# Integrated optomechanics
# and single-photon detection
# in diamond photonic integrated circuits

# Abstract

The development of quantum computers and quantum simulators promises to provide solutions to problems, which can currently not be solved on classical computers. Finding the best physical implementation for such technologies is an important research topic and using optical effects is a promising route towards this goal. It was theoretically shown that optical quantum computing is possible using only single-photon sources and detectors, and linear optical circuits. An experimental implementation of such quantum optical circuits requires a stable, robust and scalable architecture. This can be achieved via miniaturization of the optical devices in the form of photonic integrated circuits (PICs). The development of a suitable material platform for such PICs could therefore have a large impact on future technologies. Diamond is a particularly attractive material here, as it naturally offers a range of optically active defects, which can act as single-photon sources, quantum memories, or sensor elements. Besides its excellent optical properties, diamond also has a very high Young's modulus, which is important for optomechanics, and can be employed for potentially fast and low-loss tuning of PICs after fabrication.

In this work, components for future quantum optical circuits are developed. This includes the first diamond optomechanical elements, as well as the first integrated single-photon detectors on a diamond material platform. Diamond micromechanical resonators with high quality factors are realized and their actuation via optical gradient forces and electrostatic forces is demonstrated. The accomplished superconducting nanowire single-photon detectors show excellent performance in terms of low timing jitter, high detection efficiency, and low noise-equivalent power. Moreover, a novel scalable method for PIC fabrication from high quality single crystal diamond is presented. This work is therefore a promising step towards a platform for quantum optical circuits, as the demonstrated components can be used in a wide range of future applications of diamond quantum photonics.

# Zusammenfassung in deutscher Sprache

Die Entwicklung von Quantencomputern und Quantensimulatoren verspricht Lösungen für Probleme, die derzeit mit klassischen Computern nicht lösbar sind. Die Suche nach der besten physikalischen Implementierung solcher Technologien ist ein wichtiges Forschungsgebiet und die Nutzung optischer Effekte stellt einen aussichtsreichen Ansatz dar. Theoretische Forschung hat gezeigt, dass optische Quantencomputer möglich sind, welche nur aus Einzelphotonenquellen, Einzelphotonendetektoren und linearen optischen Schaltkreisen bestehen. Eine experimentelle Implementierung derartiger quantenoptischer Schaltkreise erfordert eine stabile, robuste und skalierbare Architektur. Diese kann durch die Miniaturisierung optischer Geräte in Form von photonischen integrierten Schaltkreisen (PICs) erreicht werden. Die Entwicklung einer geeigneten Materialplattform für PICs kann insofern eine große Bedeutung für zukünftige Technologien haben. Diamant ist hier ein besonders attraktives Material, da eine Vielzahl an optisch aktiven Defekten in Diamant existiert. Diese können als Einzelphotonenquellen, Quantenspeicher oder Sensorelemente fungieren. Neben exzellenten optischen Eigenschaften besitzt Diamant außerdem einen sehr großen Elastizitätsmodul. Dieser ist wichtig für optomechanische Elemente, die das schnelle und verlustarme Durchstimmen von PICs ermöglichen können.

Die vorliegende Arbeit beschreibt die Entwicklung optischer Komponenten für zukünftige integrierte quantenoptische Schaltkreise. Sowohl die ersten integrierten optomechanischen Elemente als auch die ersten integrierten Einzelphotonendetektoren auf Diamant werden demonstriert. Mikromechanische Diamantresonatoren mit hohen Qualitätsfaktoren werden realisiert und ihr Antrieb, sowohl mit optischen Gradientenkräften als auch mit elektrostatischen Kräften, wird gezeigt. Die entwickelten supraleitenden Einzelphotonendetektoren zeigen exzellente Leistungsmerkmale in Form von niedrigem Jitter, hoher Detektionseffizienz und niedriger äquivalenter Rauschleistung. Des Weiteren wird eine neuartige Methode zur Herstellung von PICs aus einkristallinem Diamant vorgestellt. Diese Arbeit stellt einen vielversprechenden Schritt in Richtung einer Plattform für quantenoptische Schaltkreise dar, weil die gezeigten Komponenten in einem breiten Spektrum zukünftiger Anwendungen der Quantenoptik mit Diamant eingesetzt werden können.

# Contents

# Publications

Parts of this thesis have already been published in scientific journals:

\* denotes equal contribution by the authors.

Additional work on related topics has been published in peer-reviewed scientific journals:

1. A. Vetter, S. Ferrari, **P. Rath**, R. Alaee, O. Kahl, V. Kovalyuk, S. Diewald, G. N. Goltsman, A. Korneev, C. Rockstuhl, and W. H. P. Pernice, "Cavity enhanced and ultrafast superconducting single-photon detectors", *Nano Letters*, vol. 16, no. 11, 2016.

2. S. Khasminskaya, F. Pyatkov, K. Słowik, S. Ferrari, O. Kahl, V. Kovalyuk, **P. Rath**, A. Vetter, F. Hennrich, M. Kappes, G. Gol'tsman, A. Korneev, C. Rockstuhl, R. Krupke, and W. H. P. Pernice, "Fully integrated quantum photonic circuit with an electrically driven light source", *Nature Photonics*, vol. 10, p. 727–732, 2016.

3. N. Gruhler, T. Yoshikawa, **P. Rath**, G. Lewes-Malandrakis, E. Schmidhammer, C. Nebel, and W. H. P. Pernice, "Diamond on aluminum nitride as a platform for integrated photonic circuits", *Phys. Status Solidi A*, vol. 213, no. 8, 2016.

4. T. T. Tran, J. Fang, H. Zhang, **P. Rath**, K. Bray, R. G. Sandstrom, O. Shimoni, M. Toth, and I. Aharonovich, "Facile self-assembly of quantum plasmonic circuit components", *Adv. Mater.*, vol. 27, no. 27, 2015.

Parts of this thesis have already been personally presented at national and international workshops and conferences:

1. June 2016: CIMTEC 2016 – Smart and Multifunctional Materials, Structures and Systems, Perugia, Italy: Oral presentation on "Nanophotonic structures made from diamond".

2. April 2016: Bristol Young Scientist Conference, Bristol, UK: Oral presentation on "Travelling-wave single-photon detectors integrated with diamond photonic circuits - Operation at visible and telecom wavelengths with a timing jitter down to 23 ps".

3. February 2016: SPIE Photonics West, San Francisco, USA: Oral presentation on "Travelling wave single-photon detectors integrated with diamond photonic circuits".

4. May 2015: International Conference on New Diamonds and Nano Carbons 2015, Shizuoka, Japan: Oral presentation on "Diamond integrated optical circuits functionalized by dip-pen nanolithography".

5. May 2015: CLEO: 2015 – Laser Science to Photonic Applications, San Jose, USA: Oral presentation on "Diamond electro-optomechanical devices with resonance frequencies above 100 MHz".

6. October 2014: CUDOS topical workshop on nano-plasmonic integrated circuits, Canberra, Australia: Oral presentation (Invited) on "Integrated superconducting single-photon detectors".

7. August 2014: QDiamond14: Workshop on diamond quantum science and technology 2014, Brisbane, Australia: Oral presentation (Invited) on "Diamond integrated circuits for photonics and sensing".

8. March 2014: Frühjahrstagung der Deutschen Physikalischen Gesellschaft, Berlin, Germany: Oral presentation on "Diamond high-Q mechanical resonators integrated in on-chip Mach-Zehnder-interferometers".

9. February 2014: Hasselt Diamond Workshop, Hasselt, Belgium: Oral presentation on "High quality factor optomechanical resonators made from polished polycrystalline diamond thin films".

10. June 2013: CLEO: 2013 – Laser Science to Photonic Applications, San Jose, USA: Oral presentation on "Integrated photonic circuits on wafer-scale diamond-on-insulator substrates".

# List of abbreviations and acronyms

Abbreviations and acronyms in alphabetical order:

| | |
|---|---|
| 3D | Three-dimensional |
| AFM | Atomic-force microscopy |
| BCS | Bardeen-Cooper-Schrieffer |
| CMP | Chemical-mechanical planarization |
| CNOT | Controlled NOT |
| CP | Cooper pair |
| CR | Count rate |
| CVD | Chemical vapor deposition |
| CW | Continuous-wave |
| DC | Direct current |
| DCR | Dark count rate |
| DOI | Diamond-on-insulator |
| EBL | Electron beam lithography |
| EDFA | Erbium-doped fiber amplifier |
| EOM | Electro-optic modulator |
| FDTD | Finite-difference time-domain |
| FEM | Finite element method |
| FSR | Free spectral range |
| FWHM | Full-width at half-maximum |
| GL | Ginzburg-Landau |
| HF | Hydrofluoric acid |
| HOM | Hong-Ou-Mandel |
| HPHT | High pressure and high temperature |
| HSQ | Hydrogen silsesquioxane |
| IC | Integrated circuit |
| ICP | Inductive coupled plasma |
| LOQC | Linear optical quantum computing |
| MEMS | Microelectromechanical systems |
| MZI | Mach–Zehnder interferometer |
| NbN | Niobium nitride |
| NEP | Noise-equivalent power |
| NEXAFS | Near-edge x-ray absorption fine-structure |
| NV | Nitrogen vacancy |
| OCDE | On-chip detection efficiency |
| PCD | Polycrystalline diamond |
| PhC | Photonic crystal |
| PIC | Photonic integrated circuit |
| PID | Proportional-integral-derivative |
| PL | Photo-luminescence |
| PMMA | Poly(methyl methacrylate) |
| PNR | Photon number resolving |
| PVD | Physical vapor deposition |
| QKD | Quantum key distribution |
| RF | Radio frequency |
| RIE | Reactive ion etching |
| RMS | Root mean square |
| SCD | Single crystal diamond |
| SCDMW | Single crystal diamond membrane window |
| SDE | System detection efficiency |
| SEM | Scanning electron microscope |
| SiV | Silicon vacancy |
| SNSPD | Superconducting nanowire single-photon detector |
| SOI | Silicon-on-insulator |
| SPAD | Single-photon avalanche diode |
| SPD | Single-photon detector |
| SPS | Single-photon source |
| TE | Transverse electric |
| TES | Transition-edge sensor |
| TM | Transverse magnetic |
| UV | Ultraviolet |
| VNA | Vector network analyzer |
| ZPL | Zero-phonon line |

# 1  Introduction – aim and scope of this thesis

The miniaturization of optical devices in the form of photonic integrated circuits offers a stable and scalable architecture for applications of classical and quantum optics. Exploiting quantum mechanics in quantum optical circuits, for example for quantum simulators or linear optical quantum computing promises to outperform the classical counterparts. The development of a suitable material platform for such circuits could hence have a large impact on future technologies. Diamond has excellent mechanical and optical properties, including a range of optically active defects, referred to as color centers, which can act as single-photon sources, quantum memories, or sensor elements. This motivates to investigate diamond photonic integrated circuits and optomechanical circuits. Quantum optical circuits require tunable elements and optomechanical elements are a promising solution for potentially fast and low-loss tunable photonic integrated circuits. Diamond has the highest Young's modulus of any dielectric thin films and mechanically tunable elements can be incorporated in diamond photonic integrated circuits. A further indispensable circuit element is the single-photon detector and superconducting nanowire single-photon detectors are a promising technology with excellent performance characteristics. The aim of the work described in this thesis is to develop central components of diamond photonic integrated circuits for future on-chip quantum optical experiments. This comprises the design, fabrication, and experimental examination of individual diamond photonic components and photonic integrated circuits. On-chip quantum optical experiments on a diamond platform rely on the emission and interaction; the routing and manipulation; and the efficient detection of single photons. While the emission and interaction of single photons from sources in diamond is being extensively studied by other research groups, the scope of this thesis is to show the manipulation of light using optomechanics and its efficient detection by single-photon detectors on a platform of diamond waveguides.

This thesis is structured in the following manner:

**Chapter 2** explains the fundamentals of integrated optics and the working principle of the photonic components, which are used in this thesis. Furthermore, a brief introduction into both the statistics of photon sources and quantum information science are given, which is necessary in order to understand the motivation and long-term goals of the presented experiments. Moreover, chapter 2 explains the choice of diamond as the material platform in this work.

**Chapter 3** describes optomechanics as a means to manipulate light in photonic integrated circuits, for example its phase and amplitude, using mechanical degrees of freedom. Two options for the control of the position of a mechanical oscillator, within photonic integrated circuits, are the use of electrostatic forces and the use of optical gradient forces. These options enable the control of the motion of micromechanical components either via electric voltages or via light. In Chapter 3 both approaches are explained and the first experimental demonstrations of integrated optomechanical circuits on a diamond substrate are reported.

**Chapter 4** presents the detection of single-photons in diamond photonic integrated circuits. The working principle of superconducting nanowire single-photon detectors is explained and the experimental results for the first demonstration of single-photon detectors on diamond photonic circuits are reported. The devices exhibit high efficiency, low dark count rate and low timing jitter.

**Chapter 5** presents a novel method for the fabrication of photonic circuits from single crystal diamond and shows its suitability for on-chip photonics. This includes a proof-of-principle demonstration showing that the device designs and the experimental results presented in chapter 2–4 can generally be transferred from polycrystalline diamond to single crystal diamond, which then allows the integration of single-photon sources into the presented photonic integrated circuits.

# 2 Diamond photonic integrated circuits and integrated quantum optics

*This chapter provides an overview over the research field of integrated optics and particularly integrated quantum optics. The basic elements of photonic integrated circuits which are used within this thesis are introduced and their properties explained. Furthermore the material properties of diamond which motivate its use for integrated optics are discussed.*

## 2.1 Fundamentals of integrated optics

Up to date mainly bulk optical components, such as bulk mirrors and lenses, are used as components for assembling optical experiments and commercial optical systems. While for many applications this solution works well, for others bulk optics does not offer a feasible solution. Limiting factors can be, for example, a lack of long-term stability, which necessitates realignment of components, or the limited scalability once thousands or more optical components are needed. Furthermore the size of an advanced optical setup can easily amount to several square meters, which prevents the production of cheap or handheld devices. In the last few decades technological improvements have allowed to address such problems by shrinking the size of optical components. Optical fibers with diameters on the order of tens of micrometers have been established very successfully as a method for transmission of light over long distances, for example for fast and reliable data transmission[1] and are nowadays replacing many bulk optical components. The next step of miniaturization is the photonic integrated circuit (PIC) with device dimensions in the nanometer to micrometer range.

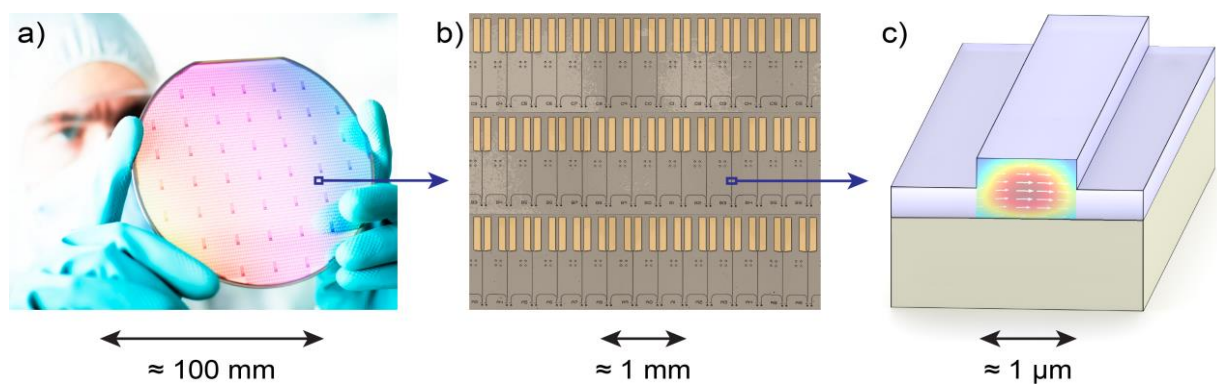

**Figure 1 - Schematic of photonic integrated circuits:** a) A wafer containing hundreds of dies which can contain hundreds of PICs each. Reprinted from Siemens AG[3,12]. b) Colorized scanning electron microscope micrograph of a die, showing tens of PICs. c) Schematic of the cross-section of a waveguide. The distribution of the electric field of an optical waveguide mode is shown as an overlay.

A PIC, or integrated optical circuit, is a solid state device, which integrates multiple photonic functions in analogy to electronic integrated circuits (IC). While in an IC electrical currents are guided along electric conductors, namely thin metal strips, in a PIC light is guided along optical waveguides made from an optically transparent material. The base material for PICs is typically a dielectric thin film of several hundred nanometers in thickness on top of an oxidized silicon wafer, as shown in Figure 1 a), from which PICs can be fabricated. For fabrication of PICs rectangular pieces are cut from the wafer, which are commonly referred to as die or chip. A wafer die with a size on the order of $15 \cdot 15 \, mm^2$ can host hundreds of PICs and Figure 1 b) shows an array of such PICs. The dimensions of the cross-section of PIC components, such as the waveguide shown in Figure 1 c), are on the order of hundreds of nanometers and hence are much smaller than bulk or fiber optical components. PICs have many advantages over free space optics and fiber optical implementations, as complex designs consisting of many optical elements can be fabricated in a scalable manner with a small footprint. If produced in large numbers this leads to low cost devices, while they are also alignment free and are long-term stable. Prime wavelength regions of interest for PICs are light which is visible to the human eye, covering a wavelength range[2] from about 380 nm to 800 nm, and light in the near-infrared which is used for telecommunication. Near-infrared light, including the C- and L-bands (1530 – 1625 nm), is used for data transmission over long distances[3] due to the fact that optical glass fibers have minimal absorption and dispersion in the near-infrared, down to 0.15 dB/km. The corresponding wavelengths are therefore referred to as *the telecom wavelengths.* Glass fibers are used in current technology for light transmission and it is expected that PICs which interface with such optical fibers will play a leading role in future technology[4–7]. The components from which many PICs, and especially those described in this thesis, can be assembled can be assigned to one of the categories presented in the following subsections.

## 2.1.1  Optical waveguides

A waveguide is the most basic and most central building block of every PIC, as it is the component which routes light across the photonic chip and which connects the various optical devices. A waveguide consists of a dielectric strip, typically with a rectangular cross-section, which confines light in two dimensions and enables light guiding along the third dimension. It is the integrated optical equivalent of an optical fiber. The widths and heights of optical waveguides are typically several hundred nanometers each. Figure 2 a) illustrates the geometry of a waveguide. The depicted geometry is called a rib waveguide, as a rectangular profile of width $w$ and height $h$ is located on top of a continuous layer of the same material, in this case diamond[1] with a refractive index $n = 2.39$ at a wavelength $\lambda = 1550$ nm.[2] This is surrounded by materials of lower refractive indices, namely air ($n = 1.0$) on the top side and oxidized silicon ($n = 1.44$) at the bottom side. Light propagates in the diamond waveguide in the $z$-direction. The refractive index contrast between diamond and the surrounding materials confines the light within the $x/y$-plane at the location of the rectangular profile. Using classical ray optics this can be intuitively understood as total internal reflection prohibiting the transmission of light from the waveguide into the surrounding material. This simple ray optics description does not explain important details concerning waveguides which are relevant for PICs, such as the transfer of light between waveguides in close proximity and the change in speed of light inside a waveguide due to objects in close proximity. Such effects can be understood using Maxwell's equations for describing light as an electro-magnetic wave, as will be explained in the following sections. An extensive discussion, of the mathematical background and the derivation of the equations presented here, is beyond the scope of this thesis and can be found in textbooks, such as *Photonic Devices* by Liu[3].

---

[1] Within this thesis diamond is used as the base material for the PICs, as will be motivated in section 2.3. Without loss of generality we will therefore use diamond as an example material in illustrations.

[2] Throughout the thesis when it is specified *at what wavelength* a property holds or an experiment has been performed, this number refers to the light's wavelength in vacuum $\lambda$, unless explicitly stated differently. The standard wavelength of $\lambda = 1550$ nm is assumed, unless explicitly noted differently. For the ease of reading this information is only repeated where clarification is needed.

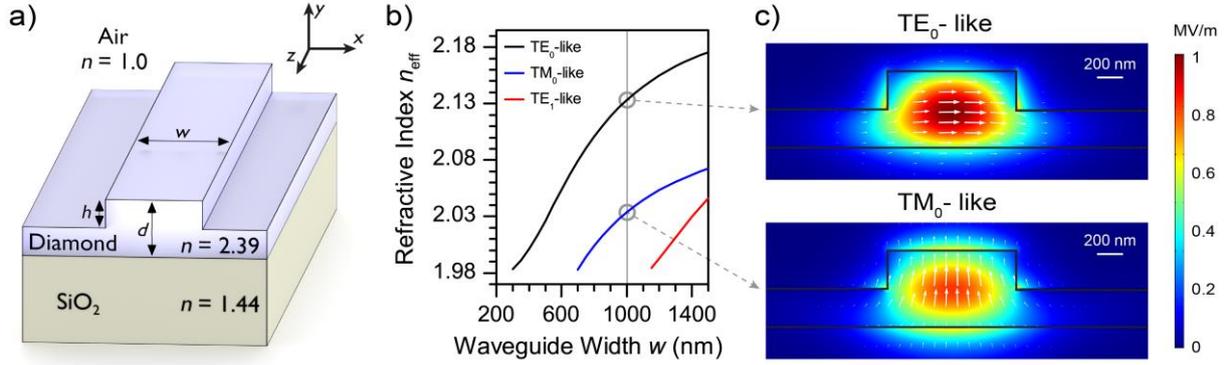

**Figure 2 - Diamond rib waveguides:** a) Schematic of a diamond rib waveguide of width $w$ and relative etch depth $h/d$. The indicated refractive indices of the materials are given for a wavelength of 1550 nm. b) Simulated effective refractive indices of all guided modes of a diamond waveguide ($d = 600$ nm, $h = 300$ nm) at 1550 nm for varying waveguide widths $w$. c) Simulated distributions of the electrical field (time averaged absolute value) of both the fundamental TE-like (upper panel) and TM-like mode (lower panel) for 1 mW optical power. The white arrows represent the direction and the relative strength of the electric field at the location of the tail of each arrow.

In the waveguide geometry illustrated in Figure 2 a) light with a frequency $\omega$ propagates along the $z$-direction. Note that the waveguide geometry is invariant under translation in the $z$-direction. Assuming homogeneous material properties, the solutions for the time-dependent spatial distributions of the electric field $\vec{E}(\vec{r},t)$ and the magnetic field $\vec{H}(\vec{r},t)$ are plane waves:

$$\vec{E}(\vec{r},t) = \vec{E}_m(x,y) \cdot e^{i(k_m z - \omega t)}$$
$$\vec{H}(\vec{r},t) = \vec{H}_m(x,y) \cdot e^{i(k_m z - \omega t)}$$

(2.1)

The solutions, called optical modes, are self-consistent electric field distributions which can be classified by their spatial distribution and their polarization. A discrete number of solutions exists, which can be enumerated using an integer mode index $m$. $\vec{E}_m(x,y)$ and $\vec{H}_m(x,y)$ are vector fields of mode $m$ which denote the distributions of the electric and magnetic fields in the $x/y$-plane. $k_m$ denotes the propagation constant of mode $m$, which is proportional to the effective refractive index or mode index $n_{\text{eff},m}$ defined by

$$k_{\text{m}} = \frac{2\pi}{\lambda_0} \cdot n_{\text{eff},m} \,,$$

(2.2)

where $\lambda_0$ denotes the optical wavelength in vacuum. Light which propagates along a waveguide is attenuated due to absorption and scattering, which can be described by a complex effective refractive index. The corresponding complex propagation constant can be written as[3]

$$k_{\text{m}} = \frac{2\pi}{\lambda_0} \cdot \text{Re}(n_{\text{eff},m}) + \frac{2\pi}{\lambda_0} \cdot \text{Im}(n_{\text{eff},m}) = \beta_m + i \cdot \alpha_m \cdot \frac{\ln(10)}{20} \,,$$

(2.3)

where $\beta_m$ is the real propagation constant und $\alpha$ the attenuation constant of the mode. The complex electric field is then given by

$$\vec{E}(\vec{r}, t) = \vec{E}_m(x, y) \cdot e^{i(\beta_m z - \omega t)} \cdot 10^{-\frac{\alpha_m}{20} z} \tag{2.4}$$

with a phase which varies sinusoidally along $z$ with a period of $1/\beta_m$. Furthermore the amplitude decreases exponentially with $z$. The optical power $P$ which propagates along the waveguide therefore decreases according to Beer's law as

$$P(z) = P(0) \cdot 10^{-\left(\frac{\alpha_m}{10} z\right)}. \tag{2.5}$$

This attenuation of optical power along waveguides is referred to as propagation loss. Curved waveguides furthermore suffer from bending losses, which increase for smaller bend radii. We will quantify the propagation loss for diamond rib waveguides in section 2.4.4.

The group refractive index, which is important in the context of pulses and propagating wave packages, is defined as $n_g = \frac{c_0}{v_g}$, where $c_0$ denotes the speed of light in vacuum and $v_g$ is the group velocity. If the dispersion is small, the group refractive index $n_{g,m}$ of mode $m$ can be calculated from the effective refractive index as[8]

$$n_{g,m}(\lambda) = n_{\mathrm{eff},m}(\lambda) - \lambda \cdot \frac{\partial n_{\mathrm{eff},m}}{\partial \lambda}. \tag{2.6}$$

For three-dimensional waveguides the field distributions $\vec{E}_m(x, y)$ and $\vec{H}_m(x, y)$ and the effective refractive index $n_{\mathrm{eff},m}$ can generally not be derived analytically and therefore numerical methods are employed. We use finite element methods (FEM), using the software COMSOL Multiphysics, for determining the waveguide modes. Waveguide modes can show different directions of polarization. If the direction of the electric field is mostly parallel to the planar layers and mostly perpendicular to the direction of propagation, the mode is called transverse electric (TE)-like[3]. If the direction of the magnetic field is mostly parallel to the planar layers and mostly perpendicular to the direction of propagation, the mode is called transverse magnetic (TM)-like[3].

For a rib waveguide, as shown in Figure 2 a), two geometric parameters exist: the waveguide width $w$ and the relative etch depth, which is defined as the ratio of the height of the rectangular profile $h$ to the total height of waveguiding material $d$. Within this thesis, diamond with a thickness of $d = 600\,\mathrm{nm}$ is employed. From this material waveguides are fabricated with methods that will be explained in section 2.4.2. A rectangular diamond strip of height $h = 300\,\mathrm{nm}$ is located on a 300 nm thick continuous diamond layer. The relative etch depth is hence $\frac{h}{d} = 50\%$. We simulate the existing guided modes at a wavelength of 1550 nm for the described waveguide geometry with varying waveguide widths $w$ and extract the effective refractive indices. Figure 2 b) shows the dependence of the effective refractive indices of all existing guided modes on the waveguide width. For widths below 300 nm no guided rib waveguide modes exist.[3] For 300 nm $\leq w < 700$ nm exactly one guided mode exists, with TE-like polarization. For widths 700 nm $\leq w < 1.15\,\mu$m exactly one TE-like and one TM-like mode exist, called the fundamental modes. For $w \geq 1.15\,\mu$m a second TE-like mode exists, resulting in a multi-mode waveguide concerning TE-like polarization. Within this thesis a waveguide width $w = 1\,\mu m$ is used.[4] This width is small enough such that, even accounting for potential width variations due to variations in fabrication, exactly one TE-like and one TM-like mode exists. This avoids unwanted potential multi-mode-interference effects within the PICs. Furthermore this width is large enough, such that the modes are well-guided with $n_{eff}$ far above the value of $\approx 1.97$ at which guided modes exist in the continuous diamond slab. Otherwise losses would occur when light couples from the rectangular waveguide into the continuous diamond layer. Figure 2 c) shows the simulated distributions of the absolute value of the electric field for the two guided modes for $w = 1\,\mu m$ with $n_{eff,TE0} = 2.133$ and $n_{eff,TM0} = 2.034$. While for both modes most of the electric field is located within the diamond waveguide, the electric field is also non-zero outside of the waveguide boundary. The electric field amplitude decreases exponentially with distance from the waveguide and is called the evanescent field. This field enables waveguides to interact with objects which are placed outside of the waveguide boundary, but in close proximity on the order of a few nanometers to a few micrometers. We will discuss this in the following chapters, in the context of evanescent coupling to mechanical oscillators and to absorptive elements of waveguide-integrated single-photon detectors. White arrows in the electric field distributions of Figure 2 c) indicate the direction and the relative electric field strength. In the upper panel the electric field points mainly in the $x$-direction, identifying the mode as TE-like polarization, while in the lower panel the electric field points mainly in the $y$-direction, identifying the mode as TM-like polarization. In the experiments within this thesis exclusively TE-like polarization is used.

---

[3] It should be noted that light can be guided in the continuous diamond layer and can propagate in any direction within the $x/z$-plane with an effective refractive index of $\approx 1.97$. In this case the rectangular profile does not fulfill its purpose.

## 2.1.2 Fiber-to-chip couplers

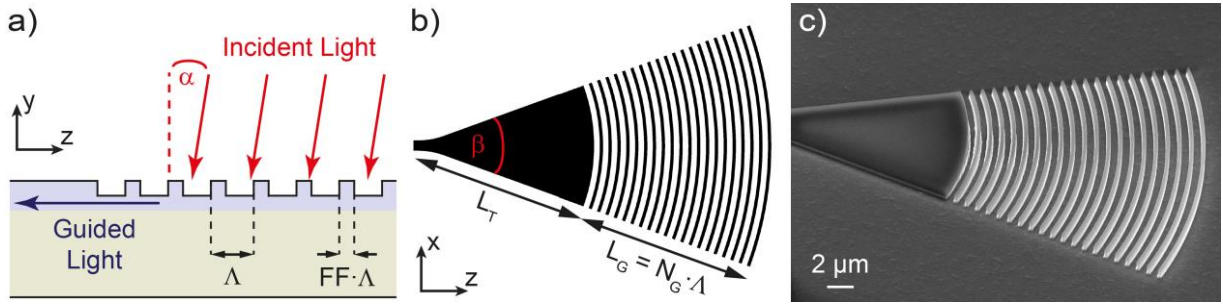

**Figure 3 - Fiber-to-chip grating couplers:** a) Schematic of a cut through a Bragg grating coupler (side view, $x/z$-plane) with a period $\Lambda$ and filling factor $FF$. Light is incident onto the grating from the top under an angle $\alpha = 8°$. Constructive interference leads to light being coupled into the waveguide (blue arrow on the left side). b) Schematic of a focusing grating coupler (top view, $x/z$-plane) indicating the opening angle $\beta$, taper length $L_T$ and grating length $L_G$. c) Scanning electron microscope image of a diamond focusing grating coupler for 1550 nm wavelength, taken under an angle of 45°.

Fiber-to-chip couplers allow transferring light from an optical fiber into and out of waveguides (called in- and out-coupling, respectively). This enables the connection of PICs to off-chip optical

---

devices such as light sources and light detectors. Fiber-to-chip couplers are often either implemented in-plane of the dielectric thin film by coupling an optical fiber to the facet of a polished waveguide[9,10] or as Bragg gratings couplers[11,12], which enable out-of-plane access to waveguides. Compared to coupling through the facets of a photonic chip, the use of focusing grating couplers increases the density of PICs on the chip, as many devices in close proximity can independently couple out-of-plane to optical fibers. Coupling efficiencies for Bragg grating couplers at telecom wavelengths up to 87% on silicon[13] and up to 30% on diamond[14] have been shown. Within this thesis focusing grating couplers are exclusively used as the method for fiber-to-chip coupling. Figure 3 a) shows the schematic cross-section of a Bragg grating, which consists of a periodic series of grating lines, with a period $\Lambda$ and a line thickness $b = \Lambda \cdot FF$, where $FF$ is the filling factor of the grating. Light incident under an angle of $\alpha = 8°$ from the top (as indicated by red arrows) is diffracted by the grating. For certain wavelengths the diffracted light undergoes constructive interference in the direction of the waveguide (located on the left side of the grating in the schematic) and hence light is guided by the waveguide. While this brief description explains the basic principle of a grating coupler, a full analysis of a three-dimensional grating coupler necessitates coupled mode theory in order to derive the spectral and polarization dependence of the coupling efficiency. A detailed description can be found in textbooks[3]. Figure 3 b) shows a schematic of the top view of a focusing grating coupler, consisting of circularly curved grating lines with opening angle $\beta$, which focuses the diffracted light into the waveguide. An optical fiber has a core diameter of about 10 µm, while photonic waveguides have diameters on the order of $0.5 - 1$ µm. The focusing grating coupler enables to match the size of waveguide mode and fiber mode within a compact photonic component size. For half-etched diamond waveguides the optimized coupler geometry for 1550 nm consists of $N_G = 20$ curve grating lines with an opening angle $\beta = 40°$ and a taper length $L_T = 20$ µm. Figure 3 c) shows the experimental implementation of a diamond grating coupler for 1550 nm. A grating coupler is typically optimized for the coupling between one mode of the optical fiber and one mode of one polarization of the waveguide.

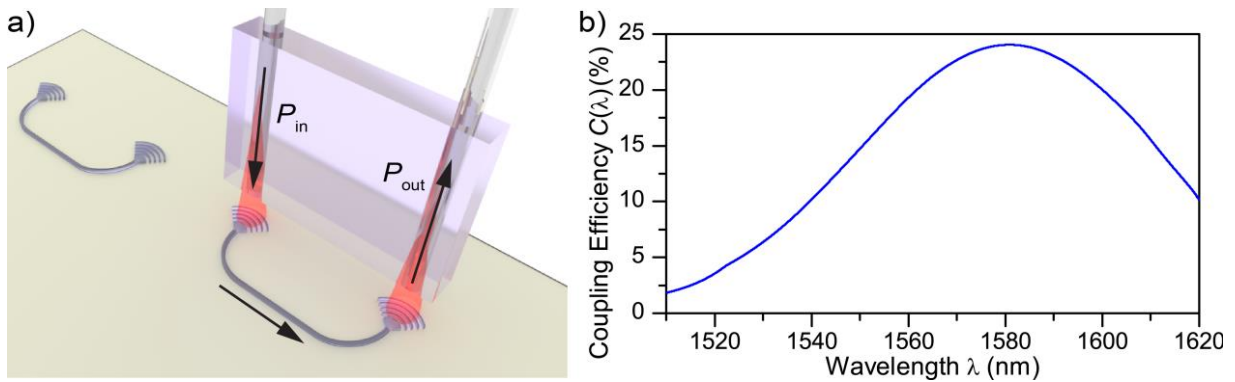

**Figure 4 - Grating coupler transmission:** a) Schematic of the coupling of light between optical fibers at the glass tip of a fiber array and the photonic waveguide via two focusing grating couplers. The flow of light is indicated with black arrows. b) Transmission spectrum of a PIC containing two diamond focusing grating couplers optimized for telecom wavelengths, with a maximum transmission of 4.4% at 1580 nm. This corresponds to a coupling efficiency of 24% for a single coupler.

Figure 4 a) shows a schematic of the coupling of light between grating couplers and optical fibers. One fiber launches light of power $P_{in}$ via a grating coupler into the PIC and a second fiber collects light of power $P_{\text{out}}$ via a second grating coupler. The details of such transmission measurements will be explained in section 2.4.3.

We vary the different geometry parameters of the grating couplers and optimize the coupling efficiency for PICs made from 600 nm of diamond at a relative etch depth of 50% for wavelengths around 1550 nm. The transmission is maximal for a filling factor of 35%. For a period of $\Lambda = 850$ nm the wavelength of maximum transmission is 1550 nm. Figure 4 b) shows a typical wavelength dependent coupling efficiency $C(\lambda)$, which quantifies the percentage of optical power which a coupler transfers between fiber and waveguide. The coupling efficiency has a maximum value of 24% at 1580 nm and a 3 dB-bandwidth of 70 nm. The coupler geometry will be adjusted in the experiments depending on the desired wavelength of maximal coupling efficiency.

### 2.1.3 Integrated beam splitters

Beam splitters[5] distribute light from one input channel into several output channels or more generally from several input channels into several output channels. Beam splitters are often implemented as 50/50 beam splitters, which divide light from one input channel equally into two output channels. A beam splitter can also act as a combiner, where light from multiple input channels is combined into a common output channel. In integrated optics, a beam splitter can for example be implemented in the form of a Y-splitter[15], as illustrated in Figure 5 a). The splitter of length $L_s$ consists of circular bends. The input waveguide (in the schematic on the left side) is tapered from its initial value $w$ to a width $2w$ and separated into two waveguides. This leads to a 50/50 splitting ratio due to the symmetry of the geometry. At the end of the curved parts the two waveguides are separated by the splitter width $w_s$. Figure 5 b) shows a scanning electron microscope (SEM) micrograph of a Y-splitter as a part of a PIC in the experiment. Other beam splitter geometries for PICs, such as multimode interference splitters[16] or directional couplers[17] exist, but within this thesis exclusively 50/50 Y-splitters are used, as the splitting ratio shows no spectral dependence and is robust concerning fabrication tolerances. Y-splitters are also used in the backward direction as combiners, with two input- and one output-waveguide.

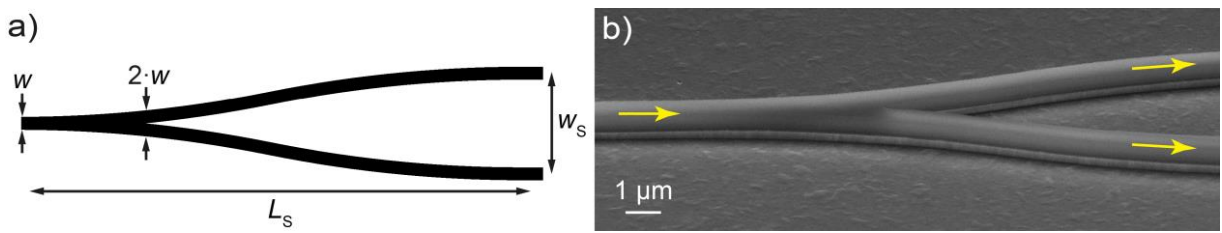

**Figure 5 - Beam splitters:** a) Schematic of the geometry of a 50/50 Y-splitter for a wavelength of 1550 nm. A waveguide is tapered from its original width $w$ to $2 \cdot w$ and separated into two waveguides via circular bends with a bend radius of 40 μm along a splitter length $L_s = 40$ μm to a splitter width $w_s = 8$ μm. b) SEM micrograph of a Y-splitter for diamond rib waveguides, taken under an angle of 45°.

---

[5] It should be noted that the term *beam* splitter is not only used in the context of ray optics but is also common when light is described as electromagnetic wave or within a quantum optical description. We will therefore use this term throughout the thesis, independent of the theory used to describe light.

## 2.1.4 On-chip Mach-Zehnder interferometers

On-chip interferometers can be assembled from aforementioned waveguides and beam splitters. They can for example be used for measurements concerning changes in the phase or amplitude of light. Such changes can be caused by absorption of light or changes in the effective refractive index due to the mechanical motion of a PIC component, as will be explained in detail in section 3.3. Figure 6 a) shows a schematic of the PIC version of a Mach-Zehnder interferometer (MZI): Light of power $P_{\text{in}}$ enters a waveguide and passes a 50/50 Y-splitter. Half of the light propagates along the upper/lower waveguide, which form the upper/lower interferometer arm. The direction of light is changed within the $x/z$-plane by changing the direction of the waveguides along circular curves of large radii $r$. Interference between both electromagnetic waves occurs at the second beam splitter, leading to an output power $P_{\text{out}}$, which depends on the phase difference $\phi$ between propagation along the upper compared to the lower arm. Assuming waveguides with no propagation losses and perfect 50/50 splitting ratios, the MZI transmission can be calculated as

$$T(\phi) = \frac{P_{\text{out}}}{P_{\text{in}}} = \cos^2\left(\frac{\phi}{2}\right), \tag{2.7}$$

which is illustrated in Figure 6 b). The phase difference $\phi$ can be calculated as

$$\phi = \int_{r=0}^{l_2} \beta_2(r)dr - \int_{r=0}^{l_1} \beta_1(r)dr = \frac{2\pi}{\lambda_0}\left[\int_{r=0}^{l_2} n_{\text{eff},2}(r)dr - \int_{r=0}^{l_1} n_{\text{eff},1}(r)dr\right], \tag{2.8}$$

where $\beta_j(r) = \frac{2\pi}{\lambda_0} n_{\text{eff},j}(r)$ is the absolute value of the wave vector and the indices $j \in [1,2]$ refer to the lower and upper waveguide, $\lambda_0$ denotes the wavelength in vacuum and $n_{\text{eff},j}(r)$ denotes the effective refractive index at location $r$ along a waveguide $j$. The integration is executed along the path of each waveguide, where $l_j$ denotes the length of each waveguide.

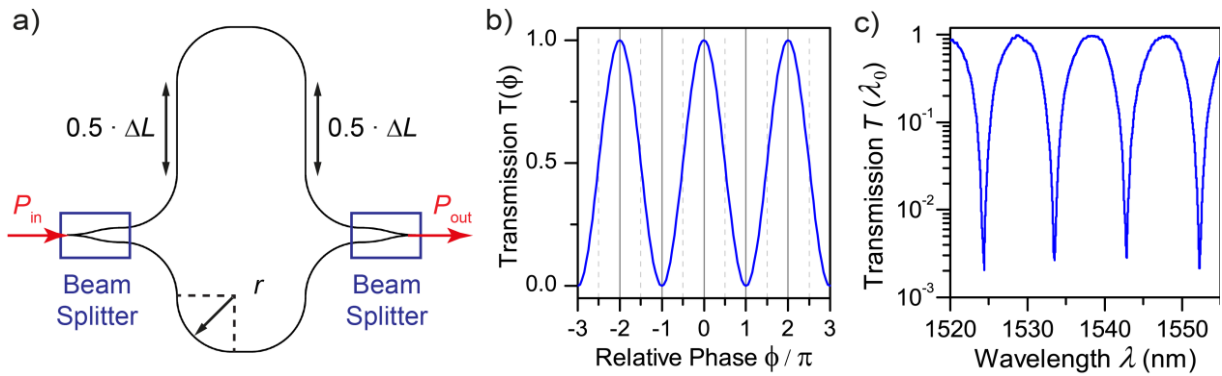

**Figure 6 - Mach-Zehnder interferometer (MZI):** a) Schematic of an integrated MZI consisting of two Y-splitters and waveguides with circular curves (curve radius $r = 30$ μm for diamond MZIs within this thesis). The upper interferometer arm is $\Delta l = 100$ μm longer than the lower arm. b) Theoretical dependence of the MZI transmission on the phase difference $\phi$ of the electromagnetic waves of both arms at the second beam splitter. c) Measured dependence of a MZI transmission on the wavelength, shown on a logarithmic scale.

Within this thesis asymmetric MZIs are used, where the upper waveguide is a distance $\Delta l = l_2 - l_1$ longer than the lower arm. For a homogeneous effective refractive index $n_{\text{eff}}(\lambda_0)$ along both waveguides the transmission spectrum of an MZI can be expressed as

$$T_{\mathrm{MZI}}(\lambda_0) = \cos^2\left(\frac{\phi}{2}\right) = \cos^2\left(\frac{\pi}{\lambda_0}(n_{\mathrm{eff}}(\lambda_0) \cdot \Delta l)\right). \tag{2.9}$$

Figure 6 c) shows the transmission spectrum of a diamond integrated MZI with $\Delta l = 100\,\mu\mathrm{m}$ path difference, realized within this thesis. The extinction ratio of an interferometer $r_{\mathrm{ext}}$ is defined as

$$r_{\mathrm{ext}} = 10 \cdot \log_{10}\left(\frac{T_{\mathrm{max}}}{T_{\mathrm{min}}}\right), \tag{2.10}$$

where $T_{\mathrm{max}}$ and $T_{\mathrm{min}}$ are the transmission values for adjacent maxima and minima. A MZI with no propagation losses and perfect 50/50 splitting ratios would have $T_{\mathrm{min}} = 0$ and hence an infinite extinction ratio. In practice the additional propagation loss along the length difference $\Delta l$ leads to a finite extinction ratio, on the order of $r_{\mathrm{ext}} = 25\,\mathrm{dB}$ for the experimental implementations within this thesis (see Figure 6 c).

## 2.1.5 Integrated optical resonators

An optical cavity or optical resonator in free space optics typically consists of two mirrors facing each other with light being reflected back and forth between them, with constructive interference for certain wavelengths, called the resonance wavelengths. Corresponding devices in PICs are for example disk resonators[18], photonic crystal cavities[19] and ring resonators[20]. In diamond PICs, optical resonators have for example been used as enhancers of the relative emission from single-photon sources into the cavity modes via the Purcell effect[21–26]. Within this thesis ring resonators are used for determining propagation losses and as elements of optomechanical circuits, as will be explained in chapter 3. This section provides the necessary background concerning integrated ring resonators for the following chapters.

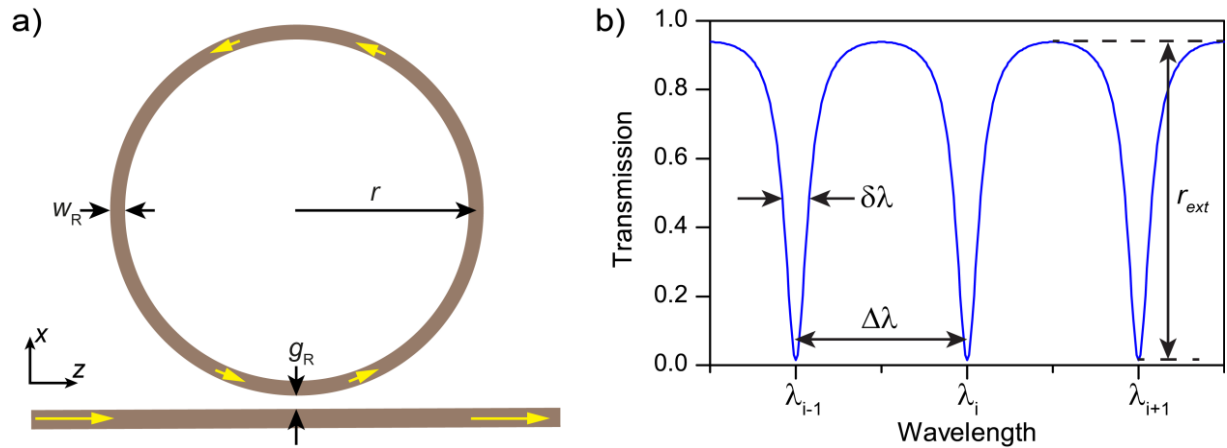

**Figure 7 - Ring resonator geometry and transmission:** a) Schematic of an integrated optical ring resonator of width $w_{\mathrm{R}}$ and radius $r$, evanescently coupled to a waveguide via a gap of size $g_{\mathrm{R}}$. b) Transmission spectrum of a ring resonator with resonances which are separated by the free spectral range $\Delta\lambda$ and have a FWHM $\delta\lambda$ and an extinction ratio $r_{\mathrm{ext}}$.

A ring resonator which is coupled to a waveguide in close proximity is shown in Figure 7 a). The ring resonator is defined by its width $w_{\mathrm{R}}$ and radius $r$ and evanescently coupled to a waveguide via a gap of size $g_{\mathrm{R}}$. The evanescent field outside of the rectangular waveguide core enables light to couple between the ring resonator and the waveguide. When the circumference of the ring is equal to an integer multiple of the wavelength of the optical mode, in-coupling light interferes constructively with light which is propagating around the ring. This is referred to as a resonance wavelength of the ring resonator. Only considering one mode of one polarization with an effective refractive index $n_{\text{eff}}$, constructive interference occurs for light with vacuum wavelengths $\lambda_i$ given by

$$\text{Re}(n_{\text{eff}}(\lambda_i)) \cdot 2\pi \cdot r = i \cdot \lambda_i, \tag{2.11}$$

where $i$ is an integer number. Figure 7 b) shows the transmission spectrum of a waveguide coupled ring resonator. The distance between two ring resonances in terms of their wavelengths, called the free spectral range (FSR) $\Delta\lambda$, can (for $2\pi \cdot r \gg \lambda$) be approximated as[8]:

$$\Delta\lambda = \lambda_i - \lambda_{i-1} \approx \frac{\lambda_i{}^2}{n_{\text{g}}(\lambda_i) \cdot 2\pi \cdot r}. \tag{2.12}$$

This enables to determine the group refractive index for a certain waveguide mode from the transmission spectrum of a ring resonator. The coupling between waveguide and ring depends on $g_{\text{R}}$. For a specific distance the transmission at resonance becomes zero. This is referred to as critical coupling. The extinction ratio $r_{\text{ext}}$ of a ring resonance quantifies the depth of a resonance and is defined as

$$r_{\text{ext}} = 10 \cdot \log_{10}\left(\frac{T_{\text{max}}}{T_{\text{min}}}\right), \tag{2.13}$$

where $T_{\text{max}}$ and $T_{\text{min}}$ are the transmission values for adjacent maxima and minima. The quality factor $Q$ of a system is a measure for the energy loss during one period $T$, compared to the total energy $W$ which is stored in the system, as

$$Q = 2\pi \frac{W}{\frac{dW}{dt}T}. \tag{2.14}$$

For ring resonators of large enough radius bending losses can be neglected compared to the propagation loss which is also present for straight waveguides. For ring resonators which are not coupled to a waveguide, the relation between propagation loss and the quality factor of a ring resonator in decibels (dB) per unit of length is given by[8]

$$Q = 10 \cdot \log_{10}(e) \cdot \frac{2\pi \cdot n_g(\lambda_i)}{\alpha \cdot \lambda_i}, \tag{2.15}$$

where $n_g$ is the group refractive index and $\lambda_i$ the vacuum wavelength at resonance. Coupling to a waveguide leads to a quality factor, which is lower than the intrinsic quality factor defined above, but for weak coupling the difference is negligible. This enables quantification of the propagation loss of waveguides by determining the quality factor of weakly coupled ring resonators, as will be shown in section 2.4.4. The quality factor of a ring resonance at wavelength $\lambda_i$ is related to the full-width at half-maximum (FWHM) of the ring resonance $\delta\lambda$ and for $Q \gg 1$ can be approximated as

$$Q = \frac{\lambda_i}{\delta\lambda}. \tag{2.16}$$

The optical power in the ring resonator can be much larger than that in the waveguide, as the wave traveling in the ring resonator at resonance interferes constructively with the input wave and thus building up the amplitude. For critical coupling at resonance the steady state power $P_{\text{back}}$ at the backside of the ring (half a circle away from the waveguide) amounts to[27]

$$P_{\text{back}} = P_{\text{in}} \cdot \frac{10^{-\left(\frac{\alpha}{10} \cdot \pi \cdot r\right)}}{1 - 10^{-\left(2 \cdot \frac{\alpha}{10} \cdot \pi \cdot r\right)}}, \tag{2.17}$$

where $P_{\text{in}}$ is the continuous-wave (CW) power in the waveguide before interacting with the ring. The intensity enhancement inside the ring increases for a decreasing attenuation coefficient $\alpha$.

## 2.1.6  Photonic crystals

Photonic crystals (PhC) are nanostructures which consist of a periodic modulation of the refractive index. This modulation can occur in one, two or three dimensions, leading to PhC of the corresponding number of dimensions. A one-dimensional PhC consists of a periodic series of layers of different refractive indices, like those used in Bragg reflectors. Reflection and transmission at each material interface and interference between the corresponding waves can lead to constructive or destructive interference, depending on the wavelength of light. Destructive interference for a certain wavelength can occur, such that light of the corresponding frequency cannot be transmitted through the periodic series of layers. The dispersion relation, which is the relation between frequency $\omega$ and wave vector $k$ (in one dimension) of the electromagnetic wave, reveals if for certain frequencies no allowed mode exists, regardless of $k$[28]. This is called a photonic bandgap. The reader is referred to Joannopoulos *et al.*[28] for a detailed description of PhCs.

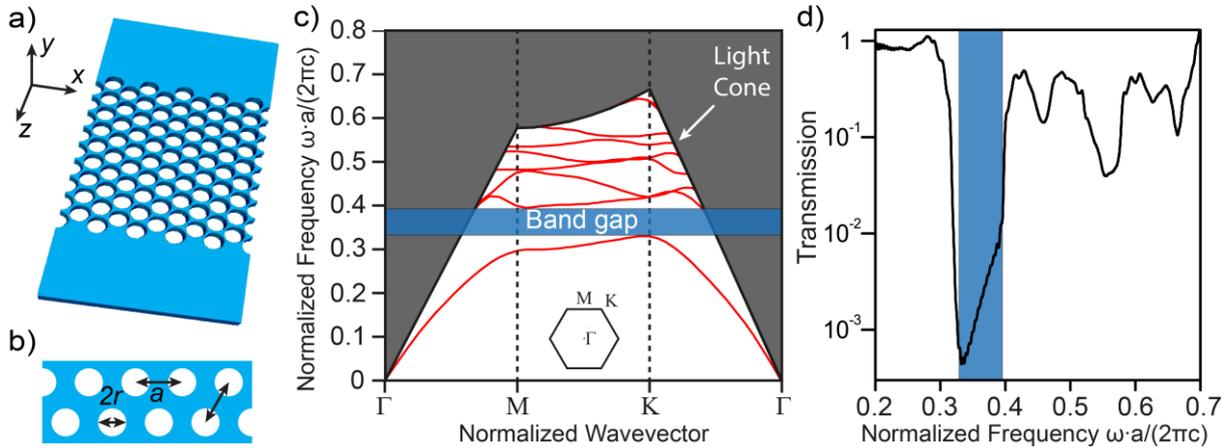

**Figure 8 - Two-dimensional photonic crystal:** a) Schematic of the 3D model of the hexagonal PhC slab. b) Top-view of a hexagonal PhC with lattice constant $a$ and a hole radius $r$. b) Simulated band structure of the TE-like modes in a PhC slab with $a = 600$ nm and r $= 180$ nm (the bands are indicated in red color). The $x$-axis shows the normalized wave vector, along directions of the irreducible Brillouin zone. The Brillouin zone and points of special interest are shown in the inset. Guided modes are bounded by the light cone for index guiding in the slab. The shaded blue region indicates the band gap for the TE-like modes. c) Simulated frequency dependent transmission through a PhC slab with nine rows of holes (As shown in a). The transmission for frequencies within the photonic band gap is reduced by two orders of magnitude.

A two-dimensional PhC slab, as schematically shown in Figure 8 a), combines two-dimensional periodicity (in the $x/z$-plane) with vertical index-guiding in the $y$-direction ($k_y = 0$). It consists of a lattice of air holes in a layer of dielectric medium. Figure 8 b) indicates the geometry parameters of the PhC: Holes of radius $r$ are arranged in a hexagonal lattice and the distance between two adjacent holes is given by the lattice constant $a$. We are interested in determining the geometry parameters $a$ and $r$, such that light with TE-like polarization at wavelengths around 1550 nm cannot propagate in a 600 nm thick diamond layer with a hexagonal lattice of air holes. This will allow for interfacing of diamond waveguides with PhC slabs.

We consider a hexagonal lattice of air holes in a 600 nm thick diamond slab, which is surrounded by air on the top and bottom sides. The periodic hole structure shows discrete translational symmetry along the two vectors which define the hexagonal lattice. Just like the propagation of electrons in solids, the propagation of light within periodic media can be described with Bloch states[28]. Due to the periodicity of the structure, only wave vectors within the first Brillouin zone need to be considered, which contains all non-redundant values of $k_z$ and $k_x$. We simulate[6] the dispersion relation of the PhC slab, referred to as band structure, using the software MIT Photonic Bands. For $a = 600$ nm and r = 180 nm, the bandgap corresponds to vacuum wavelengths from 1519 nm to 1820 nm. Figure 8 c) shows the simulated dispersion relation of the TE-like modes for the aforementioned geometry. The $x$-axis shows the normalized wave vector $\frac{k}{2\pi} \cdot a$, along different directions in the irreducible Brillouin zone. A sketch of the Brillouin zone is shown in the inset. The $y$-axis shows normalized frequencies $\frac{\omega}{2\pi c_0} \cdot a$. The bandgap for TE-like modes is indicated in blue. We confirm the band structure by simulations in the time domain via finite-difference time-domain (FDTD) methods[29] using the software MEEP. The device geometry is indicated in Figure 8 a): Nine rows of holes in a hexagonal lattice with $a = 600$ nm and r = 180 nm. The simulated transmission of light in the diamond slab from one side of the lattice to the other side is shown in Figure 8 d). For normalized frequencies corresponding to the TE-like bandgap (indicated by a blue area) a transmission of less than $T = 10^{-2}$ is found, showing that such a PhC slab enables optical isolation of the two sides of the hexagonal lattice, though they are physically connected via the remaining diamond. This PhC slab design will be applied for optomechanical resonators, as will be explained in section 3.3.6.

Summarizing the section on components for PICs, it can be said that PICs for a specific purpose can be built by combining several of the explained building blocks and adding application specific additional components, as will be shown in the following chapters. While PICs find applications in classical optics, such as telecommunications, they are nowadays also being explored for applications where the properties of light cannot be described anymore by Maxwell's equations, but where a quantum mechanical description is needed. The next section gives a brief introduction into quantum optics in combination with PICs.

---

## 2.2 Quantum optics and photonic integration

*Quantum optics studies the effects and the nature of light as quantized photons where a photon is defined as a single elementary excitation of a quantized field mode. The energy of a photon for visible or near-infrared light is on the order of $10^{-19}$ J. The concept of light being composed of quanta with discrete units of energy was first introduced by Max Planck who used it successfully for explaining the spectrum of blackbody radiation in 1899 and used by Albert Einstein in 1905 to explain the photoelectric effect. Many optical applications of quantum information science depend on single-photon sources and single-photon detectors. Hence a basic understanding of these concepts is important for understanding the aim of this thesis and will be given in the following section.*

### 2.2.1 Photon statistics and single-photon sources

Within a quantum mechanical treatment, light can be described using Fock states or number states, which are eigenstates of the number operator and correspond to a well-defined number of energy quanta.[z] When considering only a single mode, a state $|n\rangle$ describes a state of light containing exactly $n$ photons. The single-photon state is hence $|1\rangle$, the state which contains exactly one photon. The output of a laser can be described as a coherent state $|\alpha\rangle$, which can be expressed as a superposition of number states. It can be shown that when detecting light from a laser, the probability $P(k, \bar{n})$ of detecting $k$ photons follows a Poissonian distribution, with an average photon number $\bar{n}$, given by

$$P(k, \bar{n}) = |\langle k|\alpha\rangle|^2 = e^{-\bar{n}} \cdot \frac{\bar{n}^k}{k!}. \tag{2.18}$$

This result has to be considered when characterizing single-photon detectors using laser light, as will be explained in chapter 4.

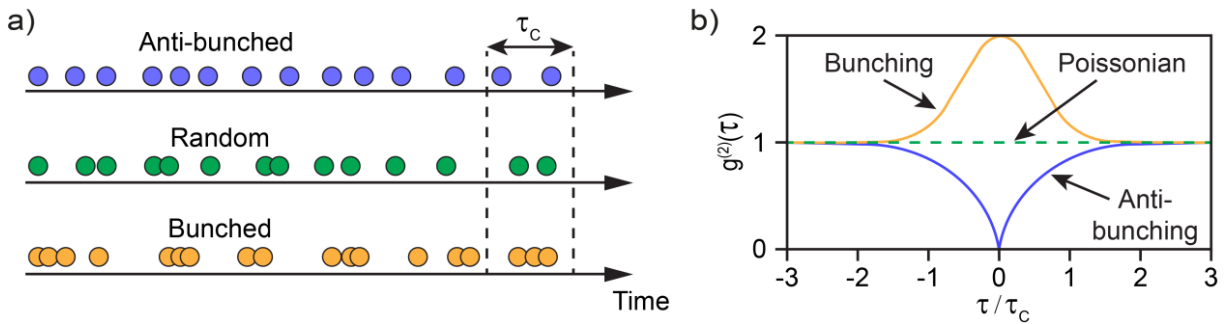

**Figure 9 - Photon statistics:** a) Schematic of the distribution of photon as a function of time for anti-bunched, random, and bunched light (bottom row). b) Schematic of the second-order correlation function of three ideal sources: a chaotic, thermal-like source (solid orange curve), a coherent Poissonian source (dotted green curve), and a single-photon source (solid blue curve), illustrating bunching and anti-bunching. The time scale is given by the coherence time $\tau_C$ of the bunched and anti-bunched sources. Both schematics are adapted from Migdall *et al.*[31].

Different light sources, such as thermal emitters, lasers, and single-photon sources, differ in the way that the emitted photons are distributed in time, as illustrated in Figure 9 a). The statistics of the distribution of photons in time can be quantified using the second-order correlation function $g^{(2)}(\tau)$,

---

as shown in Figure 9 b). $g^{(2)}(\tau)$ describes the normalized joint probability of counting a photon at time $t$ and another at $t + \tau$.[30] The value at zero delay $g^{(2)}(\tau = 0)$ hence describes the probability of simultaneously detecting two photons.

If $g^{(2)}(\tau) < g^{(2)}(0)$ for $\tau \neq 0$, the probability of detecting a second photon decreases with the time delay $\tau$, indicating a bunching of photons, as illustrated in the bottom row of Figure 9 a). As stated above, a laser follows Poissonian statistics, which results in a random distribution of photons, as illustrated in the middle row of Figure 9 a), and yields $g^{(2)}(\tau) = 1$ for all time delays $\tau$. If on the other hand $g^{(2)}(\tau) > g^{(2)}(0)$ for $\tau \neq 0$, the probability of detecting a second photon increases with the time delay, as illustrated in the top row of Figure 9 a). This is characteristic of photon anti-bunching[30]. An ideal single-photon source which never emits two or more photons at a time would show anti-bunching and yield $g^{(2)}(0) = 0$. An in-depth discussion on photon sources and their statistics can be found in Migdall *et al.*[31].

Single-photon sources play an important role in optical quantum information science, as will be explained in the subsequent section. An intuitive example of a simplified (theoretical) single-photon source can be given by considering a single atom with exactly two energy levels and exactly one electron, as schematically shown in Figure 10 a). The electron could be either in the state with lower energy, the ground state $|g\rangle$, or in the state with higher energy, the excited state $|e\rangle$. Upon transition from the excited to the ground state the atom emits a photon with a photon energy being equal to the energy difference between the two atomic states. The atom can only emit a second photon after the electron has been re-excited to the state of higher energy, so it is apparent that this idealized atom with only one electron could never emit two photons at the same time and would therefore show a second-order intensity correlation function with $g^{(2)}(0) = 0$ as a signature for a pure ideal single-photon source. Figure 10 b) shows a second-order intensity correlation function for the example of a color center in diamond, a single-photon source which motivates this work, as will be discussed in section 2.3.1. The deviation of $g^{(2)}(0)$ from zero has experimental reasons.

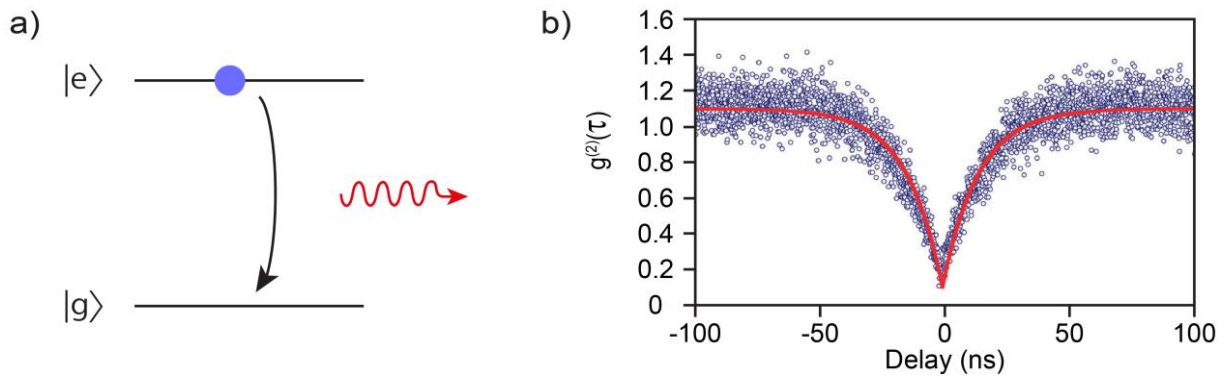

**Figure 10 - General concept of a single-photon source:** a) Electronic levels of a simplified atom with one ground state $|g\rangle$ and one excited state $|e\rangle$ and with one electron (blue circle). Upon transition to the ground state a photon is emitted (red arrow). b) Second-order intensity correlation function $g^{(2)}(\tau)$ of a color center in diamond, showing the anti-bunching of the single-photon source. The correlation function is reproduced from Naqshbandi *et al.*[343].

It is important to note that the term *single-photon source* does not simply refer to *faint light*, such that one could for example decrease the laser intensity until one receives only one photon per time slot on average in order to get single photons in the sense of a single-photon source. While for some practical applications this approach of a *faint light source* is sufficient, it is important to note that attenuation does not change the photon statistics, which means that faint laser light will still follow Poissonian statistics, while a real single-photon source is fundamentally different.

In the last decades single-photon emission from different emitters such as single molecules[32], quantum dots[33,34] and diamond color centers[35,36] has been shown. The goal of research concerning single-photon sources is to develop the best platform which is able to provide several single-photon sources of the same type which emit indistinguishable single photons on-demand at high rates. For a discussion on these goals and requirements and the state of research concerning different single-photon sources the reader is referred to Migdall *et al.*[31].

## 2.2.2 Quantum information science and quantum optical circuits

Efficient and high-quality single-photon sources play a prominent role in quantum information science, which is concerned with using quantum effects and quantum states for information science, for example in the form of quantum key distribution and quantum computation.[37] Classical information science uses classical computers, which use *bits* as the basic unit of information, where a bit can have a value of either 0 or 1. The term *qubit* was introduced as a unit of quantum information. It refers to a two-state quantum-mechanical system, whose two levels can be denoted as states $|0\rangle$ and $|1\rangle$. Unlike a classical bit which can only be in either exactly state 0 or state 1, the quantum system can be in any normalized superposition $|\Psi\rangle = \alpha|0\rangle + \beta|1\rangle$, where $\alpha$ and $\beta$ are complex numbers. The two basis states could for example refer to the polarization of a photon, which can be in the vertical or the horizontal direction, as illustrated in Figure 11 a), or in a superposition of the two polarization states, a fact that is being used for quantum key distribution (QKD). In QKD single photons are exchanged between a transmitter and a receiver and encoding information in the polarization of the photons allows the users to securely exchange a secret cryptographic key, the basis for secure communication on a public channel.[38]

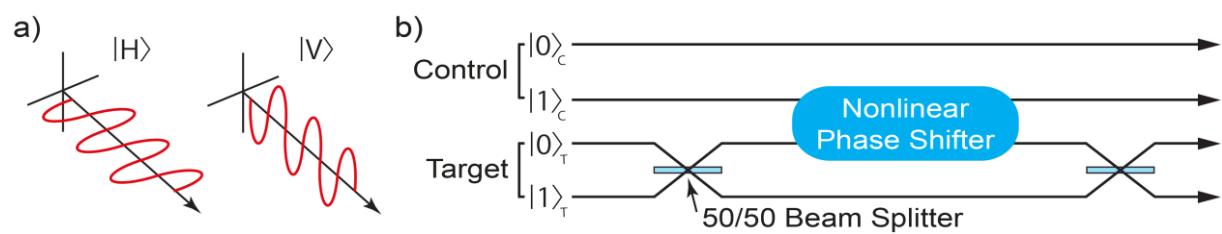

**Figure 11 - Polarization and spatial modes as basis states for photonic qubits:** a) Schematic of the horizontal $|H\rangle$ and vertical $|V\rangle$ polarization of a single photon, which can be used as the two states of a qubit. b) Schematic of the dual rail scheme, which uses the spatial waveguide modes of two waveguides to spatially encode a qubit. The states $|0\rangle$ and $|1\rangle$ refer to the photon being in the upper or the lower waveguide for each qubit. The schematic shows a possible realization of an optical CNOT gate, which requires the control photon in state $|1\rangle$ to induce a $\pi$ phase shift on the target photon. Reproduced from O'Brien *et al.*[37].

The two basis states can alternatively refer to two spatial modes, such as the fundamental modes of two waveguides. Figure 11 b) illustrates a quantum circuit with two qubits, called control and target qubit. Each black line symbolizes one waveguide and for each pair of waveguides, the states $|0\rangle$ and $|1\rangle$ refer to the photon being in the upper or the lower waveguide of the corresponding set of two waveguides. Besides QKD, photons might also be used for building a quantum simulator[39,40] or an optical quantum computer[41,42]. Quantum computers are computation systems based on scalable physical qubits. Successfully building a quantum computer promises exponentially faster computation for certain tasks, such as the factorization of integer numbers. This would for example enable to break current cryptographic systems which rely on the intractability of integer factorization for large numbers[43,44]. The qubits of a quantum computer require that they can be well isolated from the environment, while it is also necessary that the qubits can be initialized, measured, and controllably interacted. Quantum computers can for example be described using a gate model, which is analogous to the classical circuit model of computation.[43] In order to be able to carry out all possible operations on a set of $n$ qubits, a universal set of gates is needed. A universal set[44,45] can be achieved using a universal set of 1-qubit gates, together with any entangling 2-qubit gate[46], such as the *controlled NOT gate* (CNOT gate). The CNOT gate is a 2-qubit gate which flips the state of a *target qubit* conditional on the control qubit being in state $|1\rangle$. When using photons as qubits a CNOT gate could theoretically be implemented using non-linear optics.[37] This would require that one single control photon exerts a $\pi$-phase shift on a target photon, as illustrated in Figure 11 b). A very strong nonlinearity, which is not achievable in available materials, would be needed for this scheme. In 2001 it was shown that scalable quantum computing is possible without the need for a non-linear optical effect[42], but can instead be achieved using single-photon sources, single-photon detectors and optical circuits consisting of beam splitters, a scheme called linear optical quantum computing (LOQC). The initial scheme for LOQC called KLM-scheme, after its inventors Knill, Laflamme and Milburn, requires intermediate measurements on some single photons, called the ancilla photons. This imparts a type of effective Kerr nonlinearity on the system, induced through the measurements.[42,47] Furthermore feed-forward of intermediate measurement results is needed.[48,49]

At the heart of LOQC and many other quantum optical experiments and applications is the nonclassical interference and its manifestation in the Hong-Ou-Mandel (HOM) effect.[50] The HOM effect is the two-photon interference effect which occurs when two identical single-photons enter a 50/50 beam splitter, one in each input port. The probability amplitudes of detecting a photon at different output ports A and B interfere destructively. Using the notation of number states, the output state after the beam splitter can be written as $|\Psi\rangle \sim |2\rangle_A |0\rangle_B + |0\rangle_A |2\rangle_B$, a superposition of both photons being in output port A and both being in output port B. The HOM effect is at the heart of the basic entangling mechanism in linear optical quantum computing, as the beam splitter converts two single photons, potentially from different single-photon sources, into an entangled two-photon quantum state. The degree to which this operation works in an experiment critically depends on how indistinguishable the interfering photons are, in all degrees of freedom. If the photons for example differ in terms of their spectrum or their polarization, their indistinguishability degrades. Tremendous research efforts are therefore being undertaken for finding the best physical system for implementing indistinguishable single-photon sources[31,51] and Hong-Ou-Mandel in PICs has been shown in recent years.[52–54]

The motivation for implementing LOQC using PICs is generally the same as for classical optical applications, namely stability, scalable fabrication and small footprint of the components. Such quantum optical circuits promise to enable devices with thousands of components and might on the long term enable optical quantum computers. For LOQC using the KLM-scheme active elements within the PICs are needed, for example for measurements on ancilla photons with feedforward of information. While building a general purpose linear optical quantum computer is not within reach in the near future, a special purpose subset of LOQC called *boson-sampling* has attracted a lot of attention in recent years. The reason for this is that for a certain circuit configuration no active optical elements are needed. Building a boson-sampler is hence possible with current technology and boson-sampling in PICs has been shown in recent years[55–58]. A detailed description of boson-sampling can be found in a review paper by Gard *et al.*[47].

For boson-sampling $n$ indistinguishable photons are sent into $i$ input ports of a network of cascaded waveguides and beam splitters (shown as blue lines in Figure 12). At each input port either one or zero photons are inserted (hence $n \leq i$). The photons can interfere at each beam splitter, as explained above for the HOM effect. The task is to determine how many photons are exiting at each of the $j$ output waveguides at the end of the network. If this is done experimentally, then a projective measurement on a quantum state is performed and the experiment needs to be repeated many times in order to determine the probability distribution for the experiment's outcome. In 2013 it was shown by Aaronson and Arkhipov that simulating the probability distribution for the output of such a network cannot efficiently be done using a classical computer.[59] While it is possible to sample the output distribution in an experiment using the passive network of waveguides and beam splitters, single-photon sources at the input ports and photon-number resolving detectors at the output, it is not possible to predict the outcome with a classical computer without an exponential overhead in time or resources. Building a boson-sampler can therefore be considered as a step towards an optical quantum simulator or an optical quantum computer. By making the waveguide network reconfigurable in the sense that phase shifts and splitting ratios of beam splitters can be actively adjusted after device fabrication, it is possible to reconfigure the boson-sampler and hence solve many boson-sampling problems using the same PIC.

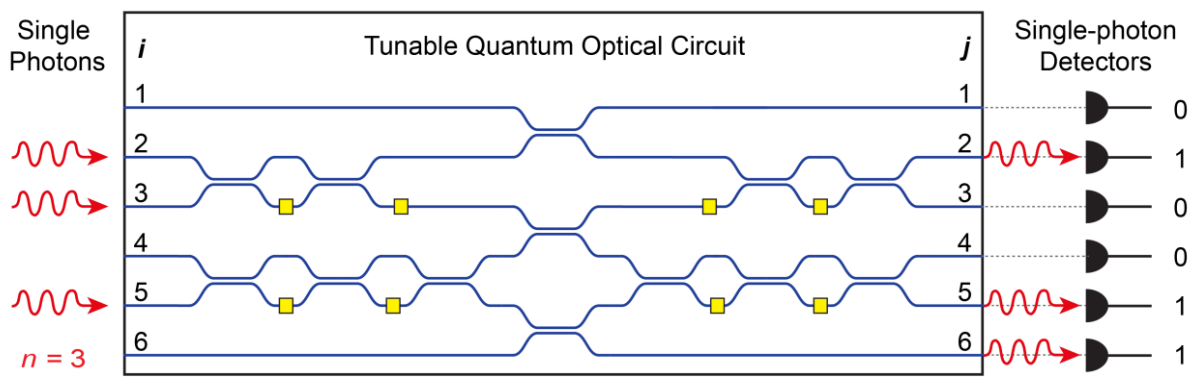

**Figure 12 - Quantum optical circuits:** Schematic of a tunable quantum optical circuit. In this example $n = 3$ photons are sent into a circuit with $i = 6$ input and $j = 6$ output waveguides. The yellow squares symbolize tunable elements. Schematic adapted from Shadbolt *et al.*[63].

To be able to build PICs which implement boson-sampling or optical quantum computing the following elements are needed:

1) photonic integrated circuits,
2) single-photon sources,
3) tunable elements,
4) single-photon detectors.

A schematic of a device which contains all elements listed above is shown in Figure 12. Single-photons (red arrows) are sent into the input ports of the photonic integrated circuits (blue lines) which consists of waveguides and beam splitters. Single-photon detectors at the output ports of the network determine the distribution of photons. Tunable elements (yellow squares) enable to reconfigure the photonic integrated circuit, which enables the solution of various boson-sampling problems using the same PIC.

While the first demonstrations of boson-sampling using PICs used off-chip single-photon sources and single-photon detectors, current research aims to implement all components on a single material platform. There are different materials which are promising for the monolithic integration of all components and we use the diamond thin films material platform, which will be motivated in the following section. While many research groups are working on optimizing the single-photon sources in diamond, the work presented in this thesis is concerned with the development of the three remaining elements, out of the four needed for a boson-sampler and quantum optical circuits in general: diamond PICs (this chapter), tunable elements using optomechanics (chapter 3) and on-chip single-photon detectors (chapter 4). The following section explains the advantages of diamond for classical and quantum integrated optics.

## 2.3 Diamond as a material for integrated optics

*In current research different materials are being studied for their use as substrates for photonic integrated circuits for quantum optics, such as silica[60–64], silicon[16,65–69], gallium arsenide[70,71], aluminum nitride[72–74] and silicon nitride[54,75–78] among others. In recent years research on diamond integrated optics has gathered momentum[79–81] due to its favorable material properties both concerning classical and quantum optical properties. The following section describes these properties and briefly explains different diamond substrates and how they can be used for fabricating PICs.*

### 2.3.1 Material properties

Diamond has a combination of outstanding material properties and some of them have been known to mankind since ancient times. The origin of the word *diamond* is the ancient Greek word ἀδάμας, which can be translated as *unbreakable* or *unalterable*[82] referring to the hardness of the diamond, as it is the hardest known natural material. Diamond is an allotrope of carbon in which the atoms are arranged in a covalent network of sp$^3$-hybridised orbitals which form a crystal structure called the *diamond lattice*, a variation of the face-centered cubic lattice. Table 1 shows a comparison of material properties which are relevant for integrated optics and optomechanics for diamond and other suitable materials. Pristine diamond has a large electrical bandgap of 5.5 eV and is therefore an electrical insulator. Diamond has a relatively high refractive index of about n = 2.4 for visible to near-infrared light and has an extremely wide wavelength range of transparency from 226 nm in the ultraviolet (UV)[83] to beyond 500 μm in the far infrared[84], with only one absorption region between about 2.5 μm and 6.5 μm. It could therefore be used for PICs in a large range of wavelengths. Diamond also has an extremely high thermal conductivity up to 2200 W/mK and one of the lowest thermal expansion coefficients, which enables it to efficiently dissipate locally generated heat[85,86] leading to a large power handling capacity. Diamond also features a large Raman frequency shift (≈ 40 THz) and a large Raman gain (≈ 10 cm/GW at 1 μm wavelength[87]) which is among the highest of available materials for PICs and enables diamond Raman lasers[88,89].

**Table 1 - Relevant material properties** for integrated optics and optomechanics for a variety of suitable materials: The bandgap $E_g$, transparency range $TR$, refractive index $n$, thermal conductivity $k$, Young's modulus $E$, density $\rho$ and the velocity of sound $c$. The materials are sorted by their Young's modulus in descending order. The table is adapted from Rath *et al.*[80].

|  | $E_g$ (eV) | $TR$ (μm) | $n$ | $k$ (W/m K) | $E$ (GPa) | $\rho$ (g/cm³) | $c$ (m/s) |
|---|---|---|---|---|---|---|---|
| Diamond | 5.47 | $0.22 - 500$ | 2.4 | 2200 | 1100 | 3.52 | 17 700 |
| Si₃N₄ | $\approx 5$ | $0.3 - 5.5$ | 2.0 | 33 | 800 | 3.24 | 15 800 |
| **3**C-SiC | 2.39 | $0.2 - 5$ | 2.6 | 1.4 | 390 | 3.21 | 11 000 |
| Sapphire | 9.9 | $0.17 - 5.5$ | 1.8 | 24 | 340 | 3.98 | 9200 |
| AlN | 6.14 | $0.2 - 13.6$ | 2.1 | 150 | 294 | 3.26 | 9500 |
| GaN | 3.44 | $0.36 - 7$ | 2.4 | 130 | 294 | 6.1 | 6900 |
| TiO₂ | 3.5 | $0.42 - 4$ | 2.5 | 10 | 250 | 4.26 | 7700 |
| Si | 1.12 | $1.1 - 6.5$ | 3.5 | 140 | 162 | 2.33 | 8300 |
| GaP | 2.26 | $0.54 - 10$ | 3.2 | 100 | 140 | 4.13 | 5800 |
| Ge | 0.66 | $1.8 - 15$ | 4.6 | 60 | 132 | 5.35 | 5000 |
| GaAs | 1.42 | $0.9 - 17.3$ | 3.7 | 52 | 116 | 5.32 | 4700 |
| ZnO | 3.4 | $0.37 - 8.85$ | 2.0 | 30 | 110 | 5.6 | 4400 |
| SiO₂ | $\approx 9$ | $0.38 - 2.2$ | 1.5 | 10 | 95 | 2.65 | 6000 |
| InP | 1.34 | $0.93 - 14$ | 3.5 | 68 | 89 | 4.8 | 4300 |

Due to the large bandgap, two-photon absorption, which is a common problem for other materials for PICs such as silicon[90,91], does not occur in diamond for infrared and visible light down to 440 nm. Overall diamond's favorable optical properties already lead to its use as a material for bulk optical components for a range of applications such as diamond Raman lasers and windows for high-power lasers[92,93]. Diamond is therefore a prime candidate for integrated optics at a broad range of wavelength, for on-chip non-linear optics[87,94,95], mid-infrared sensing of chemicals[96,97] and quantum optics, as will be explained in the following.

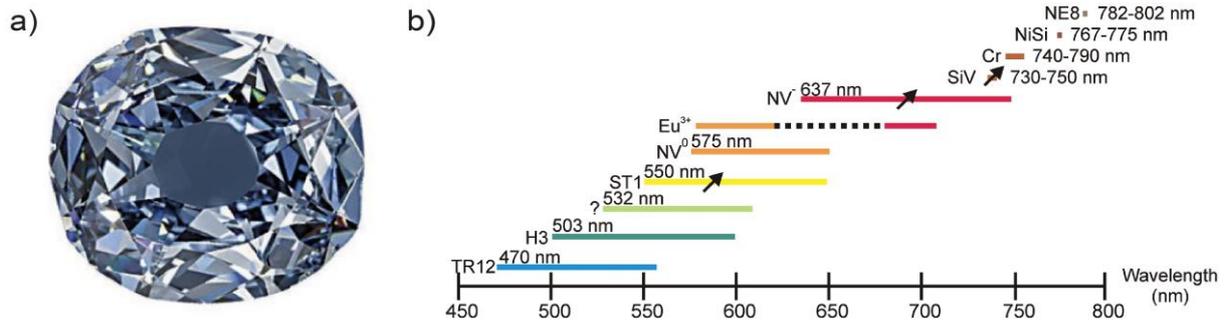

**Figure 13 - Color centers in diamond:** a) A diamond of blue color, which arises from optically active defects in the carbon lattice. b) Spectral map of various color centers in diamond which show single-photon emission. The wavelength label of each emitter indicates the observed zero-phonon-line positions while the length of the colored line represents the approximate width of the emission spectrum (for the centers with emission wavelengths below 730 nm). A black arrow indicates that the center's spin states can be manipulated. Reprinted from Aharonovich and Neu[79].

The diamond lattice can host more than 500 known optically active defects[98], referred to as *color centers*, which give diamonds their specific colored appearance, as shown in Figure 13 a). More than ten color centers have been demonstrated to show single-photon emission[79] and Figure 13 b) provides an overview of their emission spectra. Color centers are crystalline defects, which are hosted in an otherwise potentially perfect carbon lattice and can be combinations of missing carbon atoms (vacancies) and non-carbon atoms which substitute or displace carbon atoms. These modifications to the carbon lattice lead to spatially localized energy states within the electronic bandgap of bulk diamond. These optically active defects effectively act like artificial atoms which are isolated in a solid state system. The extremely high Debye temperature of 2219 K[99] leads to a low phonon population at room temperature, which in turn leads to low probabilities for phonon-induced relaxation of the electronic states of color centers. Practically no free electrons are present in diamond at room temperature, due to the large bandgap, and as 99% of natural diamond is composed of the $^{12}C$ isotope[100], which has zero nuclear spin, diamond provides ideal conditions for long coherence times for the electronic and spin states of the color centers. The color centers that currently show the most interesting properties for quantum information applications are the nitrogen vacancy (NV) and the silicon vacancy (SiV) centers, both showing interesting spin properties and, as opposed to many other emitters such as quantum dots[101–103], which need cryogenic temperatures, single-photon emission from color centers in diamond can be observed at room temperature.

The NV center in particular has been extensively studied in the last few decades[104,105], as the negatively charged defect NV$^-$ allows us to optically access the associated electron spin states, a process called optically detected magnetic resonance. The spin states of the NV center can show long coherence times[106], which enables their use as a quantum memory[107,108]. Due to the strong sensitivity to electric fields[109], magnetic fields[110–112] and strain[113], the NV center can be used for sensing small fields, often at room temperature and down to the level of a single nuclear spin[111,114]. Further details on the NV center can be found in a review by Doherty *et al.*[105]. The NV center is a bright single-photon emitter, with high photon generation rates exceeding $2 \cdot 10^6 \frac{1}{s}$[115], and can therefore be utilized for quantum information applications such as QKD[38,116]. The NV emission is typically excited off-resonance with a green laser at 532 nm. The disadvantage of the NV center is that besides the emission of photons at the zero-phonon line (ZPL) at 637 nm the NV center has a broad phonon sideband and only 3% of the photons are emitted into the zero phonon line, even at cryogenic temperatures[117], which limits its usability for quantum information applications such as boson-sampling and LOQC where indistinguishability of all involved photons is a key requirement, as described in section 2.2.2.

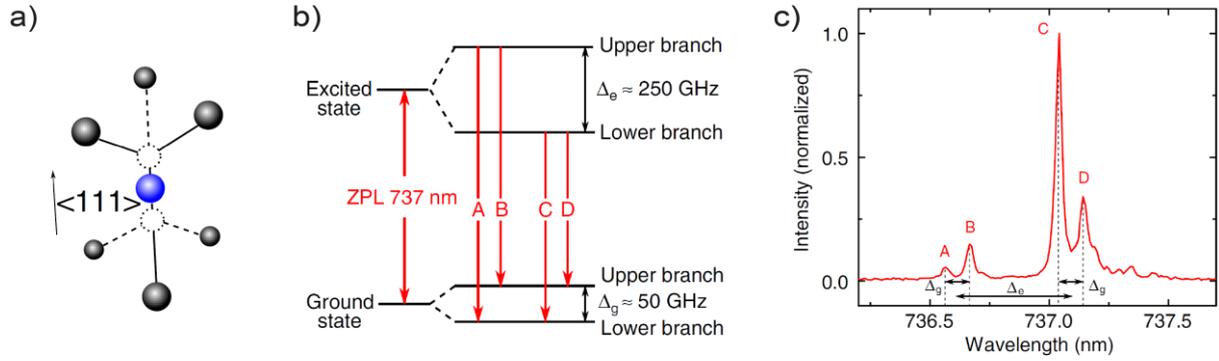

**Figure 14 - Silicon vacancy center in diamond:** a) Crystal structure of the silicon vacancy center. With carbon atoms in gray and the silicon atom (blue) in split-vacancy configuration between two unoccupied lattice sites (dashed circles). b) Energy level scheme of the SiV⁻ center. The split ground and excited states enable four optical transitions, labelled from A to D. c) Fluorescence spectrum at a temperature 4 K for a SiV ensemble obtained by non-resonant excitation at 700 nm. All three figures are reprinted from Müller *et al.*[120].

The SiV on the other hand naturally emits about 75% of the photons into the ZPL[116] and compared to the NV shows an even higher high photon generation rates exceeding $6 \cdot 10^6 \frac{1}{s}$.[118] The SiV center, as schematically shown in Figure 14 a), consists of a silicon atom and a split vacancy, which shows inversion symmetry[119]. The negatively charged NV center is an effective spin-1/2 system[120] with an energy level scheme as shown in Figure 14 b). A transition between excited and ground states leads to emission of a photon at the zero phonon line around 737 nm. At low temperatures the fine structure of the levels and its four transitions can be resolved, as can be seen in the emission spectrum shown in Figure 14 c). The narrow line width of the SiV emission is limited by the excited state lifetime[121]. The defect's symmetry results in a very weak coupling to charge fluctuations in the SiV environment, which in turn leads to the absence of spectral diffusion[121] and to a narrow inhomogeneous distribution[122]. These are important features in order to get indistinguishable photons from different defects, a requirement for many quantum information applications, and HOM interference has already been shown for photons emitted by two spatially separated SiV[119]. Besides being a single-photon source with astonishing performance at room temperature, the spin states of the SiV can be optically accessed[120,123], so the SiV might find a broad range of applications, motivating the work on diamond as a platform for quantum optics and quantum information processing. It is therefore worthwhile to develop PICs operating at the wavelength of SiV emission and especially fast single-photon detectors which work efficiently at the emission wavelengths of color centers in diamond, as will be explained in chapter 4.

When inserting color centers into diamond PICs micro- and nano-optical structures can be used for improving the performance of color centers. While solid immersion lenses[124,125] and cylindrical or cone-shaped vertical waveguides[126–128] enhance the collection efficiency of single photons from NV and SiVs by external photodetectors in the far field, optical resonators, such as photonic crystal cavities, can be fabricated in diamond at the location of color centers in order to enhance the photon emission into the ZPL via the Purcell effect[21,22,24,129]. Several research groups are working on such photonic structures and the status of research in this area is described in two recent review articles[80,81]. A further challenge for integrating single-photon sources (SPS) in PICs for quantum information applications is that, when optically exciting the SPS, it is necessary to filter out the pump light within the PIC. As the pump is many orders of magnitude brighter than the SPS this is a very challenging endeavour and it is therefore important to note that single-photon emission from NVs and SiVs can also be electrically excited.[130–133] By p- and n-doping diamond it is possible to build diamond diodes which excite electro-luminescence from color centers. When using such diode-driven color centers not only are optical filters for pump suppression not needed, but no pump light sources are needed either. Scalable electrically driven on-demand single-photon source in a solid state system might therefore be achievable in diamond. The discovery that the NV center has infrared transitions[134] and the fact that many of the more than 500 color centers are yet to be studied in detail means that color centers in diamond with properties comparable to the NV and SiV, but with emission at telecom wavelengths, might be found.[135] Having single-photon emitters at telecom wavelengths would be very desirable, as this would enable a straightforward integration into current optical communication technology. Hence it is worthwhile to study the properties of diamond PICs both for visible light and at 1550 nm wavelength, as will be done within this thesis.

Besides the optical properties which are favorable for integrated optics, diamond also offers exceptional mechanical properties, as can be seen in comparison to other materials in Table 1. It is the hardest known natural material on both the Mohs and the Vickers scale of hardness, it is mechanically stable and features an exceptionally high Young's modulus of 1100 GPa and the highest sound velocity[136]. These mechanical properties lead to diamond being used for applications ranging from tools for cutting and surface polishing[82] to diamond spheres as fuel containers for nuclear fusion reactors[137]. Diamond can also be superconducting when doped with boron[138], which enables to combine mechanical and superconducting elements on a monolithic diamond platform[139]. Due to its combination of superior mechanical and optical properties, diamond is a prime candidate for integrated optomechanics, as will be shown in chapter 3. Besides these advantages in terms of classical physics, it should be noted that mechanical strain can also be used for driving transitions between spin states in color centers.[113,140,141] This means that diamond microelectromechanical systems (MEMS)[142–144] might be used for coherent spin state manipulation, which is another motivation to study diamond optomechanical systems. Diamond has a range of further properties, which make it attractive for a range of applications of diamond integrated optics, such as its biocompatibility and its chemically inert surface, which can be useful for sensing in harsh environment[97] and for applications on or inside animal or human tissue[145].

Summarizing, it can be said that the combination of high refractive index, large transparency range and especially the color centers, which show unique properties, together with the exceptional mechanical properties makes diamond a promising platform for integrated optics and optomechanics. The following section explains how PICs can be fabricated from different diamond substrates.

## 2.3.2 Poly- and single-crystalline diamond

Diamond PICs can be either implemented in single crystalline diamond or polycrystalline diamond, which differ both in terms of material properties and availability as thin films. As explained in section 2.1 diamond waveguides have cross-sections with widths and heights on the order of hundreds of nanometers and they need to be surrounded by material of lower refractive index. Hence thin diamond films on a suitable substrate with areas as large as possible are desired for scalable PICs.

Single crystalline diamond (SCD) can only be grown homoepitaxially on existing diamond plates[82,146]. The grown diamond layer afterwards need to be cut off the original diamond. This results in relatively small but thick slabs with areas well below 1 cm$^2$, which are challenging as a template for PICs. Polycrystalline diamond (PCD) films can on the other hand be grown on a variety of substrates which are suitable for integrated optics[147,148]. PCD growth is possible on the scale of entire wafers with up to at least 6 inch ($\approx$ 15,24 cm) in diameter, which translates into diamond thin films larger than 100 cm$^2$. The small size of SCD pieces limits the scalability of both PIC designs and their fabrication, as much larger surface areas are needed for planar fabrication techniques as applied in semiconductor industry. A variety of approaches for fabricating PICs from SCD have been explored, each having a set of advantages and disadvantages. A detailed overview of different methods can be found in recent review articles[80,81]. A main problem is how to thin down a SCD slab from tens or hundreds of μm thickness to a thin film or membrane with a thickness of a few hundreds of nanometers, such that the diamond quality is preserved and the film thickness is homogeneous over the full membrane area. PCD thin films can show thickness variations below 5 nm/mm[149], while variations of more than 300 nm/mm are not unusual for SCD slabs[87], making PICs from PCD potentially more reproducible, as a variation in thickness directly translates into a variation in the performance of integrated optical components.

Some of the advantageous material properties of diamond, such as the refractive index and the Young's modulus, are largely preserved when diamond is not single-, but polycrystalline. Other properties, such as the transparency range and the thermal conductivity are deteriorated, mainly due to grain boundaries which incorporate sp$^2$-carbon and larger amounts of impurities than bulk diamond[85,98]. Synthetic diamond has become available since the development of high pressure and high temperature (HPHT) and especially chemical vapor deposition (CVD) methods through which diamond can be grown from carbon plasma on a suitable substrate[150]. A low concentration of color centers is needed for many applications and SCD can be grown with low concentrations of color centers below 0.1 part per billion[133]. While the pristine crystal structure of SCD with almost no impurities might be needed for low decoherence times of color center spins, when using them as quantum memories, this might not be as critical for other applications such as using the emitted single photons without caring about the underlying spin states.

Concerning diamond as a material for integrated optics it can be summarized that diamond has a unique combination of attractive properties which has motivated research on this topic in the last decade. While most conducted research has focused on the single-photon sources, namely the NV and SiV center which could find application as single-photon sources, sensors or quantum memory, less work has been dedicated to develop the other two crucial parts for boson-sampling or LOQC with diamond, namely tunable PICs and single-photon detectors, as explained in section 2.2.2. These are the two elements of the needed photonic toolbox for a general integrated quantum optical circuit that this thesis is concerned with, as will be explained in the following chapters.

## 2.4 Experimental methods for diamond photonic integrated circuits

### 2.4.1 Diamond deposition and polishing

Within this thesis we study diamond integrated optomechanics (chapter 3) and single-photon detectors on diamond PICs (chapter 4). We employ polycrystalline diamond thin films due to their compatibility with established planar device fabrication. In chapter 5 we will successfully demonstrate the transfer of PIC device geometries from PCD to SCD, which will in the future enable integrated quantum optics and quantum information processing on a diamond platform.

The research on diamond integrated optics using PCD is performed in collaboration with the Fraunhofer Institute for Applied Solid State Physics IAF, where polycrystalline diamond is initially deposited via plasma-enhanced chemical vapor deposition[151,152] in an ellipsoidal 2.54 GHz microwave plasma reactor, as illustrated in Figure 15 a). The growth takes place at a temperature of 850 °C using 1% methane in hydrogen at a pressure of 55 mbar.

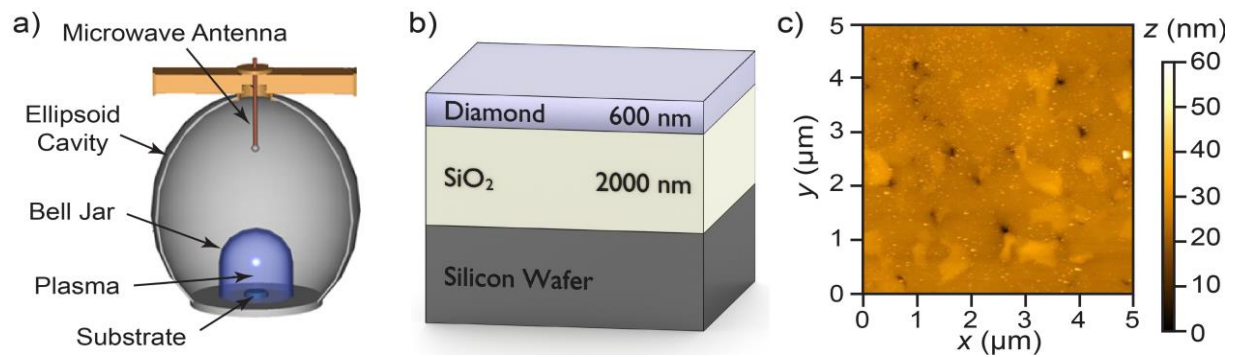

**Figure 15 - Polycrystalline diamond wafers:** a) Schematic of an ellipsoidal microwave plasma reactor for diamond CVD. b) Schematic of the layers of a diamond-on-insulator wafer before PIC fabrication. c) Atomic-force microscopy scan of the surface of a polished 600 nm thick PCD thin film revealing a small rms surface roughness of 3 nm.

To achieve diamond-based photonic circuits, the diamond thin film needs to be surrounded by a cladding material with lower refractive index, as explained in section 2.1. For this purpose, PCD is deposited onto a silicon carrier wafer covered with 2 μm of oxidized silicon, as illustrated in Figure 15 b). 1 μm of diamond is deposited and subsequently polished to a thickness of 600 nm by chemo-mechanical polishing with a soft cloth[153] in order to reduce surface roughness. The resulting wafer is sometimes referred to as a diamond-on-insulator (DOI) in analogy to silicon-on-insulator (SOI) which is a common material platform for PICs. Figure 15 c) shows an atomic-force microscopy (AFM) scan of a $5 \cdot 5 \ \mu m^2$ large area of the polished PCD surface. The root mean square (rms) surface roughness amounts to 3 nm, which is a factor of five smaller than the unpolished PCD surface. Details concerning diamond deposition, the polishing process and the material quality can be found in appendix A1.

## 2.4.2 Fabrication of diamond photonic integrated circuits

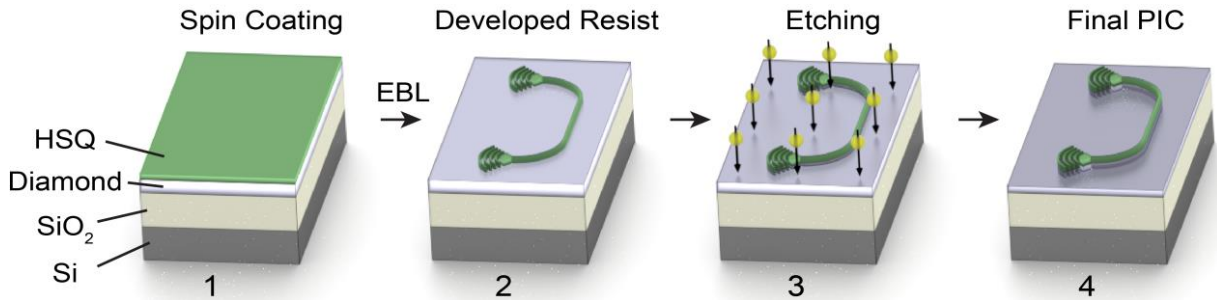

**Figure 16 - Fabrication of diamond photonic integrated circuits:** The process consists of spin coating HSQ resist, electron beam lithography (EBL), resist development and finally pattern transfer into diamond via reactive ion etching.

We fabricate diamond PICs from the polished diamond-on-insulator wafers with 600 nm of PCD, introduced in the previous section. Figure 16 illustrates the device fabrication process. We cut dies of $15 \cdot 15 \ mm^2$ size from the diamond-on-insulator wafer and deposit 5 nm of $SiO_2$ onto the dies via electron beam evaporation. The $SiO_2$ layer promotes adhesion of the hydrogen silsesquioxane (HSQ) resist (Dow Corning FOx-15) which is spin coated onto the diamond layer at a thickness of 500 nm (step 1). Electron beam lithography (EBL) is used for patterning the two-dimensional layout of the PICs into the negative tone resist. The exposure occurs in an EBL system (Jeol 5300) at 50 kV beam voltage using an area dose of $300 \ \mu C/cm^2$. Additionally a proximity effect correction is applied to the pattern, resulting in doses that locally vary between 70% and 110% of the area dose of $300 \ \mu C/cm^2$.

After 10 min of resist development in commercial developer (Microposit MF319) and rinsing in water, the HSQ structures resemble the device layout (step 2). The HSQ resist then acts as a mask for etching of the diamond. The layout is transferred into diamond via dry etching (step 3) in a capacitively coupled reactive ion etching (RIE) chamber (Oxford 100 system), using gas flows of 17 $cm^3$/min argon and 33 $cm^3$/min oxygen. The forward power is set to 200 W, which results in a direct current (DC) bias voltage between the electrodes of $\approx 535$ V. This leads to a diamond etch rate of $\approx 25$ nm/min, with a selectivity of $\approx 2:1$ for etching diamond versus etching the HSQ resist. The etching step gives rise to nearly vertical sidewalls, translating into rectangular waveguide cross-sections. For a diamond layer of 600 nm an etching time of 12 min results in a relative etch depth of 50% for the final PIC (step 4). The HSQ resist either remains on top of the diamond structure during device characterization or can be removed in hydrofluoric acid. The resist does not cause additional propagation losses, which was confirmed by transmission measurements of ring resonators (see section 2.4.4). Appendix A5 provides a detailed description concerning all device fabrication methods and experimental parameters.

### 2.4.3 Transmission measurements

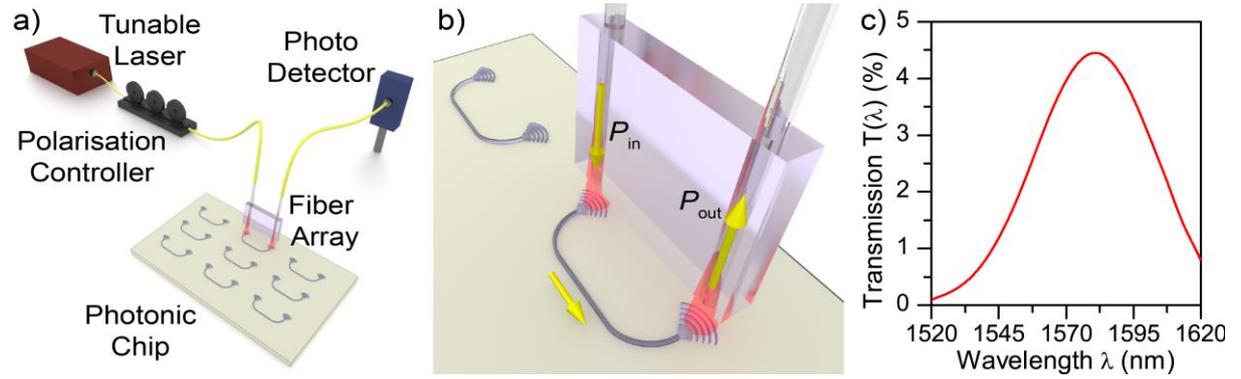

**Figure 17 - Transmission measurement:** a) Sketch of the setup for transmission measurements of PICs, consisting of a tunable laser, a polarization controller, a fiber array for coupling light between optical fibers and the PIC and a fiber-coupled photodetector (components are not to scale). b) Sketch of the coupling of light between optical fibers and the photonic waveguide via two focusing grating couplers. The flow of light is indicated with yellow arrows. c) Transmission spectrum of a PIC containing two diamond focusing grating couplers optimized for telecom wavelengths.

PICs are characterized in the setup, which is schematically depicted in Figure 17 a). The entire setup is installed on an optical table for vibrational damping. Light from a tunable laser at telecommunication wavelengths (New Focus TLB-6600 or Santec TSL-510) is coupled to an optical fiber. The photonic chip is mounted on a translation stage below the glass tip of a fiber array. A fiber polarization controller is used to adjust the polarization of the incident light in order to match the TE-like waveguide mode, as explained in section 2.1. Figure 17 b) shows a detailed view of the optical fibers of the fiber array and the photonic chip. The PIC consists of two focusing grating couplers and a connecting waveguide. The distance between the grating couplers is designed such that it matches the spacing between adjacent fibers of 250 µm. The input fiber launches light of power $P_{in}$ into the PIC, while a second fiber collects light of power $P_{out}$ coming from the PIC. We align the grating couplers of the PIC to the fiber array in the $x$-, $y$- and $z$-direction via piezo actuators for maximum transmission. After transmission through the PIC light is guided to a fiber-coupled low-noise InGaAs photodetector (New Focus 2011/2117). The transmission spectrum $T(\lambda)$ of the device can then be determined as the ratio of output to input power as

$$T(\lambda) = \frac{P_{out}}{P_{in}} = C_{in}(\lambda) \cdot C_{out}(\lambda) \cdot A_{prop} = C^2(\lambda) \cdot A_{prop} \,, \qquad (2.19)$$

where $C_{in}(\lambda)$ and $C_{out}(\lambda)$ are the wavelength dependent efficiencies of the in- and the out-coupling of the two involved couplers and $A_{prop}$ is the attenuation of light by propagation loss in the waveguide ($0 \leq A_{prop} \leq 1$, where $A_{prop} = 1$ means no attenuation). Both coupling efficiencies are typically assumed to be equal and therefore $C(\lambda)$ denoted the coupling efficiency for in- and out-coupling. Figure 17 c) shows a transmission spectrum $T(\lambda) = \frac{P_{out}}{P_{in}}$ for a simple PIC, consisting of a waveguide and two grating couplers, as shown in schematic Figure 17 b), with a maximum transmission of 4.4% centered at 1580 nm. All measurements of PICs within this thesis are performed in fiber-coupled setups, based on the setup schematically shown here. Depending on the specific experiment more fibers may be attached to the fiber array, in order to access more grating couplers at a time. Furthermore the fiber array and the photonic chip may be placed in a vacuum chamber (chapter 3) or a liquid helium cryostat (chapter 3 and 4) in order to control the environment of the photonic chip concerning pressure and temperature. Throughout this thesis, transmission spectra of PICs always include the multiplicative factor $C^2(\lambda)$ of the two grating couplers involved in any PIC transmission measurement.

### 2.4.4 Quantification of propagation losses in diamond waveguides

As explained in section 2.1.5 the propagation loss in waveguides can be estimated from the quality factors of ring resonators. We fabricate a chip containing PICs with 50% relative etch depth, according to the fabrication procedure explained in section 2.4.2. Each PIC consists of two grating couplers and a ring resonator of width $w_R = 1\,\mu m$ and radius $r$ which is evanescently coupled to a waveguide via a gap of size $g_R$, as shown in the schematic of Figure 7 a). The ring width is chosen as 1 μm, as we want to determine the propagation loss for the corresponding waveguides of the same width. Between different PICs we vary the radius of the ring resonator and the gap size $g_R$. Figure 18 a) shows a SEM micrograph of a fabricated PIC. For each device we measure the transmission for the TE-like mode around 1550 nm, as explained in the previous section.

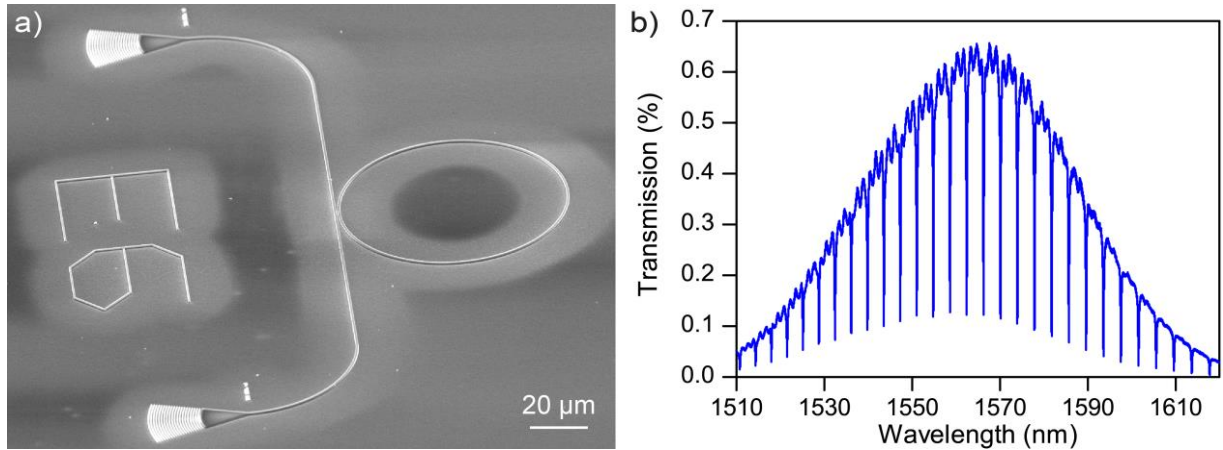

**Figure 18 - Ring resonator geometry and transmission:** a) SEM image showing a photonic integrated circuit with an optical ring resonator of $r = 40\,\mu m$ and $g_R = 600\,nm$. b) Transmission spectrum of a ring resonator ($r = 40\,\mu m$ and $g_R = 500\,nm$).

Figure 18 b) shows the transmission spectrum for a ring resonator ($r = 40\,\mu m$ and $g_R = 500\,nm$). The envelope is given by the grating couplers and the resonances are visible as dips in the transmission spectrum with a free spectral range of $\Delta\lambda = 3.76\,nm$ around 1550 nm. According to equation (2.12) this free spectral range corresponds to a group refractive index of $n_g(1550\,nm) = 2.54$, in agreement with FEM simulations. For each device we extract the quality factor and the extinction ratio for the resonances within the bandwidth of the grating coupler. Figure 19 a) shows the average quality factor for rings with $r = 70\,\mu m$ in dependence of gap size, while Figure 19 b) shows the corresponding extinction ratios. The extinction ratio is largest with $r_{ext} = 18.9 \pm 4.7$ for a gap size of 300 nm, which means that at this gap size the coupling condition is close to critical coupling. For increasing gap size, the coupling to the waveguide decreases and the quality factor increases and approaches its intrinsic value.

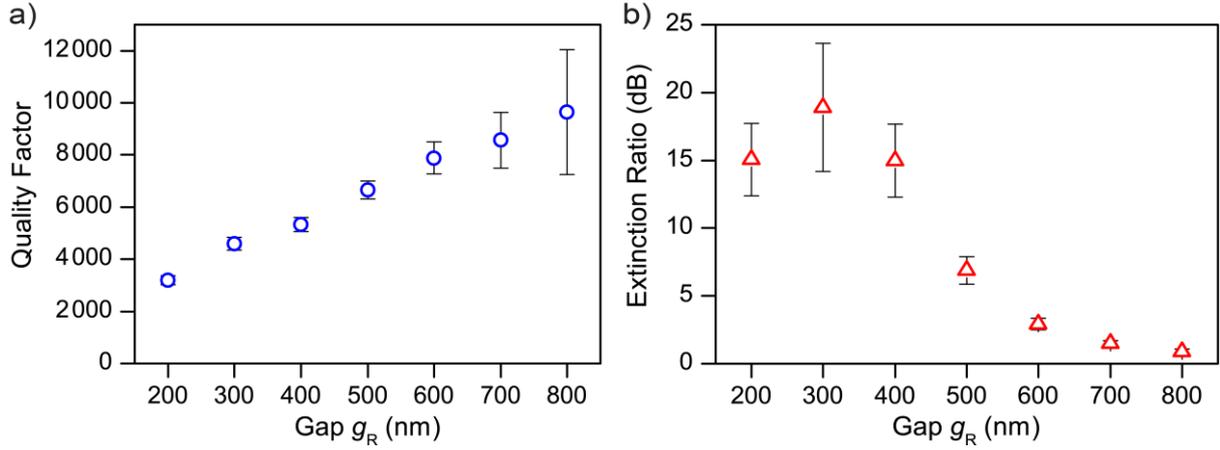

**Figure 19 - Experimental quality factor and extinction ratio**: For ring resonators with $r = 70\ \mu m$. a) Quality factor in dependence of gap size. b) Extinction ratio in dependence of gap size.

The largest quality factor is found for rings with $r = 90\ \mu m$ with $Q = 11\,800 \pm 2080$ for the weakest coupling at a gap size of $g_R = 800\ nm$. We use this quality factor to estimate the propagation losses. According to equation (2.15) a quality factor of $Q = 11\,800$ corresponds to an attenuation coefficient of $\alpha = 3.79\ dB/mm$, for propagation losses of $1\ \mu m$ wide PCD waveguides with 50% relative etch depth at a wavelength of 1550 nm. This attenuation coefficient is 28% smaller compared to waveguides of the same geometry fabricated from unpolished PCD layers[147] and we attribute this improvement to the reduction in scattering, due to the reduction in surface roughness after polishing, as explained in section 2.3.2. Scattering and absorption at grain boundaries, related to sp²-carbon and doping atoms[154,155] might be the dominating source of propagation loss in the presented PCD rings and the corresponding waveguides. In chapter 5.5 we will compare the quality factors and the propagation loss from PCD ring resonators to SCD ring resonators of the same geometry.

# 3 Diamond integrated optomechanics

*Using mechanical elements in photonic integrated circuits allows tunability in otherwise passive materials. Combining micromechanical components with photonic integrated circuits leads to optomechanical systems. This chapter motivates the use of diamond for integrated optomechanical circuits and presents proof-of-principle demonstrations. First, the fundamentals of micromechanical resonators, the detection of their motion, and their active actuation via optical gradient forces and electrostatic forces are explained. Subsequently, diamond integrated optomechanical circuit designs are explained and the experimental results are presented.*

*This chapter is partially based on results which were published previously in four publications[80,156–158], where the author of this thesis was first author or had equal contribution with the first author.*

## 3.1 Introduction to integrated optomechanics

Using mechanical degrees of freedom in the context of PICs has been established as a new research field in recent years[159–161]. Optomechanical systems find application in the study of fundamental physics, as mechanical motion in the quantum regime[160] can be studied using light. For example, the cooling of a mechanical oscillator to its quantum ground state[162–164] has been demonstrated. In the classical regime, on the other hand, mechanical components in PICs enable mechanically variable photonic systems, enabling to tune their optical properties via mechanical displacement. This enables PICs which incorporate mechanical elements such as tunable beam splitters[165], phase shifters[166,167], and transducers[168–171]. Micro- and nano-mechanical components can nowadays be fabricated and controlled at small sizes, such that their masses can be on the order of picograms or less[161] and they therefore show high responsivity to changes in their local environment. Integrated optomechanics in diamond could satisfy both purposes: On one hand it enables new active components for PICs, and on the other hand new effects arise, such as the manipulation of color center spin states in diamond, as will be described in the following section.

## 3.2 Motivation for diamond integrated optomechanics

Diamond might seem like an obvious choice as a material for optomechanics, due to its outstanding optical and mechanical properties, as explained in section 2.3. Nevertheless, only during the course of this thesis, the first integrated optomechanical circuits in diamond have been shown[18,156,172,173], as scalable device fabrication in diamond has proven to be difficult, as explained in section 2.3.

Potential applications for diamond integrated optomechanics range from sensors and the readout as well as manipulation of spin states to devices which enable the tuning of PICs, for example in the form of phase shifters as are needed for quantum optical circuits (see section 2.2.2). These three options are shortly outlined in the following:

**1. Sensors:** Diamond is both biocompatible and chemically inert and therefore diamond components are compatible with long-term use both inside sensitive living tissue[145,174,175] and in harsh chemical environments[176–178]. Micromechanical resonators can be fabricated with small masses and can be used as mass sensors by operating at their mechanical resonance frequencies. Adsorption of a particle of small mass can lead to a measurable shift of the resonance frequency[179], enabling mass detection down to single molecules[180] and single atoms[181]. A small linewidth of the resonance is crucial for high mass sensitivity and hence high mechanical quality factors are needed. The mechanical quality factors are a measure for damping and will be introduced in the following section. Excitation at resonance at gigahertz frequencies enables the operation of mechanical oscillators at ambient conditions without suffering from significant air damping[182–184]. Hence for mechanical sensors a high frequency and high quality factor are desirable. Functionalization of a sensor's surface enables the detection of specific substances and surface functionalization of various diamond devices have been shown[185–188]. Hence, all important ingredients for mechanical sensors based on diamond micromechanical elements have been demonstrated. Micromechanical and optical elements can also be combined to allow for new sensor architectures. If the sensing scheme takes advantage of the absorption or emission of light at specific wavelengths, then diamond's broadband transparency allows the operation in a large range of wavelengths. This includes the mid-infrared molecular fingerprint region, which is of key importance for applications such as trace gas sensing[189,190] and early cancer diagnostics[191]. Transmission of mid-infrared light through PCD waveguides has been demonstrated recently[96,97]. Summarizing, it can be said that optical, mechanical, and optomechanical sensors from diamond could find use in a broad range of applications, which motivates the development of PICs and optomechanical circuits from diamond.

**2. Readout and manipulation of spin states:** Single color centers, such as the nitrogen vacancy defect center, can be integrated in diamond micromechanical structures. The optical initialization and readout of color center spin states[104,192] combined with spin manipulation via pulsed microwaves allows local strain sensing through strain-mediated coupling of a single color center spin to a mechanical resonator[113]. This enables the detection of mechanical motion through the spin states[141]. Color centers in freestanding diamond elements have also been demonstrated as sensors for local electric and magnetic fields[109,193] and provide sensitivities down to the level of a single nuclear spin[114]. In the future, optical initialization and readout could be integrated on a chip with the mechanical element incorporating the single color center. Hence besides sensing schemes presented in the previous section, additional sensors based on the quantum states in color centers make diamond an important candidate for exploring optomechanical circuits. It has been suggested by Stannigel *et al.*[194] that a conversion between stationary and photonic qubits could be mediated by a mechanical resonator. As spin states in diamond are attractive as stationary qubits[26,107,108,195,196] with coherence lifetimes exceeding one second at room temperature[107] and single photons emitted by color centers in diamond can act as flying photonic qubits[196,197], diamond might be the ideal material platform for such a mechanical interface between spin states and photons. Potential applications include spin-induced oscillator sideband cooling[162], ultrafast mechanical spin driving[140] and spin squeezing[198].

Despite the promising demonstrations of diamond mechanical elements in combination with color centers, for both local sensing and quantum information processing, up to date such experiments have been performed, relying on bulk optics. The integration with PICs would take advantage of the miniaturization and the scalability of the resulting devices and hence motivate the development of diamond PICs which incorporate mechanically variable elements.

**3. Tunable elements / phase shifters:** Tunability in PICs can be achieved by altering the effective refractive index and hence the phase of electromagnetic waves which propagate in the modes of a waveguide. For semiconducting materials, such as silicon, the effective refractive index can be changed by injecting electrical carriers into the waveguides[199,200]. For electrical insulators such as silica, silicon nitride, and diamond this is not possible. A common device architecture is to place a heater[201–203] close to a waveguide. The temperature dependence of the refractive indices implies that by local heating the effective refractive index of a waveguide mode can be changed, resulting in a tunable phase shift. Such thermo-optic phase shifters are commonly used in current PICs for quantum optics[53,58,66,67,204]. The disadvantages of thermo-optic phase shifters include thermal crosstalk between closely spaced devices and a large power dissipation on the photonic chip, on the order of 10 mW per phase shifter for a $\pi$-phase shift.[201] If the development of PICs is supposed to follow the development of electronic ICs (which following Moore's law doubled the component density every two years[205]), then the use of thermo-optic phase shifters cannot be a long-term solution for thousands of tunable elements on one photonic chip. Especially for quantum optical circuits which use superconducting single-photon detectors[206] at cryogenic temperatures around 4 K, a large power dissipation for thermal tuning is unsuitable. Employing movable parts enables phase shifters which operate using mechanical displacement.[167] Opposed to heaters, optomechanical phase shifters do not require power dissipation for maintaining a static phase shift. Instead of having to maintain a certain temperature, which is larger than the temperature of the surrounding material, only a static mechanical displacement is maintained, which stores energy without notable dissipation. Besides low power consumption and avoided cross-talk, optomechanical phase shifters can also be faster than thermal phase shifters ($< 0.4\,\mu s$ compared to $\approx 3\,\mu s$)[167,202] and therefore provide a set of important advantages for tunable elements.

In summary, optomechanical devices promise advantages for tunable PICs and in particular as phase shifters. These could find applications within diamond PICs in classical applications, as well as for quantum optics. High-speed phase shifters in Mach-Zehnder interferometers have been shown to be useful for tunable feed-forward operations of a single-qubit gate of path-encoded qubits[207] and tunable phase shifters are needed for PICs, such as boson-samplers, introduced in section 2.2.2. This motivates the development of integrated optomechanical circuits from diamond.

The goal of the work presented in this chapter is to show the first proof-of-principle devices for diamond integrated optomechanics. For sensors and for manipulation of spin states high resonance frequencies and small linewidths (i.e. low dissipation) might be favorable, while for phase shifters and tunable elements comparably larger device sizes and larger displacements would be advantageous (translating to smaller resonance frequencies). We therefore design mechanical oscillators with resonance frequencies in the range of 1 MHz to 120 MHz and interface them with PICs. The resulting design could in the future be adjusted in one of the two opposite directions to match the specific requirements for one of the applications outlined above.

## 3.3 Fundamentals of optomechanics

In recent years, micromechanical oscillators have been applied in several applications, such as cooling to the quantum ground state, for which a quantum-mechanical description of the mechanical oscillator is required[159–161]. In the context of classical sensors and tunable PICs, as used within this thesis, a treatment with classical mechanics is sufficient and the following section provides the theoretical description for the following experimental implementation of diamond integrated optomechanics.

### 3.3.1 Driven harmonic oscillator

To understand the general dynamics of many optomechanical systems, a description within classical mechanics as a harmonic oscillator is sufficient. We consider a mass $m$ with only one degree of freedom, its movement in one dimension $x$, schematically shown in Figure 20 a).

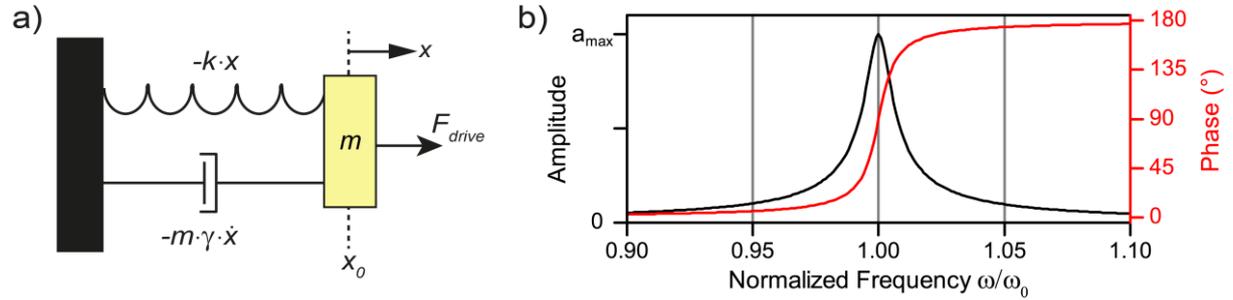

**Figure 20 - Driven harmonic oscillator:** a) Schematic of the model of a damped driven harmonic oscillator. b) Amplitude and phase response for the underdamped case ($Q = 100$) of a damped harmonic oscillator, depending on the frequency of the sinusoidal driving force.

The attachment of the mass via a spring to an anchor point leads to a restoring force $F_{\text{res}}$ which linearly depends on the displacement $x$ from its rest position, as

$$F_{\text{res}} = -k \cdot x \, , \tag{3.1}$$

where $k$ is the linear spring constant. Furthermore, a damping force $F_{\text{damp}} = -m \cdot \gamma \cdot \dot{x}$, which is proportional to the speed of the mass, and an external driving force $F_{\text{drive}} = m \cdot A \cdot \cos(\omega t)$ are considered. Here $\omega$ is the angular frequency at which the motion is driven by the external force. The motion of the mass can be described by the ordinary differential equation

$$m \cdot \ddot{x} + m \cdot \gamma \cdot \dot{x} + m \cdot \omega_0^2 \cdot x = m \cdot A \cdot \cos(\omega t) \, , \tag{3.2}$$

where $\omega_0 = \sqrt{k/m}$ is the resonance frequency of the undamped system. For our purpose only the underdamped case ($\gamma \ll \omega_0$) is relevant and the corresponding solution of the differential equation is presented in the following, as far as needed within this thesis. The solution of the differential equation consists of the inhomogeneous solution $x_{\text{inh}}(t)$ due to the driving force and the homogeneous solution $x_{\text{hom}}(t)$, which is damped away:

$$x_{\text{hom}}(t) = C \cdot e^{-\frac{\gamma}{2}t} \cdot \cos(\omega_r t - \varphi) \,. \tag{3.3}$$

With the resonance frequency $\omega_r$ defined as

$$\omega_r \equiv \sqrt{\omega_0{}^2 - \frac{\gamma^2}{2}} \,. \tag{3.4}$$

The initial amplitude $C$ and the phase $\varphi$ depends on the initial conditions. We will observe the motion under the influence of an external driving force, a long time after the motion has been excited and hence we are interested in the steady-state of the system, described by the inhomogeneous solution $x_{\text{inh}}(t)$:

$$x_{\text{inh}}(t) = a(\omega) \cdot \cos(\omega t - \varphi) \tag{3.5}$$

with a frequency dependent amplitude $a(\omega)$:

$$a(\omega) = \frac{A}{\sqrt{(\omega_0{}^2 - \omega^2)^2 + \gamma^2 \cdot \omega^2}} \tag{3.6}$$

and a phase difference between the motion and the driving force $\varphi(\omega)$:

$$\varphi(\omega) = \arctan\left(\frac{\gamma \cdot \omega}{\omega_0{}^2 - \omega^2}\right) \tag{3.7}$$

Figure 20 b) schematically shows the amplitude $a(\omega)$ and phase $\varphi(\omega)$ for frequencies around the resonance frequency. An important figure of merit for systems with low damping is the mechanical quality factor $Q$, defined as

$$Q \equiv \frac{\omega_0}{\gamma} \,. \tag{3.8}$$

The $Q$-factor is a measure for how much energy is dissipated in one oscillation period. Using $Q$ the amplitude as a function of frequency can be expressed as

$$a(\omega) = \frac{A}{\sqrt{(\omega_0{}^2 - \omega^2)^2 + \left(\frac{\omega_0 \cdot \omega}{Q}\right)^2}} \,. \tag{3.9}$$

The maximum amplitude $a_{\text{max}}$ of the driven harmonic oscillator occurs at the resonance frequency $\omega_r$, as defined in equation (3.4), and is given by

$$a_{\text{max}} = a(\omega_r) = Q \cdot \frac{A}{\omega_0{}^2} \cdot \frac{1}{\sqrt{1 - \frac{1}{4 \cdot Q^2}}} \,. \tag{3.10}$$

For high quality factors ($Q \gg 1$) this can be approximated as $a_{\text{max}} \approx Q \cdot \frac{A}{\omega_0{}^2}$. Note that the oscillation amplitude at resonance is both proportional to the amplitude of the driving force $A$ and to the mechanical quality factor $Q$. Therefore, higher mechanical quality factors enable higher transduction between driving force and oscillation amplitude.

It is possible to investigate the oscillator in the frequency domain and determine the mechanical quality factor from the dependence of $a^2$ on the frequency $\omega$ of the driving force. The energy stored in the oscillation is proportional to $a^2(\omega)$ and for a weakly damped oscillator ($\gamma \ll \omega_0$, $Q \gg 1$), starting from equation (3.6) $a^2(\omega)$ can be approximated with a Lorentzian curve, with a FWHM of $\Delta\omega = \gamma$. Hence one can extract the mechanical quality factor from a Lorentzian curve fit as $Q = \frac{\omega_r}{\Delta\omega}$.

The quality factor can also be investigated in the time domain. So far we considered the transient solution for the driving force $F_{\text{drive}} = m \cdot A \cdot \cos(\omega t)$. If we turn the driving force off at a certain time $t_{\text{off}}$ the damping leads to an exponential decay of the oscillation, as described by the homogeneous solution (see equation (3.3)). The envelope of the decaying oscillation is the oscillator's time-dependent amplitude $a(t)$, which behaves with time $t > 0$ as

$$a(t_{\text{off}} + t) = a_{\text{max}} \cdot e^{-\frac{\gamma}{2}t} = a_{\text{max}} \cdot e^{-\frac{t}{\tau}}, \tag{3.11}$$

with the characteristic decay time of the oscillation amplitude $\tau$, which can be expressed in terms of oscillation frequency and quality factor as

$$\tau = \frac{2}{\gamma} = \frac{2 \cdot Q}{\omega_0}. \tag{3.12}$$

This enables to determine the quality factor $Q$ of a mechanical oscillator in the time domain by fitting an exponential curve to the decay of the oscillator's amplitude. This is referred to as ring down measurement and will be applied for diamond micromechanical oscillators in section 3.4.3.2. The energy stored in the oscillator decays as

$$E(t) = \frac{1}{2}k \cdot a_{\text{max}}^2 \cdot e^{-\frac{\omega_0 \cdot t}{Q}}, \tag{3.13}$$

which illustrates that the quality factor is a measure for the dissipation of the system and the characteristic time of the decay of the stored energy.

## 3.3.2 Continuum mechanics

To describe the dynamics of a solid three-dimensional (3D) oscillator, in principle, the motion of all particles which make up the oscillator should be taken into account. For large, macroscopic objects this is needlessly complicated and materials can be accurately described as a continuum. This implies that the dynamics of the individual particles is irrelevant and instead the dynamics of the oscillator can be described by deflections and deformations which depend on a small set of parameters such as the elasticity tensor. The microscopic details do, however, determine the macroscopic material properties such as the Young's modulus and the Poisson ratio. A detailed description of this concept of continuum mechanics and its application to micromechanical resonators can for example be found in a review paper by Poot and van der Zant[160].

At the heart of continuum mechanics lies the relation between strain and stress in a material. Strain describes how the material is deformed with respect to its relaxed state. After a deformation of the material, the mass element which was initially located at position $\vec{x}$ is displaced to a new location $\vec{x} + \vec{u}$. The strain describes how much an infinitesimal line segment is elongated by the deformation $\vec{u}(x, y, z)$ and is given[208] by the strain tensor§

$$\gamma_{ij} = \frac{1}{2}\left(\frac{\partial u_i}{\partial x_j} + \frac{\partial u_j}{\partial x_i} + \frac{\partial u_m}{\partial x_i} \cdot \frac{\partial u_m}{\partial x_j}\right). \tag{3.14}$$

The diagonal elements are the normal strains in the coordinate directions, whereas the off-diagonal elements are the shear strains. While equation (3.14) is an exact description, the last, non-linear term is only relevant when the deformations are large[209] and will hence not be considered in this work. External forces deform the material and in turn give rise to forces inside the material. When considering the material to be composed of small volume elements, each element feels the force applied to its faces by all neighboring elements.

---

§ We use the Einstein notation for the elements of vectors and tensors: When indices appear only on one side of an equal sign, one sums over them, without explicitly writing the summation sign. The index runs over the three cartesian coordinates, where $x_1 = x$, $x_2 = y$ and $x_3 = z$.

The force $\delta\vec{F}$ on a small area $\delta A$ of the element is given by

$$\delta F_i = \sigma_{ij} n_j \delta A \, , \tag{3.15}$$

where $\vec{n}$ is the vector perpendicular to the surface and $\sigma$ is the stress tensor. The stress tensor describes the forces acting inside the material, whereas the strain tensor $\gamma$ describes the local material deformations. When the deformations are not too large, the stress and strain tensors are related linearly via the elasticity tensor with elements $E_{ijkl}$:

$$\sigma_{ij} = E_{ijkl} \cdot \gamma_{kl} \, . \tag{3.16}$$

In general the fourth-rank elasticity tensor is described by 81 elements. For an isotropic material, whose properties are the same in all directions, only two independent parameters are needed: the Young's modulus $E$ and the Poisson's ratio $v$.

For simplified geometries, such as a thin beam subjected to lateral loads, it is possible to analytically calculate the displacement $\vec{u}(x, y, z)$ and the corresponding strain. Figure 21 a) shows a schematic of such a thin beam, which is oriented along the $z$-direction, fixed in position at both ends and subjected to a force $F$.

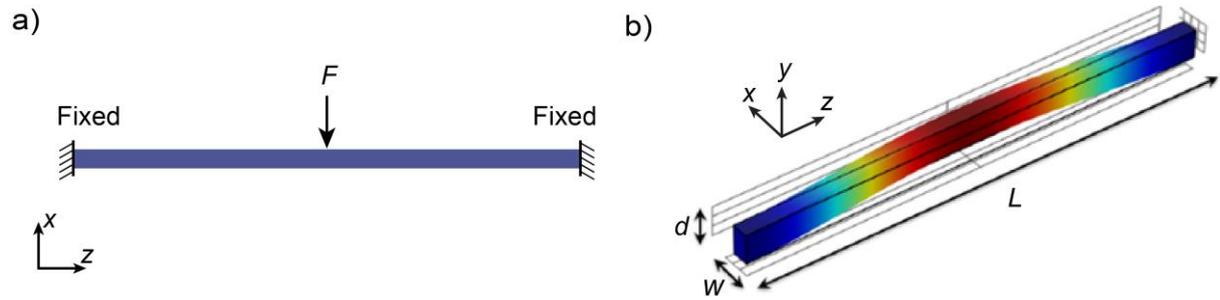

**Figure 21 - Three-dimensional mechanical oscillator:** a) Schematic of a doubly clamped mechanical beam under external force $F$. b) Displacement field of the fundamental in-plane mode of oscillation (in the $x/z$-plane) for a doubly clamped beam of width $w$, height $d$, and length $L$. The displacement is shown as a deformation of the 3D beam, as well in color, where blue denotes no displacement and red denotes maximum displacement.

Figure 21 b) shows a 3D schematic of the beam, which is defined by its length $L$, width $w$, and height $d$. If such a flexural beam is long and thin ($L \gg w, d$) and made of an isotropic material with no damping, the dynamic behavior can be treated as a one-dimensional problem using the Euler-Bernoulli beam theory[210]: Instead of having to consider the displacement field $\vec{u}(x, y, z)$ for each point in space within the beam, it is sufficient to consider one point for each cross-section perpendicular to the $z$-direction: the points along the beam's centerline. Boundary conditions for the motion are typically the fixation points to the substrate, in this case at both ends of the beam. This geometry is known as doubly clamped beam. If the beam moves in the $x$-direction we refer to it as in-plane motion, as the $x/z$-plane is defined by the thin film of the wafer. If the beam moves in the $y$-direction we refer to it as out-of-plane motion, as the motion is perpendicular to the thin film layers. In-plane motion is more relevant to integrated optomechanics, as an in-plane displacement can lead to a much larger change in the effective refractive index of a waveguide, as we will discuss in section 3.3.4.

In the following we therefore describe in-plane motion, while out-of-plane motion can be treated in the same fashion.

For in-plane motion, only the $x$-component of the displacement vector $\vec{u}$ is considered to be non-zero. For an isotropic material with no damping, the beam's centerline can show transverse displacements in the $x$-direction $u_x(z,t) \equiv u(z,t)$, which obeys the differential equation[160]

$$\rho \cdot A \cdot \frac{\partial^2 u}{\partial t^2}(z,t) = -D\frac{\partial^4 u}{\partial z^4}(z,t) + T\frac{\partial^2 u}{\partial z^2}(z,t) + F \;, \tag{3.17}$$

where $\rho$ denotes the material density, $A = w \cdot d$ denotes the cross-sectional area, $D$ is the bending rigidity, $T$ is the tension in the material in the $z$-direction, and $F$ is an external force. We consider the tension $T$ to be negligible. The bending rigidity $D$ can be written as the product of the Young's modulus $E$ and the bending moment of inertia with respect to the $y$-axis $I_y$ as

$$D = E \cdot I_y \cdot \frac{1}{1-v^2} \;, \tag{3.18}$$

where $v$ is the Poisson ratio. The Poisson ratio is typically small ($v \approx 0.06$ for diamond[211]) and is therefore often neglected. For a beam as shown in Figure 21, the bending moment of inertia with respect to the $y$-axis can be calculated as

$$I_y = \iint x^2 \, dx dy = \frac{d \cdot w^3}{12} \;, \tag{3.19}$$

where $d$ denotes the height and $w$ denotes the width. The clamped ends at $z = 0$ and $z = L$ impose the boundary conditions $u(0) = u(L) = 0$ and $\frac{\partial u}{\partial z}(0) = \frac{\partial u}{\partial z}(L) = 0$. The solutions for zero external force have the form[210]

$$u_n(z,t) = \chi_n(z) \cdot U \cdot e^{-i(\omega_n t - \varphi)} \;, \tag{3.20}$$

with the eigenfunctions

$$\chi_n(z) = C_{1n} \cdot [\cos(k_n z) - \cosh(k_n z)] + C_{2n} \cdot [\sin(k_n z) - \sinh(k_n z)] \;, \tag{3.21}$$

with eigenvectors $k_n$ satisfying

$$\cos(k_n L) \cdot \cosh(k_n L) = 1 \;. \tag{3.22}$$

The eigenfunctions $\chi_n(z)$ are mutually orthogonal and build a basis, such that any beam motion can be described as a superposition of these eigenfunctions as $u(z,t) = c_n \cdot u_n(z,t)$. The angular eigenfrequencies $\omega_n$ are given by

$$\omega_n = \sqrt{\frac{E}{\rho}} \cdot \sqrt{\frac{I_y}{A}} \cdot k_n{}^2 \;. \tag{3.23}$$

The numerical solution of the fundamental in-plane mode corresponds to the first non-zero eigenvector $k_1 \approx \frac{4.730}{L}$, with an eigenfunction $\chi_1(z)$ defined by $C_{11} \approx -1.000 \cdot L$ and $C_{21} \approx +0.983 \cdot L$. The fundamental frequency is given by

$$f_1 = \frac{\omega_1}{2\pi} = 1.028 \sqrt{\frac{E}{\rho}} \cdot \frac{w}{L^2} \;. \tag{3.24}$$

Note that the frequency is proportional to the width $w$ and proportional to $\frac{1}{L^2}$. For diamond beams with widths of several hundred nanometers and tens of micrometer length, as relevant to this thesis, fundamental frequencies are on the order of Megahertz.

For the presented simplified case of a thin beam, analytical solutions for eigenfunctions and eigenfrequencies can be calculated, which is instructive and often provides good approximations for the motion of micromechanical oscillators. More accurate treatments require numerical methods. Within this thesis 3D FEM simulations are used, which enables realistic boundary conditions and geometries with more design parameters than simple beams to be taken into account. The simulations for the specific oscillator geometry used in this thesis will be presented in section 3.3.6.2.

It is useful to introduce a mapping between the 3D beam and a hypothetical point-mass with a scalar displacement $U$, such that the harmonic oscillator model can be applied. We can write the displacement field $\vec{u}_n$ for mode $n$ as

$$\vec{u}_n(x, y, z, t) \equiv \vec{\chi}_n(x, y, z) \cdot U(t) , \qquad (3.25)$$

where $\vec{\chi}_n(x, y, z)$ is the normalized shape of the mode and $U(t)$ is the time-dependent amplitude. We choose the normalization of $\vec{\chi}_n(x, y, z)$ such that for its maximum absolute value is equal to one ($\max_{\vec{r} \in V} |\vec{\chi}_n| = 1$). This implies that $U(t)$ refers to the maximum displacement of any mass element of the 3D oscillator. The average kinetic energy of a harmonic oscillator of a mass $m_{\text{eff},n}$ and amplitude $U(t) = \bar{U} \cdot \sin(\omega_n t)$ amounts to

$$\langle E_{\text{kin}} \rangle = \left\langle \frac{1}{2} m_{\text{eff},n} \cdot \left( \frac{dU}{dt} \right)^2 \right\rangle = \frac{1}{2} m_{\text{eff},n} \cdot \omega_n^2 \cdot \frac{1}{2} \bar{U}^2 . \qquad (3.26)$$

On the other hand, the energy of a 3D oscillator with density $\rho$ within its volume $V$ under oscillation with a displacement field $\vec{u}_n(x, y, z, t) \equiv \vec{\chi}_n(x, y, z) \cdot \bar{U} \cdot \sin(\omega_n t)$ can be calculated as[209]

$$\langle E_{\text{kin}} \rangle = \left\langle \frac{1}{2} \int \rho \cdot \left( \frac{\partial \vec{u}_n(x,y,z,t)}{\partial t} \right)^2 dV \right\rangle$$
$$= \frac{1}{2} \left( m_0 \frac{1}{V} \int \vec{\chi}_n^2(x, y, z) dV \right) \omega_n^2 \cdot \frac{1}{2} \bar{U}^2 , \qquad (3.27)$$

where $m_0$ is the physical mass of the oscillator. We define the effective modal mass of mode $n$ as[183]

$$m_{\text{eff},n} \equiv m_0 \frac{1}{V} \int \left( \frac{\vec{\chi}_n(x,y,z)}{\chi_{\max}} \right)^2 dV , \qquad (3.28)$$

such that the energy of the 3D oscillator is equivalent to the energy of the point mass $m_{\text{eff},n}$ under oscillation with an amplitude $U(t)$, which corresponds to oscillation of the mass element with the largest amplitude. The spring constant which relates the amplitude $U(t)$ for the oscillation at resonance frequency $\omega_n$ to an external force is given by

$$k_n = m_{\text{eff},n} \cdot \omega_n^2 . \qquad (3.29)$$

The effective modal mass $m_{\text{eff},n}$ for a micromechanical oscillator is a constant, which solely depends on the oscillator geometry, material properties, and the mode number $n$. We can therefore determine it from simulated displacement fields $\vec{\chi}_n(x, y, z)$. While in the three-dimensional description presented in this section damping was not considered, any experimental implementation will show damping. Damping will shift the resonance frequency of each mode, as discussed for a one-dimensional point mass system (equation (3.4)), but the mode shape is typically unaffected[212] by the damping. It is hence feasible to simulate the normalized mode shapes $\vec{\chi}_n(x, y, z)$ for the undamped system and use them for mechanical oscillators with damping.

The connection presented here between a three-dimensional mass distribution with a displacement field and the harmonic oscillator model of a point mass is important, as it justifies to explain the behavior of diamond micromechanical oscillators within this thesis based on the harmonic oscillator model.

### 3.3.3 Damped oscillator driven by thermal motion

Damping implies that energy can be exchanged with the environment. Due to the fluctuation-dissipation theorem there is consequently mechanical noise associated with the damping in the system[209]. The oscillator therefore experiences a random noise force $F_N(t)$ with a zero average value $\langle F_N(t) \rangle = 0$. The dynamics of the oscillator thus follows the differential equation

$$m \cdot \ddot{x} + m \cdot \gamma \cdot \dot{x} + m \cdot \omega_0^2 \cdot x = F_N(t) \, . \tag{3.30}$$

For a completely uncorrelated force noise, the spectral density of the force $S(\omega)$ is frequency independent (white noise). It can be shown that the noise force produces a thermal equilibrium of the oscillator and its environment.[213] For an oscillator with several modes $n$ the equipartition theorem implies that the oscillator thermalizes with its environment at temperature $T$, such that the mean energy $\langle E_n \rangle$ of each mode $n$ of the oscillator[214] is given by $\langle E_n \rangle = k_B T$, where $k_B$ is the Boltzmann constant. For a 3D oscillator, such as a doubly clamped beam, each point on the oscillator experiences a noise force which acts with the same spectral density, but fluctuates independently from the force at other points. The noise at any two points on the oscillator is uncorrelated[209]. The noise-force can be expanded in terms of the eigenfunctions of the oscillator and a noise force associated with a mode $n$ is thus uncorrelated with the noise for any other mode $n' \neq n$[209]. We therefore consider only the fundamental mode of the oscillator and the white noise which leads to thermal equilibrium with the environment.

As white noise corresponds to an excitation at all frequencies with a constant spectral density, the resulting spectral density of the amplitude of the thermally driven damped harmonic oscillator resembles the curve of the frequency dependent amplitude of a harmonic oscillator driven at a single frequency (see equation (3.6)). The spectral density $S_x(\omega)$, which is associated with the kinetic energy of the fundamental mode hence resembles a Lorentzian curve. Using the average kinetic energy associated with the fundamental mode at thermal equilibrium $\langle E_{\text{kin}} \rangle = \frac{1}{2} k_B T$, it is possible to derive[209] that the spectral density of the displacement noise at resonance frequency $\omega_0$ amounts to[179]

$$\sqrt{S_x(\omega_0)} = \sqrt{\frac{4 \cdot k_B T \cdot Q}{m_{\text{eff}} \cdot \omega_0^3}} \, , \tag{3.31}$$

where $Q$ is the mechanical quality factor and $m_{\text{eff}}$ the effective modal mass (as introduced in the previous section). Note that the units of $\sqrt{S_x}$ are m/$\sqrt{\text{Hz}}$. If a signal is measured which is proportional to the displacement, then by analyzing the signal in the frequency domain, the spectral density of the displacement noise can be determined. The measured signal amplitude can then be calibrated using the theoretical thermal displacement noise (using equation (3.31)), a method which was first applied for the calibrations in atomic force microscopy[215] and gets routinely applied for the calibration of the measurement sensitivity concerning mechanical displacement in PICs[183,216,217]. We will perform this calibration for diamond micromechanical oscillators in section 3.4.2.

### 3.3.4 Integrated optomechanical system and optical phaseshift

Mechanical oscillators, such as doubly clamped beams described above, can be combined with waveguides, resulting in optomechanical integrated circuits. This enables sensitive detection of mechanical displacement, as well as active manipulation of optical properties via the control of the mechanical displacement, as will be explained in the following. Figure 22 a) shows a schematic of such a system. It consists of a waveguide at a fixed position and a mechanical element in close vicinity of a waveguide. The mechanical element can move in the $x$-direction and has internal damping and restoring forces, as considered in the previous sections. Without external forces, the rest position is such, that waveguide and mechanical element are separated by a gap of size $g_0$. The system shows translational invariance in the $z$-direction and we therefore consider the waveguide mode in a two-dimensional cut in the $x/y$-plane, as shown in Figure 22 b). The effective refractive index $n_{eff}$ of the waveguide depends on the gap size $g = g_0 + u$, as shown in Figure 22 c), and hence a displacement $u$ of the mechanical element, due to an external force, will lead to a change in $n_{eff}$ and hence a change of the phase of light which propagates in the waveguide.

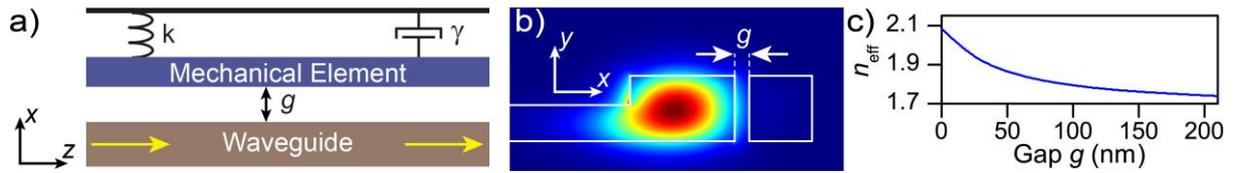

**Figure 22 - Optomechanical system and effective refractive index:** a) Schematic of an optomechanical system, composed of a waveguide, coupled to a mechanical element by a gap $g$. b) Cross section through the system, showing the waveguide mode. c) Exemplary dependence of the effective refractive index $n_{eff}$ of a waveguide mode on the size of the gap between optical waveguide and mechanical element.

We consider systems for which the gap size and hence the displacement is a function of the $z$-coordinate, for example if the mechanical element is not simply displaced, but bends in the $x/z$-plane, as considered in the previous section for doubly clamped beams. The phase acquired by photons during propagation in the $z$-direction along a waveguide section of length $l_{int}$ amounts to

$$\phi(\lambda_0) = \frac{2\pi}{\lambda_0} \int_{z=0}^{l_{int}} n_{eff}(\lambda_0, g) dz \,, \qquad (3.32)$$

where $\lambda_0$ denotes the wavelength of light in vacuum and $l_{int}$ is the interaction length of waveguide and mechanical oscillator. In-plane motion of the movable part in the $x$-direction leads to a displacement $u(z)$ from the rest position. This leads to a $z$-dependent change in the effective refractive index and the results in an overall change in phase $\Delta\phi$, which amounts to

$$\Delta\phi(\lambda_0, u(z)) = \frac{2\pi}{\lambda_0} \int_{z=0}^{l_{int}} \big(n_{eff}(\lambda_0, g_0 + u(z), z) - n_{eff}(\lambda_0, g_0, z)\big) dz \,. \qquad (3.33)$$

The effective refractive index $n_{eff}(g)$ for such a system typically shows an exponential dependence on the displacement $n_{eff}(g = g_0 + u) = n_{eff}(g_0) \cdot e^{-\frac{u}{A}}$ with a characteristic spatial decay constant $A$. For small displacements $u \ll A$, it is appropriate to Taylor expand the exponential function and only consider the linear term[218].

The change in phase then shows a linear dependence on the displacement as

$$\Delta\phi\big(\lambda_0, u(z)\big) \approx \frac{2\pi}{\lambda_0} \int_{z=0}^{l_{\text{int}}} \frac{\partial n_{\text{eff}}(\lambda_0, x)}{\partial x}\bigg|_{x=x_0} \cdot u(z)\, dz \approx \frac{2\pi}{\lambda_0} \cdot \frac{\partial n_{\text{eff}}(\lambda_0, x)}{\partial x}\bigg|_{x=x_0} \cdot \bar{u} \cdot l_{\text{int}} \,, \qquad (3.34)$$

with the average in-plane displacement of the mechanical oscillator $\bar{u}$ defined by

$$\bar{u} = u_{\text{max}} \frac{1}{l_{\text{int}}} \int_{z=0}^{l_{\text{int}}} \chi(z)\, dz \,, \qquad (3.35)$$

which is determined by the geometry of the mechanical element, the maximum displacement $u_{\text{max}}$ and the shape $\chi(z)$ of the mechanical mode under consideration.

Note that hence, according to equation (3.34), the phase shift increases with increasing interaction length $l_{\text{int}}$, with a stronger dependence of $n_{\text{eff}}$ on the lateral position $x$ and with a larger average displacement $\overline{\Delta u}$. Tunable phase shifters are for example needed for tunable quantum optical circuits, such as boson-samplers, as introduced in section 2.2.2. For such an application a phase shift $\Delta\phi$ which is tunable between 0 and $2\pi$ would be required. This means that for an optomechanical phase shifter of a given geometry, the phase shift should be tunable in this range by changing the displacement $u(z)$ via a controllable external force. We will discuss external forces in section 3.3.5 and present the corresponding experiments in section 3.4.

It is important to note the difference between two variables within optomechanics, which are generally referred to as phase:

1) the phase $\varphi(\omega_m)$ between a force which excites a mechanical oscillator and the resulting mechanical motion at frequency $\omega_m$, as defined by equation (3.7). This can be referred to as the phase of the mechanical oscillator. A static force ($\omega_m = 0$) leads to a displacement $u(z)$, but the mechanical phase $\varphi(\omega_m)$ remains zero.

2) the phase $\Delta\phi(u(z))$, as defined by equation (3.33), which the light acquires additionally, due to a displacement $u(z)$, during propagation along the interaction length $l_{\text{int}}$. This is the phase imprinted by a mechanical displacement on the electromagnetic wave, which can be non-zero for a static displacement.

Keeping in mind the distinction between both types of phase is important for the understanding of the following descriptions.

## 3.3.5 Excitation of micromechanical motion via optical gradient forces

Light can exert forces on matter[159] which act on mechanical degrees of freedom[218]. In this way light-driven mechanically variable systems can perform trapping[219–221] and actuation[169,222–224] of objects on the micro- and nanoscale. Optical forces can occur as radiation pressure and optical gradient forces. Radiation pressure[225] stems from the momentum transfer from electromagnetic radiation to matter, which for example occurs upon reflection of light from a mirror. Radiation pressure has been extensively studied in the context of optical cavities and interferometers. Such forces are relevant for large structures on kilometer length scales, such as gravitational wave detectors[226,227], as well as for small structures on the micro- and nanoscale[164,228–230]. Optical dipole forces, also called optical gradient forces, can occur when a dipole is induced[231] in a polarizable micro-particle. This effect finds application as optical tweezers[232,233] for the manipulation of micro-particles[234,235]. The optical gradient forces can also arise from internal illumination by light traveling in waveguides. The optical gradient force acts perpendicular to the propagation direction of light and for laterally varying electromagnetic fields, such as in waveguide modes, the optical gradient force can be comparably large and the mass and the dimensions of components for PICs have been miniaturized to the degree that device tuning via optical actuation is possible at micro- to milli-watt power levels[216,222,223,230].

Using Maxwell's equations and the Lorentz force, which connects the electromagnetic field and mechanical motion, the time average of the force exerted by an electromagnetic field on a rigid body within a volume $V$ can be calculated[231] as

$$\langle F \rangle = \oint_{\delta V} \langle \hat{T}(\vec{r},t) \rangle \cdot n(\vec{r}) \cdot ds \,, \tag{3.36}$$

where $\oint_{\delta V} ds$ denotes the integral over the surface of volume $V$ which encloses the geometry and $\hat{T}(\vec{r},t)$ is the Maxwell stress tensor whose components $T_{ij}$ are related to both the electric field distribution $\vec{E}(\vec{r},t)$ and the magnetic field distribution $\vec{B}(\vec{r},t)$ as

$$T_{ij}(\vec{r},t) = \varepsilon_0 \varepsilon_r \left( E_i E_j - \delta_{ij} \frac{1}{2} \vec{E}^2 \right) + \frac{1}{\mu_0 \mu_r} \left( B_i B_j - \delta_{ij} \frac{1}{2} \vec{B}^2 \right) \,, \tag{3.37}$$

where $\varepsilon_r$ and $\mu_r$ denote the dielectric constant and magnetic susceptibility, $\varepsilon_0$ and $\mu_0$ denote the permittivity and permeability of free space and $\delta_{ij}$ is the Kronecker delta.

An equivalent[218,222,236] but potentially more intuitive way to derive the force is via the change in the total energy $E_{tot}$ which is stored in the propagating optical field in the waveguide mode. Povinelli *et al.*[222] derived the force $\vec{F}_{opt}$ by arguing that the work done by the mechanical displacement $\int \vec{F}_{opt} \, d\vec{s}$, should equal the change in energy $dE_{tot}$.

If light at a well-defined frequency $\omega$ propagates in the form of a guided waveguide mode along a waveguide of length $l$, then the energy $V_\omega$ associated with the propagating wave is given by

$$V_\omega(g) = P \cdot l \cdot \frac{n_{\text{eff}}(\omega,g)}{c} \,, \tag{3.38}$$

where $P$ is the optical power, $c$ is the speed of light in vacuum, and $n_{\text{eff}}$ is the effective refractive index of the mode, which depends on the gap size $g$. Note that $\frac{c}{n_{\text{eff}}}$ is the phase velocity. The optical force is then given as the gradient of the energy as

$$\vec{F}_\omega(g) = -\nabla V_\omega(g) \,. \tag{3.39}$$

The spatial dependence of the energy $V_\omega$ is given by the spatial dependence of $n_{\text{eff}}$, which changes with displacement in the $x$-direction and hence with a change in gap size $g$, which results in an optical force in the $x$-direction of

$$\vec{F}_\omega(g) = -P \cdot \frac{l}{c} \cdot \frac{\partial n_{\text{eff}}(\omega,g)}{\partial g} \vec{e}_x \,. \tag{3.40}$$

The absolute value of this force can be expressed using the energy $V_\omega$ as

$$F_\omega(g) = \frac{1}{n_{\text{eff}}(\omega,g)} \cdot \frac{\partial n_{\text{eff}}(\omega,g)}{\partial g} \cdot V_\omega(g) \,. \tag{3.41}$$

For an electromagnetic wave, such as a wave package, which does not consist of a single frequency $\omega$, the total energy associated with light guided by the waveguide mode is given by[236]

$$E_{\text{tot}}(g) = P \cdot l \cdot \frac{n_g(g)}{c}, \qquad (3.42)$$

where $P$ is the optical power, $l$ is the length of the waveguide, $c$ is the speed of light in vacuum, and $n_g(g)$ is the group refractive index. Note that $E_{\text{tot}}$ is associated with the group velocity $\frac{c}{n_g}$. The resulting optical force $F_{\text{opt}}$ can be expressed as[236]

$$F_{\text{opt}}(g) = \frac{1}{n_{\text{eff}}(g)} \cdot \frac{\partial n_{\text{eff}}(g)}{\partial g} \cdot E_{\text{tot}}(g) = \left( \frac{n_g(g)}{n_{\text{eff}}(g)} \cdot P \cdot \frac{l}{c} \right) \cdot \frac{\partial n_{\text{eff}}(g)}{\partial g}. \qquad (3.43)$$

As can be seen from equation (3.43), waveguide geometries for which the derivative $\frac{\partial n_{\text{eff}}}{\partial g}$ is larger compared to other waveguide geometries lead to larger optical gradient forces. $F_{\text{opt}}$ is proportional to both the optical power $P$ and the length of the device $l$. It is therefore useful to define a normalized optical force $F^{n}_{\text{opt}}$ as

$$F^{n}_{\text{opt}}(g) = \frac{F_{\text{opt}}(g)}{P \cdot l} = \left( \frac{n_g(g)}{n_{\text{eff}}(g)} \cdot \frac{1}{c} \right) \cdot \frac{\partial n_{\text{eff}}(g)}{\partial g}. \qquad (3.44)$$

The absolute value of normalized optical force in integrated optomechanical circuits is typically on the order of $\frac{\text{pN}}{\mu\text{m} \cdot \text{mW}}$.[216,223] In section 3.4.3 it will be shown how optical forces drive the motion of diamond micromechanical oscillators.

## 3.3.6  Design of the mechanical resonator

The goals for our design of a mechanical resonator for tunable diamond integrated PICs are:

1) that the mechanical resonator can be fabricated in close proximity to a waveguide, with which it interacts,

2) that the resonance frequency can be easily chosen via the geometric parameters,

3) low losses concerning the photons which propagate in the waveguide,

4) the possibility to spatially isolate the waveguide from the force which drives the mechanical motion.

A design which enables to satisfy these goals is the H-resonator, which has been employed in silicon nitride as a broadband phase shifter for PICs[167]. Figure 23 a) shows a schematic of the H-resonator geometry. It consists of two doubly clamped beams of length $L$, on the order of tens of micrometers, and width $w_H$, on the order of hundreds of nanometers. Both beams are joined by a central block of length $b_{\text{cen}} = 7\,\mu\text{m}$ and width $h_{\text{cen}} = 5\,\mu\text{m}$ (block dimensions are chosen arbitrarily), forming a structure which resembles the letter "H". The central block includes an array of holes, which form a two-dimensional photonic crystal slab, which will be further explained in section 3.3.6.1. The H-resonator is evanescently coupled to a curved waveguide which supports a single mode for TE-like polarization. Along an interaction length of waveguide and mechanical oscillator $L_{\text{int}} = 12\,\mu\text{m}$ the H-resonator and waveguide are parallel to each other. Figure 23 b) shows a cross-section through both the waveguide of width $w_{\text{WG}} = 1\,\mu\text{m}$ and the H-resonator arm. They are separated by a small gap $g =$ 150 nm. The material of both structures is diamond of thickness $d = 600$ nm. The H-resonator consists of fully etched diamond, while the waveguide is partially etched and hence connected to a continuous diamond film (continuing on the left side of the schematic in the $x$-direction). The underlying silicon oxide layer is removed below both the H-resonator and the waveguide. Hence diamond and air are the only materials present. The position of the waveguide is fixed due to its connection to the surrounding diamond layer, but the H-resonator can move in the $x$-direction (as it is freestanding in the $x/y$-plane and only fixed at the end of the arms in the $z$-direction). A displacement of the H-resonator in the $x$-direction leads to a change in gap size $g$.

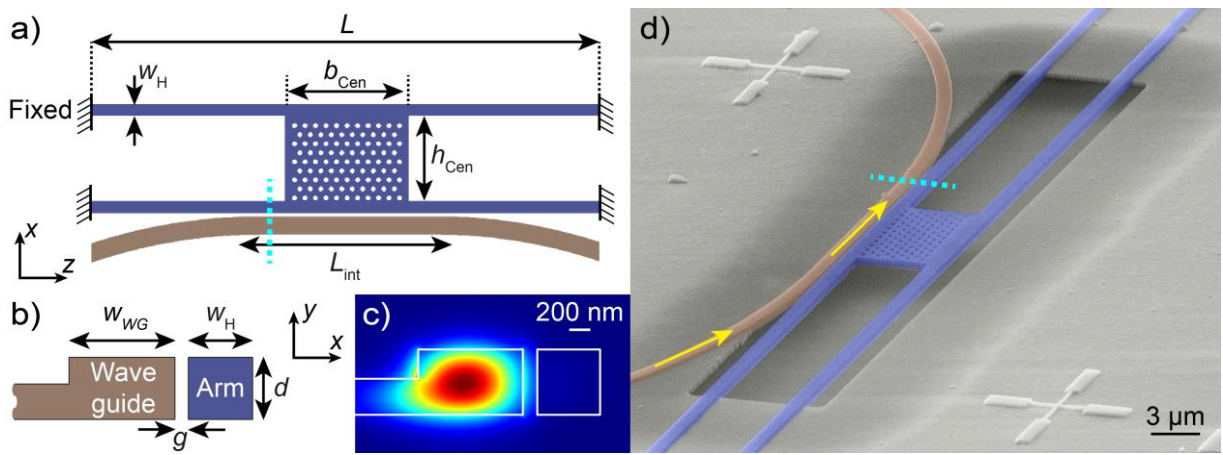

**Figure 23 - H-resonator device geometry:** a) Sketch of the H-resonator, coupled to a waveguide (top-view in the $x/z$-plane). The arm width $w_H$ and the arm length $L$ are the two geometric parameters which are varied between different circuits. b) Cross-section through the H-resonator arm of width $w_H$ and the waveguide of width $w_{WG}$, separated by a gap of size $g$. The location of the cross-section in the $x/y$-plane is indicated as a cyan dashed line in a) and d). c) Simulated distribution of the electric field for the TE-like optical mode for $w_H = 600$ nm, $w_{WG} = 1$ μm and $g = 150$ nm. d) Colorized SEM micrograph of a freestanding diamond H-resonator, which is clamped by the diamond layer and evanescently coupled to a half-etched waveguide.

Figure 23 c) shows the simulated distribution of the electric field for the TE-like waveguide mode for $w_H = 600$ nm, $w_{WG} = 1$ μm. The gap between H-resonator and waveguide is $g = 150$ nm and the evanescent field of the waveguide extends into the H-resonator arm. Figure 23 d) shows a SEM micrograph of the H-resonator and the adjacent waveguide. The underetched area, the clamping points of the beams, and the close proximity to the waveguide are clearly visible.

### 3.3.6.1 Optical isolation in nanomechanical resonators

The central block of the H-resonator contains a hexagonal array of holes, which leads to a two-dimensional photonic crystal slab, as explained in section 2.1.6. We make use of the results of the simulations presented in that section, which showed that for a hole radius r = 180 nm and a lattice period $a = 600$ nm, a photonic bandgap exists for the propagation of light with TE-like polarization at telecom wavelengths in the diamond layer. Figure 24 shows the distributions of the electric field, resulting from 3D FDTD simulations, for light propagation along a waveguide coupled to an H-resonator without (a) and with PhC slab (b). Light is launched into the waveguide on the left side and, if no air holes are present, light couples evanescently into the H-resonator and propagates into the central block and the upper arms. Such a design would lead to a significant loss for the waveguide, introduced by the H-resonator in close proximity. When replacing the central block with a PhC slab, it acts as a mirror and restricts light propagation to one side of the mechanical resonator.

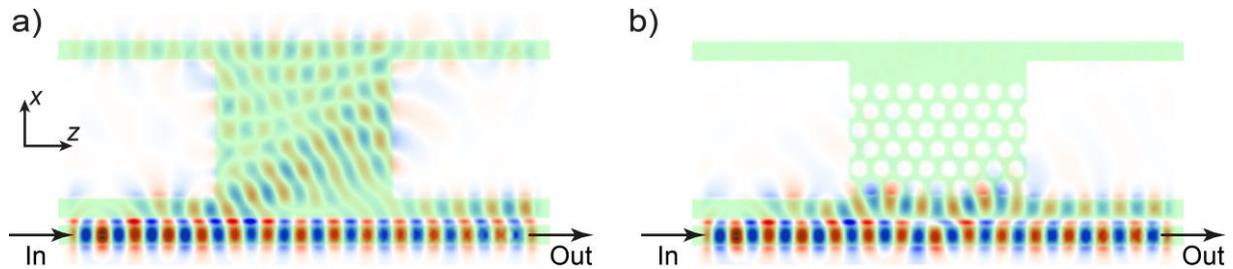

**Figure 24 - H-resonator with a photonic crystal slab:** Simulated mode pattern for an H-resonator which is evanescently coupled to a waveguide at the bottom (a) without and (b) with a PhC slab inside the H-resonator, which acts as a mirror. The PhC slab optically isolates the bottom and top side of the resonator.

This design not only supresses the losses due to light propagating into the mechanical resonator, but also optically isolates the bottom and top side of the resonator. This can for example be used for driving the mechanical oscillation by applying force on the upper arm of the H-resonator without leading to cross-talk or additional losses concerning the light in the waveguide at the bottom arm. We will make use of this concept in section 3.4.4.

### 3.3.6.2 Simulation of the in- and out-of-plane motion

We design mechanical resonators with fundamental frequencies in the radio frequency (RF) range, specifically in the range from 1 MHz to 120 MHz for which comparable devices from standard materials such as silicon and silicon nitride exist in the literature[23]. For this purpose we simulate the displacement fields and eigenfrequencies for the motion of H-resonators of various beam widths and lengths using 3D FEM in COMSOL Multiphysics. The boundary conditions are given by the H-resonator arms, which continue into a 300 nm thick diamond film. (The diamond film contains structures of 50% relative etch depth, etched into an initially 600 nm thick diamond layer.) Fixed boundary conditions are assumed for the bottom area of the diamond layer. Material properties for the diamond are assumed as indicated in Table 1. Figure 25 shows the displacements of an H-resonator ($d = 600$ nm, $w_H = 600$ nm, $L = 40$ μm) for its fundamental out-of-plane mode at 4.0 MHz and in-plane mode at 4.9 MHz.

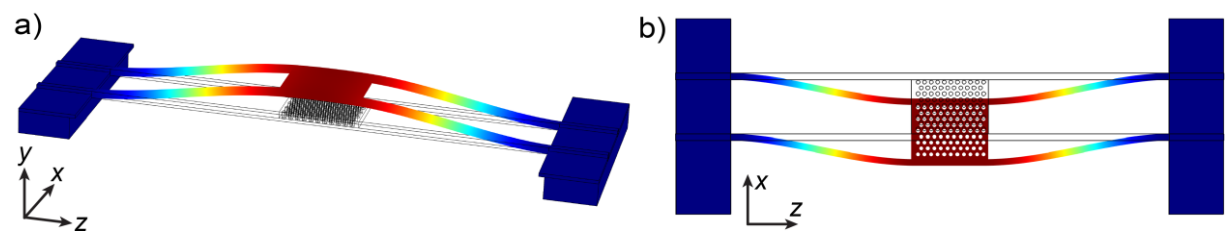

**Figure 25 - Mechanical modes of an H-resonator:** FEM simulation for an H-resonator with thickness $d = 600$ nm, arm width $w_H = 600$ nm and length $L = 40$ μm. The shown displacements are exaggerated to improve the clarity of the presentation. a) Fundamental out-of-plane mode at 4.0 MHz. b) Fundamental in-plane mode at 4.9 MHz.

We are mainly interested in the fundamental in-plane mode, which will lead to the largest phase shift in the adjacent waveguide. Figure 26 shows the resonance frequency of the fundamental in-plane mode for H-resonators of various beam widths $w_H$ (400 nm, 600 nm, 800 nm, 1000 nm) and lengths $L$ (30 µm, 35 µm, 40 µm, 45 µm). The simulated resonance frequencies are proportional to $w_H$ (Figure 26 a) and proportional to $1/L^2$, as shown in Figure 26 b). As for the simplified model of doubly clamped beams, as treated within the Euler-Bernoulli theory (see equation (3.34)), the resonance frequencies scale proportional to $w/L^2$.

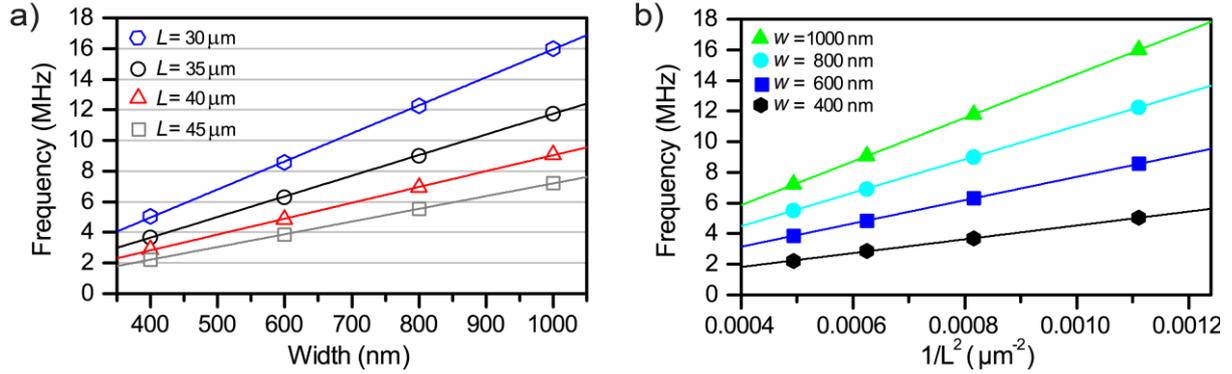

**Figure 26 - Simulated resonance frequencies:** a) Simulated fundamental in-plane resonance frequencies in dependence of width $w$, for different lengths $L$. b) Simulated resonance frequencies as a function of $1/L^2$ for different widths. As for doubly clamped beams treated within the Euler-Bernoulli theory the resonance frequencies scale proportional to $w/L^2$.

We calculate the effective modal masses from the simulated displacement fields for the fundamental in-plane mode of the various H-resonator geometries according to equation (3.28). For example for an H-resonator of length L = 40 µm and width $w$ = 800 nm the physical mass amounts to $m_0 =$ 189 pg, while the effective modal mass amounts to $m_{eff}$ = 118 pg, hence about 63% of the physical mass contribute to the energy of the oscillation.

### 3.3.7 Design of the integrated optomechanical circuits

Measuring small changes in the intensity of light with high precision is possible using commercial photodetectors. In order to detect mechanical motion it is therefore useful to make the transmission of a photonic device dependent on the position of the mechanical element. For this purpose we incorporate evanescently coupled H-resonators in a phase-sensitive PIC. Figure 27 a) shows a SEM micrograph of the integrated optomechanical circuit. Two focusing grating couplers are used for coupling of light between the PIC and off-chip light source and photodetector. The actual PIC consists of a Mach-Zehnder interferometer, as introduced in section 2.1.4, with two interferometer arms which differ in length by $\Delta l = 100$ µm. The MZI enables to translate phase changes within one interferometer arm into intensity changes. By incorporating one H-resonator in each interferometer arm we can study the dynamics of two independent mechanical resonators using one PIC. As explained in section 3.3.4, within the interaction length $l_{int}$ the effective refractive index $n_{eff}$ of the waveguide depends on the displacement $u$ of the mechanical element. Hence a mechanical motion leads to an accumulated phase shift, given by equation (3.34), which in turn translates into an intensity change via the MZI.

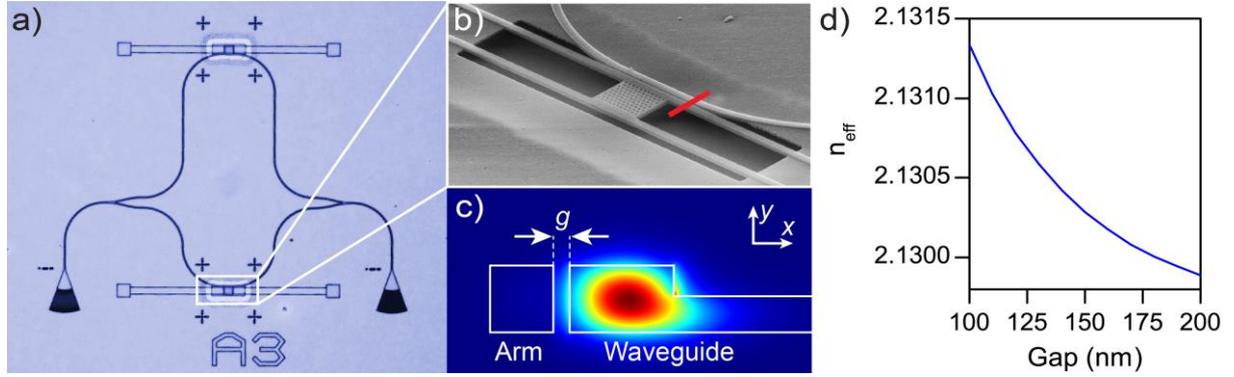

**Figure 27 - Integrated optomechanical circuit**: a) Optical micrograph of the photonic integrated circuit, incorporating H-resonators. b) Close-up SEM image of a freestanding H-resonator and the adjacent waveguide. c) Electric field distribution of the TE-like mode of the waveguide, separated from the H-resonator arm by $g_0 = 150$ nm. d) Simulated dependence of $n_{\text{eff}}$ on the gap size $g$ for an H-resonator of width $w_{\text{H}} = 600$ nm and a waveguide of width $w_{\text{WG}} = 1$ µm.

Figure 27 b) shows a close-up SEM image of the H-resonator and the adjacent waveguide, with a gap of $g_0 = 150$ nm at rest positions. Displacements of the mechanical resonator lead to a change in gap size, resulting in a change of the electric field distribution of the waveguide, which is shown in Figure 27 c). The simulation result for the dependence of $n_{\text{eff}}$ on the gap size $g = g_0 + u$ is shown in Figure 27 d). If the H-resonator moves away from the waveguide, $g$ increases, leading to a decrease in the effective refractive index. An exponential fit yields a relation of

$$n_{\text{eff}}(u) = 2.12964 + 6.43 \cdot 10^{-4} \cdot e^{-\frac{u}{52.24\,\text{nm}}}\,. \tag{3.45}$$

For our geometry we can therefore for small displacements ($u \ll 52.24$ nm) consider only the linear term of the Taylor expansion, such that a change in $n_{\text{eff}}$ is proportional to a displacement $u$. The dependence of the transmission of a lossless Mach-Zehnder interferometer $T$ on the phase difference $\phi$ between the two interferometer arms, according to equation (2.7), is given by

$$T(\phi) = \cos^2\left(\frac{\phi}{2}\right)\,. \tag{3.46}$$

We consider a MZI with a geometric path difference $\Delta l$ between both arms and a homogeneous effective refractive index $n_{\text{eff}}$ along both interferometer arms. The phase difference $\phi_o$ which follows from the path difference $\Delta l$ is given by

$$\phi_o(\lambda_0) = \frac{2\pi}{\lambda_0}(n_{\text{eff}}(\lambda_0) \cdot \Delta l)\,. \tag{3.47}$$

A displacement $u(z)$ along the interaction length $l_{\text{int}}$ leads to an additional phase difference $\phi_n(\lambda_0, u(z))$, given by equation (3.33). Following equation (3.34), for small displacements in the $x$-direction, this phase can be approximated, using the average displacement $\bar{u}$, as

$$\phi_n(\lambda_0, \bar{u}) = \frac{2\pi}{\lambda_0} \cdot \Delta n_{\text{eff}}(\lambda_0, \bar{u}) \cdot l_{\text{int}}\,, \tag{3.48}$$

with an average change in effective refractive index along the interaction length given by

$$\Delta n_{\text{eff}}(\lambda_0, \bar{u}) \equiv \left.\frac{\partial n_{\text{eff}}(\lambda_0, u)}{\partial u}\right|_{u=0} \cdot \bar{u}\,. \tag{3.49}$$

Note that the phase difference $\phi_n$ resulting from the optomechanical interaction does not depend on the length difference $\Delta l$ and hence not on the interferometer design, but rather on the interaction length $l_{\text{int}}$ and how strongly $n_{\text{eff}}$ depends on the displacement $u$. The total phase difference for light with vacuum wavelength $\lambda_0$ can hence be expressed as

$$T(\lambda_0, \overline{u}) = \cos^2\left(\frac{1}{2}[\phi_o(\lambda_0) + \phi_n(\lambda_0, \overline{u})]\right) = \cos^2\left(\frac{1}{2}[\phi_o(\lambda_0) + \frac{2\pi}{\lambda_0} \cdot \Delta n_{\text{eff}}(\overline{u}) \cdot l_{\text{int}}]\right). \quad (3.50)$$

In order to be able to measure small displacements, it is crucial to maximize the translation from displacement $\overline{u}$ into a relative change in transmitted intensity $\frac{\Delta T(\lambda_0, \overline{u})}{T(\lambda_0, \overline{u}=0)} \equiv \frac{T(\lambda_0, \overline{u}) - T(\lambda_0, \overline{u}=0)}{T(\lambda_0, \overline{u}=0)}$. Small displacements $\overline{u}$ correspond to small phase differences ($\phi_n(\overline{u}) \ll 2\pi$) and we can therefore Taylor expand equation (3.50) as

$$T(\overline{u}) = T(\overline{u} = 0) + \left.\frac{\partial T(\overline{u})}{\partial \overline{u}}\right|_{\overline{u}=0} \cdot \overline{u} + O(\overline{u}^2) \quad (3.51)$$

with $T(\overline{u} = 0) = \cos^2\left(\frac{1}{2}\phi_o(\lambda_0)\right)$ and

$$\left.\frac{\partial T(\overline{u})}{\partial \overline{u}}\right|_{\overline{u}=0} = 2\cos\left(\frac{\phi_0}{2}\right) \cdot \left(-\sin\left(\frac{\phi_0}{2}\right)\right) \cdot \left(\pi \cdot \frac{l_{\text{int}}}{\lambda_0}\right) \cdot \left.\frac{\partial n_{eff}(\lambda_0, u)}{\partial u}\right|_{u=0}. \quad (3.52)$$

The change in transmission due to a mechanical displacement can therefore be maximized by choosing a wavelength $\lambda_m$ such that the absolute value of $\left.\frac{\partial T(\overline{u})}{\partial \overline{u}}\right|_{\overline{u}=0}$ is maximal, hence $\phi_o(\lambda_m) = \phi_{\max} \equiv (2m + 1) \cdot \frac{\pi}{2}$ for $m \in \mathbb{N}$. Note that $\phi_o(\lambda_m) = \phi_{\max}$ coincides with the wavelengths of maximum slope in the transmission spectrum $T_{\text{MZI}}(\lambda_0)$, as defined in equation (2.9).[9] This enables in the experiment to maximize the translation from mechanical displacement to intensity changes by choosing the wavelength using the transmission spectrum.

We neglect higher order terms $O(\overline{u}^2)$ and can hence express the relative change in transmitted intensity[10] for a wavelength $\lambda_m$ as

$$\left|\frac{\Delta T(\lambda_m, \overline{u})}{T(\lambda_m, \overline{u}=0)}\right| = \frac{2\pi}{\lambda_0} \cdot \left(l_{\text{int}} \cdot \left.\frac{\partial n_{\text{eff}}(\lambda_0, u)}{\partial u}\right|_{u=0} \cdot \overline{u}\right) = \phi_n(\overline{u}). \quad (3.53)$$

The relative change in transmitted intensity through the MZI can therefore be considered to be proportional to the phase $\phi_n(\overline{u})$ and proportional to the displacement of the oscillator. This linear transformation enables to study the behavior of the mechanical oscillator in frequency or time domain by studying the optical transmission.

According to the simulation results, presented in Figure 27 d) for an H-resonator with $w_H = 600$ nm, the slope at rest position amounts to $\left.\frac{\partial n_{\text{eff}}(\lambda_0, x)}{\partial u}\right|_{u=0} = -1.215 \cdot 10^{-5} \frac{1}{\text{nm}}$. With an interaction length of $l_{\text{int}} = 12\ \mu\text{m}$ we can estimate the phase change for light with $\lambda_0 = 1550$ nm for an assumed average displacement of $\overline{\Delta u} = 100$ nm, according to equation (3.48), as $\Delta\phi(\lambda_0) \approx -\frac{1}{100} \cdot 2\pi$. Displacements in the experiment are much smaller and hence the Taylor expansion **Error! Reference**

---

[9] Note that the slope $\frac{\partial T_{MZI}(\lambda_0)}{\partial \lambda_0}$ depends on the path difference $\Delta l$ (see equation (2.8)), while the relative intensity change $\frac{\Delta T(\lambda_m, \overline{u})}{T(\lambda_m, \overline{u}=0)}$ does not, as can be seen in equation (3.23). It rather depends on the interaction length $l_{\text{int}}$. Hence the measurement sensitivity for displacements is independent from the path difference, as long as a wavelength $\lambda_m$ within the range of the laser and within the bandwidth of the grating coupler can be chosen, such that $\phi_o(\lambda_m) = \phi_{\max}$.

[10] Note that the sign of the intensity change depends on the choice of the wavelength $\lambda_m$, corresponding to a negative or positive slope in the transmission spectrum $T_{\text{MZI}}(\lambda_0)$.

**source not found.** is justified. This ensures the linearity of all involved elements, such that intensity changes are proportional to the mechanical displacement. If on the other hand larger phase shifts are intended, this is possible by choosing a smaller gap size $g_0$ and by increasing the arm length $L$ and the interaction length $l_{int}$, as well as by cascading multiple H-resonators. Phase shifts as large as $\Delta\phi \approx \pi$ have been shown for long H-resonators in silicon nitride[167] and phase shifts larger than $2\pi$ have been shown recently for an alternative microbridge design in silicon nitride[237]. Tunable phase shifts from 0 to $2\pi$ are hence generally possible using optomechanical devices in PICs and here we show the first proof-of-principle devices in diamond PICs.

As explained in section 3.3.5, the dependence of $n_{eff}$ on the gap size implies, that light propagating in the waveguide leads to an optical gradient force, which attracts the H-resonator. We experimentally show in section 3.4.3 the use of these optical gradient forces to actively drive the mechanical motion.

## 3.4 Diamond integrated optomechanics: experiments

### 3.4.1 Fabrication of integrated optomechanical circuits[11]

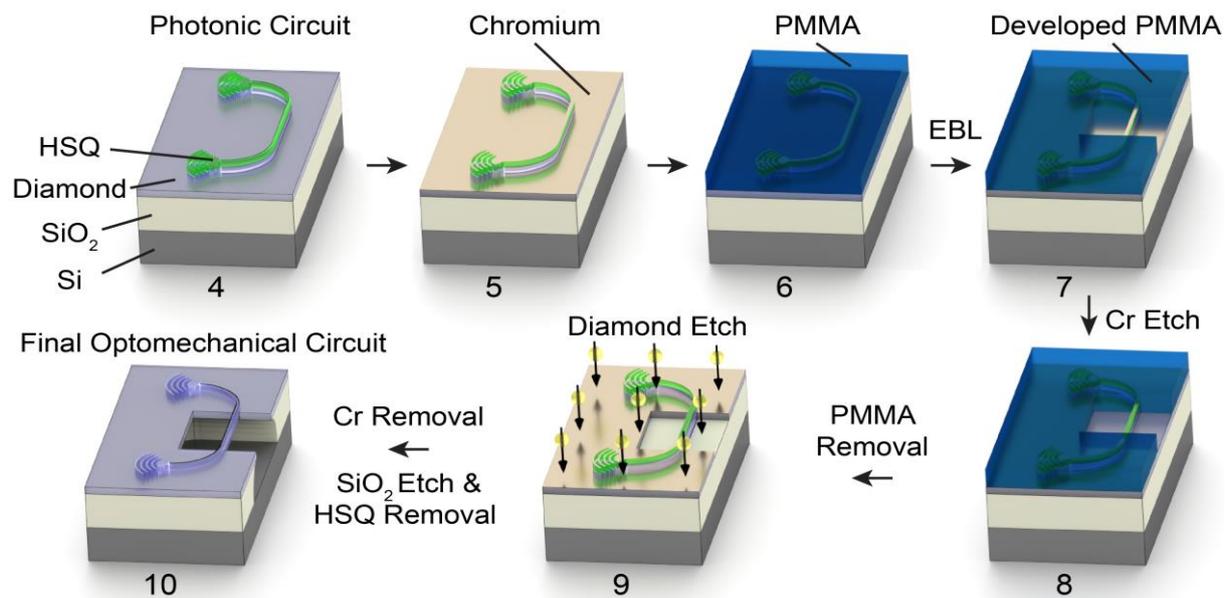

**Figure 28 - Fabrication of optomechanical circuits:** Chromium (Cr) is deposited and PMMA is spin coated on a photonic chip, containing half-etched photonic circuits, covered by HSQ resist. After electron beam lithography (EBL) and resist development, the rectangular pattern is transferred into the chromium via etching. The remaining chromium acts as hard mask for fully etching of diamond within the rectangular area. Finally the chromium and HSQ are removed and the silicon oxide within the opening window is partially removed via wet etching, resulting in the final optomechanical circuits.

The fabrication process of optomechanical components starts with a chip, containing half-etched diamond PICs, covered by HSQ resist. These PICs were fabricated using the four fabrication steps explained in section 2.4.2 (steps 1 to 4 are illustrated in Figure 16). Figure 28 a) illustrates the consecutive fabrication steps: A 25 nm thick chromium layer is deposited via electron beam evaporation onto the entire photonic chip (step 5). Then 800 nm of poly(methyl methacrylate) (PMMA) positive tone resist are spin coated on the chip (step 6). Using EBL, with an area dose of 500 µC/cm², a rectangular opening window is defined in PMMA for each PIC. The resist is developed for 15 min in a 1 : 3 mixture of methyl isobutyl ketone : isopropanol (step 7). The rectangular opening window layout is transferred from PMMA into chromium via wet etching (Sigma Aldrich chromium etchant 651826) (step 8). A reactive ion etching step in oxygen/argon plasma (plasma details are described in section 2.4.2) removes the PMMA while chromium and HSQ on the PIC act as hard mask for diamond. Hence diamond which is not covered by either metal or HSQ is fully etched down until the underlying silicon oxide is exposed (step 9). After removing the chromium hard mask via wet etching, the silicon oxide within the opening window is partially removed in a wet etching step using hydrofluoric acid (HF). This results in the final optomechanical circuits. Due to the underetching, each diamond PIC now contains a freestanding structure, which is clamped within the half-etched diamond layer at the edges of the opening window. Details concerning all lithography and etching steps are provided in appendix A2.

---

[11]The fabrication procedures for diamond optomechanical circuits presented within this section were developed by the author of this thesis within his master thesis (Diplomarbeit). The fabrication of most of the devices presented within chapter 3.4 was performed by Sandeep Ummethala within his master thesis, which was supervised by the author of this thesis.

## 3.4.2 Thermomechanical displacement measurement

Without applying an external driving force, the H-resonators are in thermal equilibrium with their environment and hence driven by a white noise force associated with the thermalization, as explained in section 3.3.3. We measure the thermal motion of the H-resonators using a vacuum measurement setup, depicted in Figure 29. The photonic chip containing the optomechanical circuits is placed on a 4-axis stage below an optical fiber array inside a vacuum chamber and one PIC is characterized at a time. The measurements are carried out at a pressure $p < 10^{-5}$ mbar, to ensure that air damping is negligible and intrinsic damping of the H-resonators is dominating their behavior. The pressure dependence of the mechanical quality factor will be presented in section 3.4.4.2.

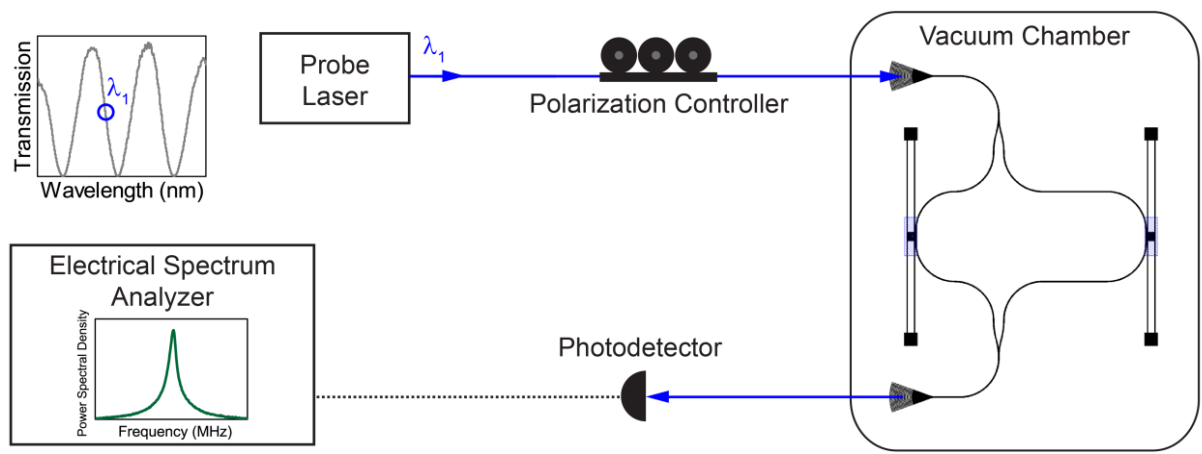

**Figure 29 - Measurement of thermomechanical motion:** Schematic of the measurement setup used for the measurement of the thermal motion of the diamond H-resonators via an on-chip interferometer. The transmitted optical power is analyzed with an electrical spectrum analyzer and reveals the spectral density of the thermal motion.

The measurement setup is an adaptation from the general transmission measurement setup explained in section 2.4.3. Infrared light from a tunable laser is coupled via optical fibers, the fiber array, and grating couplers into the input of the MZI. The relative position of fiber array and grating couplers and the polarization are adjusted for maximum transmission (corresponding to the TE-like waveguide mode). After transmission through the MZI, light is detected by a low-noise photodetector. The transmission spectrum of the PIC is acquired and the laser wavelength $\lambda_1$ is chosen such that the slope in the transmission spectrum is maximal, corresponding to a maximum concerning translation of mechanical motion into a change in transmission, as explained in section 3.3.7. We note that each PIC contains two H-resonators, one in each interferometer arm, which we can both characterize in one measurement. The output voltage of the photodetector is proportional to the transmitted optical power. For small displacements and small phase changes in the waveguide the transmitted power is proportional to the displacement. Using a swept-tuned, heterodyne[12] electrical spectrum analyzer we analyze the spectral density of the detector's output voltage, from which we infer the amplitude spectral density of the mechanical motion. The spectral analysis of the detector voltage reveals the resonance frequencies of the mechanical resonators.

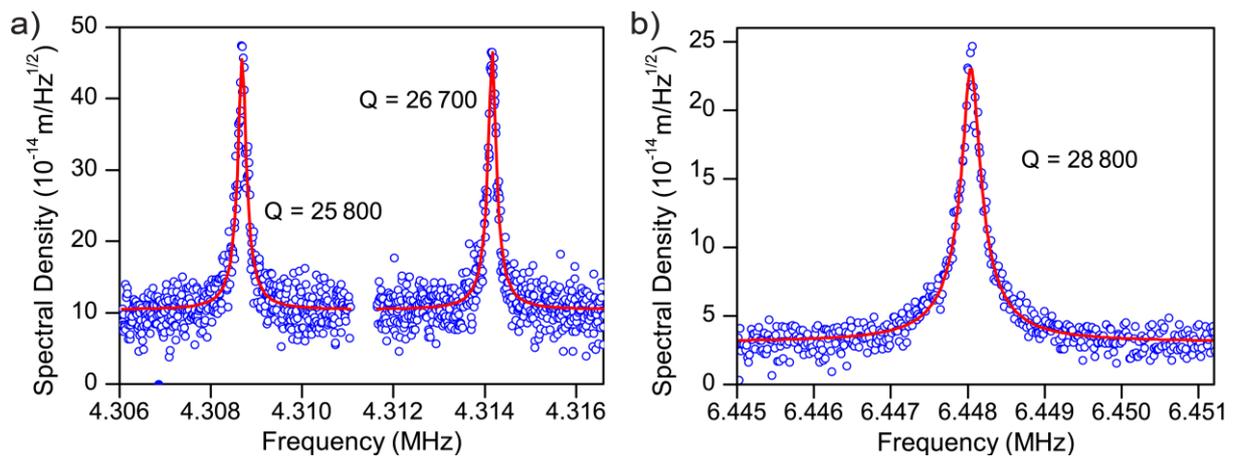

**Figure 30 - Thermal motion:** a) Calibrated spectral density of the thermal motion of an H-resonator ($L = 40\ \mu m$, $w_H = 600\ nm$). The two resonance peaks around 4.31 MHz correspond to fundamental in-plane motion of the two mechanical resonators in separate arms of the MZI. The mechanical quality factor Q is extracted from the Lorentzian fit to the power spectral density. b) The calibrated spectral density of the thermal motion of the H-resonator with the highest quality factor of $Q = 28\ 800$ at 6.448 MHz ($L = 40\ \mu m$, $w_H = 800\ nm$).

Figure 30 a) shows the measured spectral density in a frequency range around 4.3 MHz, measured for a PIC containing two H-resonators with $L = 40\ \mu m$ and $w_H = 600\ nm$. The two resonances correspond to thermal motion at the fundamental in-plane resonance frequency of the two H-resonators in the two different interferometer arms. Small variations in fabrication, such as concerning lithography and etching, lead to deviations in the device dimensions. This in turn leads to slightly different mechanical resonance frequencies for H-resonators, which by design have the same geometry. As explained in section 3.3.3, the power spectral density $S_x(\omega)$ is associated with the kinetic energy of the fundamental mode and hence follows a Lorentzian line shape. We therefore extract the mechanical quality factors by fitting Lorentzian curves to the square of the voltage spectral density. This yields values for the FWHM of the resonances of $\delta f = 167$ Hz and $\delta f = 162$ Hz, corresponding to mechanical quality factors of $Q = 25\,800$ and $Q = 26\,700$, respectively. Using equation (3.28) the effective mass is calculated, using the mode shape from the FEM simulations, as $m_{\text{eff}} = 102$ pg. We calibrate the spectral density at resonance according to equation (3.31), using the theoretical thermal energy of the resonator and the experimental value for the quality factor. The spectral density at resonance therefore amounts to $\sqrt{S_x} = 4.66 \cdot 10^{-13}$ m/$\sqrt{\text{Hz}}$, which calibrates the $y$-axis in Figure 30. The displacement sensitivity corresponds to the noise floor, above which a measurement signal can be detected. The measurement for this specific device hence shows a displacement sensitivity of 105 fm/$\sqrt{\text{Hz}}$. We measure the thermal motion for all devices on the photonic chip.

---

[12] Heterodyne refers to the down-conversion of the frequency via mixing with an internal source. Each measurement point refers to the power within the width of the passband filter.

Figure 30 b) shows the measured thermal motion for the H-resonator with the largest quality factor $Q = 28\,800$. With a length $L = 40$ µm and a width $w_{\text{H}} = 800$ nm the resonance frequency is $f = 6.448$ MHz and the FWHM amounts to $\delta f = 224$ Hz. Calibration via the thermal energy yields a measurement sensitivity of 32 fm/$\sqrt{\text{Hz}}$ at an optical power in the waveguide $P_{\text{opt}} \approx 3$ mW. Figure 31 a) shows resonance frequencies of the fundamental in-plane mode of H-resonators in dependence of their arm width $w_{\text{H}}$ for different lengths $L$, which we extract from the measurement of the thermal motion. The frequencies range from about 1 MHz to 15 MHz and scale linearly with the width $w_{\text{H}}$ and proportional to $\frac{1}{L^2}$ with the length $L$ of the H-resonator, as expected from the simulations. The measured resonance frequencies are about 15% smaller than the simulated frequencies. We attribute this to fabrication tolerances and stress in the thin film, which can lead to shifts of the resonance frequency[238]. Furthermore the Young's modulus of polycrystalline diamond layers can be as large as for single crystalline diamond (as used in the simulations) but can also be smaller, depending on deposition parameters such as power density and methane concentration[239].

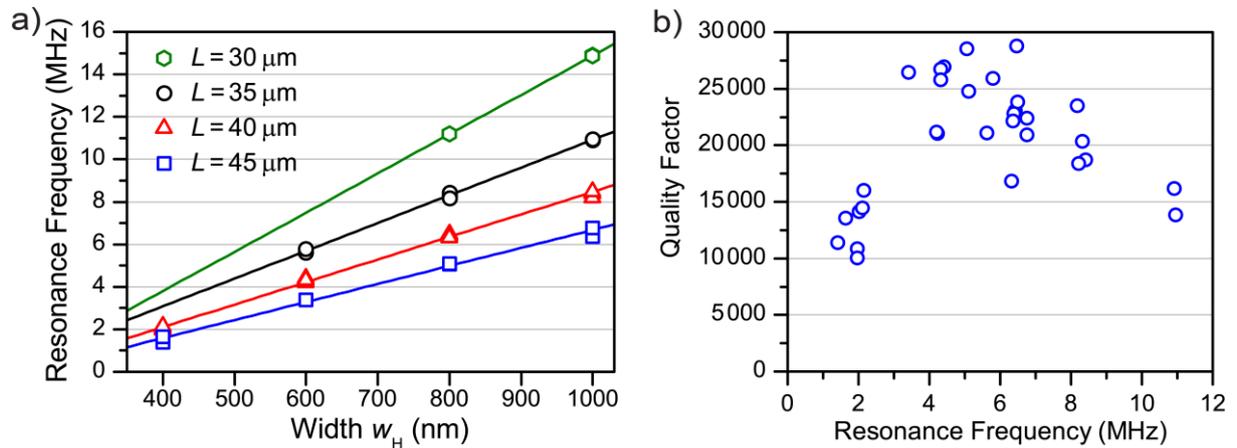

**Figure 31 - Resonance frequencies and mechanical quality factors:** a) Measured resonance frequencies of the fundamental in-plane mode of H-resonators in dependence of their arm width $w_{\text{H}}$ for different lengths $L$. b) Quality factors for the fundamental in-plane mode of H-resonators in dependence of the resonance frequency. All $Q$-factors are above 10 000, with a maximum value of $Q = 28\,800$ at 6.45 MHz.

Figure 31 b) shows the quality factors for the fundamental in-plane mode of all measured H-resonators, in dependence of the resonance frequency. All $Q$-factors are above 10 000, with a maximum value of $Q = 28\,800$ at 6.45 MHz, corresponding to the measurement curve shown in Figure 30 b). The finite value for the mechanical quality factor is a manifestation of the dissipative damping acting on the beam. The dissipation in micromechanical resonators can be attributed to aerodynamic losses, clamping losses, surface losses, and internal friction, the latter typically incorporates thermo-elastic damping, phonon-phonon interactions, and losses due to defects[240]. A decrease of $Q$ with higher frequencies has been observed for many micromechanical resonators, which can be attributed to the scaling of different damping mechanisms[241]. Therefore often the product $Q \cdot f$ is given as a figure of merit for micromechanical resonators. A high value of $Q \cdot f$ is for example critical for low phase noise oscillators[242] and cavity optomechanical mass spectroscopy[243]. An overview over demonstrations of micromechanical resonators in PCD and SCD can be found in a recent review article[80]. A comparison shows that best value of $Q \cdot f = 186$ GHz for our H-resonators is among the highest values for diamond micromechanical resonators at room temperature. Only recently a record value of $Q \cdot f \approx 1.9 \cdot 10^4$ GHz, which is sufficient for room temperature single-phonon coherence, was achieved for SCD micro-disk resonators[18]. For quantum optical circuits our resonators would be employed in SCD, and hence we do not further investigate which damping mechanism is limiting the quality factors in our H-resonators from PCD. In chapter 5 we will demonstrate the transfer of our diamond PICs from PCD to SCD.

In this section we measured the thermal motion and determined resonance frequencies and quality factors of the H-resonators. In the following sections we show the controlled actuation via optical gradient forces and electrostatic forces.

### 3.4.3 Driven motion via optical gradient forces

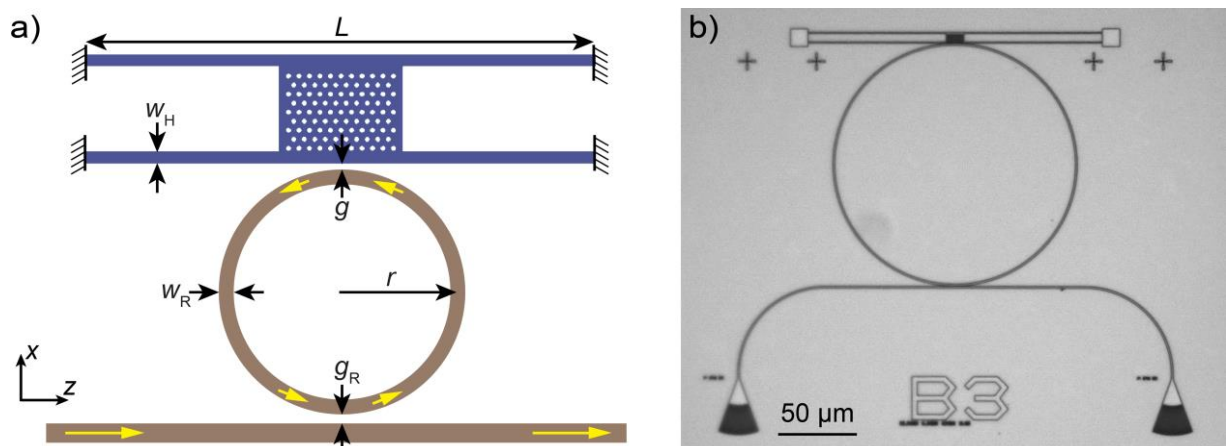

**Figure 32 - Ring resonator with H-resonator:** a) Schematic of the optomechanical system, consisting of an optical ring resonator and a mechanical H-resonator. The schematic is not to scale. b) Microscope image of a fabricated diamond integrated optomechanical circuit.

Optical gradient forces, as introduced in section 3.3.5, can be used to drive the motion of micromechanical oscillators. This allows all-optical transduction of motion on a photonic chip, without the need for on-chip electronics for the driving force. Such systems are of interest for applications in harsh environments, such as in gases or liquids[244–246] which are incompatible with electronics. Furthermore optical forces in diamond PICs are of interest for cavity optomechanics.[161] We therefore combine an optical cavity and a mechanical resonator and demonstrate actuation via optical gradient forces. Figure 32 a) shows a schematic and Figure 32 b) shows a microscope image of a fabricated optomechanical circuit. The circuit consists of a ring resonator which is evanescently coupled to a bus waveguide via a gap $g_R$ and to the H-resonator via a gap $g$. At rest position of the mechanical motion the gap amounts to $g_o = 150$ nm. Compared to the PIC design explained in section 3.3.7, the ring resonator replaces the MZI as the element which translates mechanical displacement via a phase change into a change in transmitted intensity. The ring geometry is defined by its radius $r$ and width $w_R = 1$ µm, which is equal to the waveguide width. The H-resonator geometry is equal to the one described in the previous section with arm width $w_H$ and arm length $L$. The ring resonator geometry ($r = 70$ µm) and its coupling to the waveguide ($g_R = 240$ nm) are chosen such that the coupling is close to critical coupling, while a moderate quality factor is preserved.

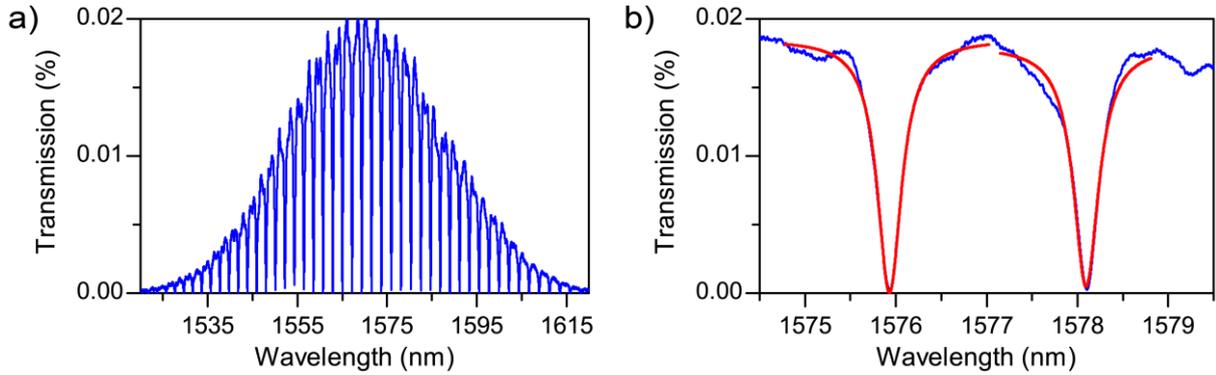

**Figure 33 - Photonic circuit transmission:** a) Transmission spectrum of a ring resonator ($r = 70$ µm, $g_R = 240$ nm, $w_R = 1$ µm), which is coupled to a freestanding H-resonator ($g = 150$ nm). The ring resonances have a free spectral range of 2.2 nm and the extinction ratio exceeds 15 dB. b) Transmission spectrum in a small wavelength range around two ring resonances. Lorentzian fits to the resonances (red curves) reveals an average optical quality factor of $Q = 4800$.

Figure 33 a) shows the transmission spectrum of one device. The envelope is given by the focusing grating couplers and the ring resonances are visible as dips with a free spectral range of 2.2 nm. The extinction ratio exceeds 15 dB over the full spectral range, showing close-to-critical coupling of waveguide and ring. Figure 33 b) shows the transmission in the spectral range around two ring resonances. From Lorentzian fits to the resonances we find an average optical quality factor of $Q = 4800$. The quality factor is consistent with the values for critically coupled optical ring resonators without mechanical H-resonators (see section 2.4.4). The evanescent coupling to the H-resonator hence does not introduce noticeable losses.

We fabricate a photonic chip which contains about 1000 of the described optomechanical circuits, using the fabrication procedure explained in section 3.4.1. We characterize the devices in the vacuum chamber setup, as explained in the previous section. We focus on H-resonators with extended lengths up to $L = 70$ µm, which have smaller spring constants and hence driving forces translate into larger amplitudes. We first measure the thermal motion and its transduction into in-tensity changes, via ring resonances. For an H-resonator with L = 70 μm and $w_H$ = 600 nm the thermal motion reveals a fundamental resonance frequency of $f$ = 1.126 MHz. A Lorentzian fit yields a FWHM of $\delta f$ = 52.7 Hz and a corresponding quality factor of $Q$ = 21350 ± 420. Accurately measuring high quality factor at comparably low frequencies is challenging in the frequency domain, because the resonance linewidth approaches the smallest filter width of the spectrum analyzer (30 Hz) and because thermal drift of the resonance frequency broadens the measured resonance shape. Temperature changes within the measurement time as small as 20 mK can lead to noticeable changes in resonance frequency (section 3.4.4.3 presents measurements concerning the temperature dependence of the resonance frequency). To avoid problems concerning thermal drifts we confirm the quality factors in the time domain via ring down of driven motion (presented in section 3.4.3.2).

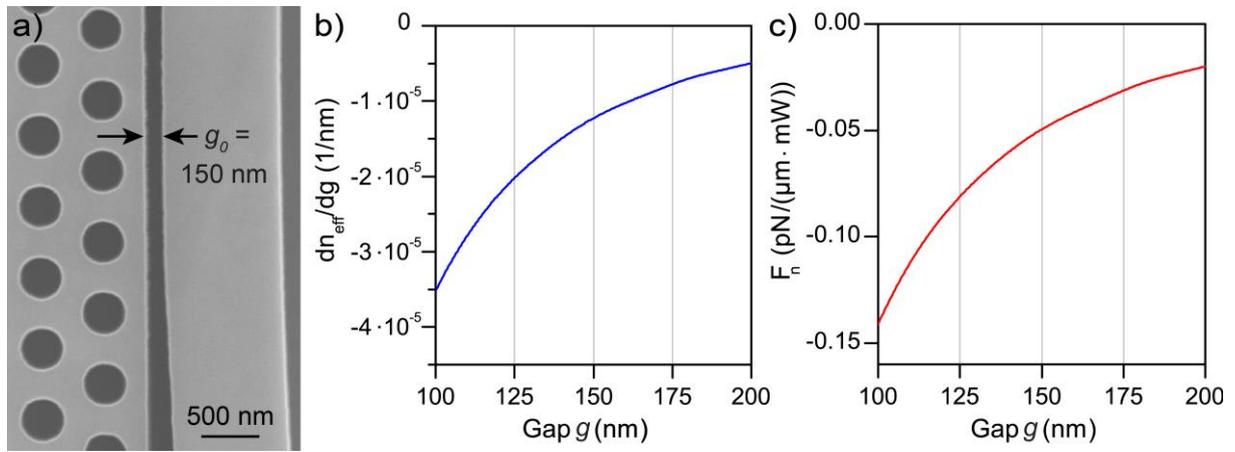

**Figure 34 - Optical gradient force between ring resonator and H-resonators:** a) SEM image of the interaction region of ring and H-resonator with a gap in rest position of $g_o$ = 150 nm. b) Simulated dependence of the derivative of $n_{eff}$ with respect to the gap size $g$ for an H-resonator with $w_H$ = 600 nm for gap sizes values around $g_o$. c) Dependence of the normalized optical gradient force on the gap size $g$, in piconewton per μm interaction length and per mW optical power in the waveguide.

Light circulating in the ring applies an attractive force on the H-resonator in close proximity. Figure 34 a) shows a close-up SEM image of the interaction region of ring and H-resonator. The gap size at the point of smallest distance at rest position of the mechanical motion is $g_o$ = 150 nm. According to equation (3.44), the optical force per interaction length $l$ and optical power $P$ can be calculated as

$$F^n_{opt}(g) = \frac{F_{opt}(g)}{P \cdot l} = \left(\frac{n_g(g)}{n_{eff}(g)} \cdot \frac{1}{c}\right) \cdot \frac{\partial n_{eff}(g)}{\partial g},$$

where $g$ is the gap size between the H-resonator and the ring resonator. Figure 34 b) shows the dependence of $\frac{\partial n_{eff}}{\partial g}$ on the gap size obtained by FEM simulation (see Figure 27 d)) for $n_{eff}(g)$. The resulting normalized optical force is shown in Figure 34 c). The force is attractive and its magnitude exponentially decreases with increasing gap size as

$$F_n(g) = -0.050 \cdot e^{-\frac{g-g_0}{54.3\,nm}} \frac{pN}{\mu m \cdot mW} \ . \tag{3.54}$$

Due to the curvature of the ring, the size of the gap between ring and H-resonator (at rest position) depends on the $z$-coordinate as $g(z) = g_0 + r - \sqrt{r^2 - z^2}$. We calculate the total force at the interface between full ring and H-resonator as

$$F_{\text{opt}}(g_0) = \int_{z=-r}^{z=r} F^n{}_{\text{opt}}(g(z)) \, dz \equiv F^n{}_{\text{opt}}(g_o) \cdot L_{\text{eff}} \ . \tag{3.55}$$

It is useful to define the effective interaction length $L_{\text{eff}}$, as the length for which a structure of parallel waveguide and H-resonator (with constant gap $g_o$) would yield the same total force. This yields an optical force per optical power of

$$F_{\text{opt}}(g_0) = -0.24 \text{ pN/mW} \tag{3.56}$$

and an effective interaction length $L_{\text{eff}} = 4.9 \text{ µm}$. Next, we actively drive the mechanical motion via optical gradient forces and verify this via a pump-probe measurement.

### 3.4.3.1 Pump-probe measurement of driven motion via optical gradient forces

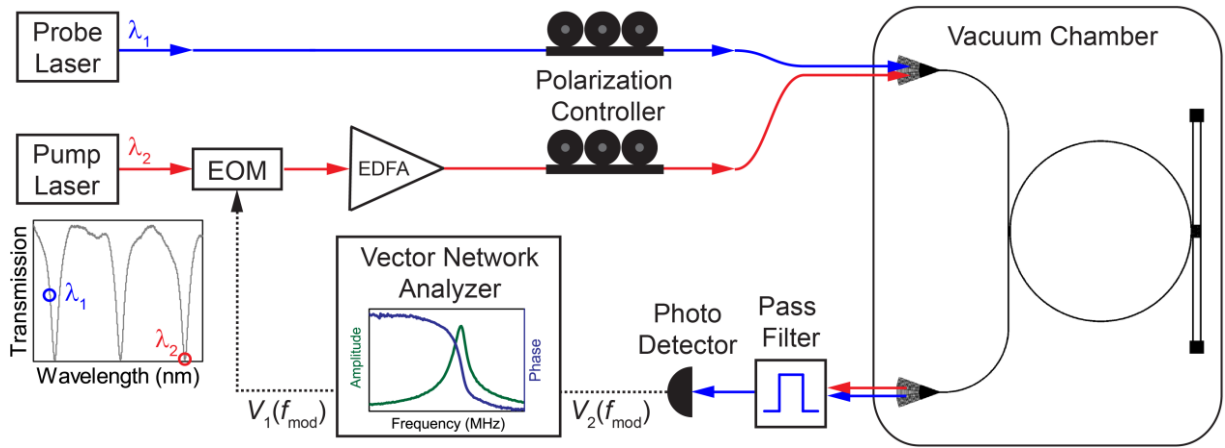

**Figure 35 - Pump-probe measurement:** Schematic of the setup for measuring mechanical motion driven by optical gradient forces. The mechanical resonator is actuated with optical forces by light from a pump laser (red color), which is modulated at the mechanical resonance frequency via an electro-optical modulator (EOM) and amplified via an erbium doped fiber amplifier (EDFA). The resulting mechanical oscillations are detected via light from a probe laser (blue color).

We drive the H-resonator motion with optical forces and measure the response in vacuum ($p < 10^{-5}$ mbar) using the pump-probe measurement setup depicted in Figure 35. The setup consists of components for the interferometric detection, shown in blue color, and the components for driving the motion via optical forces, as depicted in red color. We apply optical forces via a laser, referred to as pump laser, at a wavelength $\lambda_2$. The optical force is proportional to the power of the pump laser which we modulate via an electro-optical modulator (EOM, Lucent 2623NA) at frequency $f_{\text{mod}}$ resulting in a pump power $P_{\text{pump}}$:

$$P_{\text{pump}} = P_{\text{DC}} + P_{\text{RF}} \cdot \sin(2\pi \cdot f_{\text{mod}} \cdot t) \ . \tag{3.57}$$

Besides a static optical force due to $P_{\text{DC}}$, a sinusoidal alternating optical force acts on the mechanical oscillator, as considered in the model for a driven harmonic oscillator (see section 3.3.1). The ratio $P_{\text{RF}}/P_{\text{DC}}$ is referred to as modulation depth. We operate the EOM in its linear regime, where amplitude of the power modulation depends linearly on the amplitude of the sinusoidal voltage supply. An erbium doped fiber amplifier (EDFA, PriTel LNHPFA-33) enables comparably large optical forces with an adjustable output power up to 2.5 W.

We measure the motion of the H-resonator using a CW probe laser at a wavelength $\lambda_1$. The polarization of pump and probe light are independently adjusted for maximum transmission using two fiber polarization controllers and combined using a 50/50 directional coupler before being coupled into the PIC. The power of the pump laser is larger than the power of the probe laser ($\langle P_{\text{pump}} \rangle \approx$ 7.5 mW compared to $P_{\text{probe}} \approx 0.5$ mW, both inside the waveguide) and the probe laser power is unmodulated, which ensures that the dynamic optical force can be attributed to the pump laser. Both laser wavelengths are chosen within the bandwidth of the coupler, as shown on the left side of Figure 35. The probe wavelength $\lambda_1$ is chosen on the slope of a ring resonance for best transduction of motion into intensity changes and the pump wavelength $\lambda_2$ is chosen at the minimum of a ring resonance (and spectrally far enough separated from $\lambda_1$). This ensures the maximum achievable optical power in the ring, as well as 15 dB extinction in transmission, provided by the extinction ratio of the ring resonator. After transmission through the PIC a pass filter (PriTel TFA-1550) is applied which enables transmission of the probe light, while strongly attenuating ($> 40$ dB) the pump light. This ensures that the intensity modulations which are measured at the photodetector stem from mechanical motion and not from transmitted light from the pump laser. Transmitted light is then recorded with a low-noise photodetector (New Focus 2117) and its electrical output signal is spectrally analyzed with a vector network analyzer (VNA, Rohde & Schwarz ZVL6).

The VNA supplies the EOM with the sinusoidal modulation voltage $V_1$ at frequency $f_{\text{mod}}$ and measures the spectral density of the photodetector signal at the same frequency, which is proportional to the motional amplitude. Furthermore the phase relation of driving voltage and voltage from the photodetector is measured. This corresponds to the phase between the optical force and the mechanical motion. Hence both amplitude and phase of the driven motion at frequency $f_{\text{mod}}$ are measured. By sweeping the frequency $f_{\text{mod}}$ across the resonance of the mechanical oscillator, the frequency-dependent response in terms of amplitude and phase can be recorded.

We study the driven motion for the H-resonator ($L = 70$ µm, $w_{\text{H}} = 600$ nm) for which the measurement of its thermal motion was presented previously. We control the amplitude of the optical force via the pump power and the voltage at the EOM. The VNA supplies the EOM with an alternating voltage with an amplitude of 0.2 V, which leads to a modulation depth of 40%. Using the attenuation coefficient (see section 2.4.4) and the condition of critical coupling at resonance, we estimate the average circulating steady state pump power in the ring at closest distance to the H-resonator as $\langle P_{\text{ring}} \rangle \approx 19.4$ mW (see equation (2.17)). The resulting amplitude of the dynamic force which is acting on the H-resonator is estimated as $F_{\text{RF}} = 1.9$ pN. Using the simulated effective mass $m_{\text{eff}} = 162.4$ pg we can calculate the spring constant as $k = m_{\text{eff}} \cdot \omega_0{}^2 = 8.14$ N/m. The amplitude for the oscillator driven at resonance can be estimated (following equation (3.10)) as

$$a = Q \cdot \frac{F}{k} = 6.5 \text{ nm} . \tag{3.58}$$

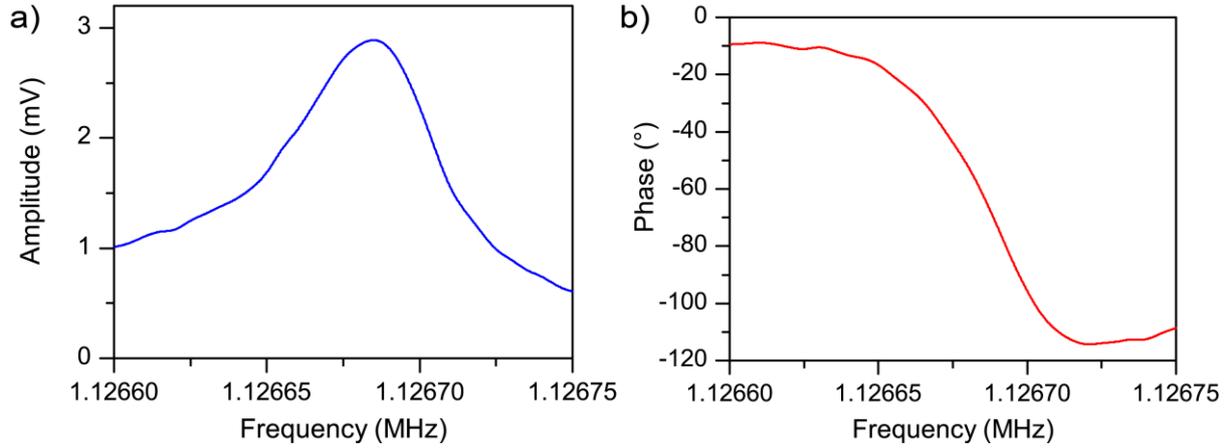

**Figure 36 - Driven motion via optical forces:** a) Voltage signal, which is proportional to the motional amplitude, in dependence of the driving frequency. A Lorentzian fit to the square of the amplitude reveals a mechanical quality factor $Q = 28\,260 \pm 1540$ at a resonance frequency of 1.127 MHz. b) Phase between the driving optical gradient force and the resulting mechanical motion.

We sweep the frequency of the driving force in a frequency range around the fundamental resonance, as determined by the thermal motion. Figure 36 a) shows the recorded voltage signal, which is proportional to the motional amplitude, in dependence of the driving frequency. Figure 36 b) shows the phase between driving force and mechanical motion. The observation of both the frequency dependence of the driven amplitude and the phase shift around the resonance shows the successful coherent drive of the H-resonators' motion via optical gradient forces. A Lorentzian fit yields a resonance frequency of $f = 1.127$ MHz and a FWHM of $\delta f = 39.9 \pm 2.2$ Hz, corresponding to a quality factor of $Q = 28\,260 \pm 1540$. This is slightly larger than the value extracted from the thermomechanical motion and we attribute this to the shorter measurement time, enabled by the better signal-to-noise ratio at the photodetector due to the larger driven amplitude compared to the thermal motion. Next we will confirm the quality factor in the time domain.

### 3.4.3.2 Ring down measurement

The quality factor can also be investigated in the time domain, by measuring the ring down of the oscillation amplitude of the driven oscillator upon turning off the driving force[247] (as explained in section 3.3.1). For micromechanical oscillators with high quality factors at relatively low frequencies, the line width of the resonance, as measured in the frequency domain in the previous section, is in the same range as the available filter widths of the employed measurement equipment. We hence measure the ring down time of the displacement amplitude in order to verify the quality factor extracted from the measurements in the frequency domain. We modify the measurement setup shown in Figure 37 by replacing the VNA with a lock-in amplifier (Zurich Instruments UHFLI). We use a phase-locked loop to ensure that we drive the motion at resonance, such that drifts of the resonance frequency do not influence the measurement results. We choose a set point for the phase between driving force and the displacement, corresponding to the phase at resonance frequency (determined via pump-probe measurement, as presented in the previous section). A proportional-integral-derivative (PID) controller ensures negligible differences between actual phase and set point.

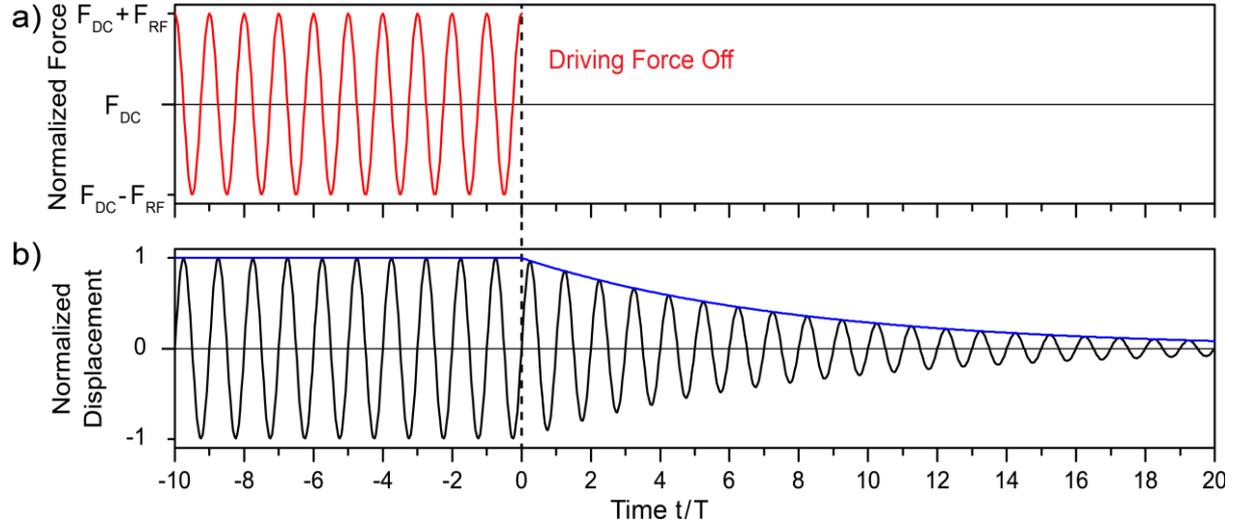

**Figure 37 - Schematic of the ring down measurement:** a) Driving force, applied via the optical gradient force, consisting of the average static force $F_{DC}$ and a dynamic sinusoidal force with amplitude $F_{RF}$. b) Displacement of mechanical oscillator. After the driving force is turned off, the displacement amplitude (shown in blue color) decays exponentially, enabling to determine the quality factor from the decay time.

Figure 37 schematically shows the measurement sequence. We apply a sinusoidal optical force at resonance frequency by applying the corresponding voltage to the EOM. The mechanical displacement follows the sinusoidal drive (besides the expected phase delay). After turning off the driving force, by turning off the alternating voltage to the EOM, the amplitude of the oscillation $a$ decays exponentially, according to equation (3.11), as

$$a(t) = a_{max} \cdot e^{-\frac{t}{\tau}}.$$

The time constant $\tau$ is related to the mechanical quality factor $Q$ as

$$Q = \pi \cdot \tau \cdot f, \tag{3.59}$$

where $f$ is the mechanical resonance frequency.

We perform the described ring down measurement for the same H-resonator for which the driven motion was presented in the previous section. We record the amplitude of the demodulated voltage[13], which is proportional to the oscillation amplitude of the H-resonator. Figure 38 shows the average of five ring down measurements. The phase-locked loop drives the motion at resonance ($f \approx 1.127$ MHz) until the force is turned off by turning off the driving voltage (at time $t = 0$ ms) upon which the amplitude exponentially decays. We extract the time constant $\tau$ from an exponential fit as $\tau = 8.1 \pm 0.04$ ms. According to equation (3.59) this corresponds into a quality factor of $Q = 28490 \pm 150$, in agreement with the estimation via the pump-probe measurement.

---

[13] We verify and ensure that the measured decay time is not limited by the bandwidth of the filter ($> 1$ kHz) used in the demodulation.

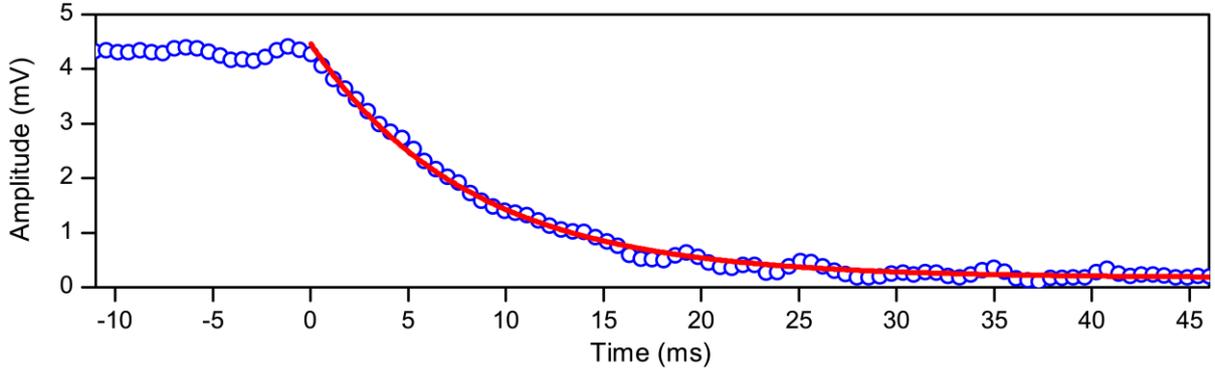

**Figure 38 - Ring down measurement:** Amplitude of the demodulated voltage, corresponding to the oscillation amplitude of the H-resonator, driven via optical gradient forces. After turning off the driving force at time $t = 0$ ms the amplitude exponentially decays. An exponential fit (red) to the data (blue circles) yields a time constant of $\tau = 8.1 \pm 0.04$ ms, corresponding to a quality factor of $Q = 28\,490 \pm 150$.

In summary, we have shown that the on-chip translation from mechanical motion into intensity modulation enables the measurement of the thermal motion with sensitivities down to $32\ \mathrm{fm}/\sqrt{\mathrm{Hz}}$. We coupled an optical cavity and a mechanical resonator within one diamond PIC and this demonstration is a step towards on-chip cavity optomechanics in diamond. Using an optical cavity enabled to enhance the optical power and hence the optical gradient force. We have shown the coherent actuation of an H-resonator via optical forces on the order of 1.9 pN, resulting in resonant motion with an estimated amplitude of about 6.5 nm. High mechanical quality factors up to 28 800 were quantified in the frequency domain, and confirmed in the time domain for the resonator with the smallest linewidth of 40 Hz. In the next section we will demonstrate an alternative actuation scheme using electrostatic forces.

## 3.4.4  Driven motion via electrostatic forces

Electrostatic forces are routinely used for the excitation and detection of micromechanical motion in MEMS devices, for example in pressure sensors and gyroscopes. In the context of PICs fabricated from electrical insulator such as diamond, it is possible to structure electrical circuits on top of the passive substrate, which enables the co-integration of electrical and photonic integrated circuits on the same chip. While optical forces, as discussed in the previous section, enable all-optical tunability of PICs, the use of electrostatic forces between on-chip electrodes has two main advantages: The applicable forces are comparably large, and under ideal conditions, no power is dissipated in static operation and the power consumption in dynamic operation is extremely small.[248] This is a striking difference to tuning by a mechanical element via optical forces, where the average optical power traveling inside a waveguide can be on the order of mW and to tuning of PICs via heaters, which necessitate local dissipation of power on the order of mW. Especially at cryogenic temperatures, where low power dissipation is crucial, electrostatic actuation of optomechanical elements is thus a promising route to tunable PICs.

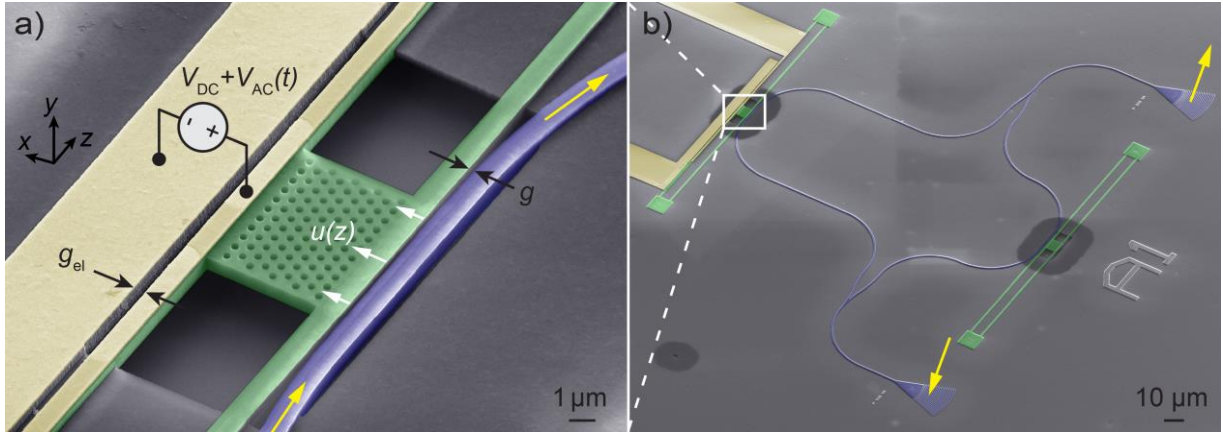

**Figure 39 - Electro-optomechanical photonic integrated circuit:** Colorized and annotated SEM images of an electrostatically driven H-resonator inside a PIC. a) Two metal electrodes (golden color) are separated by a gap of size $g_{el}$. One electrode is located on the fixed diamond layer, while the counter electrode is located on one arm of the H-resonator (green color). An applied voltage leads to an electrostatic force, which leads to a displacement $u(z)$ of the mechanical resonator (symbolized by white arrows). The mechanical resonator is separated from a waveguide (blue color) by a gap of size $g$, which changes with displacement. b) Overview of the photonic integrated circuit, consisting of electrodes, H-resonator, integrated Mach-Zehnder interferometer, and focusing grating couplers.

We modify the geometry of our optomechanical circuit by adding two gold electrodes of 100 nm thickness to the H-resonator, as shown in Figure 39 a). One electrode is located on the fixed diamond layer, while the counter electrode is located on one arm of the movable H-resonator. The electrodes are not electrically connected, but instead separated by a gap of size $g_{el} = 200$ nm, constituting a capacitor. Each electrode is connected to a large contact pad, which enables to electrically connect external electronics via micro-probes. By applying a voltage, an attractive electrostatic force between the electrodes can be applied, which leads to a displacement $u(z)$ of the mechanical resonator (as symbolized by white arrows) with a maximum displacement $U$. On the opposite side of the photonic crystal a waveguide (depicted in blue color) is evanescently coupled to the H-resonator, separated by a gap of size $g$, which changes with displacement. As in the previous sections a mechanical displacement implies a phase shift for the light which propagates within the waveguide. The device layout takes advantage of the optical isolation of both resonator arms by the photonic crystal, which enables to spatially separate the driving force and the resulting tuning of the PIC. Opposed to the use of optical forces in the same waveguide, no optical filters are needed here. This reduces the number of required optical components, especially important for quantum optical circuits which will incorporate photon sources, detectors, and all optical components in the same PIC. We incorporate the H-resonator and waveguide within a photonic circuit, shown in Figure 39 b), which consists of grating couplers and a MZI for interferometric detection of the H-resonator motion, as explained in section 3.4.2. The overall device can be referred to as an electro-optomechanical circuit.

A displacement $U$ directly changes the electrode capacitance $C$ and for an applied voltage $V$ leads to a force $F_{el}$[249]:

$$F_{el} = \frac{\partial}{\partial U}\left(\frac{1}{2}CV^2\right). \tag{3.60}$$

For a mass-spring model of a capacitive resonator[250], simplified as a parallel plate capacitor, the electrostatic elastic force $F_{el}$ is given by

$$F_{el} = \frac{\varepsilon_0 \varepsilon_r A}{2g_{el}^2}V^2 \tag{3.61}$$

where $\varepsilon_0$ is the vacuum permittivity, $\varepsilon_r$ is the relative permittivity, $A$ is the capacitive area, $g_{el}$ is the gap between the electrodes at rest position when no force is applied, and $V$ is the applied voltage. For displacements $U \ll g_{el}$ we can neglect the influence of the changing gap size on the capacitance and the corresponding force. By applying a constant voltage $U_{DC}$, a constant attractive force can be applied, thus pulling the mechanical resonator closer to the fixed electrode. We estimate the electrostatic force, assuming a parallel plate capacitor geometry and the relative permittivity as the average of air and diamond ($\varepsilon_r \approx 5.7$ at low frequencies[146]). For a voltage $V_{DC} = 5$ V this yields a force per interaction length of $F_{el}/L \approx 1$ nN/µm. This is a factor of 1000 larger than the optical force used for H-resonator actuation, which amounts to $F_{opt}/L \approx 1$ pN/µm (see equation (3.54)) for 20 mW of optical power. By applying a sinusoidal voltage $V_{RF}(t) = \hat{V}_{RF} \cdot \sin(\omega_{RF} \cdot t)$ additionally to the DC voltage, we can dynamically drive the oscillator motion with a total force $F_{el}$ which amounts to

$$F_{el} = \frac{\varepsilon_0 \varepsilon_r A}{2g_{el}^2}V_{DC}^2 + \frac{\varepsilon_0 \varepsilon_r A}{2g_{el}^2}\left[2 \cdot V_{DC} \cdot \hat{V}_{RF} \cdot \sin(\omega_{RF} \cdot t) + \hat{V}_{RF}^2 \cdot \sin^2(\omega_{RF} \cdot t)\right]. \tag{3.62}$$

For $V_{DC} \gg \hat{V}_{RF}$ we can neglect the term which is proportional to $\hat{V}_{RF}^2$ and hence we consider a dynamic force which is proportional to $V_{DC} \cdot V_{RF}(t)$ as

$$F_{el,RF}(t) = \frac{\varepsilon_0 \varepsilon_r A}{g_{el}^2}\left[V_{DC} \cdot \hat{V}_{RF} \cdot \sin(\omega_{RF} \cdot t)\right]. \tag{3.63}$$

We fabricate a photonic chip with electro-optomechanical circuits, according to the fabrication procedure explained in section 3.4.1. The fabrication process is only modified by adding additional steps for the metal electrodes (before step 1). The electrode design is patterned in PMMA resist by EBL and transferred into metal structures using deposition by electron beam evaporation (5 nm Cr, 100 nm Au, 10nm Cr) and consecutive lift-off of the metal in the unpatterned areas during PMMA removal (30 min in acetone). We note that during the HSQ lithography step for PICs the metal electrodes within the area of the future rectangular opening windows are covered in HSQ. This protects the thin metal structures from extensive O₂/Ar-etching, which is employed for fully etching the diamond within the opening windows.

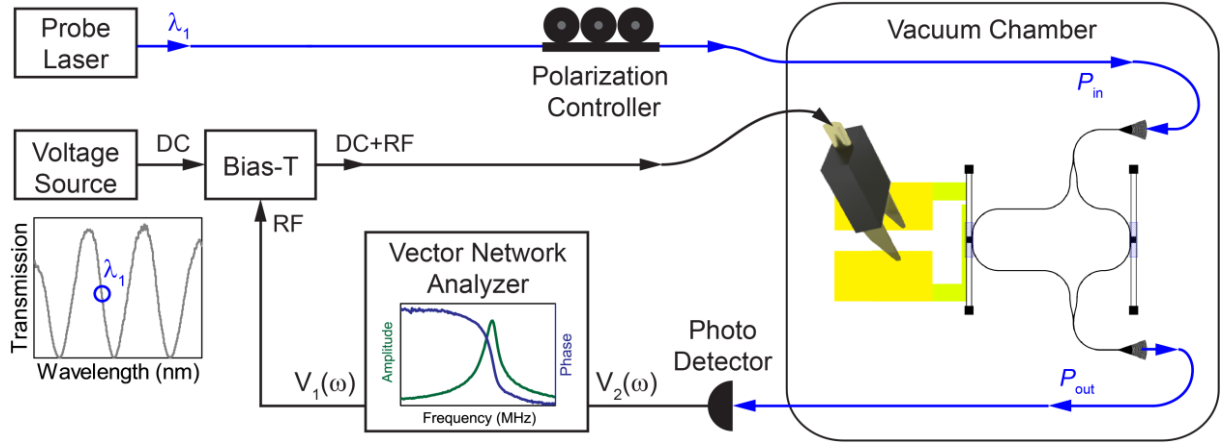

**Figure 40 - Measurement setup for electro-optomechanical circuits:** Schematic of the setup for measuring mechanical motion driven by electrostatic forces. The mechanical resonator is actuated via a voltage $V_{DC} + V_{RF}(t)$ which is applied to on-chip electrodes, resulting in an attractive force between the electrodes. A probe laser (blue color) is operated at a wavelength $\lambda_1$ on the slope of an interference fringe of the on-chip MZI, which transduces mechanical oscillations into modulations in the transmitted intensity.

We characterize the electro-optomechanical circuits in vacuum ($p < 10^{-5}$ mbar) using the setup depicted in Figure 40. As for the measurement of the thermomechanical motion we use a probe laser (Santec TSL-510) at a wavelength $\lambda_1$ on the interference fringe of the MZI, for best transduction of mechanical motion into intensity modulation. The measurements are performed with about 1 mW laser power inside each interferometer arm. The transmitted light is detected with a fast low-noise photodetector (New Focus 1811, 125 MHz bandwidth). An electrical micro-probe (Cascade Microtech Unity Probe) is mounted on a piezo stage, which enables to connect external electronics with the electrodes of an individual device in vacuum. The micro-probe is connected to a bias-T (Mini-Circuits ZFBT-6GW+) which combines the DC voltage from a tunable voltage source with the RF voltage for dynamic actuation. The sinusoidal RF voltage at frequency $f_{RF} = \omega_{RF}/2\pi$ is supplied by a vector network analyzer, which measures the amplitude and phase of the driven motion as a function of $f_{RF}$. The voltage is applied to on-chip electrodes rather than to an EOM, but besides that the measurement procedure is equivalent to the one performed with optical forces.

Due to the much larger forces we anticipate that, besides the fundamental in-plane mode investigated so far, higher order mechanical modes can be excited. Figure 41 a) shows the simulated resonator displacement for the five eigenmodes with the lowest frequencies. When sweeping $f_{RF}$ we expect peaks in the amplitude spectrum, corresponding to the resonance frequencies of the various excited modes. Figure 41 b) shows a recorded spectrum for a resonator with $L = 40$ μm and $w_H = 600$ nm. When sweeping the RF from 5 MHz to 125 MHz, more than 20 resonances can be easily detected, corresponding to the different eigenfrequencies of the oscillator. Figure 41 c) shows the quality factors for all observed resonances extracted from Lorentzian fits. For the fundamental in-plane and out-of-plane modes around 10 MHz mechanical quality factors up to 8700 are observed. For increasing frequencies the quality factor decreases. For the highest resonance frequency of 116 MHz a value of $Q = 1300$ is found. We investigate H-resonators of various lengths and width and find a maximum quality factor of 9600 at 13.8 MHz for a device with $L = 40$ μm and $w_H = 800$ nm. The observed quality factors are lower than for H-resonators of the same device geometry presented in the previous sections. We attribute this to the additional damping introduced by the gold which is deposited directly on the resonator arm.

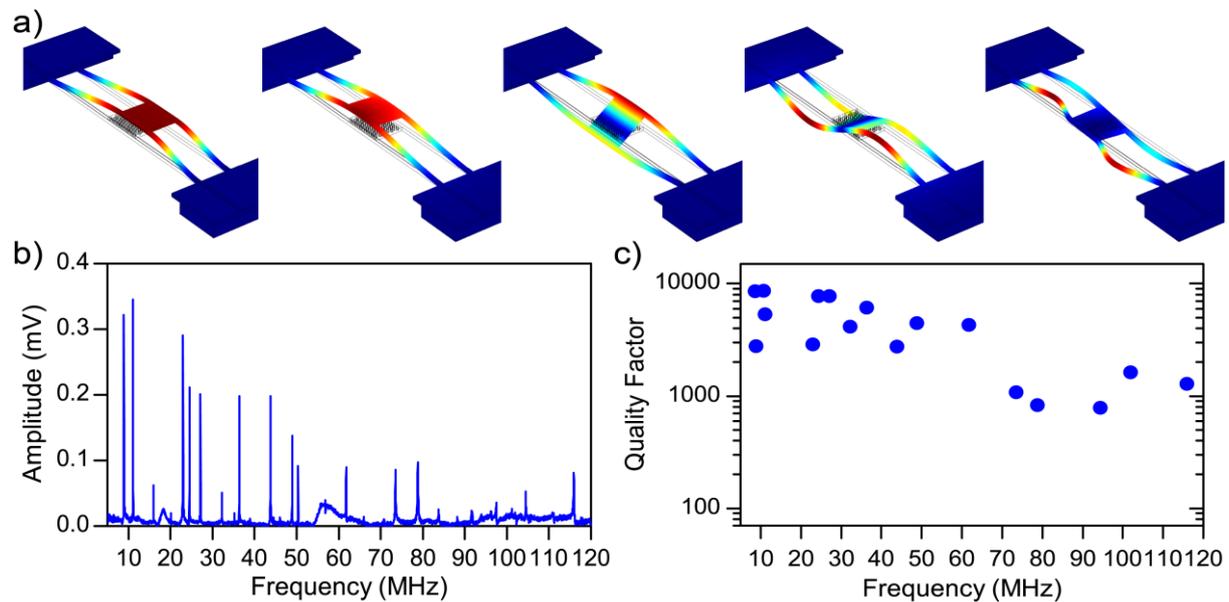

**Figure 41 - Driven higher order mechanical resonances:** a) Simulated shapes of the H-resonator, when excited at the five lowest eigenfrequencies. b) Acquired spectrum of the driven response ($V_{DC} = 5\,V$, $\hat{V}_{RF} = 3.16\,V$) for one H-resonator ($L = 40\,\mu m$, $w_H = 600\,nm$). More than 20 mechanical eigenmodes in the range of 5 MHz to 120 MHz can be excited via the electrodes and the motion detected using the on-chip interferometer. c) Mechanical quality factors for resonances of two resonators of the same geometry ($L = 40\,\mu m$, $w_H = 600\,nm$).

Besides the amplitude, we also investigate the phase between exciting force and mechanical motion. Figure 42 a) shows both amplitude and phase signal for a resonance at 10.8 MHz. A phase shift of 180° and around the resonance is observed, as expected from the driven harmonic oscillator model. Figure 42 b) shows amplitude and phase for the driven motion at the highest observed resonance frequency of 116 MHz.

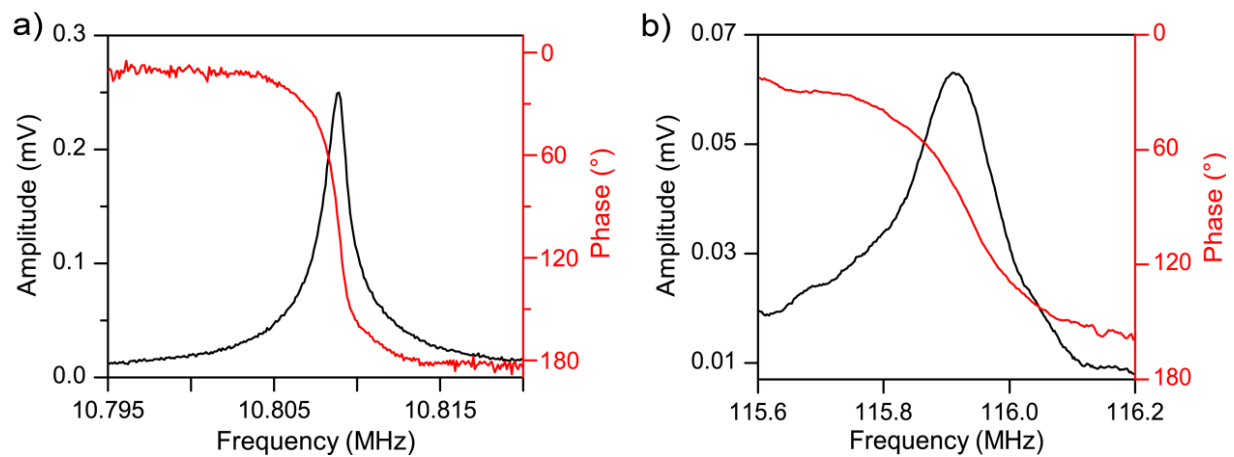

**Figure 42 - Driven motion - amplitude and phase:** a) Amplitude (black) and phase (red) signal for a resonance frequency at 10.8 MHz. By sweeping the excitation frequency across the resonance the expected Lorentzian lineshape and a 180° phaseshift are observed. b) Amplitude and phase for the mode with the highest observed resonance frequency of 116 MHz.

The measurement is limited by the bandwidth of the detector (125 MHz) and we anticipate that resonant motion at higher frequencies could be excited and detected. Higher frequencies in the GHz range would be of interest for fast tunability, as well as for the manipulation of spin states and for sensors at ambient pressures. This could be reached via the excitation of modes of even higher frequencies as well as by increasing the resonance frequencies by scaling down device dimensions.

### 3.4.4.1 Varying driving force and non-linear behavior

We vary the driving force by increasing the RF voltage amplitudes $\widehat{V}_{\mathrm{RF}}$. Figure 43 a) shows the driven response for the resonance at 10.81 MHz for $V_{DC} = 5$ V at various voltages $\widehat{V}_{\mathrm{RF}}$. For $\widehat{V}_{\mathrm{RF}} \geq 0.316\ V$ distortions from Lorentzian shape become apparent, which can be attributed to mechanical non-linearities, which manifest at higher driving amplitude[168,216,251].

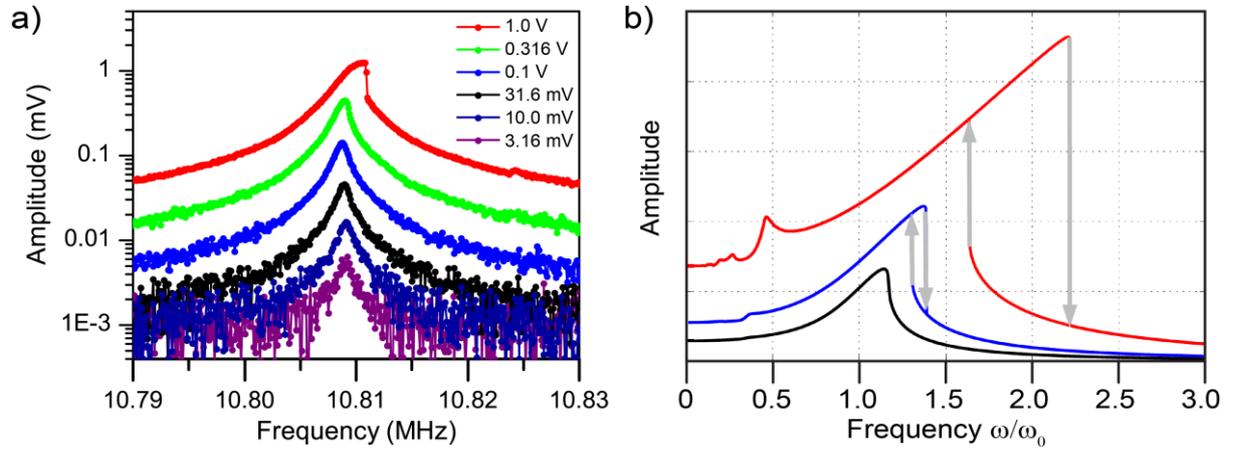

**Figure 43 - Driven motion at higher forces and non-linearity:** a) Driven mechanical resonance at $V_{\mathrm{DC}} = 5$ V and various RF voltage amplitudes $\widehat{V}_{\mathrm{RF}}$. For $\widehat{V}_{\mathrm{RF}} \geq 0.316\ V$, a stiffening Duffing non-linearity is observed, resulting from tensile stress in the diamond layer. b) Theoretical spectrum of a stiffening Duffing oscillator, showing the amplitude over the normalized oscillation frequency $\omega/\omega_0$ for increasing amplitude of the driving force (black to red). At a critical frequency the oscillation loses stability and the system undergoes a transient to another stable periodic solution as indicated by the arrows. Reproduced from Parlitz *et al.*[344].

The linear relation between driving force and amplitude, as assumed in the harmonic oscillator model (equation (3.1)), does not hold at comparably large amplitudes and higher order terms of the Taylor expansion need to be considered. The next non-zero term[14] is proportional to the third power of displacement and the restoring force can be expressed as

$$F_{\mathrm{res}} = -k \cdot x - k_3 \cdot x^3, \tag{3.64}$$

where $x$ is the displacement and $k_3$ the constant of the non-linearity. The resulting differential equation of motion, called Duffing oscillator[252], leads to an oscillator with a non-linear elasticity. The relation between force and amplitude, which in the linear case is referred to as spring constant, becomes amplitude dependent. $k_3 > 0$ corresponds to a stiffening spring, as the restoring force increases with amplitude more than in the linear case, while $k_3 < 0$ corresponds to a softening spring.

---

[14] The potential, which underlies the restoring force, is symmetric concerning displacements in positive and negative direction. Therefore the cubic term of the potential vanishes and the quadratic term of the restoring force is zero.

Figure 43 b) shows the theoretical spectrum of a stiffening Duffing oscillator for increasing amplitude of the driving force (black curve to red curve). Two branches occur in the solution. When sweeping the frequency the oscillation loses stability at a critical frequency and the system undergoes a transient to another stable solution. Depending on the direction of the frequency sweep, different transitions occur, as indicated by the gray arrows.

The occurrence of a Duffing non-linearity in micromechanical oscillators can be related to strain in the thin film and in the context of mechanical elements for PICs a softening non-linearity in silicon[216], stiffening non-linearity in silicon nitride[253], and both types of non-linearities in diamond[156] have been observed. Polycrystalline diamond thin films are grown on oxidized silicon wafers at temperatures around 850 °C. The mismatch in thermal expansion coefficients of the diamond film and the $SiO_2$/Si substrate produces compressive stress, when the sample is cooled down to room temperature. The growth process itself produces intrinsic tensile stress, which can counteract some of the thermally induced compressive stress[254]. Depending on the location on the wafer from which the photonic chip and hence the optomechanical circuit is fabricated, both compressive and tensile internal stress can occur, as shown in our previous work[156]. The nature of the internal stress translates into the sign of the non-linear constant $k_3$ of the mechanical element: for compressive stress a softening Duffing non-linearity is expected, whereas for tensile-stressed film a stiffening response occurs. As shown in Figure 43 a), a stiffening Duffing non-linearity is observed for our electro-optomechanical H-resonators. As in our measurement the frequency is swept from low to high frequencies, we record an amplitude spectrum, corresponding to the downward transition between the branches, as shown in Figure 43 b). We can hence apply large enough forces via the electrodes, such that non-linear mechanical behavior can be observed.

The H-resonator is well suited as low-loss tunable element. In particular for use at low temperature, dissipation-free electrostatic actuation provides significant advantages over currently used micro-heater devices. In following sections we present the results of testing the compatibility of our optomechanical circuit with the operation under varying pressure and temperatures from room temperature to 4 K.

### 3.4.4.2 Pressure dependence of the mechanical quality factor

The pressure dependence of the quality factor determines how well the pressure of the system in which the photonic chip is operated needs to be controlled. Furthermore for sensing applications, it is necessary to operate the electro-optomechanical resonators under non-vacuum conditions. We therefore characterize the H-resonator resonances at various pressures inside the vacuum chamber. For an exemplary device Figure 44 a) shows the driven response at the fundamental frequency of $\approx$ 10.8 MHz at different pressures. We extract the mechanical quality factor from Lorentzian fits to the power spectral density. Figure 44 b) shows the quality factor as a function of pressure over more than eight orders of magnitude. Colored squares correspond to the measurement curves of the same color in Figure 44 a).

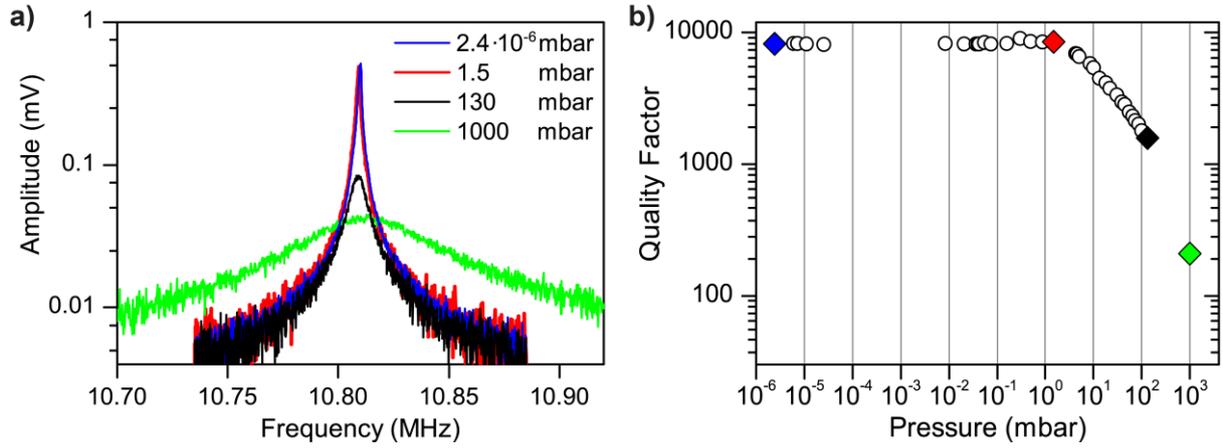

**Figure 44 - Pressure dependence of the mechanical quality factor:** a) Driven response at 10.8 MHz at different air pressures inside the vacuum chamber. b) Quality factor in dependence of the pressure. Air damping is negligible, with a quality factor of $\approx 8400$ for pressures from below $10^{-5}$ mbar up to 1.5 mbar. At higher pressures the quality factor decreases with increasing air damping and reaches a value of 210 at ambient pressure.

At the lowest pressure of $2.4 \cdot 10^{-6}$ mbar the quality factor amounts to $Q = 8400$ (in agreement with the measurements of the same device, presented in Figure 43). For pressures from $2.4 \cdot 10^{-6}$ mbar to 1.5 mbar, the resonance shapes and quality factor remain unchanged within the measurement uncertainty, as can be seen by comparing the blue and red measurement curves in Figure 44 a). At low pressures the device performance is hence insensitive to the stability of the pressure. For pressures above 1.5 mbar up to ambient pressure, the air damping is dominating the other damping mechanisms and the quality factor steadily decreases, comparable to the previous studies of mechanical resonators at varying pressures[179,255]. At ambient pressure, the driven mechanical motion can still be observed and the quality factor amounts to 210.

When used as an optomechanical phase shifter, the goal might be to operate the device as a switch which instantaneously changes the phase from one value to another one. Phase changes on a sub-µs time scale have been shown for H-resonators in silicon nitride[167]. In this context damped operation is preferred, as the transient solution to an instantaneous force leads to resonant oscillations which decay exponentially according to the quality factor. In this context high damping is intended, in order to minimize the time after which the static phase shift has stabilized. Therefore the reduction of the quality factor due to the gold electrodes, as shown in the previous section, as well as air damping are favorable for a phase shifter operated as a switch. If on the other hand a high quality factor and in turn a small linewidth are important, for example in the context of sensing, then operating the device at low enough pressure or increasing the resonance frequency, such that air damping is negligible, are two possible options.

### 3.4.4.3 Diamond optomechanics at cryogenic temperatures

For quantum optics, the PICs investigated in this thesis need to be operated at cryogenic temperatures due to constraints for single-photon sources or single-photon detectors. We therefore test the operation of our optomechanical circuits at low temperatures. For this we place our photonic chip inside a cryogenic measurement setup, which is equal to the setup shown in Figure 40, the only difference being that the vacuum chamber is replaced by a liquid helium cryostat. We drive the motion of an H-resonator ($L = 40\,\mu m$, $w_H = 600\,nm$) at its resonance frequency $f \approx 8.5\,MHz$ via the electrodes, while ensuring a linear relation between force and displacement (avoiding a Duffing nonlinearity). We sweep the excitation frequency in the range of the resonance frequency and extract the mechanical quality factor and the resonance frequency from Lorentzian fits to the power spectral density. The quality factor at room temperature is in agreement with the measurement of the same device in the vacuum chamber, showing that air damping is negligible in this measurement[15]. We continuously repeat this measurement while cooling down the sample chamber from a temperature of 300 K to 4 K, using liquid nitrogen and liquid helium.

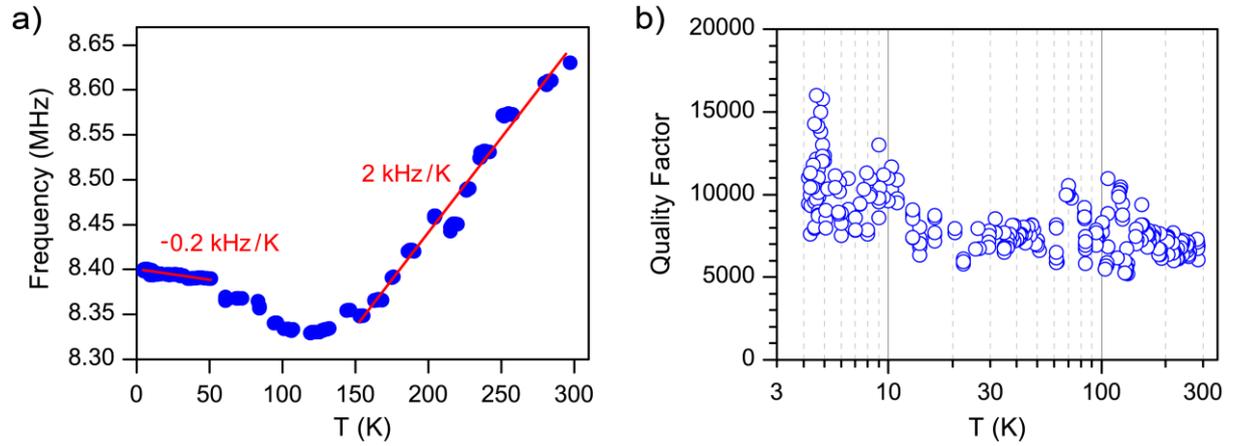

**Figure 45 - Temperature dependence:** a) Dependence of the resonance frequency of an H-resonator ($L = 40\,\mu m$, $w_H = 600\,nm$) on the temperature. b) Corresponding quality factor extracted from repetitive measurements while cooling down and heating up the cryostat. A slight increase of the quality factor towards small temperatures can be observed.

Figure 45 a) shows the dependence of the resonance frequency of an H-resonator on the sample chamber temperature. With decreasing temperature, starting at room temperature, the resonance frequency decreases linearly with 2 kHz/K until about 150 K. Around 110 K a local minimum in frequency is observed. At temperatures from 50 K to 4 K the frequency increases with 0.2 kHz/K with decreasing temperature. This means that if the resonance frequency of a micromechanical oscillator needs to be stable within its line width, the temperature needs to be stable within corresponding bounds. This is for example important for measurements of the quality factor in the frequency domain and for optomechanical sensors[179,180]. Hence a ring down measurement is, especially for high quality factors, a more accurate way to determine the $Q$-factor, as explained in section 3.4.3. H-resonators without electrodes have high quality factors of about $Q \approx 20\,000$ at $f \approx 8.5\,MHz$ (see Figure 31). This implies a FWMH of the resonance of $\delta f \approx 400\,Hz$. Therefore the temperature needs to be controlled within $\Delta T < 0.2\,K$ at room temperature and within $\Delta T < 2\,K$ at cryogenic temperatures to ensure that changes in resonance frequency are smaller than the linewidth within the time scale of relevance. Such limits can be sustained via active temperature control and our optomechanical circuits can thus be employed both at room temperature and at cryogenic temperatures. Furthermore, if the absolute value of the resonance frequency is not relevant for the specific application, then a phase-locked loop enables to drive the oscillator at resonance over long time spans, as shown in section 3.4.3.2, and hence less temperature stabilization is required. We note that if a mechanical element is used as a phase shifter, vibrations caused by the thermal energy will generally lead to a small noise in phase shift[166]. Operating optomechanically tunable circuits is therefore especially feasible for low temperature applications, as the thermal energy and its implied noise decrease with decreasing temperature.

---

[15] The measurement of the pressure inside the sample chamber of the cryostat is less accurate than the measurement of the pressure in the vacuum chamber. The sample chamber has been evacuated before the cooldown but contains some gaseous helium. The exact pressure is not relevant for our measurement, as long as air damping can be neglected compared to the intrinsic damping, which we confirm via comparing the experimental results for the quality factor for both the vacuum chamber setup and the cryogenic measurement setup.

Figure 45 b) shows the dependence of the quality factor on the temperature.[16] Generally the quality factor of the electro-optomechanical circuits is most likely limited by the gold electrode, which introduces additional damping. We observe an increase of the average quality factor by $\approx 60\%$ when comparing room temperature to 4 K. This is smaller than the increase by $\approx 200\%$ observed for micromechanical oscillators from polycrystalline diamond, as well as high quality single-crystal diamond without electrodes[256]. For our purpose a smaller quality factor is beneficial, as for an optomechanical device which applies phase shifts stepwise, the transient solution needs to decay in order for a new phase setting to stabilize. For the device studied here with a quality factor of 10 400 at 4 K and $f \approx 8.4$ MHz, the amplitude decays, according to equation (3.12), with a $1/e$-decay time of $\tau = 400$ µs, limiting the speed at which new phase settings can be applied.[166] Note that the quality factor and hence the decay time can be decreased by introducing further damping, for example by operating the device at higher pressures, as shown in the previous section.

Within this section we have shown that our H-resonator design and its active actuation via electrostatic forces can be used from room temperature to 4 K. The working principle of an optomechanical phase shifter depends on mechanical displacement and not on local heating of optical elements and hence dissipation of power. Optomechanical components are therefore energy-efficient, which ensures scalability of the PICs. They also show advantages over current solutions concerning avoided cross-talk and speed and their compatibility with cryogenic temperatures enables to combine optomechanical tunability with superconducting single-photon detectors, which we will discuss in the following section.

---

[16] We note that the extracted quality factors are likely smaller than the actual quality factors of the device, as the constant temperature change while cooling leads to a change in resonance frequency (as shown in Figure 45 a) which hence broadens the measured resonance shape.

## 3.5 Conclusions on optomechanics and outlook

In conclusion, this chapter presented the first demonstration of diamond integrated optomechanical circuits. Integrating the mechanical components within on-chip interferometers allowed for sensitive motion readout with sensitivities down to 32 fm/$\sqrt{\text{Hz}}$ using 3 mW of optical power. We demonstrated the readout of thermal motion and driven motion with high quality factors up to $Q = 28\,800$ and resonance frequencies ranging from 1 MHz to 115 MHz. Such high quality factor are important for optomechanical sensors. Depositing metal electrodes onto the resonator and air damping both lower the quality factor. Such damped operation is intended for a range of tunable elements such as phase shifters. Photonic crystals within the mechanical element allows optical isolation between driving force and the resulting mechanically induced variation of the optical waveguide properties, avoiding losses. Active control of the motion via optical gradient forces and electrostatic forces was demonstrated and non-linear mechanical behavior was observed for high driving forces.

In the future, coupling both optical and mechanical resonators, as demonstrated here for a ring resonator and an H-resonator, will enable diamond cavity optomechanics. Towards this goal the design would be adjusted for a larger overlap of optical and mechanical mode. Furthermore the demonstrated optomechanical design could either be adjusted for higher frequency to enable high quality factors at ambient pressure or for increased tunable phase shifts up to $2\pi$. While the phase shift in the demonstrated geometry is rather small, it has been shown that large phase shifts via long H-resonators can be achieved.[167] We demonstrated operation both at ambient and cryogenic temperatures, indicating that our optomechanical device can be used as fast and energy-efficient tunable phase shifters in quantum optical circuits along with single-photon sources and superconducting single-photon detectors. So far we demonstrated diamond PICs and tunable elements and the following chapter will demonstrate another key component: single-photon detectors, integrated on diamond PICs.

# 4 Superconducting nanowire single-photon detectors on diamond

*Single-photon detectors are an indispensable building block for integrated quantum optics. This chapter provides an overview over single-photon detection and figures of merit which characterize the performance of single-photon detectors. Furthermore the operation principles of superconducting nanowire single-photon detectors are explained and the experimental results for the first integrated single-photon detectors on diamond waveguides[17] are presented and an outlook for future applications of such detectors is given.*

*This chapter is partially based on results which were published previously in three publications[257–259], where the author of this thesis was first author or had equal contribution with the first author.*

## 4.1 Motivation and introduction

Since the discovery that the energy of electromagnetic radiation is quantized and therefore propagating light can be described as a stream of energy quanta called photons, it has become evident that there exists an ultimate limit for the precision of a measurement system for electromagnetic radiation. The energy of one photon of near-infrared and visible light is around $10^{-19}$ J and once a measurement device can detect such a photon, it can detect the smallest amounts of energy that exist within the electromagnetic radiation at that wavelength. Such a device is then called a single-photon detector (SPD). Since the first realization of SPDs in the form of photomultiplier tubes, there has been a growing interest in single-photon detection and various detector technologies, such as single-photon avalanche diodes and transition-edge sensors, have been developed. A detailed overview and comparison of the different single-photon detectors can be found in a publication by Hadfield[260].

The general mode of operation of an SPD is that upon the absorption of a photon an electrical output signal is generated, which can be registered by an external electrical circuit connected to the detector. The electrical output pulse corresponding to such a detection event is often referred to as *a click* or *a count* and the frequency at which these events occur on average is called the count rate. For quantum information science with photons SPDs are of central importance, as explained in section 2.2.2, and high detection efficiencies and outstanding timing characteristics are needed. For efficient linear optical quantum computing the product of photon source efficiency and detector efficiency needs to surpass a threshold[261] of 2/3. Hence the goal within this thesis is to develop SPDs, integrated with diamond PICs, with detection efficiencies surpassing this threshold, while the long-term goal for scalability of the PICs is to achieve 100% efficiency. Superconducting nanowire single-photon detectors (SNSPDs) were first introduced in 2001 by Goltsman *et al.*[206] and show outstanding properties. In the following sections we first introduce the working principle of SNSPDs and then discuss their performance characteristics.

---

[17] Within the research for this thesis the first waveguide-integrated single-photon detectors on any diamond photonic circuit were achieved.[257,258] We present here the results of the second generation of detectors, which perform better concerning basically all detector properties.

## 4.2 Superconducting nanowire single-photon detectors: theory

### 4.2.1 Detector operation principle

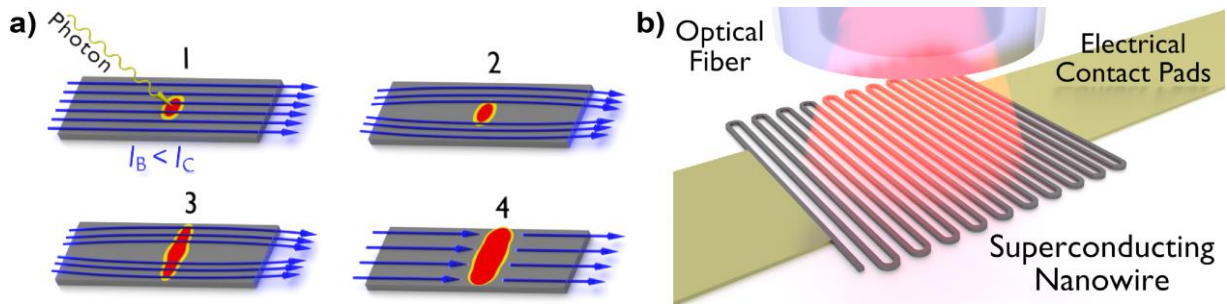

**Figure 46 - Working principle of a superconducting nanowire single-photon detector:** a) Schematic of a superconducting wire (gray) carrying a bias current $I_B$ (blue arrows) below the critical current $I_C$. Upon absorption of a photon superconductivity is locally disturbed and a resistive barrier forms, which enables the detection of a single photon. b) Schematic of a SNSPD with photon impinging under normal incidence from an optical fiber onto the superconducting nanowire meander. The nanowire is biased via gold contact pads.

A superconducting nanowire single-photon detector consists of a current-biased nanowire of superconducting material with dimensions of about 5 nm thickness and about 100 nm width. The nanowire is cooled down to a low temperature at which it is superconducting, meaning that its electrical resistance is zero[18] if it carries a small enough current density. At higher current densities the nanowire is normal-conducting and has a non-zero resistance, and the critical current density defines the threshold between superconducting and normal-conducting state. Figure 46 a) illustrates the operation of the superconducting nanowire using a phenomenological description[19], which is referred to as the hotspot model: To act as a single-photon detector, the nanowire (symbolized by a gray strip) carries a current (symbolized by blue arrows), called the bias current $I_B$. The bias current is smaller than the critical current $I_C$ above which the current density would surpass the critical current density.[20] Once a photon is absorbed in the wire, superconductivity is locally disturbed in a certain region, often called a hotspot (as symbolized in yellow and red color), and the charge carrier density in this region is decreased. This leads to an increase in current density in the outside of the hotspot and the current density surpasses the critical current density, leading to an initial normal-conducting domain across the full wire width. The non-zero resistance of the initial normal-conducting domain implies that energy is dissipated, which leads to Joule heating of the wire caused by the current. This expands the normal-conducting region and leads to a resistance on the order of several kΩ. An electrical readout circuit enables the detection of an electrical pulse on the order of a few nanoseconds, which corresponds to the detection of a single photon. Experimental implementations of SNSPDs, as shown in the schematic of Figure 46 b), consist of the aforementioned superconducting nanowire, folded into a meander shape in order to cover a larger area, which is electrically connected to a current source and read-out electronics via metal contact pads. In the first SNSPDs and in most of the SNSPDs to date, light is impinging onto the detector from the top of the detector from an optical fiber, as shown in Figure 46 b), while within this thesis waveguide-integrated SNSPDs are being investigated, which will be introduced in section 4.2.4.

---

[18] Up to date no trace of resistance in bulk superconductors was found. On the basis of the sensitivity of modern equipment it is at least smaller than $10^{-24}\,\Omega$cm. For comparison the resistivity of high-purity copper is of the order $10^{-9}\,\Omega$cm at 4.2 K[262]. It is hence feasible to argue that the resistivity of superconductors is zero and we will phrase it accordingly in this thesis.

[19] Models which explain the formation of the normal-conducting domain more accurately than the hotspot model are described in the subsequent section 4.2.2.

[20] It should be noted that the term *critical current* refers to the experimentally measured largest current at which a nanowire sustains superconducting. This experimental value cannot exceed the depairing critical current $I_{C,dep}$ at which Cooper pairs would depair in a homogeneous superconducting wire.[266] The measured critical current in a nanowire can be smaller than $I_{C,dep}$, for example due to the bias circuit[276], or due to reductions in the nanowire's cross section, which locally limit the largest current at which the nanowire sustains superconducting.

## 4.2.2 Fundamentals of superconductivity and SNSPD detection mechanism

This section gives a brief overview of fundamentals of superconductivity in terms of concepts which are important in the context of SNSPDs, such as understanding why an absorbed photon results in the creation of a normal-conducting domain in a current-biased superconducting nanowire. Detailed descriptions of superconductivity and its underlying mechanisms can for example be found in[262,263]. Superconductivity is the phenomenon of expulsion of magnetic fields and the disappearance of the DC electrical resistance, which occurs for superconducting materials when cooled below a characteristic critical temperature $T_C$. The most common material for SNSPDs is niobium nitride (NbN) with a critical temperature of about 17 K for bulk material[264] and about 13 K for a thin film[264] of 5 nm thickness. Such a moderately high critical temperature enables the operation of the devices using liquid helium which boils at 4.2 K, while SNSPDs from high critical temperature superconductors such as Yttrium barium copper oxide might in the future enable the operation of SNSPDs using liquid nitrogen.[265] Below the critical temperature and without external magnetic fields, a wire of a superconducting material sustains superconductivity as long as the electrical current density is below a critical current density. Furthermore, superconductivity is only sustained if the magnitude of a magnetic field is below the critical magnetic field $H_C$.

On a macroscopic scale, superconductivity can be described by the phenomenological Ginzburg-Landau (GL) theory in the context of second-order phase transitions. The superconducting state is more ordered than the normal-conducting state, which can be described by a suitable order parameter.[262] In the GL theory, the complex and spatially varying order parameter is chosen to be the effective wavefunction of superconducting electrons $\Psi(\vec{r})$. A variation of the order parameter due to a perturbation takes place on a length scale which defines the coherence length $\xi_{GL}$, while the length scale for the screening of static magnetic fields from the inside of a superconductor is called the penetration depth $\lambda_{GL}$. NbN superconducting nanowires can be considered as two-dimensional systems, as their film thickness on the order of $d = 5$ nm is comparable to the coherence length $\xi_{GL}$, while the wire width $w$ is at least a factor of ten larger and the length is orders of magnitude larger than the coherence length.[266] As the penetration depth is much larger than the film thickness, the supercurrent density is uniform across the thickness of the nanowire.

Two types of superconductors exist which differ concerning the ratio of penetration depth to coherence length, which causes different behaviors concerning external or self-induced magnetic fields: A type-I superconductor ($\frac{\lambda_{GL}}{\xi_{GL}} \leq \frac{1}{\sqrt{2}}$) screens external magnetic fields from the inside, up to the critical field $H_C$ above which superconductivity breaks down. Opposed to that a type-II superconductor ($\frac{\lambda_{GL}}{\xi_{GL}} \geq \frac{1}{\sqrt{2}}$) above a certain magnetic field $H_{C1}$ is penetrated by the field in the form of quantized magnetic vortices, where each vortex carries one flux quantum. The current state of research suggests that magnetic vortices play an important role in the detection mechanism of SNSPDs[266,267], as will be elaborated on at a later point of this chapter. A vortex consists of a normal-conducting core around which a supercurrent circulates, shielding the field inside the vortex from the superconducting region outside the vortex. The core has a size on the order of $\xi_{GL}$ and the supercurrent decays with distance from the core on the order of the penetration depth $\lambda_{GL}$. NbN is a type-II superconductor which furthermore belongs to the group of "dirty" superconductors. In such materials the electron mean free path is smaller than the coherence length and vortices can be also present at zero external magnetic field.[268]

On a microscopic scale superconductivity can be treated using the Bardeen-Cooper-Schrieffer (BCS) theory[269], which considers the interaction of electrons near the edge of the Fermi sphere leading to the formation of a stable bound state of two electrons with reduced energy. A weak attraction between electrons can lead to the formation of a bound state, called Cooper pair (CP), with an integer spin of 0 or 1. For metallic superconductors this attractive interaction can be mediated via phonons and a maximum interaction occurs for electrons with anti-parallel spins and momenta of equal absolute value but opposite direction. The electrons form Cooper pairs up to a point when the binding energy of an additional CP is zero. The energy gap of the resulting BCS ground state relative to the ground state of the normal-conducting material, often called condensation energy, is given by $E_{cond} = -\frac{1}{2}N(E_F)\Delta^2$, where $N(E_F)$ denotes the density of states at the Fermi energy and $\Delta$ the average potential for pairing of CPs. The condensation energy quantifies the reduction in the energy of the electron collective by transitioning into the superconducting state.

An external excitation of the superconducting ground state, for example via a thermal perturbation or the absorption of a photon in the nanowire, creates a high energy quasiparticle. Quasiparticles describe quantized excitations where the creation of a quasiparticle corresponds to the annihilation or the breaking of a Cooper pair. The energy of a photon, on the order of $E_{ph} \approx 1$ eV, is much larger than the superconducting energy gap, on the order of $\Delta \approx 1$ meV[270]. Hence the energy of a photon is large enough for breaking up many Cooper pairs, leading to a reduced Cooper pair density. The suppressed superconductivity around the location of absorption can result in a breakdown of superconductivity. The higher the initial photon energy, the higher the number of quasiparticles which are generated by this process. Hence the smaller the energy gap $\Delta$, the less energy is needed to create a sufficient amount of quasiparticles for breaking down superconductivity, making such a material more sensitive to light of longer wavelengths.[264]

The microscopic details of single-photon detection using superconducting nanowires are currently not yet fully understood and several detection models exist, each of them consistent with some of the experimentally observed properties of SNSPDs, while so far no model is capable of fully explaining the detection mechanism underlying SNSPDs.[266,271] Here we briefly describe two promising models, namely the refined hotspot and the diffusion-based vortex-entry model. A comparison of all models is for example provided by Engel *et al.*[271]. The first model for the detection mechanism, called the hotspot model, was provided in 2001 by Semenov *et al.*[272] and has been subsequently refined in order to account for more experimental observations. The refined diffusion-based hotspot model relaxes the requirement for a normal-conducting core of the hotspot. According to this model, an absorbed photon reduces the number of Cooper pairs at the absorption site and hence the current carrier density. The remaining Cooper pairs need to carry the same total current and hence their speed needs to increase. If it exceeds a critical velocity, the wire becomes normal-conducting.

In contrast to the original hotspot model, the refined model is consistent with many experimental observations[273] but fails to explain the position-dependence of detection efficiency observed by Renema *et al.*[274]. The diffusion-based vortex-entry model[271,275] on the other hand considers the entrance of magnetic vortices into the nanowire by overcoming an edge energy barrier. Overcoming the edge barrier can be induced by energy fluctuations which lower the edge barrier, for example by the absorption of a photon in the nanowire. The initially excited quasiparticles and subsequent generated quasiparticles are then considered to diffuse independently. This model is consistent with the majority of experimental observations including the position-dependent local detection efficiency. However, some aspects of the model such as the influence of external magnetic fields do not agree with the experiments. While both the refined hotspot and the diffusion-based vortex-entry model are consistent with large parts of experimental observations, a model which is consistent with all observations and which provides a detailed understanding of single-photon detection using superconducting nanowires remains to be found. While this is an interesting aspect of fundamental research, this open question has little impact on the results of the work presented in this chapter. The next section describes both the electrical operation of SNSPDs and the time scales on which the involved physical processes occur.

## 4.2.3  Electrical operation of the SNSPD

For the operation as a single-photon detector the device containing the superconducting nanowire is installed in a cryogenic setup, typically operating with liquid helium at a temperature below 4 K. The cryogenic setup used within this thesis will be explained in detail in section 4.4.1.

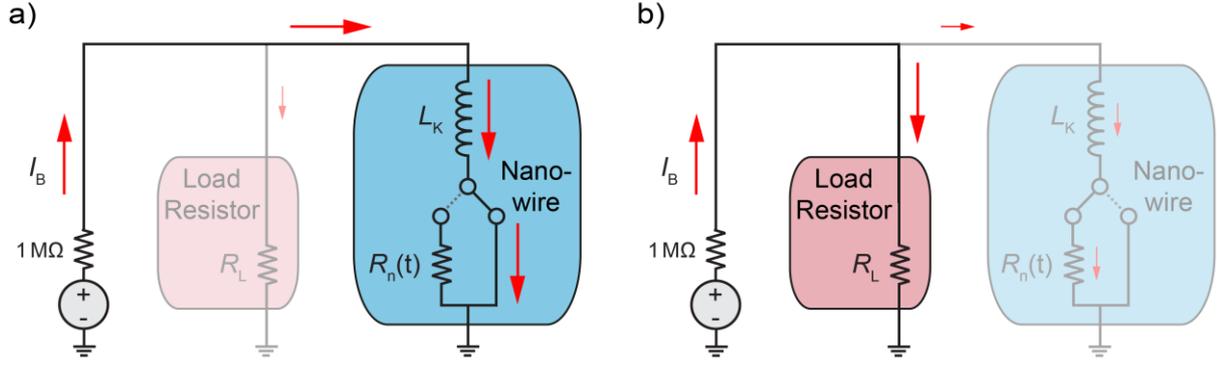

**Figure 47 - Electrical equivalent circuit diagram of a SNSPD:** in the (a) superconducting and (b) normal-conducting state. A voltage source with a 1 MΩ resistor provides a bias current $I_B$ below the critical current of the nanowire. The nanowire, represented as a blue box, can be described as a series of an inductance $L_K$ and a varying resistance which his either zero when superconducting (a) or non-zero with a value $R_n(t)$ when normal-conducting.

The electrical operation of a SNSPD can be explained using a simplified electrical equivalent circuit diagram[276], as shown in Figure 47. A voltage source with a 1 MΩ resistor provides the nanowire with a constant bias current $I_B$ below the critical current $I_C$ of the SNSPD. The nanowire, represented as a blue box, can be described as a series of an inductance $L_K$ and a varying resistance $R$. If the wire is superconducting (Figure 47 a) the resistance is zero ($R = 0$) and if a part of the nanowire is normal-conducting (Figure 47 b) the resistance has a non-zero value $R = R_n(t)$. The absorption of a photon can lead to a switching of the nanowire from superconducting to a normal-conducting state, which is represented as an electrical switch in the equivalent circuit. While in the superconducting state the current through the nanowire is essentially equal to the bias current $I_B$. Upon switching the majority of the current is shunted to the load resistor of the transmission line $R_L = 50\ \Omega \ll R_n(t)$ which is connected to the read-out circuitry. This enables to detect a photon by detecting the switching of the nanowire.

The inductance $L_K$ in the electrical equivalent circuit of the nanowire corresponds to the kinetic inductance which can be explained as follows: The well-known magnetic inductance $L_M$ of a wire or a coil is defined by the energy $E_M$ which is stored in the magnetic field that is generated by a current $I$ as

$$E_M = \tfrac{1}{2}\int \vec{H} \cdot \vec{B}\, dV = \tfrac{1}{2} L_M I^2 \ . \tag{4.1}$$

In an electric conductor charge carriers have non-zero mass and their movement at speed $v$ implies a stored kinetic energy $E_K$. The inertia of the charge carriers can cause a phase lag between voltage and current of the conductor and hence the kinetic inductance $L_K$ is introduced in analogy to the magnetic inductance as

$$E_K = \int \tfrac{1}{2} m v^2 \cdot n \ dV = \tfrac{1}{2} L_K I^2 \ , \tag{4.2}$$

where $m$ and $v$ are the mass and the velocity of a single carrier[262], $n$ denotes the number density of the current carriers and the integration is carried out over the volume of the nanowire. For normal-conducting materials the resistive impedance dominates the electrical behavior of a wire and the kinetic inductance can be neglected up to THz frequencies. For a superconductor on the other hand the DC resistance is zero and the impedance is governed by the kinetic inductance. Using the super-current density $\vec{J_s} = n_s \cdot q_s \cdot \vec{v_s}$ with the charge carrier density $n_s$, the charge $q_s = 2e$, the velocity $\vec{v_s}$ and the mass $m_s = 2m_e$ of the Cooper pairs, where $e$ and $m_e$ are the charge and the mass of one electron the kinetic inductance can be expressed as

$$L_K = \frac{1}{I^2} \int \frac{m_e \cdot \vec{J_s}^2}{2e^2 n_s} dV = \frac{m_e}{2e^2} \int \frac{\vec{J_s}^2}{n_s I^2} dV \ . \tag{4.3}$$

For a nanowire with cross section $A$, length $l$ and a uniform supercurrent density $J_s = \frac{I}{A}$, the kinetic inductance is therefore

$$L_K = \frac{m_e}{2e^2 n_s} \frac{l}{A} \ . \tag{4.4}$$

The kinetic inductance of a nanowire, which is proportional to the nanowire length $l$, limits the detector speed, as will be explained in the following. The absorption of a photon in the superconducting nanowire leads to the excitation of one electron into an unoccupied state in the conduction band[277] with high energy. Subsequent relaxation of this excitation leads to a cascade which generates further quasiparticles.[278] This locally perturbs the equilibrium of superconductivity and an initial normal-conducting domain of $\approx 100$ nm length[279] is formed[266] which gives the wire a non-zero resistance $R_n(t)$. As described in the previous section, the microscopic mechanism for the emergence of the initial normal-conducting domain is up to date not fully understood and under active investigation[266,273,280]. The initial normal-conducting domain leads to energy being dissipated in the resistance giving rise to a heating process, called Joule heating. The heating causes the normal domain to grow, leading to a growing dissipation of energy (positive electro-thermal feedback) and consequently $R_n(t)$ expands in time exponentially[279] on a picosecond timescale and very quickly exceeds the load resistance $R_L$. The increase in resistance on the other hand leads to a negative electro-thermal feedback, as an increasing portion of the bias current is diverted into the load resistance $R_L$, a process which stops the expansion of the normal-conducting domain. This diversion to the load resistor is however opposed by an induced voltage in the kinetic inductance, for a characteristic time $\sim \frac{L_K}{R_n(t)}$[276]. This enables Joule heating to increase $R_n(t)$ to a value much larger than $R_L = 50\ \Omega$.

Figure 48 a) shows the time behavior of the nanowire current during a detection event. The bias current flowing through the nanowire decreases exponentially with a time constant $\tau_1$ defined by the LR circuit, typically on the order of some hundred picoseconds, given by

$$\tau_1 = \frac{L_K}{R_L + R_n(t)} \approx \frac{L_K}{R_n(t)} \ . \tag{4.5}$$

After shunting most of the current to the load resistor, the heating is strongly reduced, and the heat dissipation to the surrounding allows the superconductivity to be restored. The different heat transfer processes involved in the recovery of superconductivity can be explained using a two-temperature model[281,282] for electrons and phonons. Figure 48 b) shows a schematic of this model where the involved time scales are indicated. As described in the previous section, the absorption of a photon leads to initially excited electrons which relax by exciting further electrons, for example by breaking Cooper pairs. The subsystem of the electrons at temperature $T_e$ is excited via this initial photon absorption and by Joule heating and relaxes by coupling to the phonon subsystem (at temperature $T_{ph}$) on a timescale $\tau_{e\rightarrow ph} \approx 10$ ps[283] and by out-diffusion of electrons. The energy of the phonons can either be coupled back to the electrons or to the substrate at a lower bath temperature $T_0$. The time scale at which energy is transferred to the substrate is given by the phonon escape time on the order of $\tau_{esc} \approx 38$ ps[283]. After the energy has been dissipated, the initial state of the superconducting material is restored.[21]

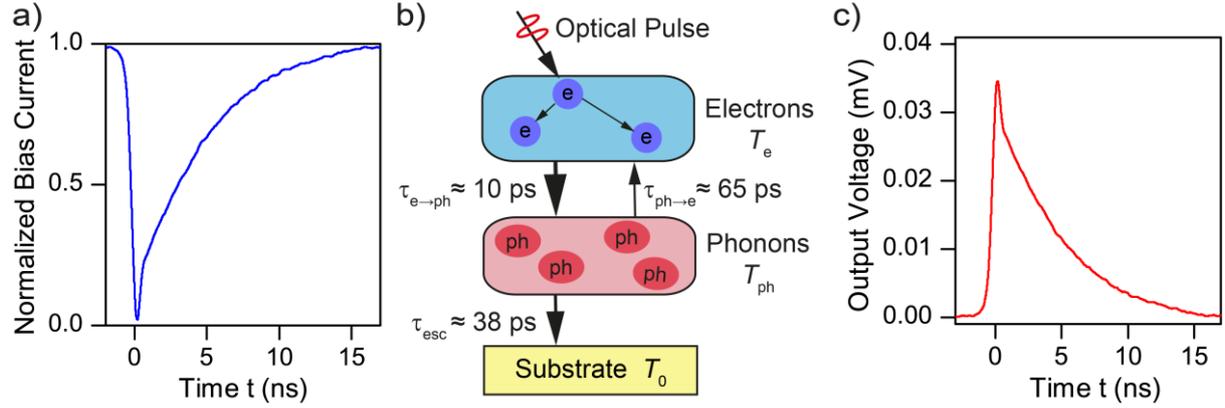

**Figure 48 - Time scales:** a) Schematic of the decrease and recovery of the bias current during a detection event. b) Schematic of the heat transfer processes involved in the recovery of superconductivity, described by a two-temperature model for electrons ($T_e$) and phonons ($T_{ph}$), and the time scales of the involved processes for 3.5 nm NbN on sapphire substrate. Schematic adapted from Il'in *et al.*[283] c) Time behavior of the electrical output pulse corresponding to a detection event. The amplitude of the output voltage depends on the bias current and the electrical amplifiers.

The current through the superconducting nanowire recovers to its initial value $I_B$ with an exponential time behavior with a time constant $\tau_2$ given by

$$\tau_2 = \frac{L_K}{R_L}. \tag{4.6}$$

It should be noted that this recovery of the current is limited by the kinetic inductance and is typically on the order of nanoseconds, which is significantly larger than the excitation timescales mentioned above (see Figure 48 b). The current at the load resistor, as shown in Figure 48 c), resembles the inverted behavior of the current through the nanowire, leading to an asymmetric pulse shape governed by the same time constants $\tau_1$ and $\tau_2$. The detector's recovery time $t_{recovery}$ describes the ability to detect a second photon subsequent to the detection of a first photon. For a SNSPD the recovery time is mainly governed by $\tau_2$ and we will discuss the relation between both times in section 4.2.5.5. It should be noted that a small $\tau_2$ is desired, but if it becomes too small, the electro-thermal feedback can become stable and lead to an effect known as latching[281]. Latching means that the detector is locked in a resistive state and can therefore no longer detect photons[279]. We will investigate in section 4.4.7 if we find latching for short and hence fast SNSPDs.

---

### 4.2.4 Waveguide-integrated SNSPDs

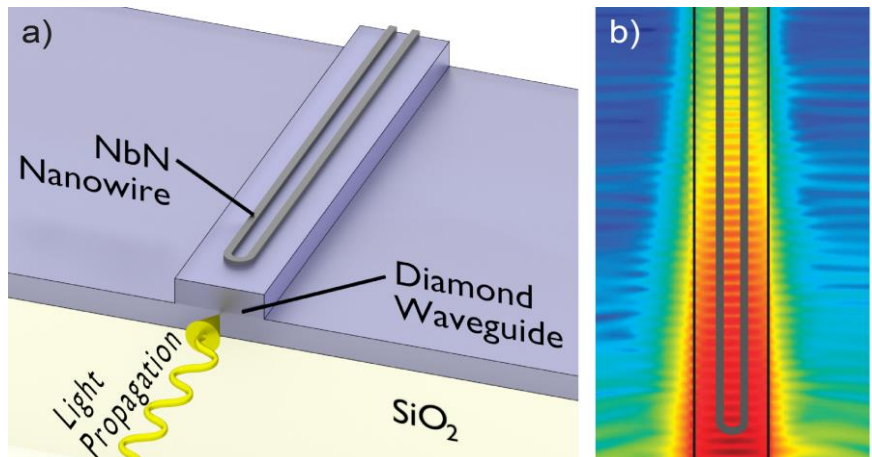

**Figure 49 - Geometry of waveguide-integrated SNSPDs:** a) Schematic of a NbN superconducting nanowire integrated on a diamond waveguide. b) FDTD simulation of the exponential decay in propagation intensity in a waveguide due to absorption of light in the NbN nanowire. The shown simulation results are for a NbN nanowire on a silicon waveguide, adapted from Pernice *et al.*[65].

A design for superconducting single-photon detectors, which makes them suitable for PICs, was proposed in 2009 by Hu *et al.*[284] and was experimentally realized by several research groups in 2011[65,71,285]. Figure 49 a) shows a schematic of such a detector: A superconducting nanowire is placed on top of an optical waveguide. The nanowire consists of two parallel NbN strips which are connected by a 180° circular turn at the side from which light is arriving at the detector. Due to the evanescent field of the waveguide mode, light is absorbed by the nanowire while it propagates along the direction of the nanowire. While in classical SNSPDs the light impinges from the top, as shown in Figure 46 b), limiting the absorption length to the thickness of the superconducting layer of about 5 nm, the absorption length of waveguide-integrated SNSPDs is determined by the length of the nanowire on the order of tens of micrometers. The absorption efficiency of SNSPDs can be improved by placing them in an optical cavity, but this has the drawback of making the absorption spectrum narrowband. Opposed to that the absorption in waveguide-integrated SNSPDs can be improved simply by elongating the nanowire. With no need for an optical cavity the large absorption efficiency is broadband.

Figure 49 b) shows the simulated intensity distribution for CW light propagating along a silicon waveguide[65]. After some tens of micrometers of nanowire length most of the light has been absorbed. We will quantify the absorption coefficient for SNSPDs on diamond waveguides in section 4.3.1. In summary it can be said that waveguide-integrated SNSPDs can feature a near unity absorption efficiency over a wide range of wavelengths, while having a small device size, which is furthermore naturally compatible with PICs.

### 4.2.5 Performance characteristics of single-photon detectors

When comparing the performance of different single-photon detectors, several figures of merit or performance characteristics should be considered. These are the detection efficiency, the dark count rate and noise-equivalent power, the spectral range, the timing jitter, the recovery time and maximum count rate, and the existance or absence of photon number resolution capability. These performance characteristics will be discussed in the following. A perfect SPD would lead to electrical detection pulses every single time when a photon arrives. It would be able to independently register two photons which arrive after each other, independent of how short the time difference is. Furthermore, this perfect detector would provide a signal which unambiguously reveals the number of photons which arrived at the detector at the same time. Moreover, the detector would never create any output signal in the absence of light. The following performance characteristics allow to compare real world detectors against this theoretical ideal detector.

### 4.2.5.1 Count rate, dark count rate and detection efficiency

Count rate and dark count rate are terms used throughout this chapter and therefore we shortly define how both terms are used within this thesis. A *count* is the event of registering an electrical pulse which heralds photodetection. This occurs when the electrical output pulse of the detector surpasses a defined threshold voltage of the counting electronics. When operating a detector for a certain time $t_{\text{meas}}$, a total number of counts $N_{\text{meas}}$ is registered and the ratio $CR = \frac{N_{\text{meas}}}{t_{\text{meas}}}$ is referred to as the count rate. Besides counts which are caused by the photons which are to be measured, there exist also false counts, referred to as *dark counts*, which are not caused by the photons of interest. Dark counts can for example be caused by electrical noise or by photons from unwanted sources, such as thermal radiation from relatively hot materials, which form part of the cryogenic measurement setup. While the term count rate refers to the number of all registered electrical pulses per unit time (irrespective of them being *real counts* or *dark counts*), the dark count rate (DCR) refers to the number of false counts per unit time. The dark count rate for a detector at certain operation conditions can be determined by disconnecting the light source of interest, covering the optical access to the detector and measuring the remaining count rate caused by all other origins. The detection efficiency $\eta$ of a SPD denotes the overall probability of registering an electrical pulse once a photon arrives at the detector.

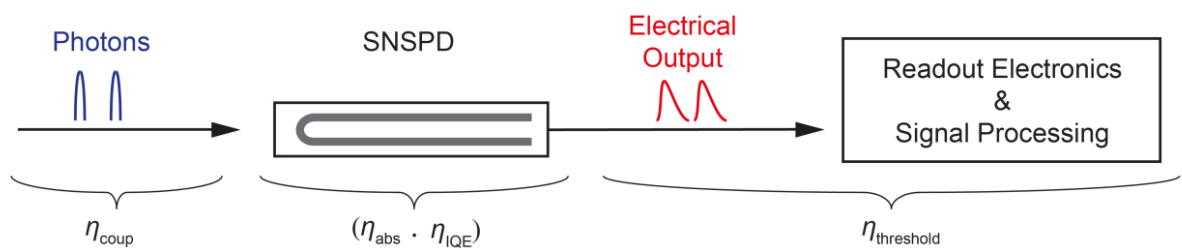

**Figure 50 - Illustration of the decomposition of the system detection efficiency**: $\eta_{\text{SDE}}$ is a product of the coupling efficiency $\eta_{\text{coup}}$, the absorption efficiency $\eta_{\text{abs}}$, the internal quantum efficiency $\eta_{\text{IQE}}$, and the threshold efficiency $\eta_{\text{threshold}}$.

The detection efficiency can be decomposed into a product of four distinct efficiencies, as illustrated in Figure 50, namely:

1) The coupling efficiency $\eta_{\text{coup}}$, which denotes the probability of a photon from a light source to arrive at the location of the detector. In the framework of PICs this consists of both the efficiency of coupling to a waveguide (for example via a focusing grating coupler) and the propagation losses along the waveguide, where photons can be absorbed or scattered.

2) The absorption efficiency $\eta_{\mathrm{abs}}$, which denotes the probability of the photon to be absorbed in the detector's active material.

3) The internal quantum efficiency $\eta_{\mathrm{IQE}}$, which for SNSPDs denotes the probability of an absorbed photon to lead to a breakdown of superconductivity and subsequently to a detectable electrical output pulse.

4) Besides these three contributions, which are determined by different physical properties of the detector, the experimental detection efficiency is also influenced by the efficiency with which the output electrical signal is registered by external electronics, referred to as threshold efficiency $\eta_{\mathrm{threshold}}$. The threshold efficiency is typically very close to unity[31]. This is also the case for the counting electronics used within this thesis and we will assume $\eta_{\mathrm{threshold}} = 1$.

The product of these four efficiencies is referred to as system detection efficiency (SDE) $\eta_{\mathrm{SDE}}$[286]:

$$\eta_{\mathrm{SDE}} = \eta_{\mathrm{coup}} \cdot \eta_{\mathrm{abs}} \cdot \eta_{\mathrm{IQE}} \cdot \eta_{\mathrm{threshold}} \ . \tag{4.7}$$

SNSPDs are operated at a certain bias current $I_{\mathrm{B}}$, and while $\eta_{\mathrm{coup}}$ and $\eta_{\mathrm{abs}}$ are independent of the bias current, $\eta_{\mathrm{IQE}}$ does depend on $I_B$. The system detection efficiency is a good figure of merit when using a stand-alone detector system, where the user is interested in how many of the photons that are sent into the fiber are registered by the detector. In the context of on-chip detectors for integrated quantum optics, the coupling efficiency is of minor importance as the challenges of minimizing both the fiber-to-waveguide coupling losses and the propagation losses of the waveguides are independent of the SPD development. Furthermore on the long-term the photon sources will be on the same chip, making fiber-to-waveguide coupling unnecessary. In the context of waveguide-integrated SNSPDs we determine the coupling losses in the measurement but do not consider them as part of the detector's efficiency. Therefore the on-chip detection efficiency (OCDE) is used as a characteristic[65], consisting of the absorption efficiency and the internal quantum efficiency:

$$\mathrm{OCDE}(I_{\mathrm{B}}) = \eta_{\mathrm{abs}} \cdot \eta_{\mathrm{IQE}}(I_{\mathrm{B}}) \ . \tag{4.8}$$

To simplify the notation throughout this thesis the term detection efficiency is therefore used as a synonym for OCDE with the notation $\eta = \mathrm{OCDE}$. The detection efficiency of a SPD at a certain wavelength can be determined using a monochromatic light source which directs light of power $P_{\mathrm{opt}}$ onto the detector. As explained above, the coupling efficiency is not considered here, hence $P_{\mathrm{opt}}$ refers to the optical power at the location where light first reaches the detector. For waveguide-integrated SNSPDs this occurs at the location of the tip with the 180°-turn. The power $P_{\mathrm{opt}}$ can be converted into the rate of photons per unit time arriving at the detector, called photon flux $\phi = \frac{P_{\mathrm{opt}}}{E_{\mathrm{photon}}}$, where $E_{\mathrm{photon}}$ denotes the energy of one photon. For clarity of the explanations we consider light as a train of single photons[22], which arrive at the detector, as shown in Figure 51 a).

---

[22] A train of single photons would be the output of an ideal triggered single-photon source, which might experimentally not be available. Considering such a source enables simpler and more intuitive explanations in this section and future quantum optical circuits are expected to include triggered single-photon sources, as explained in section 2.2.2.

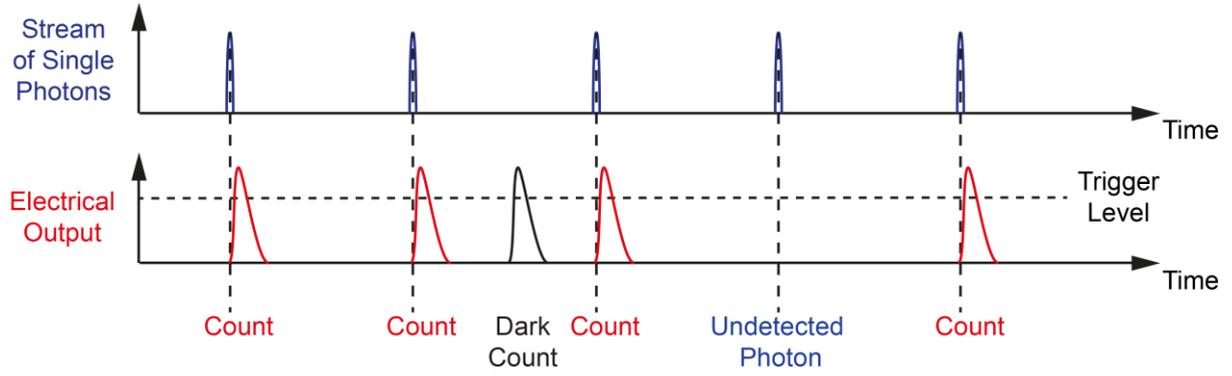

**Figure 51 - Illustration of the detection efficiency:** The detector input (top row) consists of a stream of single photons. The detector's electrical output pulses (bottom row) are registered by the readout electronics if a pulse crosses the trigger level. Output pulses, referred to as counts, can either be due to correctly registering a photon (counts in red color) or caused by noise (dark counts in black color). Undetected photons, due to a detection efficiency below 100% occur as well. Schematic adapted from Migdall *et al.*[31].

As one photon never causes more than one electrical output pulse[23], the detection efficiency $\eta(I_{\mathrm{B}})$ at a certain bias current $I_{\mathrm{B}}$ can be calculated as the ratio of the count rate $CR$, corrected for the dark count rate $DCR$, to the photon flux $\phi$ arriving at the detector:

$$\eta(I_{\mathrm{B}}) = \frac{CR(I_{\mathrm{B}}) - DCR(I_{\mathrm{B}})}{\phi} \leq 1 \ . \tag{4.9}$$

A detection efficiency smaller than unity can have different causes: A photon not being detected can be either due to the fact that (1) it was not absorbed in the nanowire, (2) the absorption did not lead to a breakdown of superconductivity or (3) the breakdown was not detected, because the pulse height was below the trigger level. If the photons do not arrive with a constant spacing in time, as in the simplified case illustrated here, then it is also possible that (4) several photons arrive within the recovery time of the detector, where the efficiency is reduced and no further detection pulse might occur, as will be explained in more detail in section 4.2.5.5, or that several photons arrive at the same time, leading to one electrical output which might not reveal that more than one photon was absorbed (the lack of photon number resolution), a potential problem that will be discussed in section 4.2.5.6. By making sure that reasons (3) and (4) are negligible in the measurement, the detection efficiency can be determined correctly.

The closer the bias current is to the critical current, the more likely it is that an absorbed photon will lead to a breakdown of superconductivity, and the general dependence of the detection efficiency on the bias current resembles a sigmoid curve, as shown in Figure 52 a). Close to the critical current, the detection efficiency saturates and becomes almost independent of the bias current. Such a saturated detection efficiency is often referred to as *the detection plateau* and is a sign of high internal efficiency[287–289]. Following the suggestion of Najafi *et al.*[287] one can introduce a saturation metric $S$ for quantifying the length of the plateau region as

$$S = \frac{I_{\mathrm{C}} - I_{\mathrm{B}}(0.9 \cdot \eta(I_{\mathrm{C}}))}{I_{\mathrm{C}}} , \tag{4.10}$$

---

[23] If more than one count can be caused by the absorption of one photon, an effect called *afterpulsing*, then this calculation of the efficiency would not be correct, but fortunately afterpulsing (opposed to single-photon detection with semiconductor single-photon avalanche diodes) is not a problem in well-operating SNSPDs[341].

where $I_B(0.9 \cdot OCDE(I_C))$ is the bias current at which the on-chip detection efficiency reaches 90% of its maximum value. As illustrated in Figure 52 a), for example a value of $S = 0.3$ expresses that for currents between $0.7 \cdot I_C$ and $I_C$ the detector's efficiency is within 90% of its maximum value.

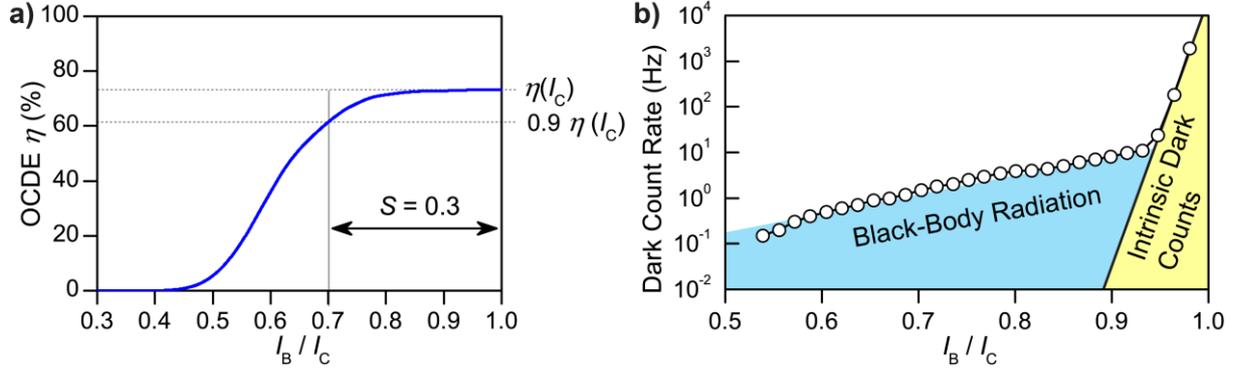

**Figure 52 - Bias current dependence of the detection efficiency and dark count rate**: a) Schematic of the on-chip detection efficiency as a function of normalized bias current, which follows a sigmoid curve, with a saturating detection efficiency at bias currents close to the critical current $I_C$. b) Schematic of dark count rate as a function of normalized bias current. At comparably low current dark count rate is dominated by counts due to blackbody radiation. At bias currents close to the critical current intrinsic dark counts, not related to thermal photons, dominate.

Figure 52 b) shows a schematic of dark count rate as a function of normalized bias current. At low bias currents the dark counts are caused by photons from blackbody radiation[290] and are hence not intrinsic dark counts, but rather unwanted real detection events caused by unwanted photons. Thermal photons from setup components at room temperature can propagate within the core and cladding of the optical fibers of the fiber array, reach the SNSPD and their absorption can trigger false counts. Following Planck's law, the thermal spectrum at $300\,K$ shows a maximum spectral emissive power at $\approx 10\,\mu m$ wavelength, but also photons at visible and near-infrared wavelength are emitted, for which the SNSPD detection efficiency is high. Dark counts caused by blackbody radiation can generally be reduced by appropriate spectral filtering of the light propagating within the optical fibers.[290]

At bias currents close to the critical current intrinsic dark counts dominate. According to recent studies comparing experimental data and different theoretical models[266], intrinsic dark counts at low temperatures are probably mainly caused by vortex-induced phase slips[29]. Magnetic vortices can cross a superconducting wire under the influence of the Lorentz force imposed by the bias current $I_B$. These leave behind a normal-conducting domain across the width of the wire, a vortex-induced phase slip, which can lead to a dark count.[267]

### 4.2.5.2 Noise-equivalent power

The noise-equivalent power (NEP) denotes the impinging optical power which is required for gaining a signal-to-noise ratio of unity after integrating the detector's output signal for one second[260]:

$$NEP(I_B) = \frac{E_{photon}}{\eta(I_B)}\sqrt{2 \cdot DCR(I_B)}\,, \tag{4.11}$$

where $E_{photon}$ denotes the energy of a photon, $\eta(I_B)$ denotes the detector's bias current dependent efficiency and $DCR(I_B)$ denotes the bias current dependent dark count rate. The NEP is often used as a figure of merit for single-photon detectors as it quantifies the smallest amount of optical power which can still be distinguished from noise. The lowest possible NEP is desirable for a detector. Besides the efficiency and the dark count rate, no further measurements are required for determining the NEP. For waveguide-integrated SPDs, typically an on-chip NEP value is given, which uses the OCDE value as the efficiency value, for the same reasons as explained in the section on detection efficiency. Throughout this thesis, NEP always refers to on-chip NEP, which is useful in combination with on-chip light sources.

### 4.2.5.3 Spectral range

The spectral range of a detector denotes the wavelength range of electromagnetic radiation for which the absorption of a photon can lead to an electrical output signal. The spectral range depends on the physical process which underlies the photodetection. A common SPD is the silicon based single-photon avalanche diode (SPAD) which can only detect light with wavelengths up to 1.1 µm, due to the electronic bandgap of silicon. The spectral range of a specific SNSPD depends on the detector material, geometry and operation conditions. It has been shown that SNSPDs generally are able to detect single photons from the ultraviolet up to wavelengths of at least 5.5 µm[292] in the mid-infrared spectral region.

### 4.2.5.4 Timing jitter

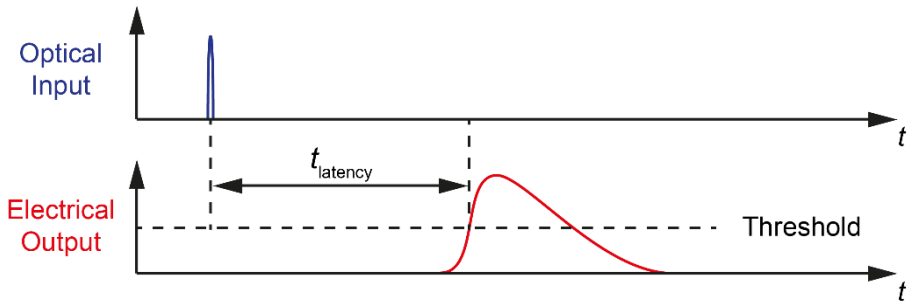

**Figure 53 - Illustration of the timing latency $t_{\text{latency}}$**: the time between optical input and when the resulting electrical output pulse crosses a given threshold level. Schematic adapted from Migdall *et al.*[31].

The timing jitter quantifies the uncertainty in how accurate it is possible to measure the arrival time of photons. The time which elapses between when a photon is incident on the detector (optical input) and when the resulting electrical output pulse crosses a given threshold level is called the timing latency $t_{\text{latency}}$, as illustrated in Figure 53. The timing latency varies for different detection events and this variation is quantified in a measure called the timing jitter. SNSPDs show very low timing jitter, which is a distinct advantage over other SPD schemes[260] and applications such as time-correlated single-photon benefit from a low jitter. The mechanism of timing jitter in SNSPDs is little understood. Recent studies show that at constrictions of the nanowire higher current densities occur. Photons that are absorbed at locations with higher current density are detected with higher efficiency and it has been shown that an absorption of photons at locations of constrictions and hence with higher efficiency leads to output pulses with shorter timing latency[293]. Thus variations in nanowire diameter lead to variations in timing latency, which in turn increases the timing jitter. Furthermore electrical noise in the readout circuitry increases the timing jitter. Therefore improving the signal-to-noise ratio, by increasing the magnitude of the output pulse or by reducing the electrical noise, improves the timing jitter. In the experimental section 4.4.6 we will explain how to measure the timing jitter of SPDs.

### 4.2.5.5 Recovery time

The recovery time $t_{\text{recovery}}$ of a detector is the time after a detection event for which the detector does not show its full detection efficiency $\eta$. As illustrated in Figure 54, it consists of the dead time and the reset time:

$$t_{\text{recovery}} = t_{\text{dead}} + t_{\text{reset}} \,, \tag{4.12}$$

where the dead time $t_{\text{dead}}$ is the duration of time during which the detector is incapable of producing an output signal in response to an additional incident photon, which means that the detection efficiency is zero during this dead time.

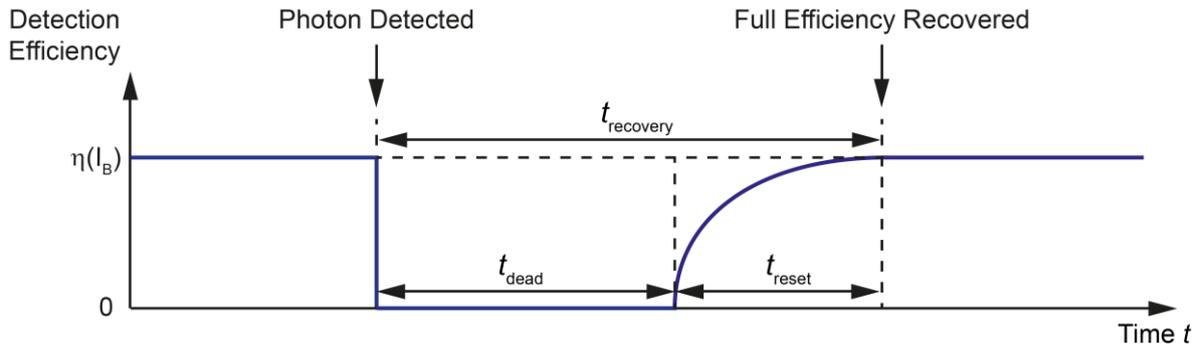

**Figure 54 - Illustration of the recovery time of a single-photon detector**, consisting of the dead time during which the efficiency is zero and a reset time during which the efficiency recovers from zero back to its initial value $\eta(I_{\text{B}})$.

The reset time $t_{\text{reset}}$ is the time during which the detection efficiency recovers from zero back to its initial value. For a superconducting nanowire, the recovery of $\eta$ is not limited by the dissipation of energy to the environment and the recovery of superconductivity. It is rather limited by the recovery of the bias current to its initial value which happens on a timescale limited by the kinetic inductance of the nanowire[294], as introduced in section 4.2.3. After *a click* of the detector, the bias current recovers back to its initial value $I_B$ and the time dependence shows an exponential behavior of

$$I(t) = (1 - e^{-\frac{t}{\tau_2}}) \cdot I_{\text{B}} \,, \tag{4.13}$$

with a time constant $\tau_2$, as introduced in section 4.2.3. As the shape of the detector's output pulse shows an exponential decay with the same time constant, it is possible to extract $\tau_2$ from an exponential fit to the output pulse and this value is often quoted as the detector's recovery time in the literature[65]. It has to be noted that after the time $\tau_2$ has elapsed, only 63% of the initial current has recovered and at this current in most cases the efficiency is significantly lower than the efficiency at the full bias current $I_B$ and therefore $\tau_2$ is not a good approximation for the recovery time of a detector. In a strict sense due to the exponential behavior there exists no value for the time after which the full current is recovered and a more practical definition for a recovery time is needed. A time of $3 \cdot \tau_2$ after which 95% of the current has recovered has been suggested as a better approximation to the SNSPDs recovery time.[279,295] A more accurate evaluation necessitates to take into account both the accurate dependence of the efficiency on the bias current $\eta(I)$ and the time dependence of the current $I(t)$ in the nanowire after a detection event. This enables to determine the time $t^*$ after which a certain fraction $A$ of the initial efficiency $\eta(I_B)$ has recovered[294]:

$$\eta\big(I(t^*)\big) = A \cdot \eta(I_B) \,. \tag{4.14}$$

The choice of $A$ is arbitrary and there seems to be no convention in the literature yet. A value of $A = 90\%$ has been suggested as a potential convention[294] and we will use this value in our data analysis. In the experimental section 4.4.7, we will estimate the detector recovery times: We will determine the time constant $\tau_2$ which governs the recovery of the bias current for different detector geometries of SNSPDs on diamond waveguides. Additionally we will estimate the time $t^*$ after which 90% of the initial efficiency has recovered and compare both approaches. The recovery times of SNSPDs are typically on the order of a few nanoseconds.

### 4.2.5.6 Energy resolution / photon number resolution

Many SPDs provide a binary output signal which enables only to distinguish between "no photons have been detected" and "photons have been detected", but the energy of the detected light or the number of photons cannot be determined from the detector signal. A scheme where the detector's output signal is dependent on the energy of the absorbed photons such that (in case of monochromatic light) the number of photons can be unambiguously determined from the detector signal, is called a photon number resolving (PNR) detector. Transition-edge sensor (TES) are superconducting detectors which are energy resolving and can therefore act as true PNR detectors. TES are microcalorimeters, where the amplitude of the output pulse is proportional to the energy of the detected light[260], which in case of monochromatic light can be converted into the number of detected photons. TES can be integrated on top of waveguides[285,296], but these detectors are relatively slow (µs recovery times) and need to be operated at millikelvin temperatures[297,298]. Opposed to that, SNSPDs do not feature photon number resolution, as the output pulse of SNSPDs is not dependent on the number of absorbed photons. From the performance characteristics of SPDs, this is the only characteristic for which SNSPDs show a disadvantage compared to other technologies.

Temporal or spatial multiplexing, for example by distributing an amount of not-PNR detectors at the output of a series of beam splitters, enables to draw some conclusions concerning the number of photons which are introduced into the series of beam splitters. Such multiplexing only provides useful information if the number of detectors is much larger than the number of incoming photons. Therefore statistical information on the distribution of photon numbers in the incoming light is needed and such multiplexing schemes[299–301] have to be clearly distinguished from real photon number resolution.

It should be noted that quantum optical applications exist for which photon number resolving detectors are often not required. For example boson-sampling, as introduced in section 2.2.2, does not necessarily require PNR detectors, as explained by Gard *et al.*, "because the number of modes scales quadratically with the number of photons, for large systems we are statistically guaranteed that all photons will arrive at different output modes"[47]. It is therefore reasonable to develop SNSPD detectors for integrated quantum optical circuits besides their lack of energy or photon number resolution.

## 4.3 SNSPDs on diamond: layout and room temperature characterization

Efficient on-chip single-photon detectors compatible with the diamond PICs presented in the previous chapters are important as a missing counterpart to single-photon sources based on color centers in diamond, as explained in section 2.3. We design superconducting nanowire single-photon detectors integrated with polycrystalline diamond PICs for two wavelength regimes, namely around 1600 nm and 765 nm. We investigate detectors at a wavelength of 765 nm, because this is close to the emission wavelength of one of the most promising single-photon sources in diamond, namely the silicon vacancy center. On the other hand efficient single-photon detection at telecommunication wavelength is more challenging than at visible wavelengths[286] and SNSPDs at 1600 nm act as a benchmark for comparison to other single-photon detectors, especially SNSPDs on other substrates. We focus on two types of detectors: on one hand short nanowire detectors, which have short recovery times and hence enable high detection rates, and on the other hand long SNSPDs which feature high absorption efficiency, a necessary requirement for a high detection efficiency.

### 4.3.1 Nanowire geometry

A central part of the design of SNSPDs is the geometry of the nanowire. A sketch of a superconducting nanowire is shown in Figure 55 a), indicating the main geometric design parameters: the NbN nanowire (gray) resides on top of the diamond waveguide (blue). It has a thickness of $d = 4$ nm and consists of two parallel stripes of width $w$ on the order of 100 nm and a length $L$ on the order of several tens of micrometers. These two stripes are separated by a gap of size $g = 100$ nm and connected by a 180° circular bend which constitutes the detector's tip at which photons arrive first. The refractive index of NbN thin films depends strongly on the wavelength, but also depends on the film thickness and the deposition conditions.[264] For a 4 nm thick NbN layer the refractive index is about $n = 5.33 + 5.90i$ at 1600 nm and $n = 3.19 + 4.15i$ at 775 nm[302], where the imaginary part corresponds to the absorption of light. For light propagating in a waveguide the absorption per length depends on the overlap of the mode with the absorptive material, which can be expressed as the effective refractive index $n_{\text{eff}}$ of the waveguide modes, as introduced in section 2.1.1.

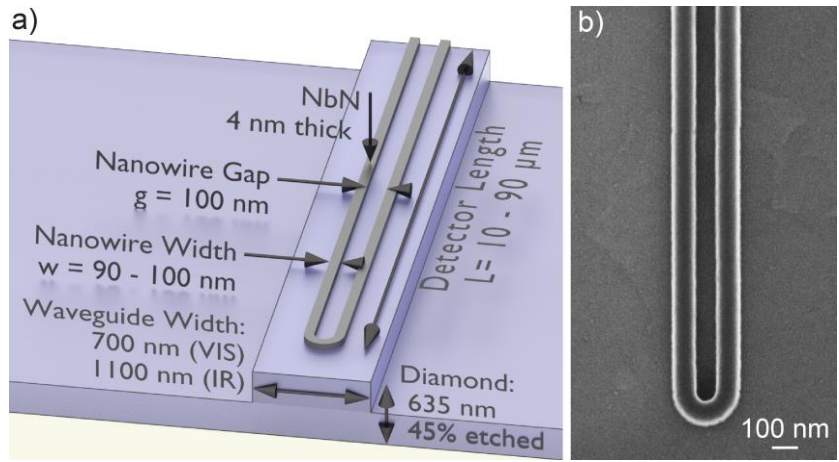

**Figure 55 - SNSPD geometry:** a) Sketch waveguide-integrated SNSPD, consisting of the NbN nanowire (gray) on top of a partially etched diamond photonic waveguide (blue). The main geometric parameters are indicated. Note that the HSQ layers which are covering the nanowire and the waveguide are omitted for clarity of the schematic. b) SEM micrograph showing the HSQ resist which defines the nanowire geometry.

Figure 55 b) shows a SEM micrograph of the HSQ resist, which defines the geometry of the underlying NbN nanowire of 100 nm width. The nanowire width is one of the parameters which determine if the energy of one photon will be large enough to lead to a breakdown of superconductivity and a subsequent detection event. Hence the nanowire needs to be narrow enough for high detector efficiency, but the trade-off is that a smaller width leads to a smaller critical current and in turn to a smaller electronic output pulse, which makes the electronic readout more challenging. Furthermore, the reliable fabrication of thin wires is more demanding, as small absolute variations of some nanometers in wire width translate into a larger relative variation, which results in incisions which limit the critical current for the entire wire. For the SNSPDs on diamond we chose nanowire widths of 90 nm and 100 nm and a gap of 100 nm, as waveguide-integrated SNSPDs from NbN with the same corresponding cross-sections showed good performance at telecom wavelengths on substrates such as silicon[65] and silicon nitride[303].

We simulate the guided modes in diamond rib waveguides for both wavelengths and choose the waveguide widths such that for TE-like polarization only one spatial mode exists. We note that after fabricating the waveguide, the nanowire and the waveguide itself are covered by HSQ resist (not shown in Figure 55 a), which we consider in the FEM simulation with a resist thickness of 400 nm and we assume the same refractive index as for the fused silica below the diamond layer. For a 635 nm thick diamond layer and 45% relative etch depth the simulations yield the following widths for which exactly one TE-like mode exists: $w = 700$ nm for 765 nm wavelength and $w = 1100$ nm for 1600 nm wavelength. The evanescent field of the waveguide mode couples to the NbN nanowire which results in the absorption of photons in the NbN strip. Figure 56 shows the mode profiles of the TE-like mode and the corresponding effective refractive indices for the waveguide at 1600 nm without NbN nanowire and with a nanowire of 100 nm width. Due to the large complex refractive index of NbN, the electric field at the nanowire is greatly enhanced, as can be seen in the magnified image (c), which leads to the desired enhanced absorption of photons in the NbN nanowire.

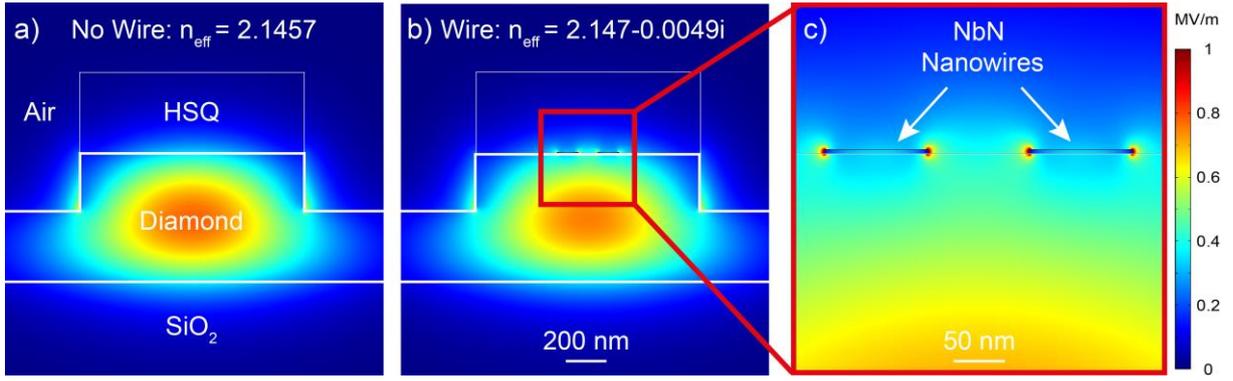

**Figure 56 - Guided TE-like modes at 1600 nm**: Spatial distribution of the electric field norm for 1 mW optical power in the TE-like mode of a waveguide a) without NbN nanowire and b) with a nanowire of 100 nm width and 100 nm gap between the two NbN strips. c) Enlarged image of the region of the NbN nanowires, as indicated by the red box in b). The waveguide dimensions are: 635 nm diamond layer, 45% relative etch depth, 1100 nm waveguide width. The color scale for the electric field strength on the right side applies to all three subfigures.

The absorption coefficient $\alpha$ (in dB per length unit) can be calculated from the imaginary part of the refractive index, according to equation (2.3), as[304]

$$\alpha = \frac{2\pi}{\lambda} \cdot \frac{20}{\ln(10)} \cdot \mathrm{Im}(n_{\mathrm{eff}}) \;. \tag{4.15}$$

Table 2 presents the simulation results for the refractive indices and the corresponding absorption coefficients for the four combinations of nanowire widths and wavelengths. Note that the absorption coefficients at 1600 nm are much larger than at 765 nm, such that for a 90 nm wide nanowire on the given waveguide geometry after 80 µm nanowire length 93% of the light is absorbed at 1600 nm, while 70% of the light is absorbed for a wavelength of 765 nm.

**Table 2 - Simulated refractive indices and absorption coefficients** for NbN nanowires of 90 nm and 100 nm width on diamond rib waveguides (635 nm of diamond, 45% etched) at two different wavelengths (1600 nm and 765 nm).

| Wavelength | Waveguide width | Detector width | Refractive index | Absorption coefficient |
|:---:|:---:|:---:|:---:|:---:|
| 1600 nm | 1100 nm | 90 nm | $2.1473 - 0.004269i$ | 0.1456 dB/µm |
| 1600 nm | 1100 nm | 100 nm | $2.1471 - 0.004943i$ | 0.1686 dB/µm |
| 765 nm | 700 nm | 90 nm | $2.3143 - 0.0009135i$ | 0.0652 dB/µm |
| 765 nm | 700 nm | 100 nm | $2.3142 - 0.0009746i$ | 0.0695 dB/µm |

Back reflection of photons at the interface between a bare waveguide and a waveguide with an NbN wire on top is a possible limiting factor for the detection efficiency of the detector. We estimate the upper bound of this reflection by assuming a sudden change in the effective refractive index of the guided mode between the bare waveguide ($n_{\mathrm{eff}} = 2.1457$) and the waveguide with a 100 nm wide nanowire ($n_{\mathrm{eff}} = 2.1471 - 0.004943i$) at 1600 nm.

Using Fresnel's equations, we find a very small reflection coefficient $R$ as

$$R = \frac{P_{\text{ref}}}{P_{\text{in}}} = 1.4 \cdot 10^{-6} \, , \tag{4.16}$$

where $P_{\text{ref}}$ and $P_{\text{in}}$ are the reflected and the incident power, respectively. The same holds for the case of a different nanowire width or a different wavelength and we therefore neglect back reflection in the following. Following the simulation, we design and fabricate a chip with PICs which contains devices for confirming the absorption coefficients, as well as devices with electrically connected nanowires which work as single-photon detectors. The following sections describe the layouts and the measurement results for both types of PICs.

## 4.3.2  NbN absorption measurement

In order to measure the absorption in NbN nanowires, we design and fabricate integrated PICs, which allow for balanced optical detection, as shown in Figure 57.

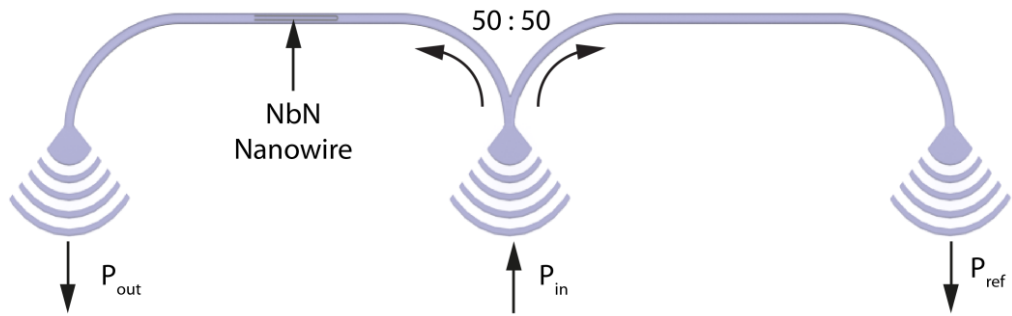

**Figure 57 - Absorption measurement:** Schematic of a PIC for balanced detection of the absorption of an NbN nanowire placed on top of a diamond rib waveguide. The unetched diamond layer is omitted for clarity of the schematic. The transmitted optical power $P_{\text{out}}$ after partial absorption by the nanowire (left side) is compared to the optical power $P_{\text{ref}}$ transmitted through a reference circuit (right side).

An input grating coupler transfers light of power $P_{\text{in}}$ from a fiber to the TE-like waveguide mode. The grating coupler is connected to a 50/50 Y-splitter, which distributes half of the light towards a reference port (right side) and half of the light towards the nanowire, where the propagating mode is attenuated by absorption in the NbN (left side). By dividing the optical power $P_{\text{out}}$ at the transmission port by the power $P_{\text{ref}}$ at the reference port, the attenuation due to the nanowire can be determined. As estimated above, scattering can be neglected and we therefore consider the relative attenuation to be purely due to absorption in the NbN. By using a symmetric design with identical waveguide lengths and grating couplers, the propagation loss and coupling loss do not contribute to the measurement of the absorption coefficient.

We fabricate PICs with nanowires of 100 nm width for various lengths, both for 1600 nm and 765 nm. We note that the length of the NbN nanowire is about twice as long as the device length, as it consists of two parallel NbN stripes which are electrically connected in series via a 180°-turn. We measure the absorption for all fabricated devices at room temperature. Light from a tunable laser source is coupled into the central waveguide, and the transmitted powers are recorded at both output ports simultaneously. Figure 58 shows the corresponding average measured absorption for 100 nm wide NbN nanowires in dependence of device length at 1550 nm and 765 nm. The absorption shows the expected exponential decay with increasing wire length. We extract the absorption in decibels per micrometer from fits to the obtained data. For devices with 100 nm width the absorption coefficient is 0.203 ± 0.009 dB/µm at 1600 nm. This is slightly larger than the value of 0.1686 dB/µm expected from simulations. We attribute this to the uncertainty concerning the thickness and the refractive index of the NbN layer and the width of the nanowire which is buried below the resist. At 765 nm (Figure 58 b) the linear fit to the data yields an absorption coefficient of 0.103 ± 0.009 dB/µm, which is slightly larger than the value of 0.0695 dB/µm expected from simulations (see Table 2). We attribute differences between simulation and measurement to the uncertainties in nanowire and waveguide geometry, due to fabrication tolerances and to uncertainties in differences between literature values and experimental values for both the refractive index of diamond and NbN.

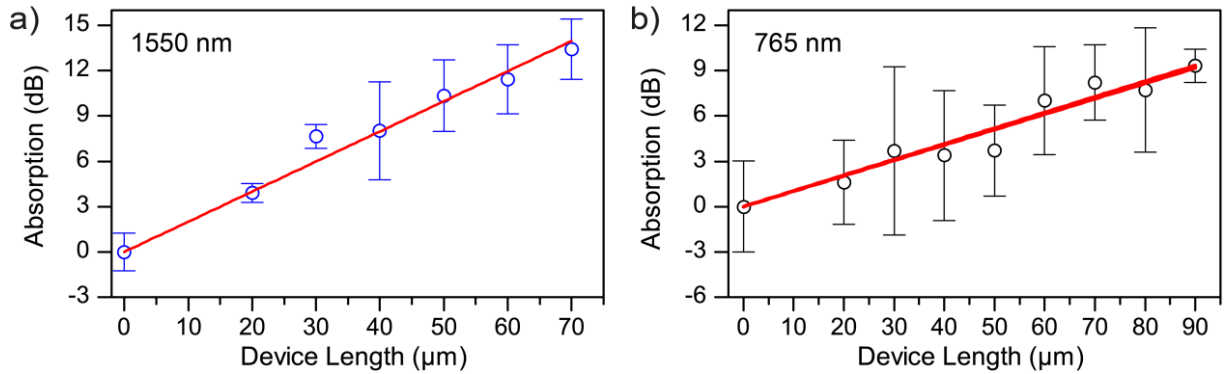

**Figure 58 - NbN absorption measurement at room temperature:** Absorption due to 4 nm thick, 100 nm wide NbN nanowires for various device lengths. a) Absorption at 1550 nm: A linear fit reveals an absorption coefficient of 0.203 ± 0.009 dB/µm. b) Absorption at 765 nm: A linear fit reveals an absorption coefficient of 0.103 ± 0.009 dB/µm.

It is important to note that the absorption coefficient at room temperature (300 K) agrees within the measurement uncertainty with the absorption coefficient at cryogenic temperatures, as we confirm by repeating the described measurement in a cryogenic setup (see appendix A4). This means that for future detector optimizations it is not necessary to repeatedly perform absorption measurements at cryogenic temperatures, as it is possible to predict the absorption efficiency of superconducting nanowire detectors from room temperature absorption measurements.

### 4.3.3 Detector circuit layout and fabrication

The design of our waveguide-integrated SNSPD is shown in Figure 59 a) and consists of the following components: A focusing grating coupler is used to couple light with power $P_{in}$ from a tunable laser source into a diamond waveguide. A Y-splitter acts as a 50/50 beam splitter where at the right output 50% of the light propagates along a waveguide to a second grating coupler which couples light out of the chip into a second optical fiber, where the output power $P_{out}$ is determined with an external photodetector. This transmission measurement enables the measurement of the coupling efficiency and hence the determination of the photon flux inside the PIC. Photons exiting at the left output of the beam splitter propagate towards the waveguide-integrated SNSPD (shown in cyan color) which is electrically connected to gold electrode pads, visible at the top side of Figure 59 a).

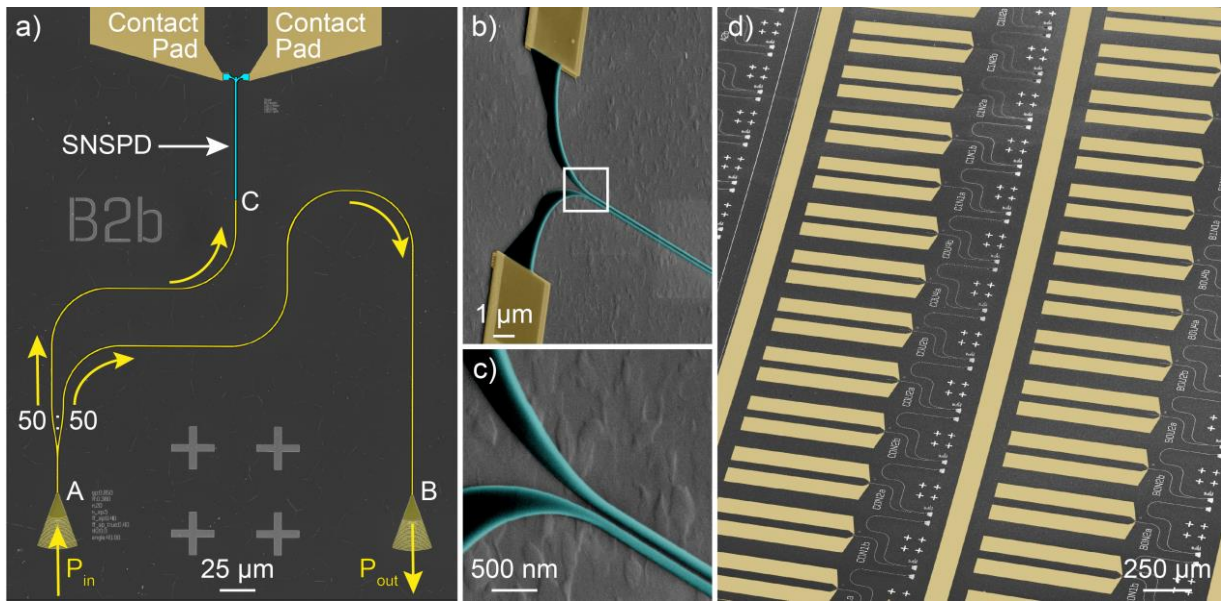

**Figure 59 - Circuit and chip layout:** a) Colorized SEM image of a photonic circuit designed for characterizing a waveguide-integrated SNSPD. b) Colorized SEM micrograph, taken under an angle of 45°, showing the nanowire geometry of the SNSPD written into negative resist (HSQ 6%) by electron beam lithography. c) Enlarged section of the SEM image of the nanowire, corresponding to the white rectangle in b). d) Colorized SEM micrograph showing a larger region of the photonic chip which contains hundreds of waveguide-integrated SNSPDs.

We fabricate a chip which contains both the PICs for absorption measurements and the PICs featuring waveguide-integrated SNSPDs via three steps of electron beam lithography: First the metal contact pads and cross-shaped alignment markers are structured from chromium and gold via a lift-off process. Next the nanowire geometries are written into HSQ (6%) negative resist, with an overlap with the metal electrode for electrical connection, as shown in Figure 59 b). The nanowire width increases at the location of the curve towards the electrode, as can be seen in the enlarged SEM image of Figure 59 c). This makes the nanowire design unsusceptible to deviations in the curve profile, which would otherwise limit the critical current of the nanowire. The nanowire geometries are then transferred from HSQ into NbN via dry etching. In the final step the PICs are written into HSQ (15%) negative resist and after resist development transferred into diamond via dry etching, resulting in hundreds of PICs containing SNSPDs, as shown in Figure 59 d). A detailed description of all fabrication steps can be found in appendix A2. The photonic chip contains two sets of waveguide-integrated SNSPDs, one each for wavelengths of 1600 nm and 765 nm. The respective sets of devices differ only in the geometry of the grating couplers and the waveguide widths ($w = 1100$ nm for 1600 nm and $w = 700$ nm for 765 nm wavelength), but are equal in all other design parameters.

While this general layout for waveguide-integrated SNSPDs has been used in most demonstrations of waveguide-integrated SNSPDs to date, one quantity which is needed for the determination of the efficiency of the detector has to be determined in a separate measurement: the propagation loss of the waveguides. As the propagation loss depends on the quality of the materials and the fabrication process, the propagation loss should be measured on the same chip as the SNSPDs. This is possible using additional PICs, such as waveguides of different lengths or PICs which include ring resonators.[147] Avoiding the need for additional PICs is especially useful when determining detection efficiencies at several wavelengths, as the propagation loss is highly dependent on the wavelength. This is possible by designing the PIC in the right fashion: If the length from input to output coupler $L_{\mathrm{ref}}$ (section A–B in Figure 59 a) is designed to be twice the distance between input coupler and detector $L_{\mathrm{det}}$ (section A–C), the propagation loss does not contribute to the calculation of the detector's efficiency, as will be explained in the following.

The on-chip detection efficiency of a detector is determined as the ratio of the number of detection events in a certain time period, called the count rate (CR), and the number of photons impinging on the detector during the same time period, called the photon flux $\phi$, as explained in section 4.2.5. The transmission $T$ through the photonic reference circuit amounts to

$$T = \frac{P_{\mathrm{out}}}{P_{\mathrm{in}}} = C^2 \cdot S \cdot \exp(-\alpha \cdot L_{\mathrm{ref}}) \, , \tag{4.17}$$

where $P_{\mathrm{in}}$ is the laser power arriving at the input coupler, $P_{\mathrm{out}}$ is the laser power measured after transmission at the output coupler, $C$ is the coupling efficiency of one grating coupler, $S = 0.5$ is the splitting ratio of the Y-Splitter, $\alpha$ is the attenuation coefficient of the waveguide, and $L_{\mathrm{ref}}$ is the length of the waveguide between the two grating couplers (section A–B in Figure 59 a). The photon flux arriving at the detector is given by

$$\phi = \frac{P_{\mathrm{in}}}{E_{\mathrm{ph}}} \cdot C \cdot S \cdot \exp(-\alpha \cdot L_{\mathrm{det}}) \, , \tag{4.18}$$

where $E_{\mathrm{ph}} = \hbar\omega$ is the energy of a photon and $L_{\mathrm{det}}$ is the length of the waveguide between input coupler and SNSPD (section A–B). Because the circuit is designed with $L_{\mathrm{ref}} = 2 \cdot L_{\mathrm{det}}$, this means that

$$\sqrt{T} = C \cdot \sqrt{S} \cdot \exp(-\alpha \cdot L_{\mathrm{det}}) \tag{4.19}$$

and hence the photon flux can be expressed as

$$\phi = \frac{P_{\mathrm{in}}}{E_{\mathrm{ph}}} \cdot \sqrt{S} \cdot \sqrt{T} = \frac{P_{\mathrm{in}}}{E_{\mathrm{ph}}} \cdot \sqrt{0.5} \cdot \sqrt{\frac{P_{\mathrm{out}}}{P_{\mathrm{in}}}} \, . \tag{4.20}$$

It is therefore possible to determine the photon flux without quantifying $\alpha$, simply by measuring the optical input and output power. Additional photonic devices for determining the propagation loss on the same chip are therefore not needed. The same holds for potential bending losses in curved waveguides. By introducing twice as many quarter circles (four compared to two) within the reference waveguide, potential bending losses cancel each other out of the calculation of the photon flux and hence do not need to be determined using additional PICs. The detector devices are fabricated on the same photonic chip as the devices for measuring the NbN absorption, presented in section 4.3.2, such that the obtained absorption coefficients are also valid for the detectors. We fabricate both detectors with long SNSPDs which feature high absorption efficiency, a necessary requirement for a high detection efficiency, and short nanowire detectors, which enable high detection rates.

## 4.4 SNSPDs on diamond: cryogenic characterization

### 4.4.1 Cryogenic measurement setup

For the characterization at low temperature, the sample chip containing the SNSPDs is mounted on a stack of nano-positioners with four axes inside a liquid helium cryostat with a base temperature of 1.8 K. All measurements within this chapter are performed at this temperature, unless indicated differently. The nano-positioners' closed-loop positon readings enable the precise localization of each detector, without the need for visual access to the sample at low temperatures. Matching the device layout shown in Figure 59 a), an optical fiber array and an electronic RF probe are mounted facing each other, such that the grating couplers can be aligned underneath the fibers for maximum transmission and, at the same time, the probe tips can be electrically connected to the contact pads of the SNSPD. Figure 60 shows a schematic of the measurement setup. The optical setup consists of a tunable CW laser (either at wavelengths around 1600 nm or 765 nm, depending on the device under investigation) combined with two variable optical attenuators and a polarization controller. Measuring both the optical input power $P_{in}$ and the transmitted power $P_{out}$ simultaneously with an optical power meter enables to calculate the photon flux arriving at the on-chip detectors. The two contact pads of a SNSPD are electrically connected to a bias-T at room temperature. The DC port is connected to a stable voltage source via a 1 MΩ resistor, which provides the bias current $I_B$. The high frequency RF port is connected to two low-noise amplifiers[24] which together amplify the SNSPD output voltage pulses by 23.3 dB. Amplified electrical output pulses, which surpass a chosen threshold voltage, are subsequently either counted with an electrical pulse counter or recorded with a fast oscilloscope.

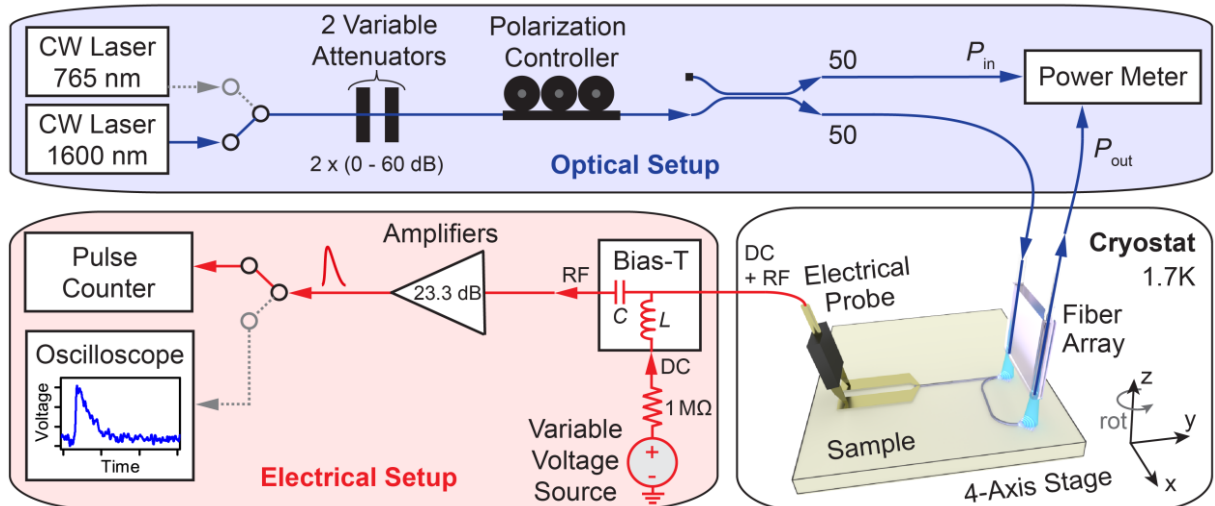

**Figure 60 - Cryogenic measurement setup:** The detector sample is mounted inside the liquid helium cryostat at temperatures down to 1 K and with fiber optical access (blue lines) and electrical access (red lines). Both the optical setup including laser sources, attenuators and polarization controller and the electrical setup including the bias voltage source for the SNSPDs and the readout electronics (bias-T, amplifiers, pulse counter and oscilloscope) are operated at room temperature.

---

[24] Mini-Circuits ZFL-1000LN+: Specification of 23.3 dB power gain (11.65 dB voltage gain) at a DC-voltage supply of 15 *V*.

### 4.4.2 Critical temperature and critical current

A high quality of the superconducting film is crucial for the SNSPD performance and the critical temperature $T_C$. The sheet resistance $R_S$ can be used as a metric for characterizing the superconducting films[287], where a high $T_C$ can be associated with high sheet current density[287] and possibly low timing jitter, while high $R_S$ could be associated with higher detector sensitivity due to larger Joule heating[305]. A high critical current $I_C$ of a fabricated nanowire on the other hand reveals that no defects or incisions are limiting the performance.

After deposition of the 4 nm thick NbN layer on the diamond film, the sheet resistance is determined by four-point probe measurement as $R_S = 400\ \Omega/\square$ at 300 K, comparable to high quality NbN layers of comparable thickness on other substrates.[264] We determine the critical temperature of the film by measuring the resistance of the film between two electrical probes while reducing the temperature. The obtained results are shown in Figure 61 a), where a distinctive transition is observed as the sample reaches the superconducting state. As the distance between the probes is arbitrary, the $y$-axis is normalized to the known sheet resistance at 300 K. The critical temperature is found to be $T_C = 11.7\ K$, comparable to values found for comparably thin NbN layers on other substrates.[287] Further details concerning the resistance characteristic can be found in appendix A3.

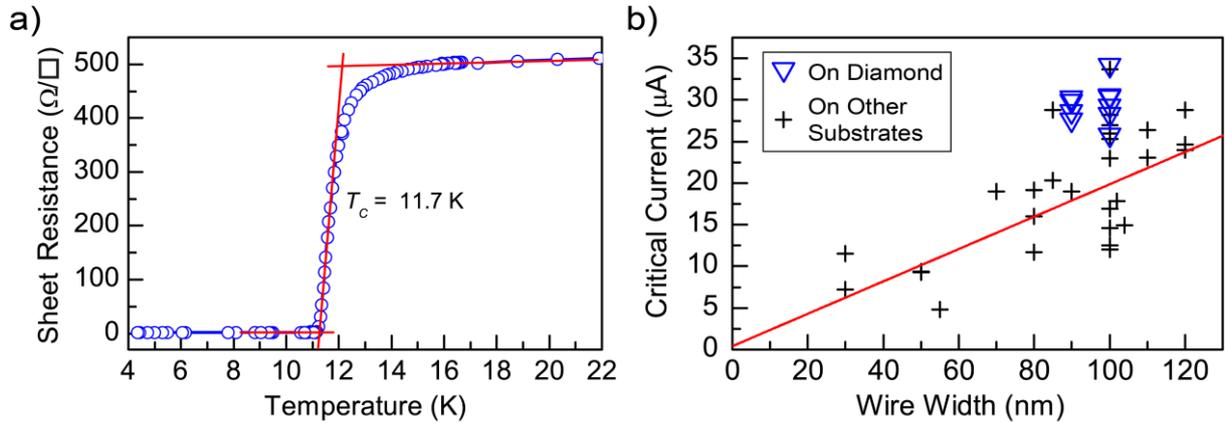

**Figure 61 - Critical temperature and critical current**: a) Dependence of the sheet resistance of the 4 nm thick NbN layer on diamond on the temperature revealing a critical temperature of $T_C = 11.7$ K. b) Dependence of the critical current of SNSPDs on the nanowire width: Comparison between highest critical current values for SNSPDs from NbN in the literature (black crosses, the red line is a guide to the eyes) to the highest critical currents measured for SNSPDs on diamond waveguides (blue triangles) showing that they are comparable to the highest values for NbN SNSPDs on any substrate. The literature comparison is adapted from Schuck *et al.*[306] and was updated with recent publications[345,346].

The photonic chip containing arrays of waveguide-integrated SNSPDs is characterized in the cryogenic setup explained in section 4.4.1, which we operate at a base temperature of 1.8 K. We contact one SNSPD at a time electrically to the electrical probe and measure the critical current in series with a 1 MΩ resistor (external at room temperature), by increasing a DC voltage until the resistance suddenly increases to a value far above 1 MΩ, revealing the transition of the nanowire from the superconducting to the normal state. During such measurements, the laser shutter is closed and the attenuators are set to 120 dB attenuation in order to avoid stray light which would disturb the superconducting state. Figure 61 b) shows the experimental values for the critical currents above 25 µA which were measured for SNSPDs on diamond waveguides (blue triangles) compared to values reported for NbN SNSPDs in the literature (black crosses). We find a maximum critical current of 34.2 µA for a nanowire cross-section of 4 nm · 100 nm and a length of 90 µm, corresponding to a critical current density of 8.55 MA/cm$^2$ which is on par with the highest values reported for NbN SNSPDs on other substrates in the literature[306].

Considering the yield of the device fabrication, it should be noted that 11 out of 21 SNSPDs designed for 1550 nm of 90 nm and 100 nm width show high critical currents above 25 µA, while the other devices show critical currents below 20 µA or cannot be electrically connected at all. We attribute this to the 3 nm rms surface roughness of the diamond layer which is comparable to the 4 nm layer thickness of NbN. This potentially leads to incisions in the nanowires which limit the critical current. As the probability of limiting the critical current with incisions increases with the nanowire length, this interpretation is supported by the fact that for 10 µm long devices five out of five devices show high critical currents above 25 µA, while for wires of at least 70 µm length this is only the case for six out of 16 SNSPDs. A further analysis of device yield depending on the length of the SNSPD and on its spatial position on the chip can be found in appendix A5. Concluding, it can be said that while the device yield could be improved, potentially by further diamond polishing, the best devices show large critical current densities comparable with the highest values on any other substrate, indicating that a high quality NbN layer on polished polycrystalline diamond was achieved. As the amplitude of the electrical output pulse of an SNSPD scales linearly with the applied bias current, high critical currents are desired as they enable a reliable detection of the output pulse with good signal-to-noise ratio. This aspect will be further discussed in the context of low timing jitter in section 4.4.6.

### 4.4.3  Single-photon detection capability

As a SNSPD is a binary detector, where the absorption of light leads either to zero or one output pulses at a time, the count rate for a given optical power depends on the probability of absorbing sufficient photons within a short enough time interval to trigger a detection event. For a laser the number of photons in a time interval follows a Poissonian distribution. For laser light with on average $\bar{n}$ photons per time interval, the probability of finding exactly $k$ photons is given by

$$P(k, \bar{n}) = \bar{n}^k e^{-\bar{n}}/k! \ . \tag{4.21}$$

For a laser with strong attenuation ($\bar{n} \ll 1$), this relation can be approximated as

$$P(k, \bar{n}) \approx \frac{1}{k!} \cdot \bar{n}^k \ . \tag{4.22}$$

The probability to find exactly $k = 1$ photon in a time interval therefore scales linearly with $\bar{n}$ and therefore with the optical power, while the probability to find $k = 2$ photons increases quadratically with the optical power. The same polynomial scaling with the power of $k$ obviously applies to higher values of $k$.

The required energy, and hence for monochromatic light the needed number of photons, for breaking down superconductivity in a SNSPD can be determined using a pulsed laser of variable pulse power. For an increasing pulse power, and hence an increasing average number of photons per pulse, the detection probability increases. For an attenuated pulsed laser ($\bar{n} \ll 1$) the probability of finding exactly $k$ photons in a pulse is given by equation (4.22), where $\bar{n}$ denotes the average number of photons per pulse. The pulse detection probability $p_{det}$ is a sum which contains the products of the probability of finding $k$ photons in the pulse $P(k, \bar{n})$ and the probability of $k$ photons triggering a detection event $\eta_k$ (the detection efficiency for $k$ photons). We note that $\eta_k \geq \eta_i$ for $k \geq i$, as more energy from more photons increases the probability of breaking down superconductivity. The probability of the photons being absorbed in the same area needs to be considered as well[310]. The pulse detection probability for a certain value of $\bar{n}$ can be dominated by the 1-photon, 2-photon, or $k$-photon events with larger numbers $k$ (sometimes referred to as detection regimes). If the energy of one photon is sufficient to trigger a detection event, then $p_{Det}$ will scale linearly with the average number of photons per pulse. If the energy of one photon is not sufficient, $p_{Det}$ will scale with the polynomial which corresponds to the smallest number of photons which is sufficient to trigger the breakdown of superconductivity.

For measuring the properties of SNSPDs concerning their interaction with light we align the two grating couplers of one PIC at a time with respect to the fiber array, such that the device transmission is maximized. Furthermore the polarization of light is optimized for maximum transmission. We employ a pulsed laser at 1550 nm, with a repetition rate of 40 MHz and a pulse duration of $\approx 1$ ps. This ensures that the pulse duration is much shorter than the recovery time of the detector (on the order of nanoseconds) and hence ensures that not more than one detector count is triggered by the same laser pulse. The photon flux towards the SNSPD is adjusted via two attenuators. The probability of detecting a laser pulse $p_{Det}$ is determined as the ratio of the detector count rate, corrected for the dark count rate, to the pulse repetition rate of the laser: $p_{Det} = \frac{CR - DCR}{RR}$. We bias the SNSPD with a current $I_B$ and record the detector count rates for varying attenuations of the pulsed laser. At a strong attenuation, each pulse contains on average much less than one photon. The average photon number per pulse $\bar{n}$ is then gradually increased by reducing the attenuation until the photon flux is too large for the SNSPD to sustain superconductivity. We note that with laser power and the corresponding average photon number we refer to their values for light inside the waveguide at the location where the nanowire starts.

Figure 62 shows the pulse detection probability $p_{det}$ as a function of the average number of photons per pulse $\bar{n}$ for various normalized bias currents on a double logarithmic scale. For a bias current of $0.85 \cdot I_C$, for a photon flux ranging from $10^{-6}$ to 1 photon per pulse a linear behavior with a slope of 1 can be observed, as expected for the operation of a binary single-photon detector[206]. At higher photon fluxes the detection probability saturates at $p_{Det} = 1$, as all pulses are being detected. The slope of 1 reveals the single-photon detection capability of the studied SNSPDs.[308,309] While for $0.65 \cdot I_C$ and higher currents $k = 1$ is dominating, for $0.55 \cdot I_C$ two regimes where single- and two-photon detection dominate can be identified. For a low bias current of $0.45 \cdot I_C$ a regime with $k = 3$ can be identified. The offset between the curves for different bias currents can be attributed to the increase of the detection efficiencies $\eta_k$ with increasing bias current, as will be explained in the following section. Throughout the rest of this thesis only $\eta_1$, the probability that one photon triggers a detection event, is considered. This is commonly referred to as *the efficiency* of an SNSPD which is commonly used as a single-photon detector and we denote it with $\eta$, as introduced in section 4.2.5.1. For waveguide-integrated detectors, $\eta$ corresponds to the on-chip detection efficiency. Summarizing, it can be said that the presented data shows the capability of the SNSPD to detect single photons.

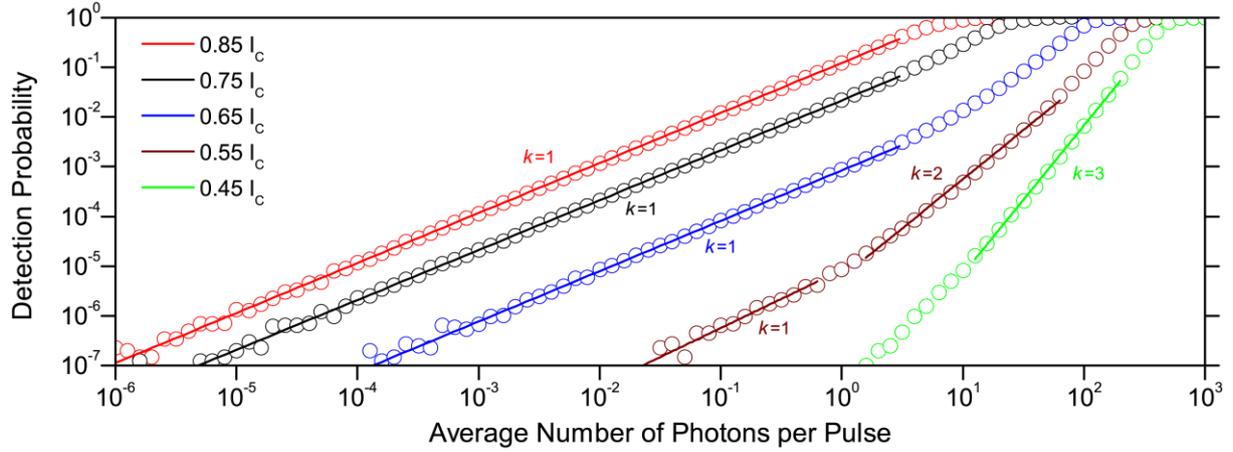

**Figure 62 - Single-photon counting capability:** Pulse detection probability in dependence of the average photon number contained in an attenuated laser pulse at 1550 nm for a 110 nm wide and 70 μm long SNSPD for various bias currents.

## 4.4.4  Detection efficiency

The goal for an ideal detector would be to feature both a short recovery time and a high detection efficiency, but for simple U-shaped SNSPDs there is a trade-off between OCDE and detector reset time. While OCDE increases with the length due to increased absorption, also the kinetic induction increases and hence the recovery time. We therefore design and fabricate short SNSPDs with a device length of 10 μm (nanowire length of 20 μm), which will be fast but have limited efficiency, and long SNSPDs with device lengths of 80 μm and 90 μm, which will show higher efficiencies but longer reset times, as will be discussed in section 4.4.7.

For the characterization the OCDE of the SNSPDs we first measure the transmission of light from a CW laser through the reference arm of the PIC, as explained in section 4.3.3, and adjust the attenuation such that the photon flux $\phi$ at the detector is either $10^6$ or $10^7$ photons per second. This ensures for detectors with recovery times of a few nanoseconds that the probability of more than one photon arriving within the recovery time is negligible. For each bias current $I_B$ we measure both the count rate $CR(I_B)$ at a given photon flux and the dark count rate $DCR(I_B)$ when the laser is disconnected and the fiber inputs are covered by metal caps to shield the fiber from light sources in the laboratory. As introduced in section 4.2.5.1, we can then calculate the on-chip detection efficiency as $\eta(I_B) = \frac{CR(I_B) - DCR(I_B)}{\phi}$ . Figure 63 a) shows the dependence of the OCDE on the bias current for detectors designed for a wavelength of 765 nm (blue circles) and 1600 nm (red triangles) on a logarithmic scale. The detectors have a width of 90 nm and device lengths of 80 μm (765 nm) and 90 μm (1600 nm) and show critical currents of $I_C = 31.2$ μA (765 nm) and $I_C = 28.6$ μA (1600 nm).

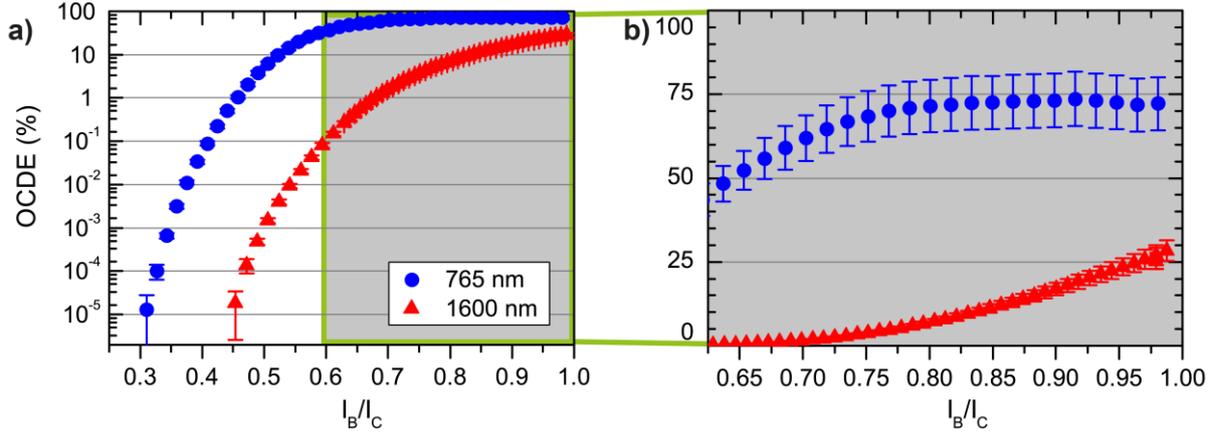

**Figure 63 - OCDE for long SNSPDs at two wavelengths:** On-chip detection efficiency for the best devices at 1600 nm (red triangles) and 765 nm (blue circles). Both nanowires have a width of 90 nm and device lengths of 80 µm (765 nm) and 90 µm (1600 nm) and show critical currents $I_C = 31.2$ µA (765 nm) and $I_C = 28.6$ µA (1600 nm), respectively. a) OCDE as a function of the normalized bias current $I_B/I_C$ on a logarithmic scale. b) Enlarged section on a linear scale showing a maximum efficiency of $28.4 \pm 3.4\%$ at 1600 nm and a plateau in efficiency for photons at 765 nm with a maximum efficiency of $73.6 \pm 11.0\%$.

For light with 1600 nm wavelength, we find that the OCDE is continuously increasing with bias current, with increasing slope up to the critical current and no efficiency plateau is reached. In this wavelength range we obtain a maximum OCDE of $28.4 \pm 3.4\%$ when biased at 98.8% of the critical current. Using the measured absorption at telecommunication wavelengths, provided in section 4.3.2, we estimate the absorption efficiency for a 90 µm long device of 90 nm width as $\eta_{abs} = 1 - 10^{(-0.203 \cdot 90/10)} = 98.5\%$. The measured detection efficiency is hence not limited by the absorption, but by the internal quantum efficiency which is in accordance with the fact that no efficiency plateau is reached. We anticipate that higher efficiencies could be reached by reducing the nanowire width, as has been shown for SNSPDs on other substrates.[303]

For detectors at a wavelength of 765 nm, the efficiency saturates at high bias currents, as can be well seen on a linear scale (Figure 63 b), with a maximum efficiency of $73.6 \pm 11.0\%$. Using the measured absorption at 765 nm we estimate the absorption efficiency for a device of 80 µm length and 90 nm width as $\eta_{abs} = 1 - 10^{(-0.103 \cdot 80/10)} = 85\%$. This is comparable to the measured OCDE, indicating that the internal quantum efficiency is likely close to unity, which is in accordance with the plateau in efficiency. SNSPD efficiencies above 90% have experimentally been shown[65,288] and higher efficiencies for waveguide-integrated SNSPDs on diamond at 765 nm could be achieved by increasing the absorption efficiency.

The presented detector for 765 nm reaches 90% of its maximum efficiency at a bias current of $0.719 \cdot I_C$. The saturation parameter, as defined in section 4.2.5.1, therefore amounts to $S = 0.281$, which means that the SNSPD can be operated 28.1% below its critical current, while losing only 10% of its maximum efficiency. This allows to operate the detector at a relatively high efficiency with low noise, as will be discussed in the following section on dark count rate and noise-equivalent power.

Concerning the fabrication yield, we have to note that from all SNSPDs of comparable nanowire length (≥ 140 μm) and width (90 nm or 100 nm) which were characterized for both wavelengths, two out of twelve devices show comparably high OCDEs around 74% at 765 nm and around 28% at 1600 nm. For the devices[25] at 1600 nm, Figure 64 provides the relation between the critical current and the maximum OCDE of the SNSPD, indicating that a high critical current is a necessary requirement for a high detection efficiency. An analysis of the influence of device length and location within the device array on the critical current is provided in Figure 84 in appendix A5. The correlation between critical current and OCDE in Figure 64 suggests that improving the proportion of nanowires with high critical current would increase the proportion of detectors with high efficiency. Low critical currents are potentially limited by defects and incisions in the nanowire. Reducing the surface roughness of the diamond layer, by extending the polishing time, might reduce the risk for incisions and defect.

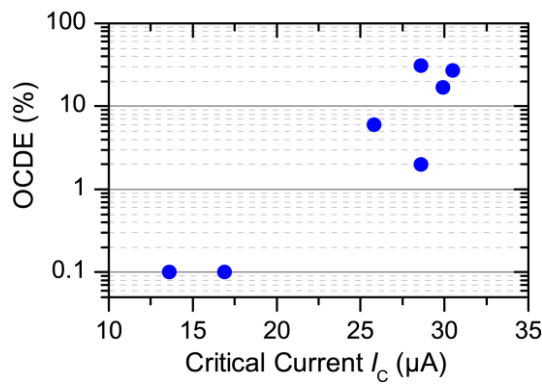

**Figure 64 - Dependence of the maximum OCDE at 1600 nm on the nanowire's critical current:** SNSPDs with lower critical currents, potentially limited by defects and incisions, show much lower efficiencies than the detectors with high critical currents.

We measure the OCDE dependence on the bias current for short SNSPDs (20 μm nanowire length), as shown in Figure 65. We find for the best device at 1600 nm a critical current of $I_C = 30.3$ μA (90 nm width) and a maximum OCDE of $2.8 \pm 0.3$%. For the best device designed for a wavelength of 765 nm we measure $I_C = 31.1$ μA (100 nm width). The OCDE curve (Figure 65 b) shows a point of inflection, revealing the onset of a plateau in efficiency[26] and a maximum efficiency of $17.3 \pm 1.9$%. As short SNSPDs have the advantage of shorter recovery times the combination of a modest OCDE of 17% at 765 nm with high possible count rates could be of interest for applications where a short recovery time and high count rate is crucial.

---

[25] We note that for some of the twelve detectors the superconductivity was not stable or the critical current so low that the detector efficiency could not be measured. Hence there are less than twelve data points in the corresponding figure.

[26] We attribute the fact that opposed to the long wires we did not find a detector with a real plateau to the fact that for each wavelength only four short devices were investigated opposed to twelve long devices, hence we expect to find short devices with a clear plateau when fabricating larger amounts of shorter devices.

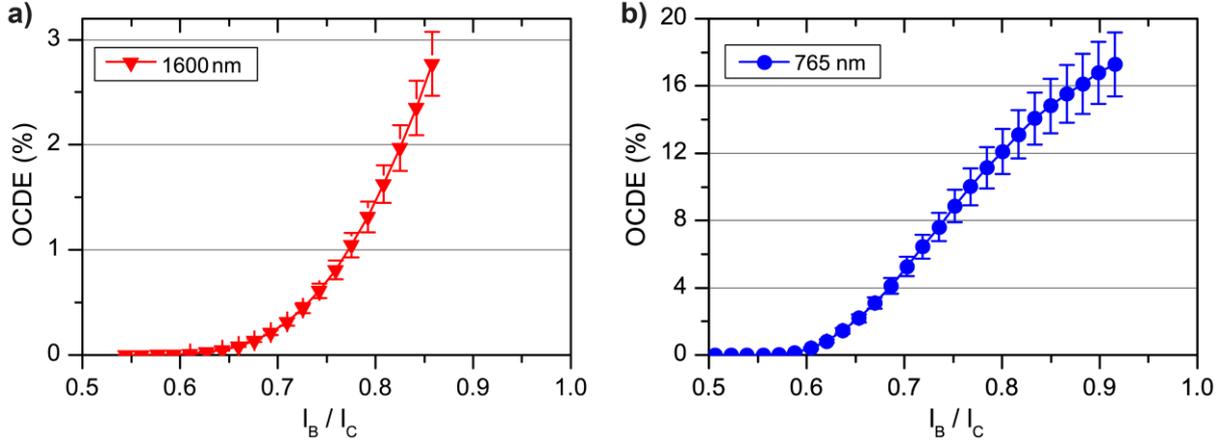

**Figure 65 - OCDE for short devices with 20 μm nanowire length**: a) SNSPD with 90 nm width showing $I_C$ = 30.3 μA and a maximum OCDE of 2.8 ± 0.3% at 1600 nm. b) SNSPD with 100 nm width showing $I_C$ = 31.1 μA and an onset of a plateau with a maximum OCDE of 17.3 ± 1.9% at 765 nm.

### 4.4.5 Dark count rate and noise-equivalent power

As mentioned in the previous section, we measure the dark count rate for each SNSPD. Figure 66 a) shows the dependence of the dark count rate (on a logarithmic scale) on the bias current for the detector with the highest OCDE each for 765 nm and 1600 nm (see Figure 63 for the corresponding OCDE measurements). For both wavelengths we find that when reducing the bias current, starting at the critical current, the dark count rate drops exponentially in the high current regime (above $0.93 \cdot I_C$) and then slowly decreases from about 10 Hz to below 1 Hz, as indicated in yellow color.

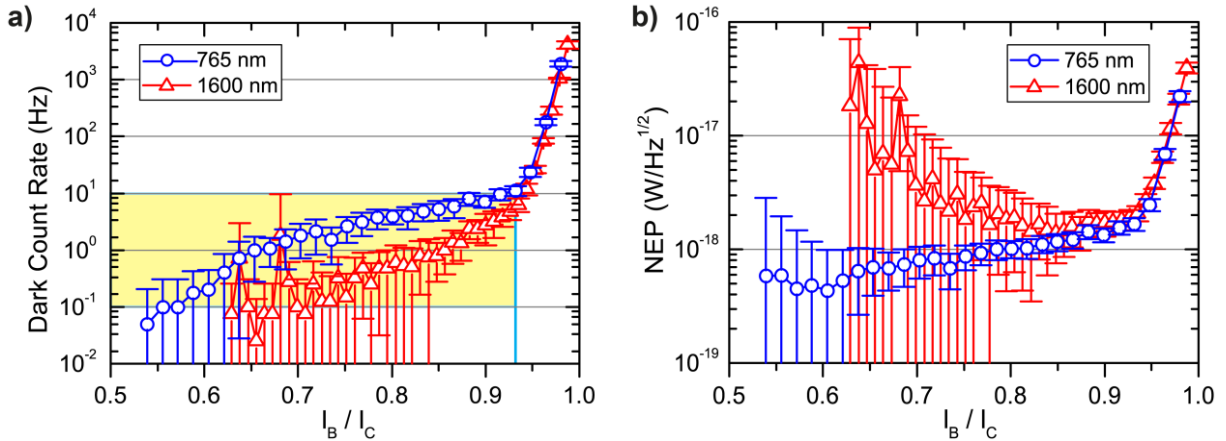

**Figure 66 - Dark count rate and noise-equivalent power for long SNSPDs at 1.8 K**: Both nanowires have a width of 90 nm and device lengths of 80 μm (765 nm) and 90 μm (1600 nm) and show critical currents $I_C$ = 31.2 μA (765 nm) and $I_C$ = 28.6 μA (1600 nm), respectively. a) Dark count rate as a function of normalized bias current for visible (765 nm) and infrared (1600 nm) photons. The yellow area indicates the regime below $0.93 \cdot I_C$ with a dark count rate between 1 Hz and 10 Hz. b) Noise-equivalent power as a function of normalized bias current for photons at 765 nm (blue) and 1600 nm (red).

As explained in section 4.2.5.1, the exponential increase in dark count rate close to the critical current is related to intrinsic dark counts, while dark counts below $0.93 \cdot I_C$ can be attributed to blackbody radiation. Dark counts due to thermal photons could be reduced by spectral filtering of the light propagating within the optical fibers.[290] We note that in future quantum photonic chips only on-chip light sources will be used, which means that no optical fibers out of the cryostat will be needed anymore and therefore the blackbody radiation and the corresponding dark counts would be largely suppressed. Hence for the development of waveguide-integrated SNSPDs within this thesis the dark count rate is of no further concern.

Figure 66 b) shows the noise-equivalent power for the same detectors. The NEP, as introduced in section 4.2.5.2, denotes the optical power which is required for gaining a signal-to-noise ratio of unity after integrating the detector's output signal for one second. For photons at 1600 nm, we obtain a minimum in NEP of $1.4 \cdot 10^{-18}$ W/$\sqrt{\text{Hz}}$ at a normalized bias current of 85%. For 765 nm, the minimum NEP value is reached at a low bias current of 60% and amounts to $4.4 \cdot 10^{-19}$ W/$\sqrt{\text{Hz}}$. The reduced NEP compared to 1600 nm is mainly due to the higher efficiency at smaller wavelengths and the plateau in efficiency furthermore moves the minimum of the curve to smaller currents. We note that the large error bars for the NEP towards low bias currents are due to the uncertainty in dark count rates below 1 Hz, due to a data acquisition time of 20 times 2s per data point in the experiment. The error bars could hence be reduced by increasing the measurement time for the dark count rate measurements.

## 4.4.6 Timing jitter

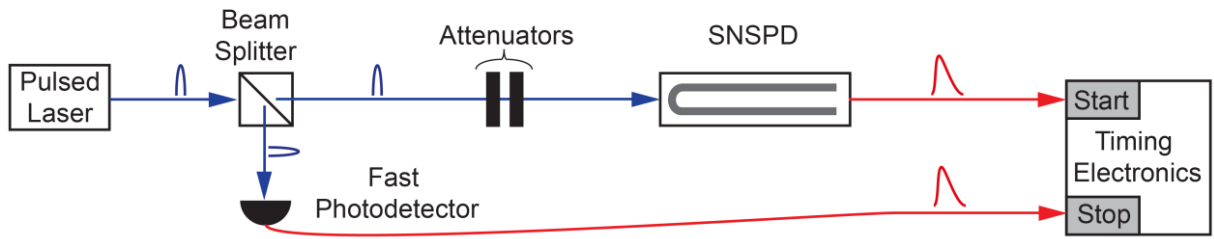

**Figure 67 - Jitter measurement setup**: Schematic of the setup for measuring the timing jitter of a single-photon detector. Schematic adapted from Migdall *et al.*[31].

The timing jitter, as introduced in section 4.2.5.4, is a measure for the variation in the time difference between optical input and electrical output. Figure 67 shows a schematic of the setup for measuring the timing jitter. Light from a pulsed laser (40 MHz repetition rate, $\approx$ 1 ps pulse width) at 1550 nm wavelength passes through a beam splitter. The portion of light which propagates towards the single-photon detector in the cryostat is strongly attenuated (far below one photon per pulse on average) in order to avoid multi-photon events occurring. The detection event at the SNSPD provides a start signal for the timing electronics. At the second output of the beam splitter the optical pulse is detected by a fast low-noise photodetector (New Focus 1611, 1 GHz bandwidth) which provides the stop signal for the timing electronics. The trigger level for the SNSPD output pulse is set at the rising slope at 50% of the pulse height. We note that all optical paths in the schematic, including the beam splitter, are implemented with optical fibers in the experiment, while the waveguide-integrated SNSPD is located inside the cryogenic measurement setup, as for all characterizations within section 4.4.

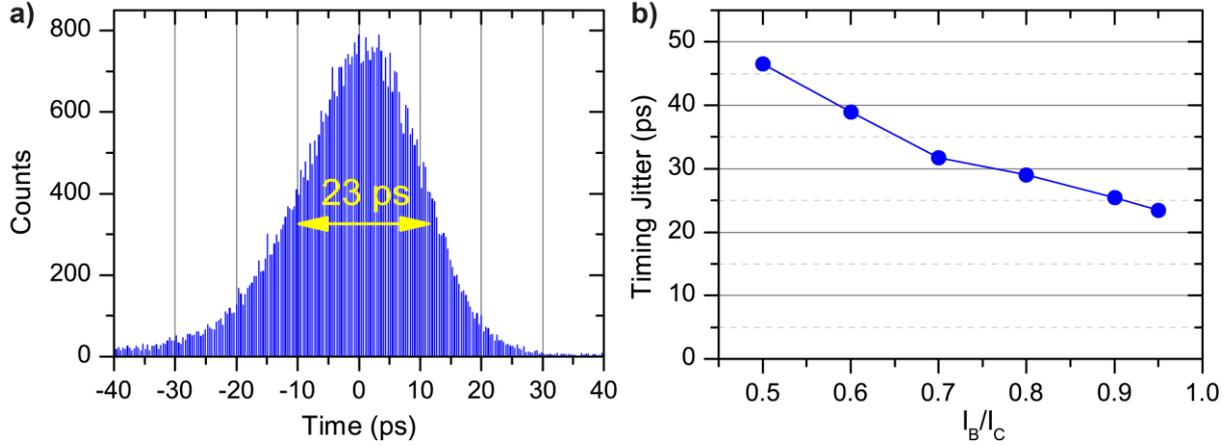

**Figure 68 - Timing jitter of a SNSPD:** $L = 90\ \mu m$, $w = 90\ nm$, $I_C = 28.6\ \mu A$. a) Histogram of the arrival times of SNSPD counts relative to a reference signal of the pulsed laser source at 1600 nm. The histogram consists of 53 000 total counts at a bin size of 0.36 ps. The SNSPD is biased at 95% of the critical current $I_C$. The Gaussian fit reveals a FWHM value of 23 ps. b) The timing jitter in dependence of the applied bias current $I_B$ relative to the critical current $I_C$ showing a decrease in jitter with increasing bias current.

We use a fast digital oscilloscope (Agilent 54855A, 6 GHz bandwidth) for the start-stop measurement and record the histogram of measured time differences, as shown in Figure 68 a). Length differences in the used electrical cables and optical fibers lead to a time difference between the detection at both detectors. Hence the absolute value of the SNSPD's timing latency cannot be measured and we set the mean value of the histogram to zero. We examine a detector with a nanowire of 180 μm length, 90 nm width and a critical current of $I_C = 28.6\ \mu A$. This detector showed up to 28% efficiency at 1600 nm (see Figure 63). Figure 68 a) shows the histogram for the detector biased at $0.95 \cdot I_C$. By fitting a Gaussian distribution we extract a FWHM of 23 ps, which is commonly referred to as the timing jitter.

The timing jitter depends on the bias current, as shown in Figure 68 b). When increasing the bias current from 50% of the critical current to 95% of the critical current, the jitter continuously decreases from 47 ps to 23 ps. The observed bias dependence of the jitter is similar to those observed for NbN and MoSi SNSPDs[311,312] and can be attributed to the improvement of signal-to-noise ratio with increasing bias current. Electrical noise in the readout circuit, for example from the electrical amplifiers at room temperature, will lead to a variation in heights of the output pulses measured at the oscilloscope. As the trigger level is set at a fixed voltage, a variation in pulse heights translates into a variation in the times at which the electrical output pulses cross the trigger level. This leads to a component in the timing jitter which is caused by the electrical noise. The magnitude of the output pulse increases linearly with the bias current, resulting in an increase in signal-to-noise ratio. A better signal-to-noise ratio results in a smaller contribution to the timing jitter and hence the jitter decreases with increasing bias current, as can be observed in Figure 68 b). We note that the timing jitter of the oscilloscope and the 1 GHz photodetector have been measured to be both less than 1 picosecond, well below the measured jitter values. We therefore attribute the measured jitter to the SNSPD jitter and to noise from the electrical amplifiers. The measured jitter value thus provides an upper bound for the intrinsic timing jitter of the SNSPD. We note that we only measured the timing jitter for detectors at 1600 nm (due to lack of a pulsed laser at 765 nm) but we anticipate that the timing jitter at 765 nm is comparable.

**Table 3 - Timing jitter for SNSPDs with different lengths and widths when biased at high currents.** The OCDE values are provided to show that SNSPDs with high efficiency at 1600 nm also show low timing jitter.

| Detector length | Detector width | On-chip detection efficiency | Critical current $I_C$ | Bias current ($I_B / I_C$) | Trigger level ($0.5 \cdot U_{max}$) | Timing jitter |
|---|---|---|---|---|---|---|
| 90 μm | 90 nm | 28% | 28.6 μA | 0.95 | 190 mV | 23 ps |
| 80 μm | 100 nm | 27% | 30.5 μA | 0.95 | 185 mV | 26 ps |
| 10 μm | 90 nm | 3% | 30.3 μA | 0.85 | 110 mV | 28 ps |

Table 3 presents the jitter for the device described above compared to the jitter of SNSPDs with wider wires and shorter lengths. At high bias currents, all detectors show a low timing jitter between 23 ps and 28 ps. The timing jitter for waveguide-integrated SNSPDs on diamond presented in this work is on par with the best results for SNSPDs on commercial substrates, such as silicon[65] and silicon nitride[303] and close to the smallest jitter for any integrated single-photon detector which to the best of our knowledge is currently 18 ps for SNSPDs on silicon waveguides[65].

## 4.4.7 Recovery time

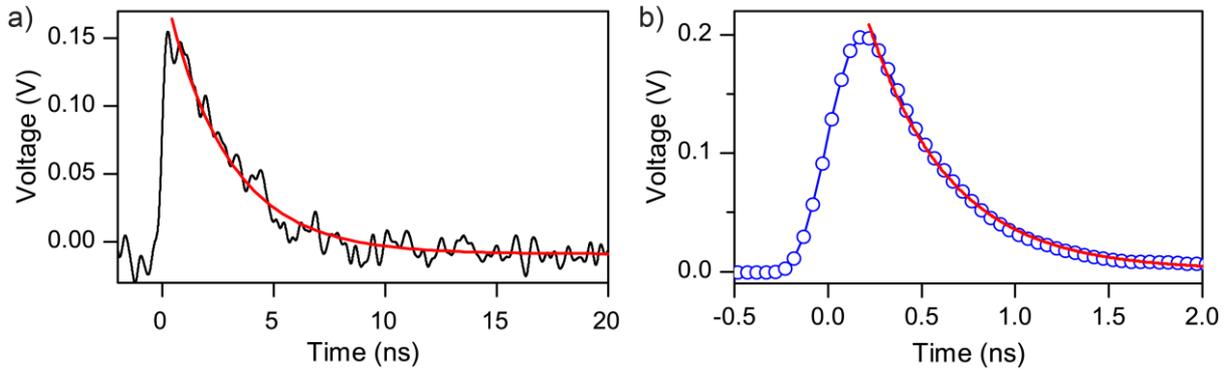

**Figure 69 - Recovery time for devices at 1600 nm:** a) One-shot time trace of the output pulse of a SNSPD with 160 μm nanowire length and 90 nm width, biased at $I_B = 14$ μA, showing a $1/e$-decay time of 2.9 ns. b) Averaged detector output pulse for a 20 μm long and 90 nm wide nanowire. An exponential fit (red) reveals a $1/e$-decay time $\tau_2 = 440$ ps.

We estimate the recovery time, after which a SNSPD can detect the next photons with the same initial efficiency, by recording the electrical pulses resulting from detection events with a 6 GHz digital oscilloscope (Agilent 54855A). As explained in section 4.2.5.5, the $1/e$-decay time of the output pulse is equal to the time constant which governs the recovery of the bias current and hence the recovery of the detection efficiency. Figure 69 a) shows the time trace of one output pulse for a SNSPD of 160 μm nanowire length and 90 nm width operated with light at 765 nm at a bias current of $I_B = 14$ μA. As explained in section 4.2.3, during a detection event the bias current is almost completely rerouted to the load resistor $R_L = 50$ Ω and hence the expected pulse height after 23.3 dB amplification is $V = 14 \text{ μA} \cdot 50 \text{ Ω} \cdot 10^{\frac{23.3}{10}} = 0.150$ V. The measured pulse height is in agreement with the expected value. An exponential fit to the pulse decay (red curve) shows a $1/e$-decay time $\tau_2 = 2.9$ ns. The corresponding kinetic inductance $L_K$ is given by $\tau_2 = \frac{L_K}{50\,\Omega}$ and amounts to 145 nH, which is comparable to the values for SNSPDs of similar geometries from NbN on other subtrates[313]. Figure 69 b) shows an average of output pulses for the shortest device (20 μm nanowire length). The exponential fit reveals a value of $\tau_2 = 440$ ps, which would correspond to a kinetic inductance of $L_K = 22$ nH. Note that this measurement is probably limited by the 1 GHz bandwidth of the electrical amplifiers and hence both $\tau_2$ and $L_K$ for this short detector are expected to be even smaller.

Latching, as explained in section 4.2.3, denotes the effect that the electro-thermal feedback can become stable, leading to a detector being locked in a resistive state. We note that we do not observe latching[279] of any detector, even for the shortest device. Single-photon detectors with decay times well below 1 ns will allow GHz count rates and are especially of interest for applications where reduced efficiency in benefit of high count rates is acceptable.

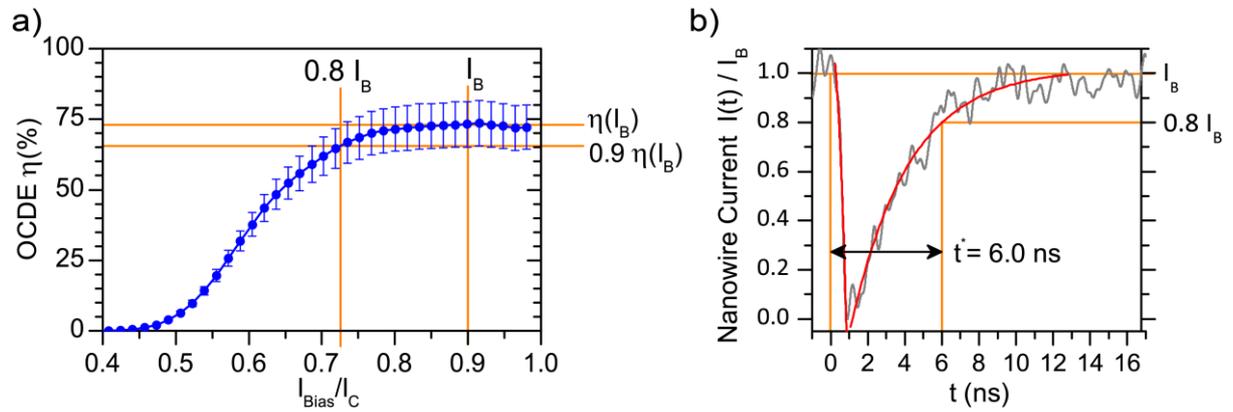

**Figure 70 - Analysis of the time after which the detector's efficiency has recovered to 90% of the initial value $\eta(I_B)$:** a) OCDE in dependence of bias current, revealing that for $I_B = 0.9 \cdot I_C$, 90% of the initial efficiency is reached at 80% of the initial bias current. b) Time trace of the bias current, estimated from the detector's electrical output pulse. 6.0 ns after the beginning of the output pulse, 80% of the initial bias current has recovered, giving an estimate of 6.0 ns for the SNSPD's recovery time.

As explained in section 4.2.5.5, a good estimation of the detector's recovery time does not only involve the $1/e$-decay time but also the dependence of the detector's efficiency on the bias current[27]. We follow the approach to define the recovery time $t^*$ as the time at which 90% of the initial efficiency has recovered: $\eta(I(t^*)) = 0.9 \cdot \eta(I_B)$. The recovery time of a SNSPD is therefore not a constant value, but depends on the choice of the initial bias current $I_B$. We analyze the recovery time for the device with the highest OCDE of 73.6% at 765 nm and a $1/e$-decay time of $\tau = 2.9$ ns (see Figure 69 a). Figure 70 a) shows the OCDE depending on the bias current. We choose a bias current of $I_B = 0.9 \cdot I_C$, which yields the full efficiency of 73.6%. We find that $0.9 \cdot \eta(I_B) = 0.66 = \eta(0.719 \cdot I_C = 0.8 \cdot I_B)$. This means that at 80% of the initial bias current 90% of the initial efficiency is reached.

---

[27] We note that only in the case of a detector with an efficiency plateau over a large enough current range it is possible that $\eta(I(\tau_2) = 0.63 \cdot I_B) \geq 0.9 \cdot \eta(I_B)$, meaning that the $1/e$-decay time $\tau_2$ would in such a case be a reasonable estimation of the recovery time. For this the saturation parameter has to fulfill $S > 0.37$, but to be able to bias below $I_C$ a larger value for $S$ is needed, such that the relation above for the efficiency holds.

Figure 70 b) shows the time trace of the bias current, estimated from the detector's electrical output pulse. We estimate the time between the start of the pulse and the recovery of 80% of the initial bias current as $t^* = 6.0$ ns. Hence 6 ns is a good estimate for the recovery time of the detector, when biased at 90% of the critical current. In combination with a pulsed single-photon source, such as a SiV center, the presented SNSPD (biased at $0.9 \cdot I_C$) could hence detect single photons with 66% efficiency at single-photon repetition rates up to 166 MHz, while the dark count rate is below 10 Hz. Up to our knowledge the largest observed single-photon creation rate for a single-photon source in diamond is $\approx 6$ MHz, measured for SiV centers.[118] Our SNSPD can thus efficiently detect single photons at count rates which can be more than 20 times larger than what is currently needed for the brightest single-photon source in diamond.

The same calculation for the detector at 1600 nm (see Figure 63) biased at $I_B = 0.9 \cdot I_C$ yields a recovery time $t^* = 16.6$ ns, which is much larger than the $1/e$-decay time $\tau_2 = 3.6$ ns due to the absence of a plateau in efficiency at 1600 nm. This detector could detect single photons from a pulsed source with 15.5% efficiency up to a repetition rate of 60 MHz. We note that up to our knowledge no single-photon sources in diamond at telecom wavelengths have been found yet[135] and hence the results at 765 nm with higher efficiency and smaller recovery time are more relevant for integrated quantum optics in diamond.

High efficiency and high count rate ($> 500$ MHz) in one SNSPD on diamond could in the future be achieved for example via one of the following strategies: Connecting several nanowires in parallel decreases the kinetic inductance while increasing the absorption and increasing the critical current and hence the signal amplitude. Such detectors, called superconducting nanowire avalanche photodetectors, have been realized on other substrates.[314,315] Optical cavities, such as one-dimensional photonic crystals, allow to increase the absorption and hence the OCDE of short nanowires[316] while preserving low kinetic inductances and hence low recovery times[317].

## 4.4.8  Maximum detector count rate

We probe the potential to operate the SNSPDs at 1600 nm at high count rates, using a pulsed measurement. The measurement setup is the same as used for the timing jitter measurement (see Figure 67). Pulses from a laser with 40 MHz repetition rate are sent to a beam splitter and an external photodetector at room temperature registers the unattenuated pulses, while the SNSPD in the cryogenic setup registers attenuated pulses. Figure 71 a) shows the time traces taken simultaneously with the reference detector (black) and with a SNSPD (red). The SNSPD with the highest OCDE of 28% was used in this measurement and biased at $0.72 \cdot I_C$. As can be seen in the enlarged section of the time trace (Figure 71 b) the detector's $1/e$-decay time of 3.6 ns is much smaller than the time between two consecutive pulses, and every pulse of the pulse train is reliably registered with a count rate of 40 MHz, matching the repetition rate of the laser.

It has to be noted that in this measurement the detector operates with relatively high input power and therefore not as a single-photon counter but as a counter of weak pulses. As explained before a pulsed single-photon source should be used instead of attenuated laser light in order to assess the limits of single-photon detection at high efficiency and at high rates.

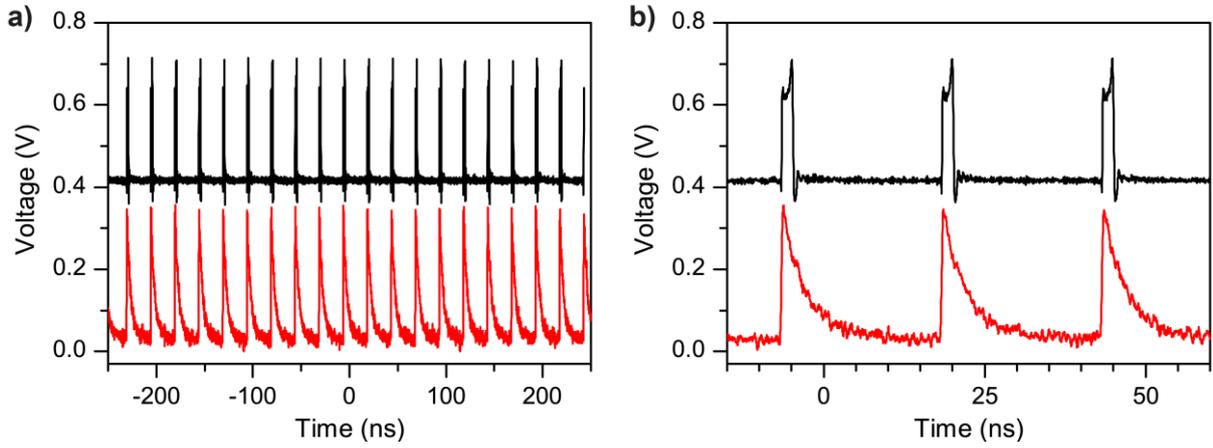

**Figure 71 - Pulsed measurements at 1600 nm:** a) Oscilloscope time trace, showing a pulse train with a period of 25 ns, registered simultaneously with a reference detector (black) and with a SNSPD (red). Note that the reference detector output was scaled to the amplitude of the SNSPD output and displaced by 0.4 V for clarity. The SNSPD ($L$ = 90 μm, $w$ = 90 nm, $I_C$ = 28.6 μA, $OCDE_{max}$ = 28%) was biased at $0.72 \cdot I_C$. b) Enlarged section of the time trace, showing four output pulses of the SNSPD with a $1/e$-decay time of 3.6 ns.

In order to experimentally estimate the highest count rate at which photons from color centers can be detected with our SNSPDs, we further test the device with the highest efficiency at 765 nm. The setup for these measurements is equal to the one employed for the OCDE measurements (see section 4.4.4). We use a 765 nm CW laser and stepwise decrease the attenuation, in order to increase the photon flux at the detector, and record the count rate. In order to keep the detector in the superconducting state (besides during short periods during detection events), we decrease the bias current while increasing the photon flux. We test the detector which showed a plateau in efficiency, a maximum OCDE of 74% (see Figure 63) and a $1/e$-decay time of 2.9 ns. We find that the count rate steadily increases with photon flux until it saturates at $\approx$ 200 MHz, limited by the employed frequency counter (Agilent 53132A).

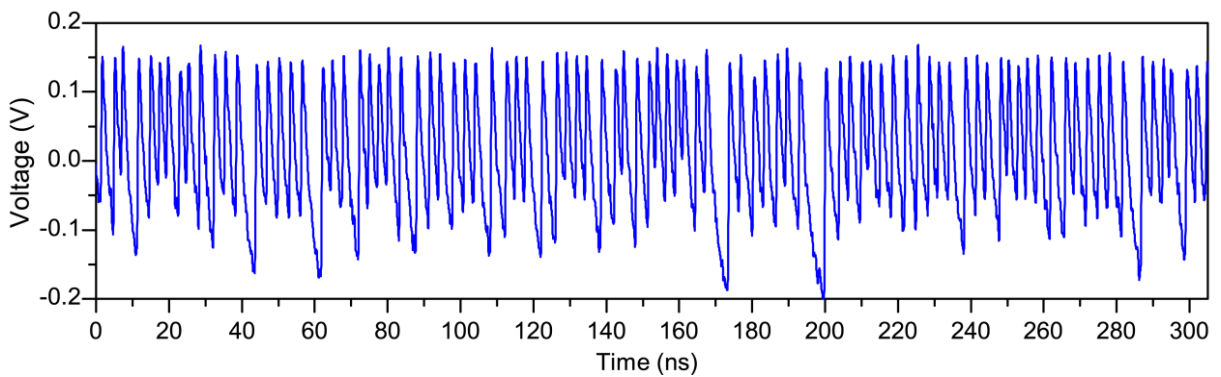

**Figure 72 - High count rate of a SNSPD at 765 nm:** One shot recording of the SNSPD's output, biased at 14 μA and operated with increased optical power of the CW laser. The time trace contains 93 pulses within a time interval of 305 ns, corresponding to a count rate of 305 MHz.

By recording time traces using a 6 GHz digital oscilloscope we are able to resolve higher count rates. Figure 72 shows a one-shot time trace of the detector's output signal recorded with the oscilloscope. The time trace contains 93 pulses within a time interval of 305 ns, corresponding to a detector count rate of 305 MHz, which illustrates that the employed SNSPDs can be used for the detection of faint light, opposed to single photons, at high count rates. It has to be noted that this high count rate was achieved at a low bias current of $I_B = 0.45 \cdot I_C = 14\,\mu A$ where the OCDE for single photons is only $\approx 1\%$. We note that CW light is not ideal for achieving high count rates simultaneous with high single-photon efficiencies, as both the photon number per time interval and the time between consecutive photons are probabilistic. It is of interest to assess the detector's performance when operating as an efficient single-photon detector with a count rate approaching the value limited by the detector's recovery time (see section 4.4.7). Such a measurement could be performed with a pulsed single-photon source with variable repetition rate (on the order of $100 - 500$ MHz). This would provide an optical input where exactly one photon impinges on the detector at a time and the time difference between consecutive photons could be adjusted to match the detector's recovery time. Such a deterministic and efficient single-photon source is currently not available and the work towards such a source is a field of research by itself[31], hence we are not able to perform such a measurement. The presented measurements are steps towards this goal and illustrate that the SNSPDs are generally able to reliably detect photons at high rates. SNSPDs therefore provide a large dynamic range, as they can be operated over at least eight orders of magnitude in count rate, from their dark count rate of $\sim 1$ Hz up to at least $2 \cdot 10^8$ Hz. Therefore the presented SNSPDs can not only be employed for applications where low count rates are expected or extremely low noise is crucial (and hence low dark count rates are needed), but also at high count rates up to hundreds of Megahertz.

### 4.4.9  Detector performance characteristics at one operation point

The best values for all performance characteristics of a SNSPD, such as high on-chip detection efficiency, low dark count rate, low noise-equivalent power and low timing jitter are typically not achieved at the same operation point. While there are tradeoffs, nevertheless excellent detector performance concerning all performance characteristics can be achieved at a common operation point. We choose the highest current below the onset of the exponential decrease in dark counts ($0.93 \cdot I_C$ at $T = 1.8$ K) and summarize the experimental results for the best SNSPD each at 1600 nm and 765 nm at this operation point in Table 4.

**Table 4 - Summary of the performance characteristics of the best SNSPD each at 1600 nm and 765 nm.** While the optimum for each characteristic (such as OCDE and NEP) is typically not achieved at the same operation conditions this summary provides a realistic detector operation point of all characteristics measured under the same conditions at the same bias current.

| Photon wavelength | Detector length | Detector width | Critical current $I_C$ | Cryostat temperature $T$ | Bias current ($I_B/I_C$) | On-chip detection efficiency | Dark count rate | NEP ($W/\sqrt{Hz}$) | Timing jitter |
|---|---|---|---|---|---|---|---|---|---|
| **1600 nm** | 90 μm | 90 nm | 28.6 μA | 1.8 K | 93% | 20% | 5 Hz | $1.9 \cdot 10^{-18}$ | 25 ps |
| **765 nm** | 80 μm | 90 nm | 31.2 μA | 1.8 K | 93% | 73% | 11 Hz | $1.7 \cdot 10^{-18}$ | -- |

# 4.5 Conclusions on SNSPDs and outlook

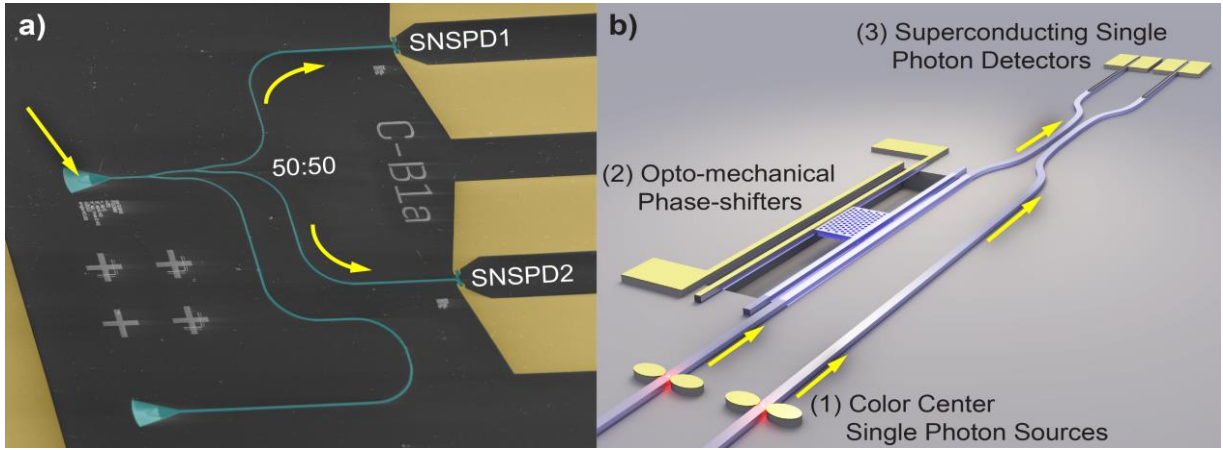

**Figure 73 - Concepts for integration of SNSPDs into advanced PICs:** a) SEM micrograph of two SNSPDs at the outputs of a 50/50-beam splitter, which could be used for on-chip correlation measurements, such as measuring the $g^{(2)}(\tau)$-function of on-chip single-photon sources. b) Schematic of a future diamond quantum optical circuit, which incorporates electrically excited color centers, acting as single-photon sources, optomechanical phase shifters, and superconducting single-photon detectors.

Within this chapter, we presented the results concerning the first superconducting nanowire single-photon detectors integrated on diamond waveguides by combining the PICs developed for polycrystalline diamond with established deposition and structuring methods for NbN nanowires. The SNSPDs feature high critical currents (up to 31 µA) and high performance in terms of high efficiency (up to 74% at 765 nm), low noise-equivalent power (down to $4.4 \cdot 10^{-19} \, W/\sqrt{Hz}$ at 765 nm) and low timing jitter (down to 23 ps). It can be expected that further improvements in the fabrication process will increase the detection efficiency close to unity as shown for waveguide-integrated SNSPDs from NbN on other substrates.[65]

The demonstrated detectors can be directly combined with all integrated optical components developed for diamond integrated optics, such as on-chip interferometers and optomechanical components, as presented in previous chapters. For example operating two SNSPDs at the two output ports of a 50/50 beam splitter, as shown in the SEM image of Figure 73 a), would allow to perform correlation measurements on-chip, with fast and efficient readout and small device footprint. Such PICs could for example be used for characterizing single-photon sources such as NV or SiV color centers, as explained in section 2.3. The implementation of waveguide-integrated SNSPDs on diamond is a promising step toward a quantum-optics-on-a-chip platform that relies on monolithically joining single-photon sources, single-photon routing and processing devices, as well as single-photon detectors. Figure 73 b) shows a schematic of how we envision such an on-chip quantum optical system in diamond: Color centers will provide efficient and controlled single-photon emission into waveguides, while this emission would be electrically excited[130–133]. Electro-optomechanical phase shifters, such as the H-resonator presented in chapter 3, will allow to control and tune the routing of single photons within the integrated photonic network, while waveguide-integrated SNSPDs at the output of the PICs will provide reliable, low-noise readout.

Such on-chip quantum optical systems might require outstanding diamond material quality in terms of low level of impurities and low propagation losses, which would require the use of single crystal diamond. We will show in the following chapter a novel method for achieving such single crystal diamond PICs, which would enable to transfer all optical and optomechanical components demonstrated within this thesis to high quality single crystal diamond.

# 5  Photonic integrated circuits on arrays of thin single crystal diamond membrane windows

*This chapter presents a novel method for fabricating photonic integrated circuits from single crystal diamond. The fabrication of arrays of thin single crystal diamond membrane is demonstrated as well as photonic integrated circuits with low loss, demonstrating the feasibility to employ previously developed photonic integrated circuit designs on high quality diamond.*

*This chapter is partially based on results which were published previously in one publication[318], where the author of this thesis had equal contribution with the first author.*

## 5.1 Motivation

The existence of single-photon sources in diamond, such as the NV and the SiV center, is one of the main motivations for developing PICs in diamond. While isolated color centers have been shown within the crystals of polycrystalline diamond (PCD)[319], it remains to be shown if PCD thin films with low enough density of photoactive defects can be grown and if they can be optimized to have low enough propagation loss to make quantum optical applications with color centers in PCD thin films possible. Single crystal diamond (SCD) on the other hand can be grown with low defect density and high crystal quality and in recent years different approaches for the fabrication of SCD photonic elements have been explored[81]. Nevertheless, up to date no mature fabrication technique exists for PICs from SCD which consist of more than just a few optical elements.

The geometry and the performance of the PICs presented in the previous chapters depend mainly on the refractive index of diamond, which is essentially the same for PCD and SCD. To transfer the developed PIC components from PCD to SCD is expected to be a straightforward process and the resulting PICs, including optomechanical elements and single-photon detectors, would find applications for integrated quantum optics.

Within this chapter, we present a new scalable fabrication method for SCD membrane windows which achieves three major goals with one fabrication method: providing high quality diamond, as confirmed by Raman spectroscopy; achieving homogeneously thin membranes, enabled by ion implantation; and providing compatibility with established planar fabrication via lithography and vertical etching. We first give a short overview over approaches and difficulties of PIC fabrication using SCD and explain our fabrication technique. Then we present the measurement results for the first proof-of-principle photonic devices which were fabricated in this way and finally give an outlook on how this new fabrication method might enable the scalable planar fabrication of SCD quantum optical circuits.

## 5.2 The challenge of fabricating PICs from SCD

The scalable fabrication of SCD thin films as template material is not a straightforward process. Single crystal diamond can only be grown homoepitaxially on existing SCD substrate which prevents wafer scale processing, as such SCD pieces are not available with sizes larger than about $10 \cdot 10 \text{ mm}^2$. While heteroepitaxial growth is possible for PCD, this is not the case for SCD. For photonic circuits SCD needs to be a thin film with a thickness of 1 µm or less, with only minor thickness variations across the film. In addition, the diamond film needs to be either free standing or must reside on top of a material with lower refractive index and negligible absorption to allow for optical wave propagation in diamond waveguides. This can be achieved by either cutting free standing structures into bulk diamond or by thinning down SCD to thin membranes before structuring PICs.

Photonic components in bulk diamond pieces have been demonstrated via structuring using angle etching[320,321], focused ion beam[322,323], and isotropic etching[173,324]. However, these methods are not compatible with established planar fabrication processes and therefore do not provide a viable path towards wafer scale fabrication, which is required for scaling up PICs (as already a reality for silicon photonics[325,326]). Fabricating thin SCD membranes can be done by thinning down SCD polished plates of 20 – 50 µm thickness using reactive ion etching.[327,328] However, these plates generally have a wedged thickness profile. When thinning down the plate, the wedge profile is transferred into the thin membrane, leading to thickness variations of about 300 nm across a millimeter-sized sample.[20,87] Thickness variations on this order limit the reproducibility of PIC components because devices on the same photonic chip with identical two-dimensional layout will have different three-dimensional geometries which results in different device properties. An alternative technique which has been developed for fabricating thin SCD membranes is ion slicing.[329,330] This approach leads to a homogeneous membrane thickness over large areas but introduces residual built-in strain.[330,331] The resulting thin membranes are therefore vulnerable to cracks and bowing which restricts the handling of large membranes. The difficulty of handling these membranes is also a major obstacle for scalable optical device fabrication.

## 5.3 SCD membrane windows – fabrication and material quality

In order to get feasible SCD templates for PIC fabrication, we use a fabrication process, developed by our cooperation partners at the University of Melbourne[28], which enables the fabrication of arrays of thin single crystal diamond membrane windows[332]. The fabrication of the SCD membrane windows, depicted in Figure 74, starts with a SCD slab with a thickness of 300 µm. Ion implantation into the top side of the SCD slab using high energy Helium ions ($5 \cdot 10^{16}$ ions/cm², E = 1 MeV) leads to a

---

thin end-of-range ion damage layer which is located 1.7 μm below the diamond surface. Next the SCD slab is annealed in vacuum ($p = 5 \cdot 10^{-6}$mbar) at a temperature of 1300℃ for one hour which converts the ion damage layer into etchable graphite-like carbon. As the ion implantation does not only damage the crystal structure at the end-of-range depth, but also damages the diamond which the ions are passing, an undamaged diamond layer needs to be regrown on top of the damaged diamond. 1 μm of undamaged high quality diamond is grown by microwave plasma chemical vapour deposition on top of the 1.7 μm thick damaged diamond layer.

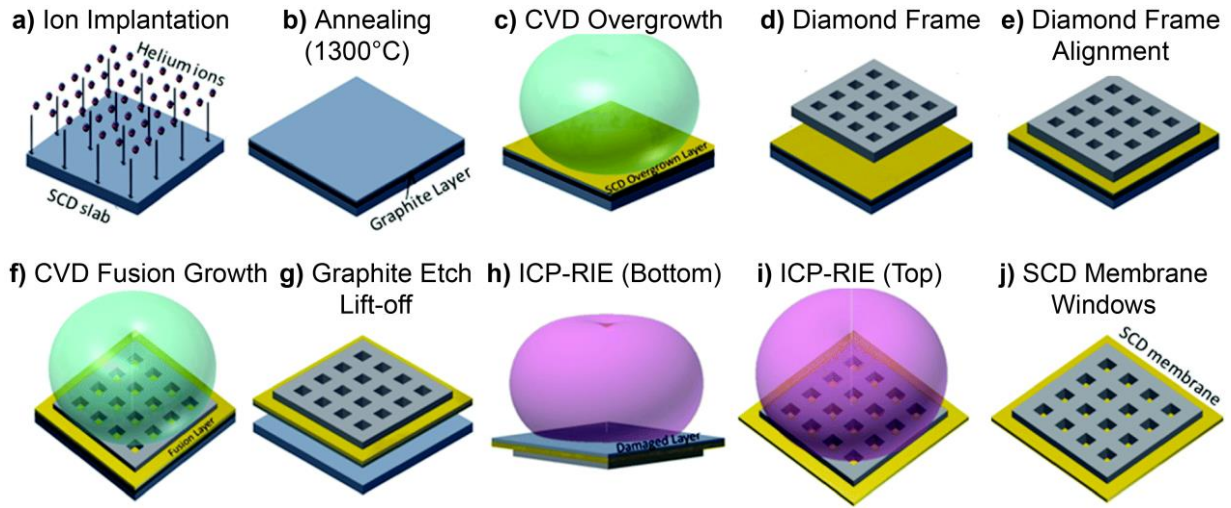

**Figure 74 - Schematic of the fabrication process of thin SCD membrane windows.** Reprinted from Piracha *et al.*[332].

A 150 μm thick diamond frame into which an array of rectangular apertures has been cut is then placed on top of the annealed and overgrown SCD slab. SCD substrate and diamond frame are then placed inside the plasma chamber and after growing diamond between both pieces, a process called fusion growth, they are permanently fused together. An electrochemical etching process is finally utilized to selectively remove the graphite-like carbon from the end-of-range ion damage layer. The thin SCD layer is now permanently fused to the diamond frame but can be detached from the original substrate. The resulting structure is an array of thin single crystal diamond membrane windows, as the diamond thin film is free standing at the locations of the rectangular apertures. After flipping the SCD membrane windows such that the lift-off membrane is on top, the 1.7 μm thick ion damaged layer is removed using inductive coupled plasma reactive ion etching (ICP-RIE) to achieve a final pristine diamond membrane with a thickness of 1 μm. This final SCD template can be used for the fabrication of photonic devices and the fusion of the thin membrane to the diamond holder enables easy handling and prevents breakage of the membrane during fabrication.

The quality of the diamond layer is assessed by Raman spectroscopy and photo-luminescence (PL) measurements. The thin membranes show a diamond Raman line centered on 1333 cm$^{-1}$ with a FWHM of 2.1 cm$^{-1}$ and free of any sign of remaining ion damage which proves that the membranes are of high quality with low strain. PL measurements using a confocal microscope and an excitation laser with a wavelength of 532 nm show optically active defects with a density of $\approx 5$ μm$^{-2}$, caused by doping atoms included in the carbon lattice during CVD growth. The PL spectra show signatures of neutral and negatively charged NV centers and the existence of single defects within the membranes is proven via autocorrelation measurements, showing anti-bunching down to $g^{(2)}(0) = 0.3$. Hence SCD membrane windows are generally suitable templates for PICs which incorporate color centers in diamond. The density of optically active defects could be reduced by avoiding traces of doping materials in the CVD chamber, which means that PICs with single-photon sources at intended locations could generally be incorporated.

Summarizing, it can be said that this method of fabricating SCD templates for PIC fabrication achieves three major goals: 1) maintaining high quality of the diamond layer throughout fabrication, as confirmed by Raman spectroscopy; 2) providing a homogeneous membrane thickness potentially over large areas, as the membrane thickness is defined by the well-defined stop range of ion implantation; and 3) providing permanent attachment of the diamond membrane to a diamond holder for ease of handling and compatibility with established planar fabrication methods. In the following section we show the fabrication of proof-of-principle PICs from such templates.

## 5.4 PIC fabrication from SCD membrane windows

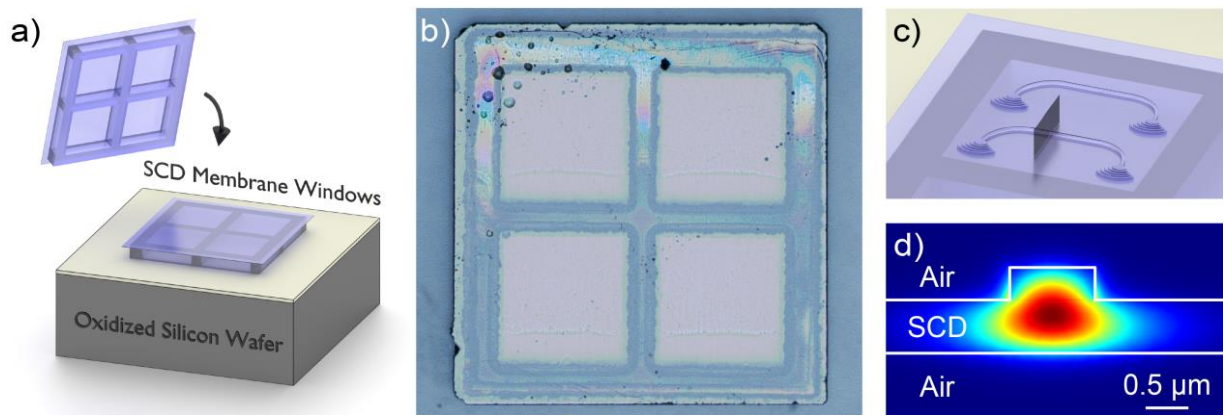

**Figure 75 - Photonic integrated circuit fabrication from SCD membrane windows:** a) Schematic of gluing the SCD template to a carrier wafer. b) Optical microscope image of the template containing four membrane windows before device fabrication. c) Schematic of PICs fabricated on top of a SCD membrane window. d) Waveguide cross-section, corresponding to the black plane in c) and simulated norm of the electric field of the TE-like guided mode at 1570 nm wavelength of the 1 μm wide diamond rib waveguide, after etching 380 nm into the 1 μm thick membrane.

We fabricate PICs starting with the SCD membrane windows (SCDMW) template described in the previous section with a membrane thickness of about 1 μm. The SCDMW is glued to an oxidized silicon wafer using PMMA resist, as shown in the schematic of Figure 75 a). The diamond frame is in contact with the PMMA, while the membranes are on the upper side, well separated from the PMMA resist and the underlying wafer. Figure 75 b) shows an optical microscope image of the SCDMW. The SCD frame has a size of $1.5 \cdot 1.5 \text{ mm}^2$ and hosts four windows of $400 \cdot 400 \text{ μm}^2$ size each. Using the template glued to the carrier wafer, we fabricate PICs on top of the membranes, as illustrated in Figure 75 c). This fabrication is performed via the fabrication procedures developed for PCD devices, as described in section 2.4.2. It consists of EBL for defining the etch mask in HSQ negative tone resist and pattern transfer into diamond via reactive ion etching in $O_2/Ar$ plasma. Further details concerning fabrication are provided in appendix A2.

We design diamond rib waveguides such that at telecom wavelengths they support exactly one mode per polarization. Figure 75 d) shows the cross-section through such a waveguide with a width of 1 μm and a relative etch depth of 38% into the 1 μm membrane thickness. Two-dimensional simulations of this cross section via finite element method (COMSOL Multiphysics) confirm that one TE-like mode with an effective refractive index of 2.26 and one TM-like mode with an effective refractive index of 2.22 are supported and no higher order modes exist. The electric field distribution of the TE-like mode at 1570 nm wavelength is shown in Figure 75 d). As no other materials besides diamond and air are involved in the waveguiding structures, PICs for a large range of wavelengths can be fabricated using the presented fabrication methods, potentially covering diamond's huge transparency range from 220 nm in the UV over visible light and covering infrared wavelengths up to 500 μm.[84]

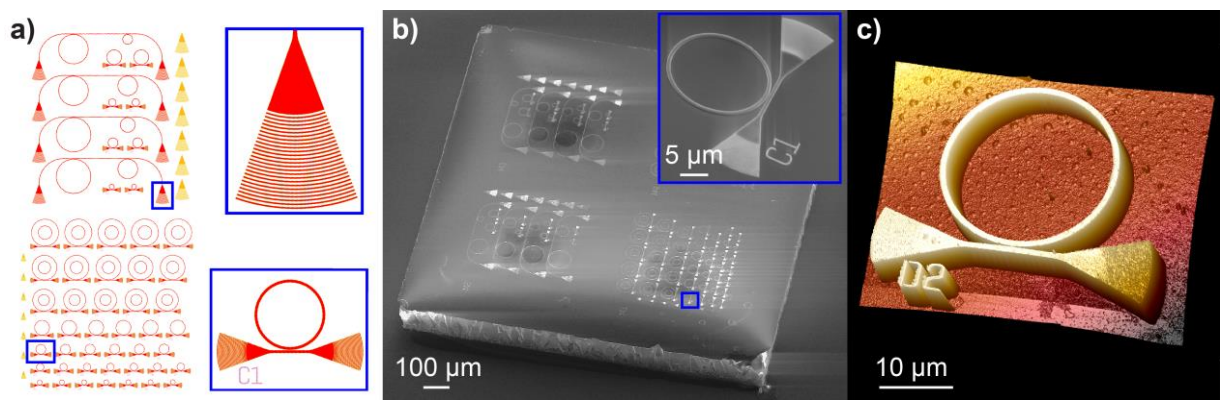

**Figure 76 - PIC fabrication results:** a) Lithography layout showing sets of ring resonators evanescently coupled to waveguides. The insets show an exemplary circuit, consisting of grating couplers, waveguide and ring resonator (bottom inset) and a zoom on a focusing grating coupler for out-of-plane coupling to optical fibers (top inset). b) SEM image of the fabricated photonic chip, hosting sets of photonic circuits on each SCD membrane. Inset: SEM micrograph of one compact device. c) Atomic force microscope scan of a device, showing 1 μm width of both ring and waveguide, 380 nm step height and 1.7 nm rms-roughness of the etched substrate.

On each of the membrane windows sets of PICs can be structured, as illustrated in the lithography layout shown in Figure 76 a). The layout at the top consists, besides some smaller test structures, of four PICs for which the distance between the two grating couplers of each device is 250 μm in order to be compatible with alignment against a fiber array in our standard transmission measurement setups. The layout at the bottom shows how more than 40 compact PICs can be fabricated on a single membrane window of 400 · 400 μm² size, illustrating the potential for dense packaging of diamond PICs. The PICs, as shown enlarged in the bottom inset, generally consist of in- and output focusing grating couplers connected by a 1 μm wide bus waveguide and one or two ring resonators which are evanescently coupled to the waveguide. The grating coupler, shown in the top inset of Figure 76 a), consists of 25 circular grating lines with a period of 820 nm and line widths of 310 nm, a geometry which was optimized for a wavelength of 1570 nm using polycrystalline diamond photonic chips (see section 2.1.2). Figure 76 b) shows a scanning electron microscope image of the experimental realization of diamond PICs implementing the layout of Figure 76 a). The enlarged SEM image shows a compact device where the gap size between waveguide and ring is 200 nm as designed, showing that small feature sizes can be realized on the SCDMW in the same way as achieved on PCD thin films.

We analyze the surface morphology of fabricated photonic structures using atomic-force microscopy (Figure 76 c). In order to correctly measure the step height in the SCD structures resulting from the RIE etching we remove the HSQ resist from the top of the structures using hydrofluoric acid (as opposed to the devices fabricated for transmission measurements where the HSQ resist remains on the devices). We find that the step height is 380 nm and that on the flat areas of the etched SCD membranes the rms-roughness amounts to only 1.7 nm in a $10 \cdot 10$ $\mu m^2$ scan area. This is comparable to the rms-roughness of 1 nm in a $5 \cdot 5$ $\mu m^2$ scan area before device fabrication, which illustrates that the fabrication procedure preserves the low surface roughness.

While here we partially etch into the diamond membranes, also fully etched diamond PICs, as for example preferable for 1D photonic crystal cavities, can be achieved by the planar fabrication process presented here. This is possible either (1) by designing anchor structures to attach all freestanding structures to the diamond frame, (2) by using a second lithography step to define regions of partial etching compared to regions of full etching, as developed for standing optomechanical structures in PCD photonic, presented in chapter 3, or (3) by depositing a layer of lower refractive index material on the bottom side of the diamond membranes such as PEVCD silicon oxide, which can be used as a cladding material for high quality diamond PICs.[87]

## 5.5 Transmission measurements

We measure the transmission spectra of PICs from SCD using the same setup used for PICs from PCD (see section 2.4.3). The photonic chip is mounted on a three-axis stage below a fiber array for alignment of the optical fibers to the grating couplers. The polarization of the input light is adjusted such that the light from the input fiber couples most efficiently to the TE-like mode of the nanophotonic waveguide, corresponding to a maximized device transmission. The wavelength of the input light is swept across the laser range from 1520 nm to 1610 nm and the wavelength dependent transmission is recorded with a fast low-noise photodetector.

We study the transmission spectrum for the device shown in the SEM micrograph of Figure 77 a). Light is coupled in and out of the PIC (symbolized by yellow arrows) via the grating couplers, which are connected by a waveguide. Two ring resonators are evanescently coupled to the ring, with a gap between ring and waveguide of 200 nm. When the wavelength of light matches a ring resonance of one resonator then light couples into the ring, as symbolized by the blue and red arrows. Figure 77 b) shows the transmission spectrum of the PIC. The envelope of the transmission spectrum is given by the product of the wavelength dependent coupling efficiencies of both grating couplers, while the resonances of the ring resonators are visible as dips with narrow linewidth. The maximum transmission is $9 \cdot 10^{-4}$, corresponding to an efficiency of one grating coupler of 3%, while coupling efficiencies up to 30% were achieved for PICs from PCD. We attribute the lower efficiency to the absence of the underlying silicon oxide and silicon layers present in the PCD stack which leads to reflection of light after initial transmission through the diamond film. This reflected light would partially couple into the diamond waveguide and increase the coupling efficiency. The coupling efficiency could hence be improved by depositing a reflecting structure on the back side of the focusing grating couplers.[333]

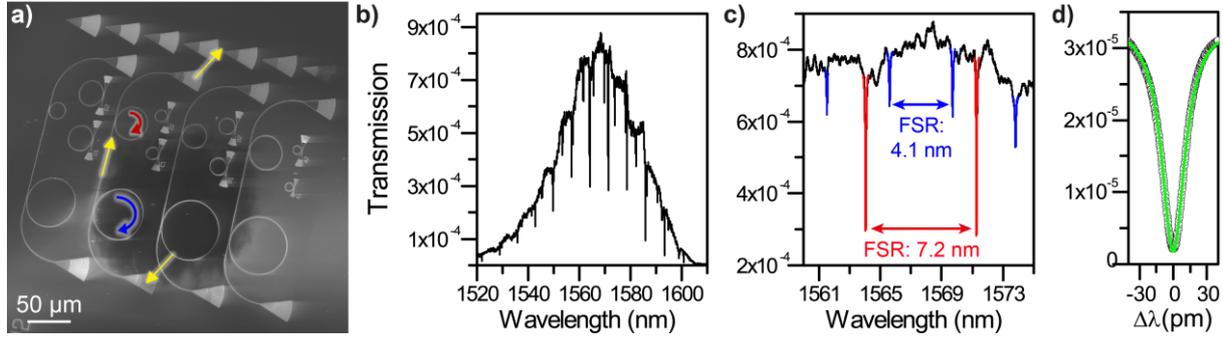

**Figure 77 - Ring resonator transmission measurements:** a) SEM micrograph of one SCD membrane window after fabrication, showing four photonic circuits designed to match the 250 μm spacing of the fiber array. The arrows indicate the direction of light during a transmission measurement. b) Transmission spectrum of a photonic circuit, showing an envelope given by two Bragg grating couplers and the resonances of the two ring resonators. c) Higher resolution transmission spectrum of the device showing resonances with a FSR of $\Delta\lambda$=4.1 nm (blue) corresponding to the ring with radius R = 35 μm and resonances with a FSR of $\Delta\lambda$=7.2 nm corresponding to the ring with R = 20 μm. d) Zoom-in on one resonance of a ring resonator with 20 μm radius. A Lorentzian fit (green) to the data (black circles) reveals an extinction ratio of 12 dB and a FWHM of 23.5 pm, corresponding to a quality factor of 66 000.

The ring resonances are visible as dips in the transmission spectrum, as can be more clearly seen in the higher resolution wavelength scan shown in Figure 77 c). One can distinguish two sets of resonances: One with a FSR of $\Delta\lambda$ = 4.1 nm, corresponding to the ring with radius R = 35 μm (blue color in Figure 77 a) and c)), and a second set of resonances with a FSR of $\Delta\lambda$ = 7.2 nm and a larger extinction ratio, corresponding to the ring with R = 20 μm (red color). From the free spectral range we can calculate[8] the group refractive index $n_g$ for the TE-like waveguide mode at wavelengths around $\lambda$ = 1565 nm according to equation (2.12) which yields a value of $n_g \approx 2.71$. By fitting Lorentzian curves to the resonances we extract the optical quality factor of each ring. For the ring with 20 μm radius a high resolution scan of one resonance is shown in Figure 77 d). A Lorentzian fit (green line) to the data (black circles) reveals an extinction ratio of 12 dB and a FWHM of 23.5 pm, which corresponds to an optical quality factor of 66 000. As this is the loaded quality factor, higher intrinsic quality factors approaching $10^5$ are expected for weakly coupled rings. From the optical quality factor we can estimate the propagation loss $\alpha$ for diamond waveguides, according to equation (2.15), as

$$\alpha = 10 \cdot \log_{10}(e) \cdot 2\pi \cdot \frac{n_g}{Q_{\text{int}} \cdot \lambda}, \tag{5.1}$$

where $Q_{\text{int}}$ is the intrinsic quality factor. As the exact value of the intrinsic quality factor is experimentally not accessible, we do a conservative estimation of the propagation loss by using the measured loaded quality factor of 66 000 as a value for $Q_{\text{int}}$. This yields a value of $\alpha$ = 0.72 dB/mm as an upper bound for the propagation loss. Ring resonators of comparable geometry from PCD show $Q$-factors of about 5300 with similar extinction ratios (see section 2.4.4). We attribute the 12-fold increase in $Q$-factor to 66 000 and the corresponding 12-fold decrease in propagation loss to the superior diamond quality, as the lithography layout and the fabrication methods (including lithography and etching) are the same for both PCD and SCD devices. Compared to other work on SCD ring resonators at telecom wavelengths in the literature, the $Q$-factor presented here is a factor of two larger than achieved using RIE thinning and lithography[20], a factor of two smaller than achieved recently with angle etching into bulk diamond[320]. The results presented in this work are hence comparable to the best results for such devices at telecom wavelengths via any fabrication method. A comparison to $Q$-factors of other optical resonator geometries and for other wavelengths are not directly possible but a detailed comparison of various diamond optical cavity geometries can be found in recent review papers[80,81]. We expect that the quality factor could be further improved by burying the diamond ring resonator in silica cladding to reduce propagation losses caused by scattering at surface or side wall roughness[20,87], by removing any potentially remaining thickness variation across the membrane and by optimizing the dry etching recipe, which was previously optimized to yield the lowest propagation losses for PICs from PCD[147] and has not been re-optimized for SCD templates.

## 5.6 Conclusions on SCD membrane windows and outlook

The fabrication method of PICs using SCD membrane windows presented here has intrinsic advantages over previously demonstrated fabrication methods: In contrast to the fabrication of photonic devices into bulk diamond via angle etching, no exotic etching procedures are needed. Furthermore, rectangular device cross-sections are achieved which automatically preserve a uniform device thickness throughout the entire photonic chip, opposed to triangular cross-sections resulting from angle etching. Device thickness variations would not only limit the degrees of freedom in device design but also impede potential multilayer integration in future large photonic chips. Opposed to this the use of SCDMW in combination with lithography and vertical etching ensures compatibility with standard planar nanofabrication.

Compared to previous demonstrations of SCD photonic components from thinned bulk diamond which is directly attached to a carrier material such as silica, in the method presented here the membranes are never in direct contact with glue or the carrier wafer itself. This prevents cracks, bowing, and contamination of the diamond which are common problems for other methods of handling thin diamond membranes. A carrier material such as silica can furthermore lead to background fluorescence and absorption, effects that are detrimental to quantum photonics where low-loss and background-free propagation of single photons through waveguides is necessary. Fusing the diamond membrane to a diamond holder using diamond itself provides the strongest possible connection via covalent bonds. This prevents peeling off during device fabrication and device lifetime. Ion implantation with defined ion energy yields a very well defined thickness of the resulting SCD membrane without needing a perfectly parallel initial substrate. In addition, the all-diamond photonic chip does not suffer from strain and cracks caused by a difference in thermal expansion coefficients, which is a commonly encountered problem for thin film devices on carrier substrates of different material. While each membrane window is limited in size to several hundred micrometers in both directions of the membrane, PICs can extend over several windows by connecting them via waveguides across micro-channels[332] which is a potential solution for large diamond PICs. The thickness of the diamond membranes can be chosen during SCDMW fabrication by adjusting the time of the RIE thin-down process and membranes as thin as 300 nm have been fabricated with the SCDMW approach. Such membranes would be suitable for PICs operating at visible wavelengths, which could incorporate color centers as single-photon sources.

In summary, we presented the first proof of principle integrated PICs based on arrays of high quality thin single crystalline diamond membrane windows, where ion implantation yields a very well defined SCD membrane thickness and CVD fusion growth ensures a permanent fusion to a diamond frame. We showed that established photonic device designs and planar fabrication methods, both tested previously for polycrystalline diamond devices, can be used to fabricate devices from SCD membrane windows, as successfully shown for PICs consisting of grating couplers, waveguides, and ring resonators. We therefore anticipate that other photonic components such as optomechanical elements and waveguide-integrated single-photon detectors, as presented in the previous chapters, can be transferred from PCD to SCD via the fabrication method demonstrated here. The ease of diamond template handling and the scalability of the membrane arrays hold promise for large scale PICs for on-chip quantum optics.

# 6  Summary and outlook

The miniaturization of optical devices in the form of photonic integrated circuits (PICs) offers a range of advantages for applications of classical and quantum optics. Diamond has excellent mechanical and optical properties, including a range of optically active defects, referred to as color centers, which can act as single-photon sources, quantum memories, or sensor elements. This makes diamond highly attractive for PICs and optomechanical circuits. The work presented in this thesis consists of the first demonstration of optomechanical components, as well as the first single-photon detectors integrated in diamond PICs, which are fast and efficient. Moreover, a novel method for PIC fabrication on single crystal diamond was demonstrated.

Using mechanical elements in photonic integrated circuits allows tunability in otherwise passive materials. We demonstrated the combination of mechanically variable elements and on-chip interferometers. This enabled the readout of motion with high sensitivity, as demonstrated by the readout of thermal motion. Our H-resonator design, which incorporates a photonic crystal for optical isolation, shows high quality factors up to 28 800. We demonstrated two different schemes for the active control of the mechanical motion, which are advantageous for different applications: actuation of a micromechanical resonator via optical gradient forces as well as electrostatic forces. We anticipate that our resonator design can be modified such that either higher frequencies can be achieved or such that large tunable phase shifts can be achieved. We successfully operated our circuits at various temperatures, including ambient and cryogenic temperatures. Our optomechanical circuits could hence be employed in future quantum optical circuits, along with single-photon sources and superconducting single-photon detectors.

Single-photon detectors are an indispensable building block for integrated quantum optics. We demonstrated the first superconducting nanowire single-photon detectors integrated on diamond waveguides. Single-photon detection both at telecommunication wavelengths as well as at wavelengths relevant for single-photon emission from color centers in diamond were shown. The demonstrated detectors feature high critical currents of several tens of μA and excellent performance in terms of low timing jitter, high efficiency and low noise-equivalent-power, comparable to established platforms. Such detectors can be integrated into large diamond PICs.

On-chip quantum optical systems require an outstanding material quality, in terms of low level of impurities and low propagation losses, which is available in single crystal diamond. We have shown a novel method for the fabrication of PICs on arrays of thin single crystal diamond membrane windows. In this method, ion implantation yields a very well-defined membrane thickness and permanent fusion of thin membranes to a diamond frame is ensured. This leads to an ease of diamond template handling as well as scalability of the membrane arrays. Thus, such a method holds promise for large scale PICs. We showed that established photonic device designs and planar fabrication methods, both tested previously for polycrystalline diamond devices, can be used to fabricate devices on single crystal diamond membrane windows, resulting in low-loss diamond PICs. This paves the way for transferring the complete set of developed photonic components to a material with superior quality, including optomechanical resonators and waveguide-integrated single-photon detectors. The demonstrated membrane arrays are suitable for PICs operating at visible wavelengths, which could incorporate single color centers.

We conclude that the presented results are a promising step towards a platform for quantum optical circuits on diamond, which relies on monolithically joining single-photon sources, single-photon routing and processing, as well as single-photon detectors. We envision devices in which color centers will provide efficient and controlled single-photon emission into waveguides. Electro-optomechanical phase shifters will allow to tune and control the routing of single photons within the integrated photonic network, while waveguide-integrated superconducting nanowire single-photon detectors at the output of the PICs will provide a reliable, low-noise readout. The building blocks for diamond quantum photonic technologies exist and a suite of functional elements is available, which opens new doors for a wide range of applications, from sensing to diamond quantum photonics.

# Appendices

## A1. Polycrystalline diamond deposition and material properties

### Chemical vapor deposition process

We employ high-quality silicon substrates, which provide an atomically flat surface. A low-refractive index buffer layer is grown by thermally oxidizing the wafer to an oxide layer thickness of 2 μm. During the oxidation process, the surface morphology is preserved, providing a smooth starting layer for later diamond growth. Microcrystalline diamond layers are then deposited directly onto the oxidized substrates. To achieve this diamond growth, a diamond nano-particle seed layer is first spread onto the $SiO_2$ film by ultrasonication for 5 min in a water-based suspension of ultra-dispersed (0.1 wt %) nano-diamond particles of typically $5 - 10$ nm size. Then the samples are rinsed with de-ionized water. After dry spinning the wafer is transferred into an ellipsoidal 2.54 GHz microwave plasma reactor. The growth of polycrystalline diamond layers takes place at a temperature of $800 - 875$ °C using 1% methane in hydrogen at a pressure of $45 - 55$ mbar. Substrate rotation was applied to avoid angular non-uniformities arising from the gas flow. Growth rates were in the range of $0.1 - 0.2$ μm/h. After growth, the samples are cleaned in concentrated $HNO_3:H_2SO_4$ to remove surface contaminations.

### Chemical mechanical planarization

The polycrystalline diamond films are polished using slurry based chemical mechanical planarization (CMP). This approach is commonly employed in the IC fabrication industry for the polishing of dielectric and metal interconnects, where softer polyester based polishing pad is used with the aid of a colloidal silica at room temperature[334]. The technique does not require the use of expensive diamond grit, or cast iron scaifes. The CMP polishing is performed with a contact force of 120 N at a rotational frequency of 90 rpm and usage of 80 ml/min of polishing liquid containing silica particles. The polishing mechanism consists of the wet oxidation of the surfaces while the polishing fluid facilitates the attachment of silica particles to the diamond film. This is followed by shear removal of the particles due to forces from the polishing pad which is employed throughout. The experimental conditions closely follow the procedure presented by Thomas *et al.*[153]. By using the CMP method, thin films can be polished without fear of film cracking. Figure 78 shows atomic-force microscopy scans before and after polishing. Before polishing (Figure 78 a), the average rms-roughness over an area of $5 \cdot 5$ μm$^2$ amounts to 15 nm. After polishing the diamond layer from 1 μm to 600 nm thickness (Figure 78 b), the roughness is reduced to 2.6 nm rms.

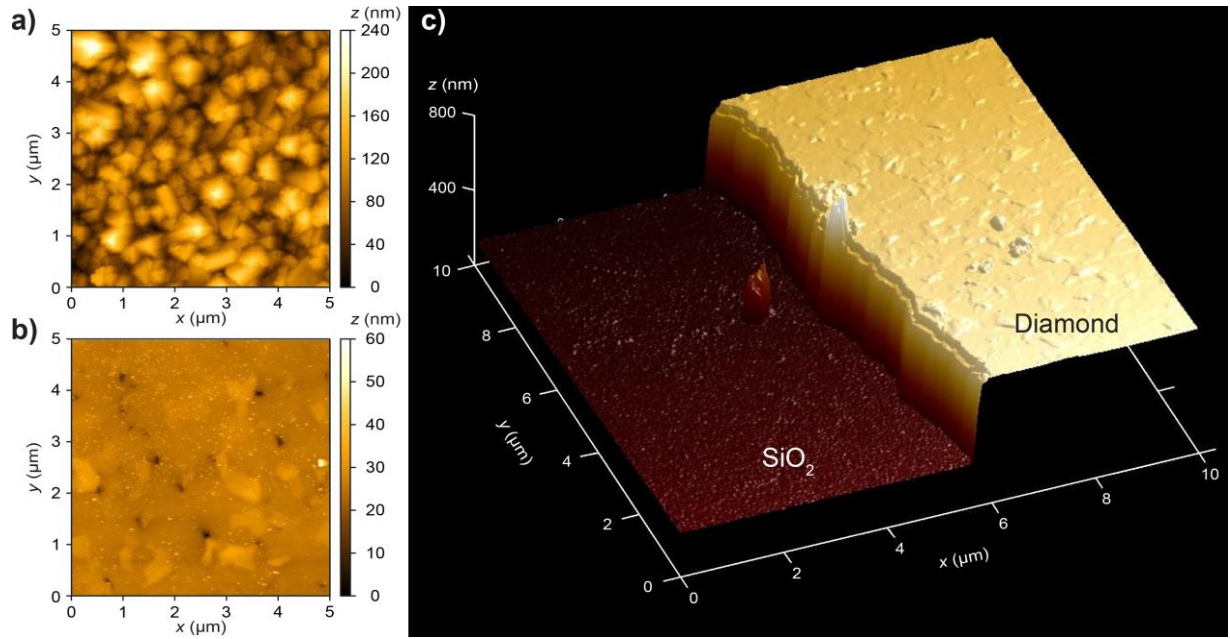

**Figure 78 - Atomic force microscopy:** a) Before polishing, showing 15 nm roughness on 5 · 5 μm² scan area. b) After polishing, showing a reduced rms surface roughness of 2.6 nm. c) AFM scan across an edge where the wafer has been cleaved.

Figure 78 c) shows an AFM scan across an edge where the wafer has been cleaved. The step height between the silicon oxide substrate and the diamond layer is clearly visible. This enables a comparison of the remaining surface roughness on top of the diamond layer to the diamond layer thickness of 600 nm. The clear reduction in surface roughness can also be observed via scanning electron microscopy before and after polishing (see Figure 79).

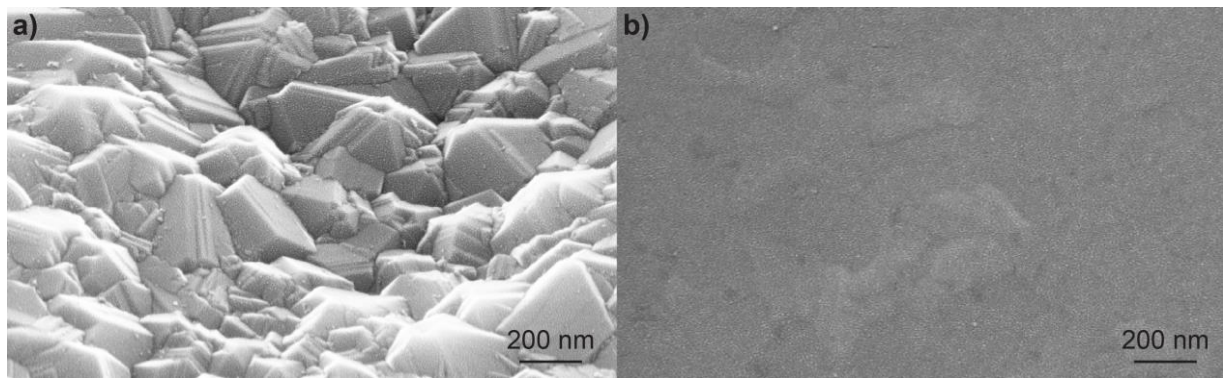

**Figure 79 - Surface roughness:** a) SEM micrograph of the unpolished diamond. b) SEM micrograph of the diamond surface after chemical mechanical planarization.

We measure the thickness of the polished diamond layers by white-light reflectometry (Filmetrics F20) and characterize the homogeneity of the deposited diamond films. Figure 80 shows a measurement across a wafer with a diameter of 3 inches (7.62 cm).The maximum thickness difference between any two points is below 100 nm and the largest occurring thickness variation within 5 mm amounts to less than 5 nm/mm. For the experiments we use wafer dies with a size of 15 mm · 15 mm. Over the area of such a die the thickness variations are negligible. This enables reproducible photonic components, including focusing grating couplers for specific wavelengths. Hence, the employed polished polycrystalline diamond films provide large device areas with comparably small thickness variations.

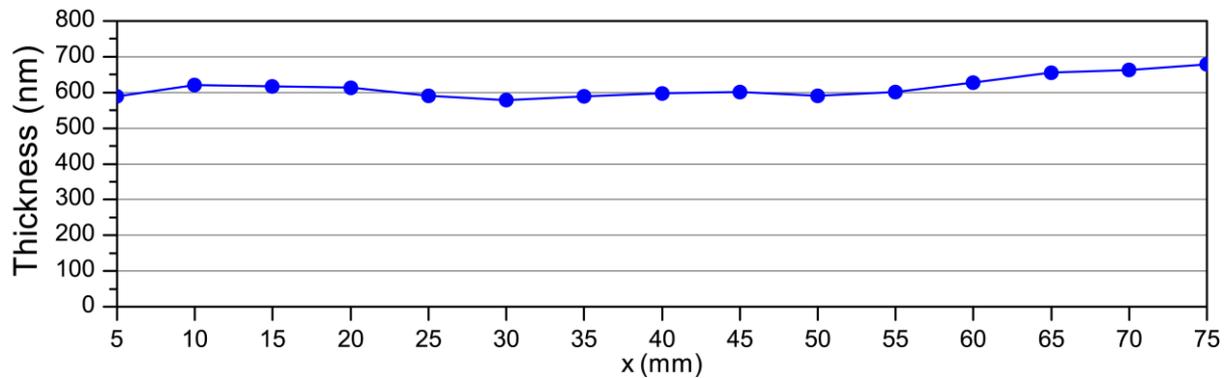

**Figure 80 - Thickness uniformity of polished diamond-on-insulator templates.** White-light reflectometry measurements of the thickness across a polished diamond thin film on a wafer with a diameter of 3 inches (7.62 cm). The direction of measurement was along a line parallel to the wafer flat and through the center.

## Raman spectroscopy

A possible reason for propagation loss in diamond waveguides is the inclusion of non-bound carbon or graphite, which can lead to absorption and scattering of light. In order to estimate the content of non-bound carbon within the PCD films, Raman spectroscopy was performed on the unpolished PCD thin film.

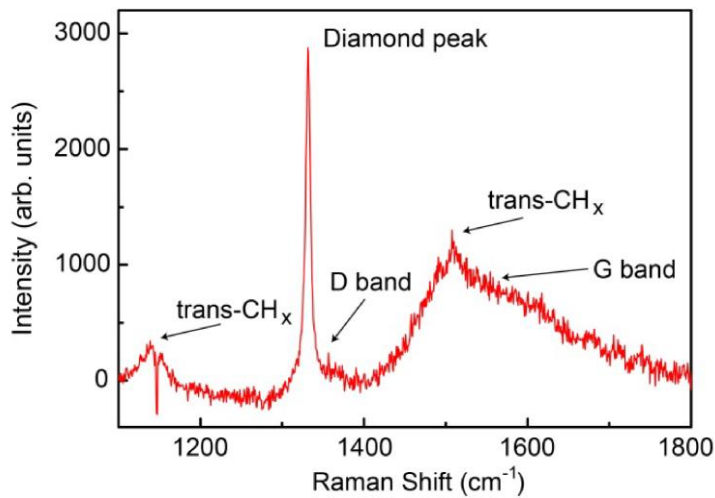

**Figure 81 - Measured Raman spectrum of a polycrystalline diamond thin film.** The data show the results of Raman spectroscopy at room temperature of an as-grown PCD thin film, with a strong diamond peak at 1332 cm⁻¹. This dataset has been published previously by Rath *et al.*[156].

The Raman spectroscopy measurements and the interpretation of the results were performed by our collaborators at the Fraunhofer IAF in Freiburg: "The sample was characterized at room temperature, using an argon-ion laser at 458 nm wavelength with a laser spot of 200 µm, focused onto the diamond surface with a laser power of 300 mW. A measured spectrum is shown in Figure 81. Generally, these spectra show a prominent Stokes shift at 1332 cm$^{-1}$, which has been identified as a parameter for high-quality single-crystalline diamond with a minimum of scattering intensity at other wavenumbers. Nano- and poly-crystalline CVD diamond films, however, show complex bands[335] with scattering intensity reaching from 1100 to 1600 cm$^{-1}$. These lines are assigned to various constituents such as sp$^2$-bonded carbons, diamond progenitors, amorphous carbon and so on. In our data, a well-defined diamond (sp$^3$) Raman peak at 1332 cm$^{-1}$ is detected. The FWHM of this peak is about 7 cm$^{-1}$, which is typical for such material.[336] The peak at 1150 cm$^{-1}$ corresponds to *trans*-polyacetylene (*trans*-CHx), which is found in nanocrystalline CVD diamond.[337] The weak peak at 1350 cm$^{-1}$ is attributed to the D-band (disordered carbon[338]) and the band around 1510 cm$^{-1}$ is due to an overlap of the G-band (graphite) and *trans*-CHx of diamond at 1480 cm$^{-1}$. These data indicate that non-diamond bound carbon is present most likely only in the nucleation layer of the film. From the Raman data we estimate that the relative part of non-sp$^3$ bound carbon will be below 2% and therefore constitute only a minor part of the measured propagation loss. Additional complications in the interpretation of the Raman spectra arise from the fact that the intensity of scattering from sp$^2$-bonded carbon is very dependent on the excitation wavelength because of resonance effects, and that scattering from diamond is quite dependent on crystallite size as would be expected from a phonon spectrum. The interpretation of Raman spectra for a discussion of sp$^2$ and sp$^3$ contents in diamond films is therefore only possible after careful calibration using near-edge x-ray absorption fine-structure (NEXAFS) experiments, a characterization technique that unequivocally distinguishes between sp$^2$- and sp$^3$-bonded carbon. From NEXFAS[339] measurements, the structural properties of high-quality nano-diamond films have been evaluated to determine the ratio of sp$^2$ to sp$^3$ bonding. The measurements reveal that no more than 1% is sp$^2$. In NEXFAS, the relative sensitivity to sp$^2$ and sp$^3$ bonding is roughly the same. In contrast, Raman spectroscopy is $50 - 100$ times more sensitive to sp$^2$ bonding. Thus, it is reasonable to assume that the actual percentage of sp$^2$ bonding in our high-quality polycrystalline diamond is likely $< 1\%$, although Raman spectra indicate somewhat poorer quality."[156]

# A2. Device fabrication

## Electron beam lithography (EBL): JEOL 5300 50 kV

**Before each EBL step:**

Clean the chip by rinsing with acetone (if structures from previous fabrication steps are present on the chip) or by ultra-sonicating in acetone for 10 min (if no structures are present), rinse with isopropanol and $H_2O$ and blow dry with a nitrogen gun.

**(A) EBL using HSQ 15% (Dow Corning Fox15) – on PCD**

1.  Make sure a 5 nm thick glass adhesion layer has been deposited on the chip.
2.  Take resist bottle from refrigerator and let it warm up to room temperature ($> 20$ min).
3.  Heat chip on hot plate at 100 °C for 2 min.
4.  Spin coat resist ($\approx 60$ µl) at 3000 rmp, 1000 rpm/s for 60 s.
5.  Verify resist thickness with reflectometer. Nominal value: $\approx 500$ nm on $15 \cdot 15$ mm$^2$ chip.
6.  EBL area dose: 300 µC/cm$^2$ at 50 kV using proximity effect correction.
7.  Development: Microposit MF-319 for 10 min, rinse with $H_2O$ and gently blow dry with a nitrogen gun.

**(B) EBL using HSQ 6% (Dow Corning XR-1541) – on PCD**

1.  Make sure a 5 nm thick glass adhesion layer has been deposited on the chip.
2.  Take resist bottle from refrigerator and let it warm up to room temperature ($> 20$ min).
3.  Heat chip on hot plate at 100 °C for 2 min.
4.  Spin coat resist ($\approx 60$ µl) at 3000 rmp, 1000 rpm/s for 60 s.
5.  Verify resist thickness with reflectometer. Nominal value: $\approx 120$ nm on $15 \cdot 15$ mm$^2$ chip.
6.  EBL area dose: 2000 µC/cm$^2$ at 50 kV for NbN nanowires ($\approx 100$ nm diameter) and 900 µC/cm$^2$ at 50 kV for NbN connections to metal electrodes ($> 200$ nm diameter).
7.  Development: Microposit MF-319 for 10 min, rinse with $H_2O$ and gently blow dry with a nitrogen gun.

**(C) EBL using HSQ 6% (Dow Corning XR-1541) – on SCD**

Do NOT use ultra-sonic cleaning. This can damage SCD membranes. Gently rinsing via pipette and blow dry with much reduced pressure compared to PCD. Otherwise membranes can crack.

1.  Make sure a 5 nm thick glass adhesion layer has been deposited on the chip.
2.  Take resist bottle from refrigerator and let it warm up to room temperature ($> 20$ min).
3.  Heat chip on hot plate at 100 °C for 2 min.
4.  Spin coat HSQ resist ($\approx 60$ µl) at 2100 rmp, 300 rpm/s for 60 s.
5.  Resist thickness nominal value: $\approx 500$ nm on top of SCD membranes, for SCD size of $1 \cdot 1$ mm$^2$ (Could not be verified by reflectometry, but was measured via AFM after successful EBL).
6.  Rotate sample in spin coater and drop about four drops of conductive polymer (Showa Denko Espacer 300) onto the sample during sample acceleration (3000 rmp, 300 rpm/s for 40 s).
7.  EBL area dose: 400 µC/cm$^2$ at 50 kV using proximity effect correction.

8. Remove Espacer by rinsing with water (use pipette with low pressure).

9. Development: Microposit MF-319 for 10 min. Use pipette with low pressure and gently rinse chip with $H_2O$. Blow dry with much reduced pressure compared to PCD. Otherwise membranes can crack.

**(D) EBL using PMMA 950K 8.0% – for metal lift-off and opening windows**

1. Heat chip on hot plate at 100 °C for 2 min.

2. Spin coat resist ($\approx$ 60 µl) at 4000 rmp, 1500 rpm/s for 90 s.

3. Bake hotplate at 120 °C for 3 min.

4. Check resist thickness with reflectometer (no UV light). Nominal value: $780 - 830$ nm.

5. EBL area dose: $500\ \mu C/cm^2$

6. Development: $1:3$ mixture of MIBK: Isopropanol for $\approx 3 \cdot 5$ min, transfer to Isopropanol to stop development. Rinse with $H_2O$ and gently blow dry with a nitrogen gun.

**(E) EBL on PMMA 950K 8.0% – for thin electrodes on H-resonator**

1. Heat chip on hot plate at 100 °C for 2 min.

2. Spin coat resist ($\approx$ 60 µl) at 4000 rmp, 1500 rpm/s for 90 s.

3. Bake hotplate at 180 °C for 2 min.

4. Check resist thickness with reflectometer (no UV light). Nominal value: 800 nm

5. EBL area dose: $320\ \mu C/cm^2$

6. Development: $1:3$ mixture of MIBK: Isopropanol for 14 min, transfer to Isopropanol to stop development. Rinse with $H_2O$ and gently blow dry with a nitrogen gun.

## Reactive ion etching (RIE)

**(1) RIE removal of $SiO_2$ adhesion layer**

Parameters on an Oxford 80 system:

Gas flow: 25 $cm^3$/min argon

Pressure: 10 mTorr

Power: 200 W

Time: 90 s to ensure removal of 5 nm of $SiO_2$

**(2) RIE of Diamond**

Parameters on an Oxford 80 system:

Gas flows: 33 $cm^3$/min $O_2$ and 17 $cm^3$/min argon

Pressure: 15 mTorr

Power: 200 W

Resulting DC-bias: $\approx$ 535 V

Low pressure strike: Start plasma at 30 mTorr.

Resulting etch rates: PCD wafer die $\approx$ 25 nm/min, SCD membranes $\approx$ 16 nm/min

**(3) RIE of NbN**

Parameters on a Sentech SI 220 system:

Gas flows: 30 cm$^3$/min CF$_4$

Pressure:1.33 Pa

Power: 100 W

Etch time: 49 s to remove 5 nm adhesion layer and fully etch 4 nm NbN

**(4) Chip cleaning in oxygen plasma**

Parameters on an Oxford 80 system:

Gas flows: 20 cm$^3$/min O$_2$

Pressure: 60 mTorr

Power: 30 W

Resulting DC-bias: $\approx 126$ V

Cleaning time: 3 min

## Wet etching

**Chromium wet etch**

Sigma Aldrich Chromium Etchant 651826. Etch rate $\approx 4$ nm/s. Round up estimated etch time to ensure full removal of chromium in the intended regions. Stop development by rinsing in water and dry blow with a nitrogen gun.

**Hydrofluoric acid etch for underetching of diamond**

J.T. Baker Buffered Oxide Etch 6:1. Etch rate $\approx 1.6$ nm/s for oxidized silicon. For not-freestanding structures: Rinse in water and blow dry with a nitrogen gun. For freestanding structures (optomechanical devices): Rinse in water, transfer to methanol bath which slightly covers the surface of the photonic chip (make sure structures never fall dry). Then let the chip dry by letting methanol evaporate on a hotplate (120°C for $\approx 2$ min). Alternatively use critical point drying.

# Workflows for the fabrication of photonic chips

Note that cross-shaped alignment markers are used for the alignment of subsequent EBL steps.

**(1) Electrode fabrication**

Either EBL Process (E) for thin electrodes on H-resonators or EBL process (D) for large area electrodes (for SNSPDs). After development: Chromium (Cr) and Gold (Au) deposition via Physical vapor deposition (PVD): First 5 nm Cr for gold adhesion, then 100 nm Au, then 15 nm Cr on top. 30 min in acetone for PMMA removal and metal lift-off of the metal in the unpatterned area. Rinse in acetone, isopropanol and dry blow with a nitrogen gun.

**(2) Chips with diamond PICs**

Deposition of 5 nm $SiO_2$ via PVD. EBL process (A) for HSQ 15%. RIE process (1) for $SiO_2$ removal. RIE process (2) for diamond etch, timed etch to reach 50% relative etch depth.

**(3) Chips with diamond PICs and optomechanical circuits**

Following workflow (II) for PICs. Deposition of 25 nm Cr via PVD. EBL Process (D) for PMMA. Cr wet etch. RIE process (2) for diamond etch, timed etch to reach 100% relative etch depth within PMMA windows.

**(4) Chips with electrodes, diamond PICs and optomechanical circuits**

Electrode fabrication following workflow (I). Cleaning by rinsing in Acetone instead of ultra-sonic. Then following workflow (III) including workflow (II), starting with deposition of 5 nm $SiO_2$.

**(5) SNSPDs on diamond PICs**

Electrode fabrication following workflow (I). Deposition of 5 nm $SiO_2$ via PVD. EBL process (B) for HSQ 6% on PCD. RIE process (3) for $SiO_2$ removal and NbN etch. Then following workflow (II) for PIC fabrication.

**(6) SCD device fabrication**

Spin coat PMMA on carrier die ($15 \cdot 15$ mm$^2$) at 2000 rmp, 1000 rpm/s for 20 s. Place SCD membrane array on PMMA and gently press it into the PMMA (Tweezers touch the sides of the thick diamond frame) → Bonding of SCD membrane array. Baking 2 min on hotplate at 180°C. Clean the bonded chip in hydrofluoric acid for 20 min. Clean chip in oxygen plasma via RIE process (4). Deposition of 5 nm $SiO_2$ via PVD. EBL process (C) for HSQ 6% on SCD. RIE process (1) for $SiO_2$ removal. RIE process (2) for diamond etch, timed etch to reach intended relative etch depth.

## A3. NbN deposition

The superconducting NbN film of 4 nm thickness, which was used for all experimental results presented in chapter 4, was deposited by the group of Prof. Goltsman at the Moscow State Pedagogical University by DC magnetron sputtering in nitrogen and argon atmosphere. A discharge current of 350 mA, a nitrogen partial pressure of $2 \cdot 10^{-4}$ mbar, and a substrate temperature of 850°C were used. We measure the dependence of the sheet resistance on the temperature, as shown in Figure 82. The temperature range around the transition from superconducting to normal-conducting is shown in Figure 82 a), while the full temperature range up to room temperature is shown in Figure 82 b). The deposited films show a critical temperature $T_C = 11.7$ K and a transition width $\Delta T_C = 0.9\ K$. The film has sheet resistance of $R_S = 400\ \Omega/\square$ at room temperature and shows critical current densities in the range of $3 \cdot 10^6 A/cm^2$ to $6 \cdot 10^6$ A/cm² at 4.2 K. This is comparable to NbN films grown on silicon or sapphire substrates[306]. The temperature dependence of the sheet resistance at higher temperatures is typically quantified by the ratio of the resistance at 300 K to the resistance at 30 K, called residual resistivity ratio $k = R_{300K}/R_{30K}$. For this layer a value of $k = 0.74 < 1$ reveals a negative temperature coefficient, consistent with the temperature dependence observed for NbN layers below 10 nm thickness on other substrates[264].

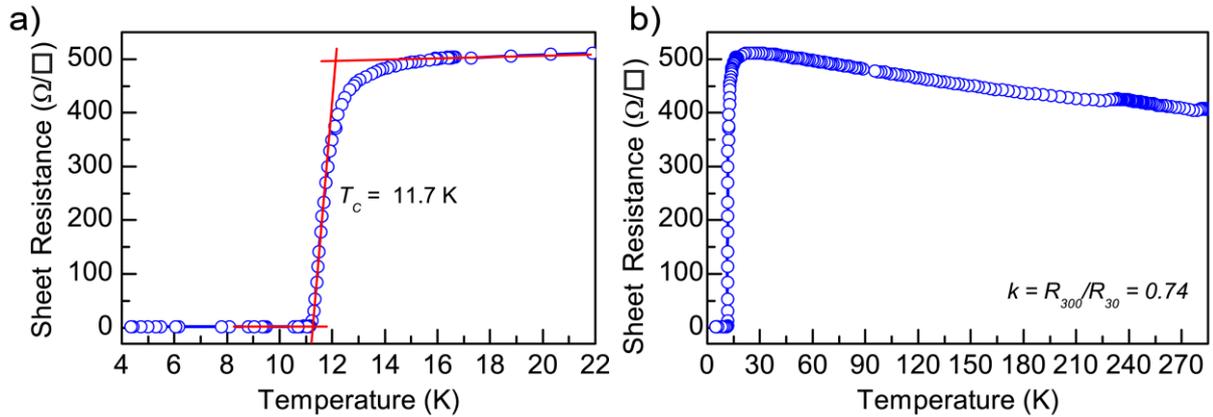

**Figure 82 - Sheet resistance of 4 nm NbN layer on diamond:** a) Sheet resistance in the temperature range from 4 K to 22 K showing the transition from superconducting to normal-conducting with a critical temperature $T_C = 11.7$ K and a transition width $\Delta T_C = 0.9$ K. b) Sheet resistance in the temperature range from 4 K to 290 K revealing a negative temperature coefficient between 30 K and room temperature.

## A4. NbN absorption measurement at room temperature and cryogenic temperature

As a preliminary study on the feasibility of SNSPDs on diamond using NbN nanowires, we fabricated devices for measuring the absorption of light by NbN nanowires from a 5.6 nm thick NbN layer, deposited by the group of Prof. Siegel at the Institute of Micro- and Nanoelectronic Systems (IMS) at the Karlsruhe Institute of Technology. The critical temperature of this layer was determined at the IMS as $T_C = 7$ K. We measured the absorption at room temperature and cryogenic temperature in order to assess if cryogenic absorption measurements are necessary for determining the absorption efficiency of SNSPDs. The device layout for the PICs and the measurement procedure are described in the main text in section 4.3.2.

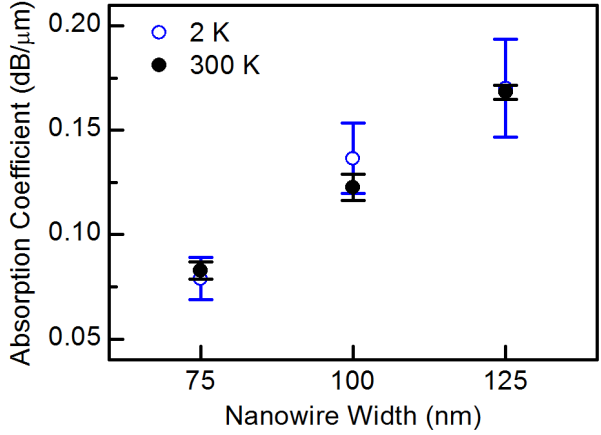

**Figure 83 - NbN nanowire absorption:** Comparison of the absorption coefficients at room temperature (≈ 300 K, black filled circles) and cryogenic temperature (2 K, blue open circles).

We measured the absorption for nanowires of widths of 75 nm, 100 nm and 125 nm for varying length and extract the absorption coefficient for each width. Figure 83 shows the absorption coefficients resulting from measurements at room temperature (≈ 300 K, black filled circles) and cryogenic temperature (2 K, blue open circles). The absorption coefficients agree within the measurement uncertainty. This suggests that measurements at room temperature are sufficient for determining the absorption efficiency of NbN waveguide-integrated SNSPDs.

We note that the NbN layer from which the devices presented in the main text have been fabricated was deposited by the group of Prof. Goltsman at the Moscow State Pedagogical University. Hence the deposition systems, deposition parameters, and exact material properties can differ between the nanowires presented in this appendix and in the main text. Within this appendix we showed that the absorption coefficients at room temperature and cryogenic temperature agree within the error bars and we assume that this generally holds, also for NbN layers deposited under different conditions, such that measurements at room temperature would be sufficient for determining the absorption efficiency of NbN waveguide-integrated SNSPDs.

## A5. SNSPD device yield

Reduced critical currents are a main limiting factor for the efficiency of SNSPDs. Both roughness of the substrate and the fabrication of nanowires can lead to defects and incisions in the nanowire width which can limit the critical current. We hence analyze the dependence of critical current on both the device length and the location on the chip, as different areas of a substrate could show an increased probability of defects and incisions. We note that the length of the U-shaped nanowire is twice as long as the device length.

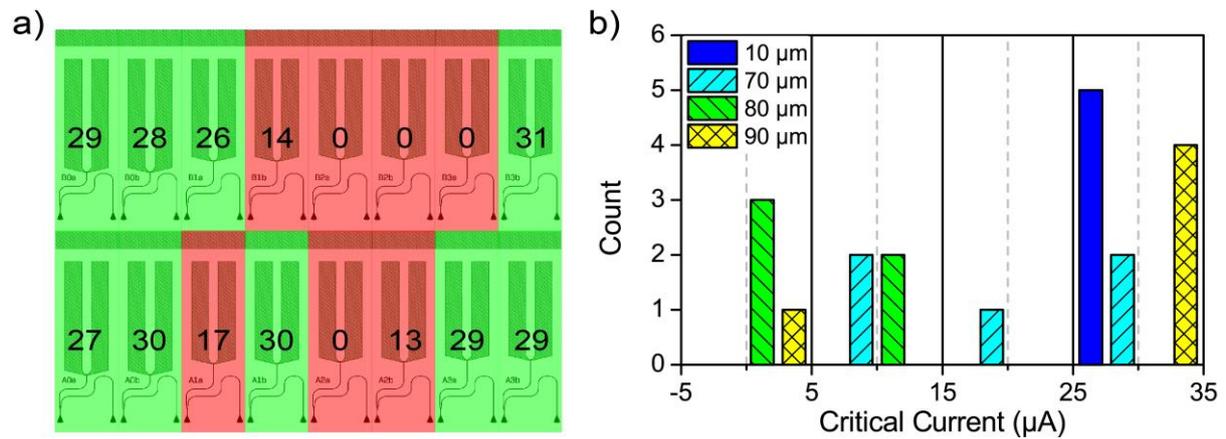

**Figure 84 - SNSPD device yield:** a) Map of device locations of 16 SNSPDs and their corresponding critical currents (in µA). The length increases from left to right with two devices per length (10, 70, 80 and 90 µm). Critical currents above 25 µA are marked in green and lower values in red color. b) Histogram of the critical currents for SNSPDs of varying device lengths (10, 70, 80 and 90 µm). 5 SNSPDs of each length were measured. The bin size is 10 µA and the bins are centered at 0, 10, 20 and 25 µA. The bin boundaries are indicated with black vertical lines. The columns of different color, corresponding to different SNSPD lengths, are displaced in the horizontal direction within each bin for clarity.

Figure 84 a) shows a map of one array of 16 waveguide-integrated SNSPDs designed for a wavelength of 1600 nm. The nanowire widths are 90 nm (bottom row) and 100 nm (top row) and the device length increases from left to right with two devices per length (10, 70, 80 and 90 µm). Each device is colorized indicating high and low critical currents (the values in µA are indicated at each location), where values above 25 µA are marked in green and lower values are marked in red. Devices which could not be electrically contacted or did not show superconductivity are indicated with a critical current value of 0 µA. It is noticeable that almost all devices with low or zero critical current are located within a continuous region, indicating that inhomogeneity in fabrication might play a role. Figure 84 b) shows a histogram of the critical currents of 20 SNSPDs, five of each length (indicated by color). The bin size is 10 µA and the bins are centered at 0, 10, 20 and 25 µA. The bin boundaries are indicated with black vertical lines. The data corresponds to the 16 devices of the array shown in the location map of Figure 74 a) and four additional devices (one of each length) located adjacent to the array.

For 10 µm length (blue color) five out of five devices show high critical currents $I_C \geq 25$ µA, while for 80 µm (green color) five out of five devices show comparably low $I_C < 15$ µA. For 90 µm though (yellow color) four out of five devices show high $I_C \geq 25$ µA, hence the distributions do not show a clear trend with increasing length. We take 25 µA as a threshold for distinguishing high from low critical currents. Aggregating the data into short (10 µm) and long ($\geq 70$ µm) devices reveals that for short devices five out of five devices show high $I_C$ while this only six out of 15 long devices show high $I_C$. The increasing probability for incisions and defects with increasing nanowire length certainly plays a major role in limiting the amount of SNSPDs with high critical currents. Larger amounts of devices need to be fabricated and characterized in order to allow a quantitative analysis concerning the influence of length and location on the probability of a SNSPD to feature a high critical current.

# Acknowledgements

I would like to thank Prof. Dr. Martin Wegener and Prof. Dr. Wolfram Pernice for enabling me to carry out my dissertation at Karlsruhe Institute of Technology. Thank you for your advice and encouragement. I would like to thank my fellow colleagues as PhD students in the Pernice group at the Institute of Nanotechnology: Simone Ferrari, Nico Gruhler, Oliver Kahl, Svetlana Khasminskaya, Anna Ovvyan and Matthias Stegmaier. Thank you for the great collaboration and the good times we had in the last 4.5 years. A special thank you to Svetlana and Matthias with whom I happily shared the office. It is funny to realize that Svetlana and I both installed the first setups and performed the last cryostat measurements on the day before the Pernice lab relocated to Münster. Many thanks to Matthias for the continuous discussions and exchange of ideas in the office.

I would like to thank the people who have contributed to my work on diamond optomechanics: Most of all Sandeep Ummethala for the great collaborative work during his Master thesis and afterwards. Thank you to Menno Poot for explanations concerning fundamental questions on optomechanics and advice concerning simulation and data interpretation concerning the H-resonators.

I would like to thank the people who have contributed to my work on SNSPDs on diamond: Simone Ferrari and Oliver Kahl for the introduction to and help with the cryostat setup. Andreas Vetter for the help with cryogenic measurements. Fabian Sproll for the first depositions of NbN layers on diamond. Vadim Kovalyuk for the deposition of NbN layers. Matthias Stegmaier and Wladick Hartmann for performing the NbN absorption measurement on my chip, after the lab had moved to Münster. A special thank you to the group of people who had immense patience with discussing my ideas and questions concerning superconductivity, SNSPDs, device designs and measurements: Simone Ferrari, Vadim Kovalyuk and Andreas Vetter. Teaming up with you guys improved my understanding of the topic a lot and made the work on SNSPDs interesting and fun.

I would like to thank the people who enabled my work via technical help and advice: Stefan Kühn and Silvia Diewald for the electron beam lithography at the JEOL at the Center for Functional Nanostructures. Christian Benz for initial assistance with EBL processes at the E-Line at Institute of Nanotechnology and for advice on many other topics. Romain Danneau for borrowing me electronic equipment. Simone Dehm for advice and assistance concerning sputtering, SEM and RIE. Matthias Blaicher for developing a much better software for making lithography designs and for the help with his expertise concerning electronics. Simone Ferrari for advice and help with his expertise in a large range of topics. Nicolás Quesada for explanations and discussions on single photons, boson-sampling and related topics.

I would like to thank everyone who proof-read parts of this thesis: Christian Benz, Pascal Casper, Simone Ferrari, Daniel Golestan, Nico Gruhler, Wladick Hartmann, Vadim Kovalyuk, Georgia Lewes-Malandrakis, Wolfram Pernice, Christoph Riedel, Matthias Stegmaier, Orlando Torres, Sandeep Ummethala and Andreas Vetter. A special thank you to Christoph, Daniel, Orlando and Andreas for their invaluable care for details. I would also like to thank Simone Rath and Klaus Hledik for advice concerning the layout of this thesis and for final corrections to the completed thesis.

I would like to thank our collaborators at the Fraunhofer-Institut für Angewandte Festkörperphysik in Freiburg: Christoph Nebel for leading the collaboration, Georgia Lewes-Malandrakis for the deposition of diamond and Dietmar Brink for the polishing of the diamond layers. I would like to thank Prof. Jason Twamley for the invitation to give a talk at the QDiamond14 conference and Prof. Steven Prawer for afterwards proposing a collaboration on single crystal diamond. Thank you to Afaq Piracha for the collaborative work on this topic. I would like to thank Prof. Brant Gibson, not only for hosting me at the Royal Melbourne Institute of Technology, but also for introducing me to Australian culture such as "Footy" at the Melbourne Cricket Ground. I would like to thank the people who helped me with measurements at RMIT, especially Desmond Lau. I want to thank Prof. Igor Aharonovich and his research group for hosting me at the University of Technology, Sydney. Furthermore I would like to thank him for his advice, both concerning science and scientific careers. Great that the research community remembers us as the two crazy guys who climbed Mount Fuji during the off season.

I want to thank the Karlsruhe School of Optics & Photonics (KSOP), Karlsruhe House of Young Scientists (KHYS) and the Deutsche Telekom Stiftung for their financial support, which financed both my dissertation and enabled me to present my research at a variety of interesting international conferences. Furthermore I would like to thank all active members of OSKar - Optics Students Karlsruhe e.V. for making the time of my PhD more diverse and fun.

Finally I would like to thank my friends and I would like to thank Mariu Hernando for the last five years that we spent together and supported each other. Der größte Dank gebührt meiner Familie: Meiner Schwester Simone und meinen Eltern Monika und Mario Rath, die mich mit Liebe und Geduld immer unterstützt haben.